%% file: arp_klt.tex
\documentclass[journal,onecolumn,11pt]{IEEEtran}
\usepackage{amsmath,amssymb,amsthm}
\usepackage{graphicx}
\usepackage{booktabs}
\usepackage{cite}
\usepackage{url}
\usepackage{bm}
\usepackage{xcolor}
\usepackage{algorithm}
\usepackage{algorithmic}
\usepackage{tikz}
\usetikzlibrary{arrows.meta,positioning,calc,decorations.pathreplacing}
\input{numbers_v27}
\input{numbers_impl}   
\input{numbers_fo3}    
\input{numbers_fo3_rf} 
\input{numbers_apps}   

\newif\ifextended
\extendedtrue

\newtheorem{theorem}{Theorem}
\newtheorem{proposition}{Proposition}
\newtheorem{corollary}{Corollary}
\theoremstyle{remark}
\newtheorem{remark}{Remark}

\newcommand{\R}{\mathbb{R}}
\newcommand{\C}{\mathbb{C}}
\newcommand{\Z}{\mathbb{Z}}
\newcommand{\bR}{\bm R}
\newcommand{\bQ}{\bm Q}
\newcommand{\bD}{\bm D}
\newcommand{\bJ}{\bm J}
\newcommand{\bI}{\bm I}
\newcommand{\bN}{\bm N}
\newcommand{\bv}{\bm v}
\newcommand{\bT}{\bm T}
\newcommand{\bH}{\bm H}
\newcommand{\bA}{\bm A}
\newcommand{\bB}{\bm B}
\newcommand{\bC}{\bm C}
\newcommand{\bP}{\bm P}
\newcommand{\bU}{\bm U}
\newcommand{\bF}{\bm F}
\newcommand{\bK}{\bm K}
\newcommand{\bGam}{\bm\Gamma}
\newcommand{\bW}{\bm W}
\newcommand{\bV}{\bm V}
\newcommand{\bDel}{\bm\Delta}
\newcommand{\bZ}{\bm Z}
\newcommand{\bLam}{\bm\Lambda}
\newcommand{\bnu}{\bm\eta}
\newcommand{\bUps}{\bm\Upsilon}
\newcommand{\bX}{\bm X}
\newcommand{\cg}{\mathsf{G}}
\newcommand{\bQx}{\bQ_{\mathrm x}}
\newcommand{\bUsev}{\bU_{\mathrm{VII}}}
\newcommand{\be}{\bm e}
\newcommand{\one}{\bm 1}
\newcommand{\norm}[1]{\lVert#1\rVert}
\newcommand{\T}{\mathsf{T}}
\newcommand{\herm}{\mathsf{H}}

\newcommand{\bx}{\bm x}
\newcommand{\bG}{\bm G}

\newcommand{\bOm}{\bm\Omega}
\newcommand{\comm}{\mathcal{C}}
\newcommand{\rep}{\Pi}


\begin{document}
\sloppy

\title{Karhunen-Lo\`eve Transforms of Autoregressive Windows Based on Matched Symmetry Groups}
\author{Mitchell~A.~Thornton,~\IEEEmembership{Senior Member,~IEEE}%
\thanks{M. A. Thornton is with the Darwin Deason Institute for Cyber Security and the Department of Electrical and Computer Engineering, Southern Methodist University, Dallas, TX 75275 USA (e-mail: mitch@smu.edu). ORCID 0000-0003-3559-9511.}}
\markboth{}{}
\maketitle

\begin{abstract}
\ifextended
The Karhunen-Lo\`eve transform of a window of a stationary autoregressive process of any order $p$, the decorrelating transform for coding and estimation, is described through its poles, extending the Algebraic Diversity framework, which selects a transform by the matched group. For simple roots, every basis vector is a sum of at most $p$ sinusoids, at the frequencies where the spectrum equals the eigenvalue, plus modes decaying from the ends. The window covariance differs from a member of the class diagonalized by the type-II discrete cosine transform by a closed-form matrix of rank $2p$, exactly when no pole is zero. The eigenvalues therefore interlace spectrum samples on the cosine grid, and the coding-gain loss of the cosine transform is an exact function of that matrix with computable bounds, below \cgSepCorrLossPctNineFive{} percent for the $8\times8$ Markov block of image coding at pole $0.95$. By Kronecker's theorem the rank is at most $2\max(p,q)$ for every rational spectrum. The matrices of the class are Toeplitz plus Hankel, that is, a matrix constant along each diagonal plus one constant along each anti-diagonal. The covariance of a moving-average window is the Toeplitz part of its class member alone, and the Hankel part it lacks is nonzero only in two small triangles at the top-left and bottom-right corners, referred to here as ``Hankel corners.'' An integrated (ARIMA) window departs by low rank from the type-VII sine class, exactly for the random walk. The exact symmetry is generically the reflection and contains a wreath product for interleaved windows. The dihedral group of the even extension is matched approximately, the sign of the dominant pole choosing cosine or sine. The two groups give the transform a fast form: a parity fold, half-length cosine cores and a cascade of at most $p$ first-order Cauchy stages, exactly $p$ for a minimal AR($p$) model. A Cauchy stage is the eigenvector basis of a diagonal matrix plus a matrix of rank one, the correction a first-order window needs in each branch. That basis is a Cauchy matrix with scaled rows and columns, whose entries are reciprocals of differences between the diagonal entries and the new eigenvalues, so fast multipole methods apply it to a window of length $M$ in $O(M\log M)$ operations to a fixed accuracy. The first-order case recovers results obtained by other means. This version carries the fast form from a factorization to a realization. Complex data has no parity fold, since the conjugate reversal is antilinear, and is carried by the cyclic group with a border departure of rank exactly $2p$. Mixed, moving-average and co-channel windows are built from the covariance side at the rank Kronecker's theorem bounds, and vector, periodic and cyclostationary windows depart from the block-circulant class by at most twice the dimension of a realization of their autocovariance. On the reference implementation released with the paper the kernel evaluation counts, which do not depend on the machine, are flat in $M\log_2(M/n_{\mathrm{leaf}})$ to \applyFlatArThree{} percent over a \countRange-fold range of length. Experiments on recorded acoustic-array, radar and vibration data, and comparisons with established methods for whitening, likelihoods and coding, state where the transform helps and where existing methods match it.
\fi
\end{abstract}


\section{Introduction}

\IEEEPARstart{T}{he} discrete cosine transform of type II (DCT-II) is used in image and video coding because it is close to the Karhunen-Lo\`eve transform (KLT) of a first-order Markov window with a pole near one~\cite{ahmed1974discrete,ahmed1975orthogonal,rao1990discrete,wallace1992jpeg,legall1991mpeg}. The exact KLT of that window is a basis of sinusoids whose frequencies depend on the pole~\cite{ray1970further,kac1953eigenvalues}, the DCT-II is its limit as the pole tends to one~\cite{clarke1981relation,flickner1982derivation} and was compared with the KLT and the Fourier transform from the start~\cite{pearl1972performance,hamidi1976comparison}, the sine and cosine transforms are the exact KLTs of Markov windows with modified boundaries~\cite{jain1976fast,jain1979sinusoidal}, the cosine and sine transforms are asymptotically equivalent to the KLT of every finite-order Markov process as the window grows~\cite{yemini1979asymptotic,unser1984approximation,sanchez1995diagonalizing}, and explicit factorizations of the first-order KLT are available~\cite{torun2013efficient,reznik2026direct,reznik2026exact}. A retrospective at the transform's fiftieth year rests the case on the same two steps, the limit of the first-order KLT as the correlation coefficient approaches one and the empirical fit of that model to images, now checked against the KLT estimated from a current image set~\cite{wang2023discrete}. To our knowledge, the following questions about the finite window of a process of order $p\ge2$ remain open: how its KLT depends on the poles, how far its covariance departs from the cosine class at a finite pole and a finite length, which symmetry group, if any, the window matches, and whether the transform of order $p$ has a fast form.

This paper treats these questions as an extension of the Algebraic Diversity framework~\cite{thornton2026ad_foundations,thornton2026ad_framework}, in which a transform is selected from the symmetry group matched to the covariance, and of the unification of the classical transforms that the framework yields~\cite{thornton2026unification}. Let a finite group $G$ act on $\R^M$ by a representation $\rep$ with the covariance $\bR$ commuting with every $\rep(g)$. When the representation is multiplicity-free, that is, when no irreducible representation of $G$ occurs in it more than once, every matrix in its commutant (the set of matrices that commute with every $\rep(g)$, $\bR$ among them) leaves each isotypic component invariant and acts on it as a scalar. When those components are one-dimensional the group therefore fixes the eigenbasis up to signs and ordering, and that eigenbasis is the matched transform of $G$. Components of higher dimension keep a basis freedom inside them. The cyclic group gives the discrete Fourier transform. The dihedral group acting on a cycle gives a real-valued Fourier basis, determined up to a rotation in each cosine-sine plane, for which the discrete Hartley transform of Hartley and Bracewell~\cite{hartley1942more,bracewell1983discrete}, whose kernel adds the cosine and the sine, is the standard choice, and which the algebraic signal processing theory obtains as the Fourier transform of the real finite time model~\cite{puschel2008time}. The dihedral group acting on the even-reflected extension of a window of length $M$, the window followed by its mirror image, gives the DCT-II~\cite{thornton2026unification,sanchez1995diagonalizing}. The dihedral symmetry of the cosine matrices themselves is identified in~\cite{egner2001automatic}, a cosine transform is obtained from the orbits of the reflection on a cycle by supercharacter theory in~\cite{garcia2018supercharacters}, and the algebraic reading of the cosine and sine types as Fourier transforms of boundary conditions is given in~\cite{puschel2003algebraic,puschel2008time,puschel2008algebraic}. The matched group of a covariance can be recovered from the covariance in polynomial time~\cite{thornton2026double_commutator}, which turns the selection of a transform into a computation. The recovery is unsupervised and works from the commutant of one covariance, as against the supervised fitting of the regular representation of a group supplied as input to labeled data in~\cite{hoyos2026}. Section~\ref{sec:groups} applies it to the window and reads the exact group, the continuous stabilizer of an interleaved window (the orthogonal mixings of its phases that leave the covariance fixed) and the carrier of a complex window (the frequency by which its poles are rotated) from sample covariances.

The group reading of the autoregressive window and the localization of its departure from the cosine class to its first and last $p$ samples, for order $p\ge2$, appeared first in~\cite{thornton2026unification}, and this paper develops them in closed form. Shortly afterward, and independently, the first-order closed form and the direct and exact fast factorizations of the first-order transform were obtained in~\cite{reznik2026direct,reznik2026exact}, starting from the tridiagonal precision of~\cite{jain1976fast,jain1979sinusoidal} and the parity of centrosymmetric matrices, those unchanged by the reversal of both indices, whose eigenvectors are even or odd~\cite{cantoni1976eigenvalues}. The two routes are complementary. The route of~\cite{reznik2026direct,reznik2026exact} needs no group theory and gives the first-order factorization explicitly, with closed forms at short lengths, while the matched symmetry groups used here and in~\cite{thornton2026unification} give the structure at every order and explain why the cores are fixed. Every first-order statement of Sections~\ref{sec:klt} to~\ref{sec:fast} on the basis, the departure and the fast form can also be found in~\cite{reznik2026direct,reznik2026exact} or in the table of~\cite{jain1979sinusoidal}, and those results are cited as the known case of each statement. The contribution of this paper is the order $p\ge2$ together with the group that organizes all orders, as well as the coding-gain bounds, the moving-average and integrated windows, and the reading of the groups from samples, which are new at every order.

The first group of results describes the Karhunen-Lo\`eve basis itself and how far the window covariance is from the cosine class. The window covariance of an AR($p$) process has a precision matrix that is banded with Toeplitz interior and two $p\times p$ corner blocks in closed form (Theorem~\ref{thm:departure}), so each eigenvector satisfies a recurrence of order $2p$ in the interior: it is a sum of at most $p$ sinusoids, at the frequencies where the spectrum crosses the eigenvalue, plus modes decaying from the two ends, with polynomial weights at a repeated root, and for $p=1$ the frequencies solve the single equation that moves them from the type-I sine grid to the DCT-II grid as the pole rises (Theorem~\ref{thm:levels}). The covariance is a perturbation of rank at most $2p$ of a member of the DCT-II class, of rank exactly $2p$ when no pole is zero, by a determinant formula for the corner block, and its eigenvalues interlace the spectrum sampled on the DCT-II grid (Corollary~\ref{cor:identity}); every window with a rational spectrum, an ARMA($p,q$) window or a co-channel sum of autoregressive windows, is a perturbation of rank at most $2\max(p,q)$ of a class member, by Kronecker's theorem on Hankel matrices of rational sequences (Corollary~\ref{cor:rational}); and the coding-gain loss of the DCT-II against the KLT is an exact function of the normalized perturbation, with two computable bounds, and multiplies over the dimensions of a separable block (Theorem~\ref{thm:gain}). A moving-average window departs from the class in its two Hankel corners, with no inversion (Corollary~\ref{cor:ma}), and an integrated window, the ARIMA model, which is not stationary and has no reflection symmetry, departs from the type-VII sine class by a matrix of low rank and lies in it exactly for the random walk, with the sixteen trigonometric types scored on it (Proposition~\ref{prop:integrated}, Table~\ref{tab:models}). Section~\ref{sec:images} assembles these statements into a finite-window account of the cosine transform for images, with the modeling step and the proved steps told apart.

The second group of results identifies the symmetry groups of the window, the exact one and the approximate one. The exact permutation symmetry of the window is the reflection whenever the autocovariances at the lags of the window are distinct. If the characteristic polynomial is a polynomial $b(z^m)$ in $z^m$, the exact group contains the wreath product $\Z_2\wr S_m$, the group that permutes the $m$ interleaved phases of the window and reflects any of them, and under the nondegeneracy conditions of Theorem~\ref{thm:exact}(c) it equals $\Z_2\wr S_m$. The dihedral group of the even extension is matched approximately, with a signed conjugation by the alternating sequence $(-1)^n$, the sinusoid at the Nyquist frequency, mapping negated poles onto the sine class (Proposition~\ref{prop:nyquist}) and the dominant real pole selecting the limit class (Proposition~\ref{prop:limits}); and a complex window with rotated poles is the real one conjugated by the modulation, so the same groups serve it exactly (Proposition~\ref{prop:modulated}). Section~\ref{sec:groups} also reads these groups from sample covariances.

The third group of results turns the two groups into a fast transform of any order. The exact group supplies the parity butterfly, the dihedral closure supplies the half-length DCT-II and DCT-IV cores, the DCT-VI and DST-VII at odd length, and the corner blocks supply a correction of rank at most $p$ per branch, equal to $p$ for a minimal AR($p$) model (one with $\phi_p\ne0$), whose eigenvalues interlace the core grid in the direction set by the inertia of the corner block (Proposition~\ref{prop:fast}); the correction is a cascade of as many first-order Cauchy stages as its rank, so that the order enters the flow graph only as the number of stages (Proposition~\ref{prop:cascade}, Fig.~\ref{fig:butterfly}), with an algorithm and its operation count, in which the merged correction is measured to be as compressible as a single stage (Algorithm~\ref{alg:fast}, Section~\ref{sec:count}). The same form holds for every rational spectrum with the folded departure in place of the corner block (Proposition~\ref{prop:rational}), and the recursive form folds every branch of length at least $4p$, with a correction of rank at most $p$ for an even-length branch and at most $p+1$ for the branch that holds the center sample of an odd-length fold. When $M=2p\cdot2^k$ every fold is even and the recursion is the dyadic tower of the Haar basis, the binary tree of nested half-windows on which that basis is built~\cite{haar1910theorie,mallat1989theory}, with a correction at every node (Proposition~\ref{prop:recursive}). Sections~\ref{sec:complexdata} to~\ref{sec:realization}, new in this version, carry the fast form from a factorization to a realization. Complex data breaks the construction at its first step, since the conjugate reversal is antilinear and supplies no invariant subspaces, so the matched group is the cyclic one and the departure from the circulant class member is border-supported of rank exactly $2p$ (Proposition~\ref{prop:complex}); mixed, moving-average and co-channel windows are built from the covariance side at the rank Corollary~\ref{cor:rational} bounds, with nothing of the window's size formed (Proposition~\ref{prop:covside}); a complex integrated window sits on the cyclic class rather than the sine class of the real case (Proposition~\ref{prop:complexintegrated}); the two cores of the odd-length fold are the only cosine families the reflection leaves invariant (Proposition~\ref{prop:oddcores}); and the polyphase route needs no mixing of its phases (Remark~\ref{rem:polymix}). Sections~\ref{sec:fmmreal} and~\ref{sec:secularreal} give the hierarchical evaluation of the Cauchy stages and the secular step it needs, including the interaction rule that removes a term proportional to the square of the stage dimension and the rational step that replaces a safeguarded Newton iteration which degenerates to bisection. Section~\ref{sec:bench} reports the accuracy and the operation count of the reference implementation released with the paper, with the order claim carried by kernel evaluation counts, which do not depend on the machine a measurement is taken on.\ifextended{} Section~\ref{sec:cara} relates the sinusoids of the eigenvectors to the sum of sinusoids in Carath\'eodory's representation of the same covariance, which the two frequency sets, one set by the magnitude and one by the phase of the characteristic polynomial, keep apart. Section~\ref{sec:block} carries the construction to vector, periodic and cyclostationary windows, Section~\ref{sec:recorded} applies it to recorded microphone-array and radar data, and Section~\ref{sec:apps} compares it with established methods for whitening, likelihoods and coding, including the comparisons in which it showed no advantage.\fi

In a basis in which the window coefficients are uncorrelated the coding gain, the ratio of the arithmetic to the geometric mean of the coefficient variances, is the largest of any orthogonal basis, and it sets the rate of a transform coder at high resolution~\cite{jayant1984digital,goyal2001theoretical}. The same basis truncates to the subspace of least mean-square error, and in it the Wiener filter, the matched filter and the Gaussian likelihood ratio of the window are diagonal operations. The transform coders of image and video standards approximate that basis by a fixed one, on the strength of a first-order Markov model of the rows and columns. The signals of sensor arrays, speech, biomedical recordings and economic series are often modeled by autoregressive, moving-average and integrated processes of higher order, where the fixed basis is less close to the KLT and the exact basis has been computed numerically from the covariance, at $O(M^3)$ operations once and $M^2$ per window.

For most uses of these models what is needed is a likelihood, a whitener or a power spectrum, and exact recursions and banded factorizations already deliver these in time linear in the window length. Some tasks need the eigenbasis itself: an exact Karhunen-Lo\`eve representation of a finite window, a decomposition into coherent modes, or the number of modes that hold a prescribed fraction of the variance. Beyond the first-order autoregressive case, and for the other model classes treated here, we are not aware of an exact eigensolver for these window covariances that also has a fast transform. This paper provides one, which applies in $O(pM\log M)$ operations per window after a design stage, and states how far the fixed cosine basis is from it with a bound on the loss. The difference between a finite window and the transform of its stationary class is confined to the edges of the window. Edge effects are usually handled with tapers, lapped transforms or an extension of the record. Where those do not suffice, as on short isolated records, the transforms here represent the edges exactly and keep a fast transform, a combination that to our knowledge has not been available before. The closed-form diagonalization of these window covariances for order $p\ge2$, the low-rank boundary structure on which it rests, and the cascade that turns it into a fast algorithm are new mathematical results.

\ifextended
\section{Background: the Cosine, Sine and Hartley Families}\label{sec:background}

The transforms compared in this paper are real orthogonal bases of $\R^M$ with sinusoidal columns. This section fixes their definitions, the boundary symmetries that generate them, the matrix classes they diagonalize, the relations among them, and the place of the KLT of a stationary window among them.

\emph{The sixteen trigonometric transforms.} A finite sequence $x_0,\dots,x_{M-1}$ can be extended to a periodic sequence by reflecting it at each end, either about a sample (whole-sample symmetry, W) or about the midpoint between two samples (half-sample symmetry, H), and either with the same sign (symmetric, S) or with the opposite sign (antisymmetric, A). The four choices at each end give sixteen extensions, and the Fourier series of each extension, restricted to the original samples, is one of eight cosine transforms (DCT-I to DCT-VIII) and eight sine transforms (DST-I to DST-VIII)~\cite{wang1984fast,martucci1994symmetric}. The first four of each are the most common, and the DST-VII is also used for small intra-predicted blocks in the HEVC standard~\cite{sullivan2012overview}. Table~\ref{tab:family} lists the kernels of DCT-I to DCT-IV and DST-I to DST-IV with their left and right symmetries. The DCT-I is the symmetric cosine transform of~\cite{kitajima1980symmetric}, and~\cite{britanak2007discrete} collects the properties of the whole family. The DCT-II, with half-sample symmetric extension at both ends, is the transform of JPEG, MPEG and the later video standards~\cite{wallace1992jpeg,legall1991mpeg,sullivan2012overview}. The DCT-III is its transpose and inverse. The DCT-IV and DST-IV are symmetric and involutory. The DST-I has whole-sample antisymmetric extension at both ends, which places virtual zeros one sample outside the window.

\begin{table}[t]
\caption{The first four cosine and the first four sine transforms. Kernel entry for row $n$ and column $k$, $0\le n,k\le M-1$, up to the scaling that makes the matrix orthogonal; the extension symmetry at the left and right ends (W whole-sample, H half-sample, S symmetric, A antisymmetric); and the period of the extension.}
\label{tab:family}
\centering\footnotesize
\begin{tabular}{@{}llll@{}}
\toprule
Type & kernel & extension (left, right) & period\\
\midrule
DCT-I & $\cos\bigl(\pi kn/(M-1)\bigr)$ & WS, WS & $2(M-1)$\\
DCT-II & $\cos\bigl(\pi k(n+\tfrac12)/M\bigr)$ & HS, HS & $2M$\\
DCT-III & $\cos\bigl(\pi(k+\tfrac12)n/M\bigr)$ & WS, WA & $4M$\\
DCT-IV & $\cos\bigl(\pi(k+\tfrac12)(n+\tfrac12)/M\bigr)$ & HS, HA & $4M$\\
DST-I & $\sin\bigl(\pi(k+1)(n+1)/(M+1)\bigr)$ & WA, WA & $2(M+1)$\\
DST-II & $\sin\bigl(\pi(k+1)(n+\tfrac12)/M\bigr)$ & HA, HA & $2M$\\
DST-III & $\sin\bigl(\pi(k+\tfrac12)(n+1)/M\bigr)$ & WA, WS & $4M$\\
DST-IV & $\sin\bigl(\pi(k+\tfrac12)(n+\tfrac12)/M\bigr)$ & HA, HS & $4M$\\
\bottomrule
\end{tabular}
\end{table}

\emph{Second differences and the classes they generate.} Each of the eight transforms is the eigenbasis of the second-difference matrix $\operatorname{tridiag}(-1,2,-1)$ with its corner entries modified by the boundary condition of its extension~\cite{strang1999discrete}: the DST-I is the eigenbasis of the plain matrix (Dirichlet ends), the DCT-II of the matrix with corner entries $1$ (Neumann ends), and so on. Since the eigenbasis of a matrix diagonalizes every polynomial in it, each transform diagonalizes an entire commutative algebra, and that algebra has a direct description. For the DCT-II it is the set of symmetric Toeplitz-plus-Hankel matrices with entries $c_{i-j}+c_{i+j+1}$ for an even $2M$-periodic sequence $c$~\cite{sanchez1995diagonalizing}, and the eigenvalue of the $k$th column is the symbol $c(\pi k/M)=\sum_m c_me^{im\pi k/M}$. The other types have the same form with the Hankel index shifted and, for the sine types, with the Hankel part subtracted~\cite{sanchez1995diagonalizing,sanchez1996generating}, and every function of a second-difference matrix is Toeplitz plus Hankel~\cite{strang2014functions}. In the matched-group reading each of these algebras is a compression of the commutant of the dihedral group acting on the corresponding extension: the extension of length $2M$ (or $2M\pm2$) carries the permutation action of the dihedral group of the cycle of that length, the original window is the subspace fixed by the reflection that defines the extension, that subspace is invariant under the symmetric circulants, which are the commutant, though not under the rotations of the group, and the compression of the commutant to it is the Toeplitz-plus-Hankel class, with the transform of that type as the matched transform~\cite{thornton2026unification}. The dihedral symmetry group of the cosine matrices was identified in~\cite{egner2001automatic} and used there to generate fast algorithms; the supercharacter construction of~\cite{garcia2018supercharacters} obtains a cosine transform from the orbits of the reflection on a cycle together with the algebra it diagonalizes; and the algebraic signal processing theory of~\cite{puschel2003algebraic,puschel2008time,puschel2008algebraic} derives all sixteen types as Fourier transforms of polynomial algebras attached to the boundary conditions, the time model in~\cite{puschel2008time} and the space models with the sixteen types in~\cite{puschel2008algebraic}.

\emph{The sine family through the cosine family.} With $\bN=\operatorname{diag}((-1)^n)$ the Nyquist sign matrix, the modulation by the sinusoid $(-1)^n=e^{i\pi n}$ at the Nyquist frequency, and $\bJ$ the exchange matrix (the matrix with ones on the antidiagonal, the simplest Hankel matrix), the DST-II columns are the DCT-II columns multiplied by $\bN$ and taken in reverse order, $\bN u_k=s_{M-1-k}$ (Proposition~\ref{prop:nyquist} below). Conjugation by $\bN$ maps the sequence $c$ onto $(-1)^kc_k$ and the Toeplitz-plus-Hankel class onto the Toeplitz-minus-Hankel class, $(-1)^{i+j}(c_{i-j}+c_{i+j+1})=c'_{i-j}-c'_{i+j+1}$ with $c'_k=(-1)^kc_k$, which is the DST-II class. The relation between the cosine and the sine families is therefore the Nyquist modulation together with the reversal of the index, and the Hankel half of the class changes sign. It is not the Hankel integral transform of Bessel analysis, which plays no role here. Every column of the types with the same symmetry at both ends, the DCT-I, DCT-II, DST-I and DST-II, is symmetric or antisymmetric under $\bJ$, $\bJ u_k=(-1)^ku_k$ and $\bJ s_k=(-1)^ks_k$ for the DCT-II and the DST-II, which is the reflection symmetry shared with the KLT of every stationary window (Theorem~\ref{thm:exact}(a) and~\cite{cantoni1976eigenvalues}). The columns of types III and IV, whose two ends differ, do not have it.

\emph{The Hartley transform.} The discrete Hartley transform has the kernel $\operatorname{cas}(2\pi kn/M)=\cos(2\pi kn/M)+\sin(2\pi kn/M)$~\cite{hartley1942more,bracewell1983discrete}. With the factor $M^{-1/2}$ its matrix is real, symmetric and involutory. The columns for $k$ and $M-k$ span the same plane as the cosine and the sine at frequency $2\pi k/M$, so the Hartley basis diagonalizes every real symmetric circulant, which are the commutant of the dihedral group $D_M$ acting on the $M$-cycle. That action is multiplicity-free with two-dimensional isotypic components, so the matched transform is determined only up to a rotation in each cosine-sine plane. The Hartley basis and the separate cosine and sine basis (the real discrete Fourier transform) are two choices of the same matched transform. The algebraic signal processing theory reaches the same two bases from the other side: over the reals the shift algebra of the finite time model splits into one- and two-dimensional components, the two-dimensional ones carrying plane rotations, and the Fourier transforms of the real finite time models are the real discrete Fourier transform and the Hartley transforms of four types~\cite{puschel2008time}. The discrete Fourier transform, with complex exponentials, diagonalizes every circulant, symmetric or not, and is the matched transform of the cyclic group alone. The DCT-II of $x$ is the discrete Fourier transform of the even extension of $x$ to length $2M$ with the half-sample phase $e^{-i\pi k/2M}$, so the cosine transforms are Fourier transforms on the orbit space of a reflection.

\emph{Where the KLT of a stationary window sits.} The KLT of a window of a stationary process shares the reflection symmetry of all of these bases, since its covariance is symmetric Toeplitz and hence centrosymmetric, so each eigenvector can be chosen symmetric or antisymmetric, and is so when its eigenvalue is simple~\cite{cantoni1976eigenvalues}. For an autoregressive process the eigenvectors are sinusoidal as well (Theorem~\ref{thm:levels}), but their frequencies are not on any of the grids of Table~\ref{tab:family}: they lie at points where the spectrum crosses the eigenvalue, set by the poles, and the boundary rows of the precision matrix fix the eigenvalues and the weight of each crossing. The first-order window makes this explicit (Theorem~\ref{thm:levels}(c) and Fig.~\ref{fig:frequencies}(a)): its eigenfrequencies solve one equation whose roots sit on the DST-I grid for a vanishing pole and move to the DCT-II grid as the pole tends to one. The matched-group reading of the same fact is that the window covariance is not in any of the classes but near the DCT-II class, by a rank-$2p$ corner matrix (Theorem~\ref{thm:departure} and Corollary~\ref{cor:identity}), and that the sign of the dominant pole decides between the cosine and the sine class (Propositions~\ref{prop:nyquist} and~\ref{prop:limits}). Carath\'eodory's representation of the same covariance (Section~\ref{sec:cara}) is a third set of sinusoids, at frequencies set by the phase rather than the magnitude of the characteristic polynomial.
\fi

\section{Setting}\label{sec:setting}

Let $x_t=\sum_{k=1}^{p}\phi_kx_{t-k}+\varepsilon_t$ be stationary, with $\varepsilon_t$ white of variance $\sigma^2$ (uncorrelated, not necessarily independent or Gaussian: every statement but the sampling calculation of Section~\ref{sec:groups}, which takes the windows Gaussian, is second-order), and let $a(z)=1-\sum_k\phi_kz^k$ have all zeros outside the unit circle. Factor $a(z)=\prod_{j=1}^p(1-\zeta_jz)$. The $\zeta_j$ are the poles of the process, the zeros of $z^pa(1/z)=z^p-\phi_1z^{p-1}-\dots-\phi_p$, so stability is $|\zeta_j|<1$, a real process has real poles or conjugate pairs, and the order $p$ is the number of poles. Here, we write $a_0=1$, $a_m=-\phi_m$, $i$ for the imaginary unit, and
\begin{equation}\label{eq:symbol}
c_k=\sum_{\ell=0}^{p-k}a_\ell a_{\ell+k},\ \ c(\omega)=\sum_{k=-p}^{p}c_{|k|}e^{ik\omega}=\prod_{j=1}^p|1-\zeta_je^{i\omega}|^2,
\end{equation}
for $0\le k\le p$; $c(\omega)$ is the symbol of the precision matrix below, the trigonometric polynomial whose coefficients fill its diagonals. A real pole $\zeta_j=r$ contributes the factor $1-2r\cos\omega+r^2$, increasing on $[0,\pi]$ for $r>0$ and decreasing for $r<0$. A pair $re^{\pm i\omega_0}$ contributes $(1-2r\cos(\omega-\omega_0)+r^2)(1-2r\cos(\omega+\omega_0)+r^2)$, with a minimum near $\omega_0$ that deepens as $r\to1$. The spectrum is $S(\omega)=\sigma^2/c(\omega)$ and the autocovariance is $r_k=\int_{-\pi}^{\pi}S(\omega)e^{ik\omega}\,d\omega/2\pi$, which for distinct poles is the sum of one damped exponential per pole,
\begin{equation}\label{eq:poles}
r_k=\sigma^2\sum_{j=1}^{p}\frac{\zeta_j^{\,k+p-1}}{\prod_{i}(1-\zeta_i\zeta_j)\prod_{i\ne j}(\zeta_j-\zeta_i)},\qquad k\ge0,
\end{equation}
in which a conjugate pair combines into a damped cosine at $\omega_0$; \eqref{eq:poles} is the residue sum of $\sigma^2z^{k+p-1}/\prod_i(1-\zeta_iz)(z-\zeta_i)$ over the poles inside the circle. The dominant correlation scale is $-1/\log\max_j|\zeta_j|$, asymptotic to $1/(1-\max_j|\zeta_j|)$ as the dominant pole approaches the circle. The window $\bm x=(x_0,\dots,x_{M-1})^\T$, $M\ge2p+1$, has covariance $\bR$ with entries $r_{|i-j|}$ and precision $\bQ=\bR^{-1}$. The exchange matrix $\bJ$ reverses the samples, $\bN=\operatorname{diag}((-1)^n)$ is the Nyquist sign matrix, so called because $(-1)^n=e^{i\pi n}$ is the sinusoid at the Nyquist frequency $\pi$, so multiplying a window by $\bN$ shifts its spectrum by $\pi$ and conjugating its covariance by $\bN$ negates the poles (Proposition~\ref{prop:nyquist}), and superscripts $\T$ and $\mathsf H$ denote the transpose and the conjugate transpose\ifextended; $\be(\omega)=(1,e^{i\omega},\dots,e^{i(M-1)\omega})^\T$ is the steering vector at frequency $\omega$\fi.

A finite group $G$ acting on the index set gives the permutation representation $\rep(g)$. Its invariant class is the commutant $\comm_G=\{\bA:\bA\rep(g)=\rep(g)\bA\ \forall g\}$, and the Reynolds projection $\bP_G\bA=|G|^{-1}\sum_g\rep(g)\bA\rep(g)^\T$ is the orthogonal projection onto it~\cite{thornton2026ad_framework}. The residual $\norm{\bR-\bP_G\bR}_F^2/\norm{\bR}_F^2$ measures how far $\bR$ is from the class. For a multiplicity-free action with one-dimensional isotypic components and matched basis $\bU$ it equals the fraction of the energy of $\bU^\T\bR\bU$ off the diagonal. The dimension of the class is the number of parameters of a symmetric $G$-invariant matrix, the number of orbits of $G$ on unordered index pairs. For the cosine and sine classes the group acts on an extension of the window, and the class is the compression of the commutant of that action to the window~\cite{thornton2026unification}. It is the set of matrices diagonalized by one fixed basis, an algebra closed under inversion of its invertible members, and it is called the commutant of the extended action where no confusion can arise. The coding gain $\cg(\bU)$ of an orthonormal basis $\bU$ is the ratio of the arithmetic to the geometric mean of the coefficient variances $(\bU^\T\bR\bU)_{kk}$~\cite{jayant1984digital,goyal2001theoretical}. The arithmetic mean is $\operatorname{tr}\bR/M$ for every orthonormal basis, so the ratio $\cg_{\mathrm{KLT}}/\cg(\bU)$, which scores the basis at the transform level, is the geometric mean of the coefficient variances of $\bU$ divided by the geometric mean of the eigenvalues of $\bR$, and it is at least one by Hadamard's inequality, with equality exactly when $\bU$ diagonalizes $\bR$.

The word matched is used in two senses, which are kept apart by name. A group is an exact matched group of $\bR$ when $\bR$ lies in its commutant, so that the group fixes the eigenbasis or its block structure. A class is an approximately matched class when $\bR$ departs from it by a matrix of low rank, as the window departs from the cosine class, and the transform of that class is then a fixed transform near the KLT, not the KLT. When a group is said to be matched approximately, it is the class of its commutant that is meant.

\section{The Karhunen-Lo\`eve Basis}\label{sec:klt}

The DCT-II basis $u_k(n)=\gamma_k\sqrt{2/M}\cos(\pi k(n+\tfrac12)/M)$ ($\gamma_0=1/\sqrt2$, $\gamma_k=1$ otherwise) diagonalizes exactly the matrices $(\bT_c+\bH_c)_{ij}=c_{i-j}+c_{i+j+1}$ for an even $2M$-periodic sequence $c$~\cite{sanchez1995diagonalizing}, a Toeplitz matrix, constant along each diagonal, plus a Hankel matrix, constant along each antidiagonal. This set is the DCT-II class, and a matrix in it is a class member. These are the compressions to the even-extension subspace of the symmetric circulants of size $2M$, the invariant class of the dihedral group $D_{2M}$ on that extension, and $u_k$ has eigenvalue $c(\pi k/M)$~\cite{thornton2026unification}. Extend the symbol of~\eqref{eq:symbol} by $c_k=0$ for $p<k\le M$.

\begin{theorem}\label{thm:departure}
$\bQ$ is banded with bandwidth $p$ and persymmetric, that is, symmetric about its antidiagonal, $\bJ\bQ\bJ=\bQ$. Outside the two $p\times p$ corner blocks $\sigma^2Q_{ij}=c_{|i-j|}$, and inside the top-left block $\sigma^2Q_{ij}=\sum_{\ell=0}^{\min(i,j)}a_\ell a_{\ell+|i-j|}$ for $0\le i,j<p$, the autocorrelation sum truncated after $\min(i,j)+1$ terms, mirrored in the bottom-right block. Hence $\sigma^2\bQ=\bT_c+\bH_c-\bD$ with $\bD$ supported on the two corner blocks, of rank at most $2p$, with top-left entries
\begin{equation}\label{eq:departure}
D_{ij}=c_{i+j+1}+\!\!\sum_{\ell=\min(i,j)+1}^{p-|i-j|}\!\! a_\ell a_{\ell+|i-j|},\quad 0\le i,j<p,
\end{equation}
and for $p=1$, $\bD=-\phi_1(1-\phi_1)(\be_0\be_0^\T+\be_{M-1}\be_{M-1}^\T)$ with $\be_n$ the $n$th unit vector. Let $\bT_a$ and $\bH_a$ be the $p\times p$ matrices with entries $a_{i-j}$ and $a_{i+j+1}$ ($a_m=0$ for $m<0$ and for $m>p$), lower triangular Toeplitz with unit diagonal and Hankel with $a_p$ on the antidiagonal and zeros below it. Then the top-left block $\bD_1$ of $\bD$ factors as $\bD_1=\bH_a(\bH_a+\bT_a)$ and
\begin{equation}\label{eq:detcorner}
\det\bD_1=(-1)^{p(p-1)/2}a_p^{\,p}\prod_{j=1}^{p}(1-\zeta_j)\prod_{i<j}(1-\zeta_i\zeta_j),
\end{equation}
so for a stable process $\bD_1$ is nonsingular, and $\bD$ has rank exactly $2p$, if and only if $a_p=-\phi_p\ne0$, that is, if and only if no pole is zero. In general $\operatorname{rank}\bD=2p_0$ with $p_0$ the number of nonzero poles.
\end{theorem}

\begin{proof}
\ifextended
\emph{The precision as a Gram matrix.} The Gaussian process with the same second moments has density equal to the density of $(x_0,\dots,x_{p-1})$ times the conditional densities of $x_t$ given its $p$ predecessors, $t=p,\dots,M-1$. Hence
\[
\sigma^2\bQ=\bA^\T\bA+\sigma^2\bigl(\bm\Sigma_p^{-1}\oplus\bm0\bigr),\qquad A_{tj}=a_{t-j}\ \ \text{for }0\le t-j\le p\text{ and }t\ge p,
\]
with $\bm\Sigma_p$ the covariance of the first $p$ samples. The second term lives in the top-left block, and the Gram entry is
\[
(\bA^\T\bA)_{ij}=\sum_{t=\max(i,j,p)}^{\min(i+p,\,j+p,\,M-1)}a_{t-i}\,a_{t-j}.
\]

\emph{The interior and the corners.} Outside the corner blocks the limits of the sum are $\max(i,j)$ and $\min(i,j)+p$, so the sum equals $c_{|i-j|}$. In the bottom-right block the upper limit is $M-1$, and with $i'=M-1-i$ and $j'=M-1-j$ the sum is
\[
\sum_{\ell=0}^{\min(i',j')}a_\ell\, a_{\ell+|i'-j'|}.
\]
Persymmetry, from Theorem~\ref{thm:exact}(a), carries this formula to the top-left block. The Hankel entry $c_{i+j+1}$ is nonzero only inside the corner blocks, and subtracting the corner formula from $c_{|i-j|}+c_{i+j+1}$ leaves~\eqref{eq:departure}. The banded Toeplitz interior with corner corrections is the classical form of the inverse~\cite{siddiqui1958inversion,galbraith1974inverses,greville1979band}. The closed form of the corners is what is used below.

\emph{The factorization of the corner block.} For $0\le i,j<p$ three sums identify the terms of $D_{ij}=c_{|i-j|}+c_{i+j+1}-\sum_{\ell=0}^{\min(i,j)}a_\ell a_{\ell+|i-j|}$:
\begin{align*}
\sum_{\ell=0}^{\min(i,j)}a_\ell\, a_{\ell+|i-j|}&=\sum_{t\le\min(i,j)}a_{i-t}\,a_{j-t}=(\bT_a\bT_a^\T)_{ij},\\
c_{|i-j|}-(\bT_a\bT_a^\T)_{ij}&=\sum_{s\ge0}a_{i+1+s}\,a_{j+1+s}=(\bH_a\bH_a^\T)_{ij},\\
(\bH_a\bT_a)_{ij}=\sum_{t\ge j}a_{i+t+1}\,a_{t-j}&=\sum_{s\ge0}a_s\,a_{s+i+j+1}=c_{i+j+1}.
\end{align*}
Since $\bH_a$ is symmetric,
\[
\bD_1=\bH_a\bH_a^\T+\bH_a\bT_a=\bH_a(\bH_a+\bT_a).
\]

\emph{The determinant.} $\bH_a$ is zero below its antidiagonal, so $\det\bH_a=(-1)^{p(p-1)/2}a_p^{\,p}$. It remains to show that $\bm\Xi=\bH_a+\bT_a$, with entries $\Xi_{ij}=a_{i-j}+a_{i+j+1}$, has
\[
\det\bm\Xi=\prod_{j=1}^{p}(1-\zeta_j)\prod_{i<j}(1-\zeta_i\zeta_j).
\]
As $a_m$ is $(-1)^m$ times the $m$th elementary symmetric polynomial of the poles, $\Xi_{ij}$ is a polynomial in the poles of degree at most $i+j+1$. So $\det\bm\Xi$ is a symmetric polynomial in $\zeta_1,\dots,\zeta_p$ of degree at most $\sum_i(2i+1)=p^2$, and it equals $1$ at $\zeta=0$, where $\bm\Xi=\bI$. Two families of factors divide it.

(i) It vanishes when some $\zeta_j=1$. Each row of $\bm\Xi$ sums to $a(1)$,
\[
(\bm\Xi\one)_i=\sum_{m\le i}a_m+\sum_{m>i}a_m=a(1)=\prod_j(1-\zeta_j),
\]
so $\one$ is an eigenvector with eigenvalue $a(1)$.

(ii) It vanishes when $\zeta_1\zeta_2=1$ with $\zeta_1\ne-1$. For $0\le i<p$, $(\bm\Xi\bv)_i$ is the $i$th coefficient of the convolution of $a$ with the half-sample even extension of $\bv$, $\tilde v_{-1-n}=\tilde v_n$, and only the entries $\tilde v_n$ with $-p\le n<p$ enter. The sequence
\[
\tilde v_n=\zeta_1^{\,n}+\zeta_1^{-n-1},\qquad n\in\Z,
\]
is half-sample even and nonzero at $n=0$, and it is annihilated by the convolution with $(1-\zeta_1z)(1-\zeta_2z)=1-(\zeta_1+\zeta_1^{-1})z+z^2$, hence by $a$. So $\bv=(\tilde v_0,\dots,\tilde v_{p-1})$ is a null vector of $\bm\Xi$.

\emph{From vanishing to divisibility.} Since $\det\bm\Xi$ is symmetric in the poles, (i) holds for every $j$ and (ii) for every pair $i<j$. Each of the $p+p(p-1)/2$ polynomials $1-\zeta_j$ and $1-\zeta_i\zeta_j$ is irreducible, and no two of them are proportional. By (i), $\det\bm\Xi$ vanishes on the whole zero set of $1-\zeta_j$. By (ii), it vanishes on the zero set of $1-\zeta_i\zeta_j$ except on the points with $\zeta_i=-1$, which form a proper closed subset of that irreducible set, with dense complement, so by continuity it vanishes on all of it. Over $\C$ a polynomial that vanishes on the zero set of an irreducible polynomial is divisible by it, so
\[
\det\bm\Xi=g(\zeta)\prod_{j=1}^{p}(1-\zeta_j)\prod_{i<j}(1-\zeta_i\zeta_j)
\]
for a polynomial $g$. The product has degree $p+p(p-1)=p^2$, which is the bound on the degree of $\det\bm\Xi$, so $g$ is a constant, and the value $1$ at $\zeta=0$ gives $g=1$. With $\det\bH_a$ this is~\eqref{eq:detcorner}.

\emph{The rank.} $\bD$ is $\bD_1$ and its mirror, which occupy disjoint rows and columns for $M\ge2p+1$, so $\operatorname{rank}\bD=2\operatorname{rank}\bD_1$. For $|\zeta_j|<1$ every factor of~\eqref{eq:detcorner} but $a_p^{\,p}$ is nonzero. If $a_p=0$ the process is the AR($p_0$) process of its nonzero poles, $\bD$ does not depend on the order declared, and the statement at order $p_0$ gives the rank $2p_0$.
\fi
\end{proof}

\begin{remark}\label{rem:firstorder}
For $p=1$ the corner value is classical in two forms. In Jain's family of boundary-modified Markov matrices $\bm J(k_1,k_2,0,0)$, with unit diagonal, off-diagonal entries $-\alpha$, $\alpha=\rho/(1+\rho^2)$, and corner entries $1-k_1\alpha$ and $1-k_2\alpha$, the window precision is $(1+\rho^2)\bm J(\rho,\rho,0,0)$ (in units of $\sigma^{-2}$) and the matrix that the DCT-II diagonalizes exactly is $(1+\rho^2)\bm J(1,1,0,0)$~\cite{jain1979sinusoidal}. Their difference is Jain's $\alpha(1-\rho)$ on the two corners, which multiplied by $1+\rho^2$ is $-\bD$. In the factorizations of~\cite{reznik2026direct,reznik2026exact} the same quantity $\rho(1-\rho)$ is the rank-one boundary term of each parity branch after the fold. Theorem~\ref{thm:departure} gives the corner block at every order; Section~\ref{sec:fast} folds it.
\end{remark}

\begin{theorem}\label{thm:levels}
Let $\mu$ be an eigenvalue of $\bR$ with eigenvector $\bv$ and let $z_1,\dots,z_{2p}$ be the zeros of $z^p\bigl(c(z)-\sigma^2/\mu\bigr)$, where $c(z)=\sum_{k=-p}^pc_{|k|}z^k$ is the Laurent polynomial with $c(e^{i\omega})$ the symbol of~\eqref{eq:symbol}. They come in reciprocal pairs. If the zeros are simple, $v_n=\sum_j\alpha_jz_j^n$ for $0\le n\le M-1$, with the $\alpha_j$ fixed by the $2p$ boundary rows of $\bQ$. A zero of multiplicity $m$ contributes the modes $z^n,nz^n,\dots,n^{m-1}z^n$ in place of $z^n$, polynomially weighted exponentials, and the statements below hold with that reading.
(a) The zeros on the unit circle are $e^{\pm i\omega}$ at the frequencies where $S(\omega)=\mu$, that is, where $\prod_j|1-\zeta_je^{i\omega}|^2=\sigma^2/\mu$. There is at least one such frequency in $(0,\pi)$ and at most $p$, and the eigenvalue equals the spectrum at every one of them. Hence $\bv$ contains at most one sinusoid at each such frequency, polynomially weighted if the zero is repeated, together with the modes of the zeros inside the circle, which decay from the two ends of the window.
(b) If all poles are real and positive, $c$ increases on $[0,\pi]$, the level set of the spectrum at height $\mu$, the set of frequencies where $S(\omega)=\mu$, is one frequency, and every eigenvector carries at most one sinusoid, whatever $p$.
(c) For $p=1$ with pole $\rho\in(-1,1)$, $\rho\ne0$, every eigenvector is a single sinusoid, $v_n=\cos\bigl(\omega(n-\tfrac{M-1}2)\bigr)$ or $\sin\bigl(\omega(n-\tfrac{M-1}2)\bigr)$, with $\mu=S(\omega)=\sigma^2/(1-2\rho\cos\omega+\rho^2)$ and $\omega$ one of the $M$ roots in $(0,\pi)$ of
\begin{equation}\label{eq:rayd}
\tan(M\omega)=\frac{-(1-\rho^2)\sin\omega}{(1+\rho^2)\cos\omega-2\rho}.
\end{equation}
As $\rho\to1$ the roots tend to the DCT-II frequencies $\pi k/M$, as $\rho\to-1$ to the DST-II frequencies $\pi(k+1)/M$, and as $\rho\to0$ to the DST-I frequencies $\pi(k+1)/(M+1)$, $k=0,\dots,M-1$.
\end{theorem}

\begin{proof}
\ifextended
\emph{The recurrence.} Rows $p\le n\le M-1-p$ of $\bQ\bv=\bv/\mu$ lie outside the corner blocks of Theorem~\ref{thm:departure} and read
\[
\sum_{k=-p}^{p}c_{|k|}\,v_{n+k}=\frac{\sigma^2}{\mu}\,v_n,
\]
a recurrence of order $2p$ with characteristic equation $c(z)=\sigma^2/\mu$. Its solution space on the window is spanned by the $z_j^n$ for simple zeros, and by $n^kz_j^n$, $k<m$, for a zero of multiplicity $m$. The remaining $2p$ rows are a homogeneous linear system for the $\alpha_j$ whose determinant vanishes exactly at the eigenvalues.

(a) \emph{The level set.} On the circle, $c(\omega)=\sigma^2/\mu$ is a polynomial equation of degree $p$ in $\cos\omega$, so it has at most $p$ solutions in $(0,\pi)$. It has at least one. For a unit eigenvector $\bv$ with trigonometric polynomial $V(\omega)=\sum_nv_ne^{in\omega}$,
\[
\mu=\bv^\T\bR\bv=\int_{-\pi}^{\pi}S(\omega)\,|V(\omega)|^2\,\frac{d\omega}{2\pi},
\]
a weighted mean of $S$. So $\mu$ lies strictly between the minimum and the maximum of $S$ when $S$ is not constant~\cite{grenander1958toeplitz,gray2006toeplitz}, and $\sigma^2/\mu$ lies strictly between the minimum and the maximum of $c$ on $[0,\pi]$, so it is attained in $(0,\pi)$.

(b) \emph{Real positive poles.} For real positive poles each factor $1-2r\cos\omega+r^2$ of~\eqref{eq:symbol} increases on $[0,\pi]$, so $c$ increases and the level set has one point.

(c) \emph{The first-order equation.} For $p=1$ write $\lambda=\sigma^2/\mu$. The interior rows and the first corner row are
\[
-\rho v_{n-1}+(1+\rho^2)v_n-\rho v_{n+1}=\lambda v_n,\qquad v_0-\rho v_1=\lambda v_0,
\]
the last row being the mirror of the first, and the two roots are $e^{\pm i\omega}$ with $1+\rho^2-2\rho\cos\omega=\lambda$. $\bR$ is centrosymmetric, so $\bv$ is symmetric or antisymmetric about $(M-1)/2$~\cite{cantoni1976eigenvalues}, which gives the two forms. With $\theta=\omega(M-1)/2$ the corner row reduces to
\[
\tan\theta=\frac{\cos\omega-\rho}{\sin\omega}\ \ \text{(symmetric)},\qquad \tan\theta=\frac{-\sin\omega}{\cos\omega-\rho}\ \ \text{(antisymmetric)},
\]
and both give
\[
\tan2\theta=\frac{2\sin\omega\,(\cos\omega-\rho)}{\sin^2\omega-(\cos\omega-\rho)^2}.
\]
The addition formula for $\tan(M\omega)=\tan(2\theta+\omega)$ then yields~\eqref{eq:rayd}. The $M$ eigenvalues are simple because $\sigma^2\bQ$ is tridiagonal with nonzero off-diagonal entries, and each gives one $\omega\in(0,\pi)$.

\emph{The limits.} As $\rho\to\pm1$ the right side of~\eqref{eq:rayd} tends to zero at fixed $\omega\in(0,\pi)$, so $M\omega\to\pi k$, with the lowest root leaving through $0$ for $\rho\to1$ and the highest through $\pi$ for $\rho\to-1$. As $\rho\to0$ the right side tends to $-\tan\omega$, so $\sin((M+1)\omega)\to0$.
\fi
\end{proof}

Theorem~\ref{thm:levels} is the trivial-group description of the window: the basis depends on the poles, but its vectors are organized by the heights of the spectrum, with at most one sinusoid for each crossing of that height by the spectrum, and the sinusoid frequencies are set by the magnitude $|a(e^{i\omega})|$ of the characteristic polynomial. Part (c) is the first-order basis of~\cite{ray1970further}, derived here from the corner rows of the precision. Its limits are the analytical derivations of the DCT-II~\cite{clarke1981relation,flickner1982derivation} and of the DST-I~\cite{clarke1984relation} as limiting KLTs, and the mode structure for $p\ge2$ is the one that Trench uses to compute eigenvalues of band matrices with Toeplitz inverses~\cite{trench1986band,trench1985eigenvalue}, whose eigenvector formulas are rederived through skew Schur polynomials in~\cite{maximenko2017cofactors}. The asymptotic spectral theory of banded Toeplitz matrices is in~\cite{bottcher2005spectral}. For a resonant pair the basis vectors at heights below the peak are carriers at the resonance under slow envelopes.

\begin{figure*}[!t]
\centering
\ifextended
\includegraphics[width=0.9\textwidth]{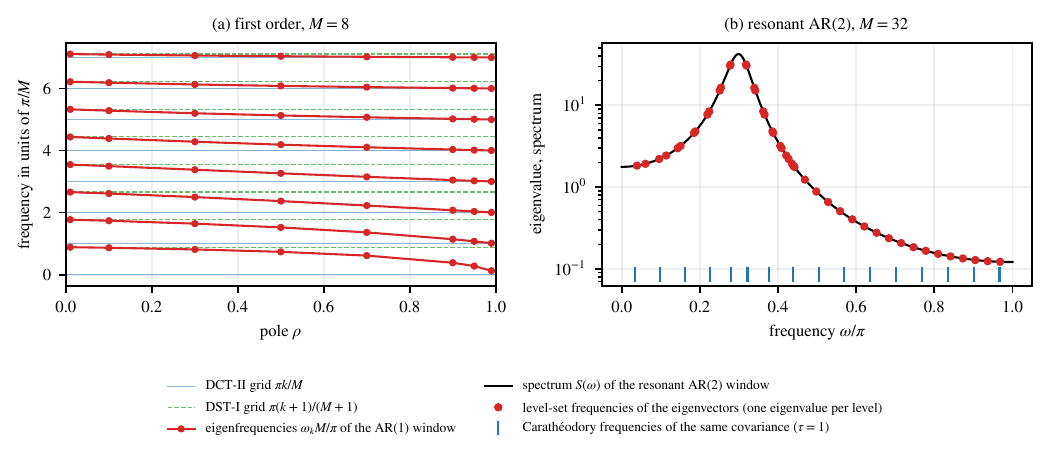}
\caption{The basis against the pole. (a) The eigenfrequencies of the first-order window at $M=8$, from~\eqref{eq:rayd}, against the pole, with the DST-I grid $\pi(k+1)/(M+1)$ and the DCT-II grid $\pi k/M$. (b) The resonant AR(2) window at $M=32$: the spectrum, the level-set frequencies of each eigenvector at the height of its eigenvalue, and the Carath\'eodory frequencies of the same covariance (Section~\ref{sec:cara}).}
\fi
\label{fig:frequencies}
\end{figure*}

\emph{The boundary system.} For the record, the $2p$ boundary rows are the first $p$ and the last $p$ rows of $(\sigma^2\bQ-\lambda\bI)\bB$, where $\lambda=\sigma^2/\mu$ and $\bB$ is the $M\times2p$ matrix with columns $z_j^n$ (for an inside pair $z,1/z$ the columns $z^n$ and $z^{M-1-n}$ are better conditioned). The determinant of this $2p\times2p$ matrix, as a function of $\lambda$ with the $z_j(\lambda)$ substituted, vanishes at the $M$ eigenvalues of $\sigma^2\bQ$, and at a level where the $z_j$ are simple it vanishes only there, since a null vector of the system then gives a nonzero eigenvector. Where two crossings merge, at an interior extremum of $c$, the columns of $\bB$ become dependent and the determinant vanishes as well. For $p=1$ it reduces to~\eqref{eq:rayd}. By Theorem~\ref{thm:levels}(a) the number of sinusoids in the eigenvector at every eigenvalue is at most the number of monotone branches of $S$ that cross the level $\mu$: one for a window with real positive poles, two between the peak of a resonant window and the larger of its end values, up to $p$ in general. It can be smaller, because the boundary system can give a crossing the coefficient zero. For the AR(3) window with the pair $0.9e^{\pm i0.3\pi}$ and the real pole $\inactR$ at $M=8$ the level of the top eigenvector is crossed at $\inactFreqA\pi$, $\inactFreqB\pi$ and $\inactFreqC\pi$, and the unit eigenvector has components of norm $\inactNorm$, $\inactNormB$ and $\inactNormC$ there. It lies in the span of the other modes to $\inactResid$, and its eigenvalue lies $\inactGap$ percent from the nearest eigenvalue of the same parity. In the six windows of Section~\ref{sec:num} every crossing carries a sinusoid. Fig.~\ref{fig:frequencies}(b) shows the resonant AR(2) window at $M=32$, where the eigenvectors with eigenvalues above the tails of the spectrum carry two sinusoids, one on each side of the peak, and those below carry one.

\begin{corollary}\label{cor:identity}
Let $\bR_c=\bQ_c^{-1}$ with $\sigma^2\bQ_c=\bT_c+\bH_c$ the DCT-II class member of Theorem~\ref{thm:departure}. Then $\bR_c$ lies in the DCT-II class, its DCT-II coefficient variances are the spectrum samples $S(\pi k/M)$, and
\begin{equation}\label{eq:identity}
\bR-\bR_c=\sigma^{-2}\bR\bD\bR_c,
\end{equation}
a symmetric matrix of rank at most $2p$, and of rank exactly $2p$ when no pole is zero, since $\bR$ and $\bR_c$ are nonsingular and $\bD$ then has rank $2p$ by Theorem~\ref{thm:departure}. Hence, with $\mu_1\ge\dots\ge\mu_M$ the eigenvalues of $\bR$ and $\nu_1\ge\dots\ge\nu_M$ the samples $S(\pi k/M)$ in decreasing order, $\nu_{k+2p}\le\mu_k\le\nu_{k-2p}$ for every $k$ (indices outside $1,\dots,M$ dropped). For $p=1$ the perturbation is $-\sigma^{-2}\rho(1-\rho)(\bR\be_0\be_0^\T\bR_c+\bR\be_{M-1}\be_{M-1}^\T\bR_c)$, of rank two.
\end{corollary}

\begin{proof}
\ifextended
\emph{The class member.} The DCT-II class is the set of matrices diagonalized by the DCT-II, an algebra closed under inversion of its invertible members, so $\bR_c$ lies in it. The eigenvalue of $u_k$ for $\bT_c+\bH_c$ is $c(\pi k/M)$, so for $\bR_c$ it is $\sigma^2/c(\pi k/M)=S(\pi k/M)$.

\emph{The identity.} With $\bQ_c-\bQ=\sigma^{-2}\bD$,
\[
\bR-\bR_c=\bQ^{-1}-\bQ_c^{-1}=\bQ^{-1}(\bQ_c-\bQ)\bQ_c^{-1}=\sigma^{-2}\bR\bD\bR_c,
\]
which is~\eqref{eq:identity}.

\emph{The interlacing.} A symmetric perturbation of rank at most $r$ moves each eigenvalue by at most $r$ places in the ordering, and a positive semidefinite one moves each eigenvalue up, by at most $r$ places~\cite{thompson1976behavior}, \cite[Thm.~4.3.6]{horn1985matrix}. The perturbation in~\eqref{eq:identity} is symmetric of rank at most $2p$, so it moves each eigenvalue by at most $2p$ places.
\fi
\end{proof}

Corollary~\ref{cor:identity} is the finite-window form of the asymptotic statements: the KLT eigenvalues of an AR($p$) window are sandwiched between spectrum samples on the DCT-II grid shifted by at most $2p$ positions, at every $M$ and every pole, and the whole departure of the covariance from the class is carried by the corner matrix $\bD$ of Theorem~\ref{thm:departure}. In the examples of Section~\ref{sec:num} the observed shift is at most \shiftMaxObserved. For $p=1$ the interlacing is in~\cite{reznik2026exact} in a sharper form, one place within each parity class and in one direction, which Section~\ref{sec:fast} recovers at every order as $p$ places within each class, with the direction and the count set by the inertia of the corner block (Proposition~\ref{prop:fast}(d)).

The rank statement does not need the banded precision, and it holds for every window with a rational spectrum. Write $\bR_c(S)$ for the member of the DCT-II class whose DCT-II coefficient variances are the samples $S(\pi k/M)$ of a spectrum $S$. For the AR($p$) window $\bR_c(S)=\bR_c$.

\begin{corollary}\label{cor:rational}
Let $\bR$ be the window covariance of a stationary process with rational spectrum $S(\omega)=\sigma^2|b(e^{i\omega})|^2/|a(e^{i\omega})|^2$, with $a$ of degree $p$ and distinct zeros, $b$ of degree $q$ (an ARMA($p,q$) process; $q=0$ is the AR window and $p=0$ the MA window), and $M>2q$. Then $\bR-\bR_c(S)$ has rank at most $2\max(p,q)$, so the eigenvalues of $\bR$ interlace the samples $S(\pi k/M)$ with a shift of at most $2\max(p,q)$. For $q\le p$ the departure lies, on both sides, in the span of the $2p$ sequences $\zeta_j^{\,n}$ and $\zeta_j^{-n}$, one pair per pole, and its rank is exactly $2p$ when the poles are nonzero and none of them is a zero of $b$. If $\bR=\bR_{\mathrm u}+\bR_{\mathrm v}$ is the window of a sum of uncorrelated AR($k_1$) and AR($k_2$) processes, a co-channel sum, two signals received on one channel, then $\bR_c(S_{\mathrm u}+S_{\mathrm v})=\bR_c(S_{\mathrm u})+\bR_c(S_{\mathrm v})$ and the departure has rank at most $2(k_1+k_2)$, which is the bound of the corollary for the ARMA representation of degrees at most $(k_1+k_2,\max(k_1,k_2))$ that the sum admits.
\end{corollary}

\begin{proof}
\ifextended
\emph{The departure.} The class member is $\bR_c(S)=\bT_c+\bH_c$ with $c$ the even $2M$-periodic sequence whose $2M$-point Fourier coefficients are the samples $S(\pi k/M)$. Sampling the spectrum aliases the autocovariance, so $c$ is the periodization
\[
c_m=\sum_{\ell\in\Z}r_{m+2M\ell}.
\]
Hence
\begin{equation}\label{eq:rational}
\bR-\bR_c(S)=-\bigl[\textstyle\sum_{\ell\ne0}r_{|i-j|+2M\ell}\bigr]_{ij}-\bigl[c_{i+j+1}\bigr]_{ij}.
\end{equation}

\emph{The exponential and the finite parts.} The autocovariance of a rational spectrum satisfies the homogeneous recursion $\sum_\ell a_\ell r_{k-\ell}=0$ for $k>q$, so
\[
r_k=\sum_j\varsigma_j\zeta_j^{\,k}+\delta_k,\qquad k\ge0,
\]
the exponential sum being the partial-fraction part of $S$ (for $q=0$ it is~\eqref{eq:poles}) and $\delta_k=0$ for $k>q-p$. Only $\delta_1,\dots,\delta_{q-p}$ enter below, so for $q\le p$ the departure is the exponential part alone.

\emph{One pole.} Put
\[
\bm u=(\zeta^n)_{n<M},\qquad \bm v=(\zeta^{-n})_{n<M},\qquad \gamma=\frac{\zeta^{2M}}{1-\zeta^{2M}}.
\]
For $0\le m<2M$ the aliases sum to
\[
\sum_{\ell\ge1}\zeta^{m+2M\ell}=\gamma\zeta^m,\qquad \sum_{\ell\ge1}\zeta^{2M\ell-m}=\gamma\zeta^{-m}.
\]
So the Toeplitz part of~\eqref{eq:rational} for this pole is
\[
-\varsigma\gamma\bigl(\zeta^{i-j}+\zeta^{j-i}\bigr)=-\varsigma\gamma\bigl(\bm u\bm v^\T+\bm v\bm u^\T\bigr),
\]
with no absolute value because the two aliases are symmetric in $m$, and the Hankel part is
\[
-\varsigma(1+\gamma)\bigl(\zeta\,\bm u\bm u^\T+\zeta^{2M-1}\bm v\bm v^\T\bigr).
\]
Both lie in $\operatorname{span}\{\bm u,\bm v\}\otimes\operatorname{span}\{\bm u,\bm v\}$, so each pole contributes rank at most two, and the exponential part contributes at most $2p$.

\emph{The rank for $q\le p$.} The exponential part is then the whole departure,
\[
\bR-\bR_c(S)=\bm U\bm C\bm U^\T,\qquad \bm U=[\bm u_1\ \bm v_1\ \cdots\ \bm u_p\ \bm v_p],
\]
with $\bm C$ block diagonal with the $2\times2$ blocks
\[
\bm C_j=-\varsigma_j\begin{bmatrix}(1+\gamma_j)\zeta_j&\gamma_j\\ \gamma_j&(1+\gamma_j)\zeta_j^{2M-1}\end{bmatrix},\qquad \det\bm C_j=\frac{\varsigma_j^2\,\zeta_j^{2M}}{1-\zeta_j^{2M}}.
\]
The $2p$ ratios $\zeta_j$ and $\zeta_j^{-1}$ are distinct for distinct stable poles, so $\bm U$ has full column rank when $M\ge2p$. The rank is therefore $2p$ exactly when every $\varsigma_j\ne0$, that is, when every pole is nonzero and uncancelled by a zero of $b$.

\emph{The finite part for $q>p$.} The sequence $\delta$ has support in $|k|\le q-p<M$, so it is not aliased and enters only the Hankel part. It does so in the entries $\delta_{i+j+1}$ with $i+j+1\le q-p$, a corner anti-triangle of rank at most $q-p$, and in its mirror image $\delta_{2M-1-i-j}$ in the opposite corner, of the same rank. The total is at most $2p+2(q-p)=2q$ when $q>p$ and $2p$ otherwise. The interlacing follows as in Corollary~\ref{cor:identity}.

\emph{The sum.} The class is a linear space and the coefficient variances add, so the departures add and their ranks add. The sum has spectrum
\[
\frac{\sigma_{\mathrm u}^2|a_{\mathrm v}|^2+\sigma_{\mathrm v}^2|a_{\mathrm u}|^2}{|a_{\mathrm u}a_{\mathrm v}|^2},
\]
with denominator and numerator of degrees at most $k_1+k_2$ and $\max(k_1,k_2)$ before any cancellation, which lowers the degrees and not the bound, and $2\max(k_1+k_2,\max(k_1,k_2))=2(k_1+k_2)$. The rank is exactly $2(k_1+k_2)$ when the poles of the two summands are nonzero, distinct and disjoint, since the partial-fraction weights are then the summands' own.
\fi
\end{proof}

Corollary~\ref{cor:rational} is a second proof of the rank statement of Corollary~\ref{cor:identity}, and it says where the departure of an AR window lives: in the modes $\zeta_j^{\,n}$ and $\zeta_j^{-n}$ decaying from the two ends, the same modes that Theorem~\ref{thm:levels} attaches to the eigenvectors. Its content for the matched-group reading is that the DCT-II class is within a low-rank departure of every window with a rational spectrum, of rank set by the larger of the two orders, so the statements of Section~\ref{sec:fast} on the fast transform hold for ARMA windows and for co-channel sums of autoregressive windows with the corner block replaced by the folded departure.

The departure also measures what the DCT-II loses against the KLT at the transform level. Write $\nu_k=S(\pi k/M)$ for the coefficient variances of the class member, $\chi_k=(\bU^\T\bR\bU)_{kk}$ for the DCT-II coefficient variances of the window, and $\cg_{\mathrm{KLT}}/\cg_{\mathrm{DCT}}$ for the coding-gain ratio of Section~\ref{sec:setting}.

\begin{theorem}\label{thm:gain}
Let $\bR$ be a window with a rational spectrum of orders $(p,q)$ as in Corollary~\ref{cor:rational}, $\bU$ the DCT-II, and
\begin{equation}\label{eq:ups}
\bUps=\operatorname{diag}(\nu)^{-1/2}\,\bU^\T(\bR-\bR_c(S))\bU\,\operatorname{diag}(\nu)^{-1/2},
\end{equation}
the departure normalized by the class member. Then $\bUps$ is symmetric of rank at most $2\max(p,q)$, its nonzero eigenvalues $\upsilon_1,\dots,\upsilon_r$ exceed $-1$, $\chi_k=\nu_k(1+\Upsilon_{kk})$, and for the AR($p$) window $\operatorname{tr}\bUps=\sigma^{-2}\operatorname{tr}(\bD\bR)$ with $\bD$ the corner matrix of Theorem~\ref{thm:departure}, which is $-2\rho/(1+\rho)$ for $p=1$.
(a) [identity] $\cg_{\mathrm{KLT}}/\cg_{\mathrm{DCT}}=\bigl[\prod_{k}(1+\Upsilon_{kk})/\prod_{i}(1+\upsilon_i)\bigr]^{1/M}$.
(b) [rank bounds]
\begin{equation}\label{eq:gainbounds}
1\le\frac{\cg_{\mathrm{KLT}}}{\cg_{\mathrm{DCT}}}\le\frac{1+\operatorname{tr}\bUps/M}{\prod_i(1+\upsilon_i)^{1/M}}\le e^{\frac1M\sum_i(\upsilon_i-\log(1+\upsilon_i))},
\end{equation}
so the loss is set by at most $2\max(p,q)$ numbers whatever $M$.
(c) [correlation bound] Let $\bX$ be the off-diagonal part of the DCT-II coefficient correlation matrix, $X_{kl}=(\bU^\T\bR\bU)_{kl}/\sqrt{\chi_k\chi_l}$ for $k\ne l$ and $X_{kk}=0$. If $\norm{\bX}_2<1$, then
\begin{equation}\label{eq:corrbound}
\frac{\cg_{\mathrm{KLT}}}{\cg_{\mathrm{DCT}}}\le\exp\Bigl(\frac{\norm{\bX}_F^2}{2M(1-\norm{\bX}_2)}\Bigr).
\end{equation}
(d) [separable block] For $\bR=\bR_{\mathrm v}\otimes\bR_{\mathrm h}$ and the two-dimensional DCT-II $\bU_{\mathrm v}\otimes\bU_{\mathrm h}$, the coding-gain ratio is the product of the two one-dimensional ratios, and each bound in (b) and (c) multiplies in the same way.
\end{theorem}

\begin{proof}
\ifextended
\emph{The identity (a).} The arithmetic means of the coefficient variances agree for all orthonormal bases, so
\[
\frac{\cg_{\mathrm{KLT}}}{\cg_{\mathrm{DCT}}}=\Bigl(\frac{\prod_k\chi_k}{\det\bR}\Bigr)^{1/M}.
\]
Since $\bU^\T\bR_c(S)\bU=\operatorname{diag}(\nu)$ (Corollaries~\ref{cor:identity} and~\ref{cor:rational}),
\[
\bU^\T\bR\bU=\operatorname{diag}(\nu)^{1/2}(\bI+\bUps)\operatorname{diag}(\nu)^{1/2},
\]
so
\[
\det\bR=\prod_k\nu_k\prod_i(1+\upsilon_i),\qquad \chi_k=\nu_k(1+\Upsilon_{kk}),
\]
which is (a).

\emph{The properties of $\bUps$.} $\bI+\bUps$ is positive definite because $\bR$ is, so $\upsilon_i>-1$, and the rank of $\bUps$ is that of $\bR-\bR_c(S)$. For the trace, with $\bR_c^{-1}=\bQ_c$ and $\bQ\bR=\bI$,
\[
\operatorname{tr}\bUps=\operatorname{tr}\bigl(\operatorname{diag}(\nu)^{-1}\bU^\T(\bR-\bR_c)\bU\bigr)=\operatorname{tr}(\bR_c^{-1}\bR)-M=\operatorname{tr}\bigl((\bQ_c-\bQ)\bR\bigr)=\sigma^{-2}\operatorname{tr}(\bD\bR).
\]
For $p=1$ the corner matrix of Theorem~\ref{thm:departure} and $r_0=\sigma^2/(1-\rho^2)$ give $\sigma^{-2}\operatorname{tr}(\bD\bR)=-2\rho(1-\rho)r_0/\sigma^2=-2\rho/(1+\rho)$.

(b) \emph{The three bounds.} The lower bound is Hadamard's inequality (Section~\ref{sec:setting}). The entries $1+\Upsilon_{kk}=\chi_k/\nu_k$ are positive, so the inequality of the arithmetic and geometric means gives
\[
\prod_k(1+\Upsilon_{kk})^{1/M}\le1+\frac{\operatorname{tr}\bUps}{M},
\]
which with (a) is the middle bound. Since $\operatorname{tr}\bUps=\sum_i\upsilon_i$, the inequality $1+x\le e^x$ together with
\[
\prod_i(1+\upsilon_i)^{-1/M}=\exp\Bigl(-\frac1M\sum_i\log(1+\upsilon_i)\Bigr)
\]
gives the last form.

(c) \emph{The correlation bound.} With $\operatorname{diag}(\chi)$ the diagonal of the coefficient variances,
\[
\bU^\T\bR\bU=\operatorname{diag}(\chi)^{1/2}(\bI+\bX)\operatorname{diag}(\chi)^{1/2},
\]
so, over the eigenvalues $x_j$ of $\bX$,
\[
\frac{\prod_k\chi_k}{\det\bR}=\frac{1}{\det(\bI+\bX)}=\exp\Bigl(-\sum_j\log(1+x_j)\Bigr).
\] These are real, sum to zero and satisfy $|x_j|\le\norm{\bX}_2<1$. For $|x|<1$,
\[
-\log(1+x)=-x+\sum_{m\ge2}\frac{(-1)^mx^m}{m}\le-x+\frac12\sum_{m\ge2}|x|^m=-x+\frac{x^2}{2(1-|x|)}.
\]
Summing over $j$ with $\sum_jx_j=0$ and $\sum_jx_j^2=\norm{\bX}_F^2$ gives
\[
\log\frac{\prod_k\chi_k}{\det\bR}\le\frac{\norm{\bX}_F^2}{2(1-\norm{\bX}_2)},
\]
and the $M$th root gives~\eqref{eq:corrbound}.

(d) \emph{The separable block.} The eigenvalues of $\bR_{\mathrm v}\otimes\bR_{\mathrm h}$ are the products $\mu_i\mu'_j$, and the coefficient variances of $\bU_{\mathrm v}\otimes\bU_{\mathrm h}$ are the products $\chi_k\chi'_l$. The geometric mean of a product set is the product of the geometric means, so the ratio factors, and a bound on each factor multiplies.
\fi
\end{proof}

Theorem~\ref{thm:gain} is the quantitative form of the matched-group reading. The DCT-II diagonalizes the whole class, the window is at most $2\max(p,q)$ ranks away from it, and the coding gain the DCT-II loses is a function of those ranks alone: through the at most $2\max(p,q)$ nonzero eigenvalues of $\bUps$, or, with no eigendecomposition, through its trace, a closed form in the corner matrix, and through the coefficient correlations, which one transform of the covariance gives. Section~\ref{sec:images} applies (d) to the image block, and Section~\ref{sec:num} measures (a) to (c) on \cgNumWindows{} windows from $M=\cgSmallestM$ to $\cgLargestM$, where the identity holds to rounding and the bounds hold in every case, the correlation bound exceeding the ratio by at most \cgCorrGapMax{} where it applies. The proof uses only that $\bU$ is orthonormal and diagonalizes a positive definite $\bR_c$ with eigenvalues $\nu_k$, so (a) to (c) hold for any orthonormal basis and any such $\bR_c$, with the rank of $\bUps$ that of $\bR-\bR_c$: the theorem scores any fixed transform against the KLT through the departure from the class it diagonalizes, and Proposition~\ref{prop:integrated} applies it to the DST-VII.

The moving-average window is the case $p=0$ of Corollary~\ref{cor:rational}, and there the departure needs no inversion.

\begin{corollary}\label{cor:ma}
Let $\bR$ be the window covariance of an MA($q$) process, with autocovariances $r_0,\dots,r_q$ and $r_k=0$ for $k>q$, and $M>2q$. Then $\bR=\bT_r$ is the banded symmetric Toeplitz matrix of bandwidth $q$, the class member of Corollary~\ref{cor:rational} is $\bR_c(S)=\bT_r+\bH_r$ with the same sequence, and
\begin{equation}\label{eq:ma}
\bR-\bR_c(S)=-\bH_r,\qquad (\bH_r)_{ij}=r_{i+j+1}+r_{2M-1-i-j},
\end{equation}
the two Hankel corners, each an anti-triangle of rank $q$ when $r_q\ne0$, so the departure has rank exactly $2q$. The AR($p$) precision and the MA($q$) covariance with the same polynomial are thus the same object on the two sides: the banded Toeplitz matrix of the polynomial's autocorrelation, $\bT_c$ in Theorem~\ref{thm:departure} and $\bT_r$ here, plus a corner matrix.
\end{corollary}

\begin{proof}
\ifextended
For $M>q$ no alias of a lag below $M$ falls within $q$, so the periodized sequence of the proof of Corollary~\ref{cor:rational} is
\[
c_m=\begin{cases}r_m,&m\le q,\\ r_{2M-m},&m\ge2M-q,\\ 0,&\text{otherwise,}\end{cases}
\]
and~\eqref{eq:rational} loses its Toeplitz part. The two corners of the Hankel part occupy disjoint rows and columns since $M>2q$. The top-left corner $[r_{i+j+1}]_{i+j+1\le q}$ is a $q\times q$ matrix that is zero below its antidiagonal with $r_q$ on it, so its determinant is $\pm r_q^{\,q}$, and the mirror corner is its reflection. Each corner therefore has rank $q$ when $r_q\ne0$, and the departure has rank $2q$.
\fi
\end{proof}

The fourth class of models, the integrated process, the ARIMA($p,d,q$) model, is not stationary, and its window is matched by a different group. Let $\nabla$ be the $M\times M$ lower bidiagonal matrix with ones on the diagonal and $-1$ below it, the first difference with the sample before the window taken as zero, and let $\bUsev$ be the DST-VII basis, with columns proportional to $\sin(\pi(n+1)(k+\tfrac12)/(M+\tfrac12))$, $0\le n,k<M$, and grid $\omega_k=\pi(k+\tfrac12)/(M+\tfrac12)$, $k=0,\dots,M-1$.

\begin{proposition}\label{prop:integrated}
(a) [the DST-VII class] For an even sequence $c$ supported on $|m|\le M-1$, the matrix with entries $c_{i-j}-c_{i+j+2}+c_{2M-1-i-j}$ is the compression to the window of the symmetric circulant of size $4M+2$ with symbol $c$, acting on the extension of the window that is symmetric about $M-\tfrac12$ and antiperiodic over $2M+1$, hence antisymmetric about the sample $-1$. These matrices are the compression of the commutant of the dihedral group $D_{4M+2}$ to that subspace, which the commutant leaves invariant and the rotations do not, and $\bUsev$ diagonalizes them with eigenvalues $c(\omega_k)$.
(b) [integrated AR window, ARIMA($p,d,0$)] Let $\bm x$ be the window of the $d$-fold integral of the AR($p$) process of Section~\ref{sec:setting} started at zero, $\nabla^d\bm x=\bm y$ with $\bm y$ the AR($p$) window, so that $\bQx=(\nabla^\T)^d\bQ\nabla^d$ is its precision. Then $\sigma^2\bQx$ is banded with bandwidth $p+d$ and Toeplitz outside its first $p$ and last $p+d$ rows, with symbol $c(\omega)(2-2\cos\omega)^d$, and it differs from the member of the DST-VII class with that symbol by a matrix supported on the first $p+d-1$ and the last $p+d$ rows and columns, so of rank at most $2(p+d)-1$.
(c) [random walk] For $p=0$ and $d=1$ the departure vanishes: $\sigma^2\bQx=\nabla^\T\nabla$ is the class member with symbol $2-2\cos\omega$, so the KLT of the random-walk window started at zero is the DST-VII, with eigenvalues $\sigma^2/(4\sin^2(\pi(k+\tfrac12)/(2M+1)))$.
(d) [no reflection] $\bJ\bQx\bJ\ne\bQx$: the integrated window has no reflection symmetry, so Theorem~\ref{thm:exact}(a) and the fold of Section~\ref{sec:fast} do not apply to it.
\end{proposition}

\begin{proof}
\ifextended
(a) \emph{The extension.} Extend the window to the cycle $\Z_{4M+2}$ by
\[
x_{2M-1-n}=x_n,\qquad x_{n+2M+1}=-x_n.
\]
The extension of the unit vector $\be_n$ has entries $+1$ at $n$ and $2M-1-n$ and $-1$ at $n+2M+1$ and $4M-n$, so the extension operator $\bm E$ satisfies $\bm E^\T\bm E=4\bI$, and $\tfrac12\bm E$ has orthonormal columns. Its range, the subspace of extensions, consists of the sequences that the half-turn $n\mapsto n+2M+1$ negates and the reflection $n\mapsto2M-1-n$ about $M-\tfrac12$ fixes.

\emph{The compression.} The half-turn is central in the cyclic group, and every symmetric circulant $\bC$ commutes with it and with the reflection, so $\bC$ maps the subspace of extensions to itself. Its restriction there, in the basis $\tfrac12\bm E$, is $\tfrac14\bm E^\T\bC\bm E$, with entries read modulo $4M+2$
\[
\bigl(\tfrac14\bm E^\T\bC\bm E\bigr)_{ij}=c_{i-j}+c_{i+j+1-2M}-c_{i-j-2M-1}-c_{i+j-4M}.
\]
With $c$ supported on $|m|\le M-1$ the third term vanishes, the second is $c_{2M-1-i-j}$ because $c$ is even, and the fourth is $-c_{i+j+2}$ because $i+j-4M\equiv i+j+2$. This gives the matrices of the statement,
\[
\bigl(\tfrac14\bm E^\T\bC\bm E\bigr)_{ij}=c_{i-j}-c_{i+j+2}+c_{2M-1-i-j}.
\]
The symmetric circulants are the commutant of the dihedral group on the cycle, so these matrices are the compression of that commutant to the subspace.

\emph{The basis.} Let $\vartheta_k=2\pi(2k+1)/(4M+2)=\omega_k$. The sequence $\sin(\vartheta_k(n+1))$, $n\in\Z_{4M+2}$, is negated by the half-turn, since $\vartheta_k(2M+1)=\pi(2k+1)$, and fixed by the reflection, since $\vartheta_k(2M-n)=\pi(2k+1)-\vartheta_k(n+1)$. It therefore lies in the subspace of extensions. It is also a combination of the harmonics $e^{\pm i\vartheta_kn}$, which every symmetric circulant multiplies by $c(\vartheta_k)$. Restricted to the window it is
\[
\sin\bigl(\vartheta_k(n+1)\bigr)=\sin\Bigl(\frac{\pi(n+1)(k+\tfrac12)}{M+\tfrac12}\Bigr),
\]
the DST-VII column $k$, with eigenvalue $c(\omega_k)$. Distinct $k<M$ give distinct pairs of harmonics, so these sequences are orthogonal on the cycle, and since $\bm E^\T\bm E=4\bI$ their restrictions to the window are orthogonal as well. They are therefore $M$ orthogonal eigenvectors, a basis.

(b) With $A_{M,\cdot}=A_{\cdot,M}=0$,
\[
(\nabla^\T\bA\nabla)_{ij}=A_{ij}-A_{i+1,j}-A_{i,j+1}+A_{i+1,j+1}.
\]
So one differencing convolves the symbol of a Toeplitz interior with $|1-e^{i\omega}|^2=2-2\cos\omega$ and widens the band by one. It leaves the rows touched by the top-left corner block of Theorem~\ref{thm:departure} where they are, and moves the rows touched by the bottom-right block and by the truncation one row further from the end. Applying it $d$ times gives the band, the symbol, and corner blocks in the first $p$ and the last $p+d$ rows. The class member has the same interior and its own corners, $-c_{i+j+2}$ in the first $p+d-1$ rows and $c_{2M-1-i-j}$ in the last $p+d$ rows, which gives the support of the difference and the rank bound.

(c) $\nabla^\T\nabla$ is tridiagonal with $2$ on the diagonal except $1$ in the last entry, and $-1$ beside it. The class member with $c_0=2$, $c_1=-1$ has $-c_{i+j+2}=0$ throughout and $c_{2M-1-i-j}=-1$ only at $(M-1,M-1)$, so the two agree. The eigenvalues of the covariance are the reciprocals of $2-2\cos\omega_k=4\sin^2(\omega_k/2)$, times $\sigma^2$.

(d) The variance of $x_t$ grows with $t$, so the diagonal of $\bQx^{-1}$ is not symmetric under $\bJ$.
\fi
\end{proof}

The mechanism is the one that selects the cosine class for a positive dominant pole and the sine class for a negative one: integration from a fixed start selects the boundary pair of the DST-VII, antisymmetric at the pinned sample and symmetric at the free end, and the group of that extension supplies the basis. Matched means here what it means for the DCT-II and the autoregressive window: the class the window departs from by a matrix of low rank, exactly for the random walk and by rank at most $2(p+d)-1$ otherwise, not exact invariance. Part (c) is the transform obtained in~\cite{han2012jointly} for an intra-prediction residual as the correlation coefficient tends to one, here as the exact KLT of the random walk. Section~\ref{sec:num} measures the proposition at $M=16$ and $32$ for ten windows with $d=1,2,3$: the departure has rank $2(p+d-1)$ in all cases but one, where it is $\intRankOneThreeZero$ against $6$. The residual of the covariance in the DST-VII class is \intResOneOneZeroThirtyTwo{} for the integrated AR(1) window with pole $0.9$ at $M=32$. Scored against all sixteen trigonometric types, the DST-VII has the largest coding-gain ratio for every once-integrated AR window measured (\intDstSevenGainOneOneZeroNineThirtyTwo{} for that window at $M=32$, at least \intDstSevenGainMinDOneThirtyTwo{} over the five), while the DST-IV, the sine type of the other boundary pair, leaves a smaller residual (\intResDstFourOneOneZeroNineThirtyTwo{} against \intResOneOneZeroThirtyTwo), the residual weighing the largest eigenvalues and the coding gain the nulls. For $d\ge2$ none of the sixteen types is near the windows measured: at $M=32$ the smallest residual over all of them is \intMinResDTwoThirtyTwo{} and the largest coding-gain ratio \intMaxGainDTwoThirtyTwo, for the twice-integrated random walk, falling below \intMaxGainDTwoArThirtyTwo{} once a pole enters; the departure of the precision keeps the rank of (b), but the pinned start of order two makes the variance grow as $t^3$ and no fixed trigonometric basis is near the KLT at these lengths. On the covariance side the integrated ARMA($p,q$) window with $d=1$, the ARIMA($p,1,q$) model, differs from the DST-VII member with coefficient variances $S_y(\omega_k)/(2-2\cos\omega_k)$, $S_y$ the spectrum of the ARMA process, finite because the grid avoids $\omega=0$, by rank at most $2\max(p,q)$ in the four cases measured, and Theorem~\ref{thm:gain} applies with the DST-VII in place of the DCT-II: the DST-VII loses \intGainLossPctOneOneZeroNineThirtyTwo{} percent of the KLT coding gain on the integrated AR(1) window with pole $0.9$ at $M=32$ (ratio \intGainRatioOneOneZeroNineThirtyTwo, rank bound \intGainJensenOneOneZeroNineThirtyTwo), more than the DCT-II loses on the stationary window, because the two eigenvalues of $\bUps$ are larger. The fast form of the DST-VII is its own, through the odd-length transforms, and is not treated here. Table~\ref{tab:models} collects the four model classes.

\begin{table*}[t]
\caption{The four model classes: the matched class, where the departure from it lives, and the rank of the departure, with the condition under which the rank is attained where that is proved}
\label{tab:models}
\centering\footnotesize
\setlength{\tabcolsep}{5pt}
\begin{tabular}{@{}llp{4.2cm}p{4.5cm}@{}}
\toprule
Model & class & departure & rank\\
\midrule
AR($p$) & DCT-II (DST-II, negative pole) & corner blocks of $\bQ$, Thm.~\ref{thm:departure} & $2p$, exactly when $\phi_p\ne0$\\
MA($q$) & DCT-II & Hankel corners of $\bR$, Cor.~\ref{cor:ma} & $2q$, exactly when $r_q\ne0$\\
ARMA($p,q$), sums & DCT-II & aliased tail and corners, Cor.~\ref{cor:rational} & $\le2\max(p,q)$; $2p$ for $q\le p$ with uncancelled nonzero poles; $\le2(k_1+k_2)$ for a sum\\
ARIMA($p,d,0$), pinned start & DST-VII & end rows of $\bQx$, Prop.~\ref{prop:integrated} & $\le2(p+d)-1$, measured $2(p+d-1)$; $0$ for the random walk\\
ARIMA($p,1,q$), covariance side & DST-VII & measured & $\le2\max(p,q)$, measured\\
\bottomrule
\end{tabular}
\end{table*}

\section{Exact and Approximate Symmetry}\label{sec:groups}

\begin{theorem}\label{thm:exact}
(a) $\bJ\bR\bJ=\bR$ for every stationary process. (b) If $r_0,\dots,r_{M-1}$ are pairwise distinct, the permutations commuting with $\bR$ are the identity and the reflection, so the exact matched group is $\Z_2$, whatever $p$. (c) If $a(z)=b(z^m)$ for some $m\ge2$ dividing $M$, then $r_k=0$ unless $m$ divides $k$, the $m$ phase subsequences $(x_{mt+s})_t$ are mutually uncorrelated copies, identical at second order, of the window of length $M_b=M/m$ of the process with polynomial $b$, and after the polyphase permutation~\cite{bellanger1976digital,vaidyanathan1993multirate}, which lists the samples phase by phase ($x_s,x_{m+s},x_{2m+s},\dots$ for $s=0,\dots,m-1$), the permuted covariance becomes $\bI_m\otimes\bR_b$. The permutations that permute the phases and reflect any subset of them commute with $\bR$ and form the wreath product $\Z_2\wr S_m$, of order $2^mm!$. If the autocovariances of the $b$-process at lags $0,\dots,M_b-1$ are pairwise distinct and nonzero, the group is exactly $\Z_2\wr S_m$. Every eigenvalue of $\bR$ has multiplicity $m$ times its multiplicity in $\bR_b$, and a KLT of the window is an orthonormal mixing of the phases tensored with a KLT of $\bR_b$, with a separate mixing allowed in each eigenspace.
\end{theorem}

\begin{proof}
\ifextended
(a) A symmetric Toeplitz matrix is persymmetric.

(b) A permutation $\varpi$ commutes with $\bR$ exactly when
\[
r_{|\varpi(i)-\varpi(j)|}=r_{|i-j|}\qquad\text{for all }i,j,
\]
which with distinct autocovariances forces $|\varpi(i)-\varpi(j)|=|i-j|$. An isometry of $\{0,\dots,M-1\}$ for the absolute difference is $\varpi(n)=\varpi(0)\pm n$, and surjectivity leaves $\varpi(n)=n$ and $\varpi(n)=M-1-n$.

(c) \emph{The phases.} The recursion
\[
x_t=\sum_j\beta_jx_{t-mj}+\varepsilon_t,
\]
with $\beta_j$ the coefficients of $b$, connects only samples congruent modulo $m$, and the innovations are uncorrelated across phases. So the phases are mutually uncorrelated AR processes with the same polynomial $b$ and innovation variance. This gives $r_{mk}=r^{(b)}_k$ and $r_k=0$ otherwise, and $\bR=\bI_m\otimes\bR_b$ in phase-major order.

\emph{The wreath product.} A permutation acting as an element of $S_m$ on the phase index and as an independent reflection or identity on each phase preserves both the phase-equality pattern and the within-phase lags, so it commutes with $\bR$. These permutations compose as $\Z_2\wr S_m$.

\emph{Equality under the nondegeneracy conditions.} With nonzero within-phase autocovariances, $R_{ij}\ne0$ exactly when $i$ and $j$ share a phase, so a commuting permutation maps phases to phases. With distinct within-phase autocovariances it acts within each phase as an isometry of the lags, by the argument of (b), and so lies in $\Z_2\wr S_m$.

\emph{The spectrum.} The spectrum of $\bI_m\otimes\bR_b$ is that of $\bR_b$ with every multiplicity multiplied by $m$.
\fi
\end{proof}

The polyphase case is the composition rule for wreath products of~\cite{thornton2026unification} realized by a stationary process: the symmetric group mixes the phases and the shorter window supplies the multiplicity space. The group is not multiplicity-free, so the matched transform fixes the phase split and leaves the basis of $\bR_b$ to the shorter window, which is again an AR window, now of order $\deg b$, to which the rest of this paper applies. The dimension of the symmetric commutant drops from $M(M+1)/2$ to $\dimGenericEight$ for $\Z_2$ at $M=8$, and to $\dimPhaseTwo$ and $\dimPhaseFour$ at $M=8$ for the two- and four-phase wreath groups and $\dimPhaseThree$ at $M=9$ for the three-phase group of Table~\ref{tab:groups}.

The matched-group map of the window is therefore fixed by the poles, not by the order. The exact group is $\Z_2$ for distinct autocovariances. If $a(z)=b(z^m)$, the exact group contains $\Z_2\wr S_m$, and under the nondegeneracy conditions of Theorem~\ref{thm:exact}(c) it equals $\Z_2\wr S_m$. The window then separates into $m$ shorter windows, and the matched transform of the wreath product is the polyphase split, the polyphase permutation followed by a transform of length $M/m$ on each phase. With the DCT-II or the DST-II of length $M/m$ on each phase it is the polyphase DCT-II or DST-II of Section~\ref{sec:num}, and we call a window with $a(z)=b(z^m)$ a polyphase pattern. The approximate group is the dihedral group of the even extension, in its cosine representation for a dominant positive real pole and its sine representation for a dominant negative one (Propositions~\ref{prop:nyquist} and~\ref{prop:limits} below), at a distance given by the corner matrix of Theorem~\ref{thm:departure}, for every order. The order sets the rank of the departure and the regime of the poles, nothing else. A resonant pair is matched by the cyclic group on the window when it lies on the window's grid (Proposition~\ref{prop:limits}(c)) and otherwise, when the values $\cos(\omega_0k)$ are distinct, by the reflection alone (Theorem~\ref{thm:exact}(b)). The modulated representations hold its limit, and the resonance frequency is the continuous parameter of a discovery procedure on the complex window, as measured at the end of this section.

\begin{proposition}\label{prop:nyquist}
Let $\bR'$ be the window covariance of the process with polynomial $a(-z)$, the poles negated. Then $\bR'=\bN\bR\bN$, and $\bN u_k=s_{M-1-k}$ where $s_k(n)=\gamma'_k\sqrt{2/M}\sin(\pi(k+1)(n+\tfrac12)/M)$, with $\gamma'_{M-1}=1/\sqrt2$ and $\gamma'_k=1$ otherwise, is the DST-II basis. Hence the DST-II residual and coding-gain ratio of $\bR'$ equal the DCT-II residual and coding-gain ratio of $\bR$, and the representation $g\mapsto\bN\rep(g)\bN$ of the dihedral group, a signed permutation representation equivalent to $\rep$, has the DST-II class as its commutant.
\end{proposition}

\begin{proof}
\ifextended
$a(-z)$ has autocovariance $r'_k=(-1)^kr_k$, so $R'_{ij}=(-1)^{i-j}R_{ij}$, that is, $\bR'=\bN\bR\bN$. For the basis, since $\cos(\pi(n+\tfrac12))=0$,
\[
(-1)^n\cos\Bigl(\frac{\pi k(n+\tfrac12)}{M}\Bigr)=\sin\bigl(\pi(n+\tfrac12)\bigr)\cos\Bigl(\frac{\pi k(n+\tfrac12)}{M}\Bigr)=\sin\Bigl(\frac{\pi(M-k)(n+\tfrac12)}{M}\Bigr).
\]
Conjugation by $\bN$ is unitary and preserves diagonals, which gives the residual and coding-gain statements, and it conjugates the commutant of $\rep$ onto the commutant of $\bN\rep\bN$.
\fi
\end{proof}

\begin{proposition}\label{prop:limits}
Let $a(z)=f(z)\varphi(z)$ with $\varphi$ fixed and stable. (a) If $f(z)=1-\rho z$ and $\rho\to1$, then $r_k/r_0\to1$ for every fixed $k$, $\bR/r_0\to\one\one^\T$, which lies in the DCT-II class, and the DCT-II residual tends to zero. (b) If $\rho\to-1$, then $\bR/r_0\to\bN\one\one^\T\bN$, which lies in the DST-II class. (c) If $f(z)=(1-\rho e^{i\omega_0}z)(1-\rho e^{-i\omega_0}z)$, $0<\omega_0<\pi$, and $\rho\to1$, then $r_k/r_0\to\cos(\omega_0k)$ and $\bR/r_0\to\bC(\omega_0)$ with entries $\cos(\omega_0(i-j))$, a matrix of rank two whose range is spanned by $(\cos\omega_0n)$ and $(\sin\omega_0n)$, on which the translations act by rotation. $\bC(\omega_0)$ is a symmetric circulant, hence in the commutant of the dihedral group of the $M$-cycle, with the Hartley basis as a matched transform, if and only if $\omega_0\in(2\pi/M)\Z$.
\end{proposition}

\begin{proof}
\ifextended
Write $S_\varphi(\omega)=\sigma^2/|\varphi(e^{i\omega})|^2$, continuous and even.

(a) $(1-\rho^2)/|1-\rho e^{i\omega}|^2$ is the Poisson kernel, an approximate identity at $0$, so for every $k$
\[
(1-\rho^2)\,r_k=\int_{-\pi}^{\pi}\frac{1-\rho^2}{|1-\rho e^{i\omega}|^2}\,S_\varphi(\omega)\,e^{ik\omega}\,\frac{d\omega}{2\pi}\ \longrightarrow\ S_\varphi(0).
\]

(b) follows from (a) and Proposition~\ref{prop:nyquist}.

(c) The kernel is now centered at $\pm\omega_0$, with the other factor continuous there, so
\[
(1-\rho^2)\,r_k\ \longrightarrow\ \frac{S_\varphi(\omega_0)\cos(k\omega_0)}{2\sin^2\omega_0}.
\]
The identity
\[
\cos(\omega_0(i-j))=\cos\omega_0i\,\cos\omega_0j+\sin\omega_0i\,\sin\omega_0j
\]
gives the rank and the range, and translation by one sample rotates that plane by $\omega_0$. A Toeplitz matrix with entries $\cos(\omega_0(i-j))$ is circulant exactly when $\cos(\omega_0(k-M))=\cos(\omega_0k)$ for $k=1,\dots,M-1$, which for $M\ge3$ and $0<\omega_0<\pi$ holds exactly when $\omega_0M\in2\pi\Z$.
\fi
\end{proof}

Proposition~\ref{prop:limits} is the content of the AR(1)-DCT relation and its extension. For $p=1$ the selection by the sign of the pole is in Jain's ordering of the distances from the window precision to his boundary-modified matrices, nearest to the DCT-II form for $\rho>0.5$, to the DST-I form for $|\rho|\le0.5$ and to the DST-II form for $\rho<-0.5$~\cite{jain1979sinusoidal}, and in the observation that negating the pole exchanges the two parity branches of the factorization in~\cite{reznik2026exact}; Proposition~\ref{prop:nyquist} is the statement for every order, as a conjugation of the representation. The relation does not depend on the order: any window whose dominant pole is real and positive tends to the rank-one member of the cosine class, any window whose dominant pole is real and negative tends to the rank-one member of the sine class, and the rate is set by the pole, with the corner departure~\eqref{eq:departure} measuring the finite-radius distance. For $p=1$ the departure itself vanishes as $\phi_1\to1$. For $p\ge2$ it does not (its relative Frobenius norm at radius $0.999$ is $\departRealPosNNNEight$ for the poles $0.999$ and $0.5$ at $M=8$), and only the rank-one limit survives. A resonant pair is matched by the cyclic group on the $M$-cycle when it sits on the window's grid, which is the statement that a sinusoid of period dividing $M$ is a circulant process. Off the grid the rotation representation of the translations at $\omega_0$ has no realization by a permutation of the window, and the extension classes, whose frequencies lie on the grid $\pi k/M$ of the doubled window, hold one quadrature of an on-grid resonance each and neither quadrature off the grid.

\emph{The modulation family.} Conjugating the dihedral representation by $\bm E_\omega=\operatorname{diag}(e^{i\omega n})$ gives a family of equivalent representations whose commutants are the modulated classes $\bm E_\omega(\bT_c+\bH_c)\bm E_\omega^*$; $\omega=0$ is the cosine class and $\omega=\pi$ the sine class of Proposition~\ref{prop:nyquist}, the two members with real signed-permutation realizations. For complex windows the conjugation is exact.

\begin{proposition}\label{prop:modulated}
Let the process be complex with polynomial $a(z)=a_0(e^{i\omega}z)$, $a_0$ real and stable: its poles are those of the real process with polynomial $a_0$ rotated by $\omega$. Then its window covariance is $\bR=\bm E_\omega\bR_0\bm E_\omega^*$ with $\bR_0$ the real window, and every complex first-order window, with pole $\rho e^{i\omega}$, is of this form. The conjugated analogues of Theorems~\ref{thm:departure} to~\ref{thm:exact}, Corollaries~\ref{cor:identity} and~\ref{cor:rational} and the fast form of Section~\ref{sec:fast} hold for $\bR$, with every group operator $\rep(g)$ replaced by $\bm E_\omega\rep(g)\bm E_\omega^*$, the DCT-II class by the modulated class $\bm E_\omega(\bT_c+\bH_c)\bm E_\omega^*$, which is the commutant of the conjugated representation, the DCT-II by the modulated DCT-II $\bm E_\omega\bU$, its matched transform, every matrix $\bm X$ of a statement on $\bR_0$ by $\bm E_\omega\bm X\bm E_\omega^*$, transposes by conjugate transposes, and the window $\bm x$ by the demodulated window $\bm E_\omega^*\bm x$. For instance $\bJ\bR\bJ=\bR$ fails for the complex window, and $\bJ_\omega\bR\bJ_\omega^*=\bR$ holds with $\bJ_\omega=\bm E_\omega\bJ\bm E_\omega^*$.
\end{proposition}

\begin{proof}
\ifextended
Let $y$ be the real process with polynomial $a_0$ and white innovations. Then $x_t=e^{i\omega t}y_t$ satisfies
\[
\sum_\ell a_\ell x_{t-\ell}=e^{i\omega t}\sum_\ell a_{0,\ell}\,y_{t-\ell},\qquad a_\ell=a_{0,\ell}\,e^{i\omega\ell},
\]
with $a_\ell$ the coefficients of $a_0(e^{i\omega}z)$, and its innovations $e^{i\omega t}\varepsilon_t$ are white. So $x$ is the complex process of the statement, with
\[
r_x(k)=e^{i\omega k}r_y(k),\qquad \bR=\bm E_\omega\bR_0\bm E_\omega^*.
\]
A single pole $\rho e^{i\omega}$ is the rotation of the real pole $\rho$. Conjugation by the unitary $\bm E_\omega$ carries every statement about $\bR_0$, its precision, its class member, its eigenvectors and its parity split to $\bR$, and carries the commutant of $\rep$ to the commutant of the conjugated representation. For the reflection,
\[
\bJ_\omega\bR\bJ_\omega^*=\bm E_\omega\bJ\bR_0\bJ\bm E_\omega^*=\bR.
\]
\fi
\end{proof}

For real windows the natural member at a resonance is the two-quadrature class spanned by $\cos(\omega_0n)$ and $\sin(\omega_0n)$ times half-length envelopes, which contains $\bC(\omega_0)$ for every $\omega_0$. Section~\ref{sec:num} measures it: it holds the limit (residual $\limQuadEightOff$ at $M=8$ where every class of the catalog leaves at least $\limMinPlainEightOff$) but recovers little coding gain, because the envelopes of the KLT have frequencies set by the corner blocks and a fixed envelope set does not reproduce the nulls that the coding gain rewards.

\emph{Discovery from samples.} The groups of this section are read from sample covariances, which makes the matched-group map a computation on data and not only on the model. For this sampling calculation take the windows Gaussian, that is, the innovations Gaussian. For other innovations the floor below acquires a fourth-cumulant term, not treated here. Let $\hat\bR$ be the covariance of $N$ windows drawn from $\bR$. For a permutation matrix $\bP$ the commutator energy $\lVert[\hat\bR,\bP]\rVert_F^2$ is the quadratic form of the double-commutator pencil~\cite{thornton2026double_commutator} on the single candidate $\bP$, and when $\bP$ commutes with $\bR$ its expectation is the Gaussian sampling floor $\mathcal{N}(\bP)=\tfrac2N[(\operatorname{tr}\bR)^2-\operatorname{tr}(\bR\bP)^2+\operatorname{tr}\bR^2-\operatorname{tr}(\bR\bP\bR\bP)]$, by Isserlis' theorem. The statistic $\lVert[\hat\bR,\bP]\rVert_F^2/\mathcal{N}(\bP)$ has expectation one on the group when the population floor is used. With $\hat\bR$ in place of $\bR$ in the floor, which is what the data allow, the plug-in statistic is calibrated to one up to a relative error of order $N^{-1/2}$, provided $\mathcal{N}(\bP)>0$, the floor being a smooth function of $\hat\bR$. Off the group it grows as $N/N^\star(\bP)$, where $N^\star(\bP)=N\mathcal{N}(\bP)/\lVert[\bR,\bP]\rVert_F^2$, the resolving sample size, does not depend on $N$. Sorting the statistics of the $M!-1$ nontrivial permutations and reading the group as the identity with the permutations below the largest ratio between consecutive statistics, statistics below one being taken as one, returns the exact group of Theorem~\ref{thm:exact} from $N=\discAllExactFromN$ windows in every trial of Section~\ref{sec:num} for all \discNumPatterns{} patterns measured, the wreath groups of orders \discOrderPhaseTwo, \discOrderPhaseThree{} and \discOrderPhaseFour{} included: the interleaving of a window is found from its data, since the wreath product is a permutation group and the search runs over permutations. The continuous symmetry of an interleaved window, the orthogonal mixing of the phases of Theorem~\ref{thm:exact}(c), is the stabilizer $O(m)^{M_b}$, one orthogonal group for each eigenspace of $\bR_b$, with Lie algebra of dimension $M_bm(m-1)/2$: the pencil over skew-symmetric candidates has a kernel of that dimension (\skewDimPhaseTwo, \skewDimPhaseThree{} and \skewDimPhaseFour{} for the three interleaved patterns, $0$ for the generic ones) and its calibrated form counts it from $N=5000$ windows. For a family of interleaved windows with different inner poles the common kernel is $\mathfrak{so}(m)\otimes\operatorname{span}\{\bI,\bJ_b\}$, with $\bJ_b$ the exchange matrix of length $M_b$, the phase mixing composed with the inner reflection, of dimension $m(m-1)$, and the common permutation group is the wreath product. Finally, the carrier of a complex window with rotated poles (Proposition~\ref{prop:modulated}) is the continuous parameter of the discovery: the residual of the modulated class $\bm E_\omega(\bT_c+\bH_c)\bm E_\omega^*$ as a function of $\omega$ is minimized at the carrier, with the residual of the demodulated real window in the DCT-II class (\carResEight{} at $M=8$ and \carResThirtyTwo{} at $M=32$ for the pole $0.9e^{i\carWzero\pi}$), and from $N$ sample windows the minimizer has root-mean-square error $\carRmseFirstEight\pi$ at $N=\carNfirst$ and $\carRmseLastEight\pi$ at $N=\carNlast$ ($M=8$). For a real window the same family is minimized at $\omega=0$, because the real window carries the untwisted reflection exactly, and the resonance of a real window is located only by the quadrature class, which is not a commutant. The transform found on the complex window is the modulated DCT-II, which is to that window what the DCT-II is to the real window. The exact basis remains the data-dependent one of Theorem~\ref{thm:levels}.

\begin{table*}[t]
\caption{Residual (energy outside the class) / coding-gain ratio to the KLT at $M=8$, by pole pattern and class; the last column is the polyphase construction of the wreath group or the quadrature class at the resonance}
\label{tab:classes}
\centering\footnotesize
\setlength{\tabcolsep}{5pt}
\begin{tabular}{@{}lcccccl@{}}
\toprule
Pattern & DCT-I & DCT-II & DST-I & DST-II & cyclic (Hartley) & polyphase or quadrature\\
\midrule
\input{class_table.tex} 
\bottomrule
\end{tabular}
\end{table*}

\begin{figure*}[!t]
\centering
\includegraphics[width=0.9\textwidth]{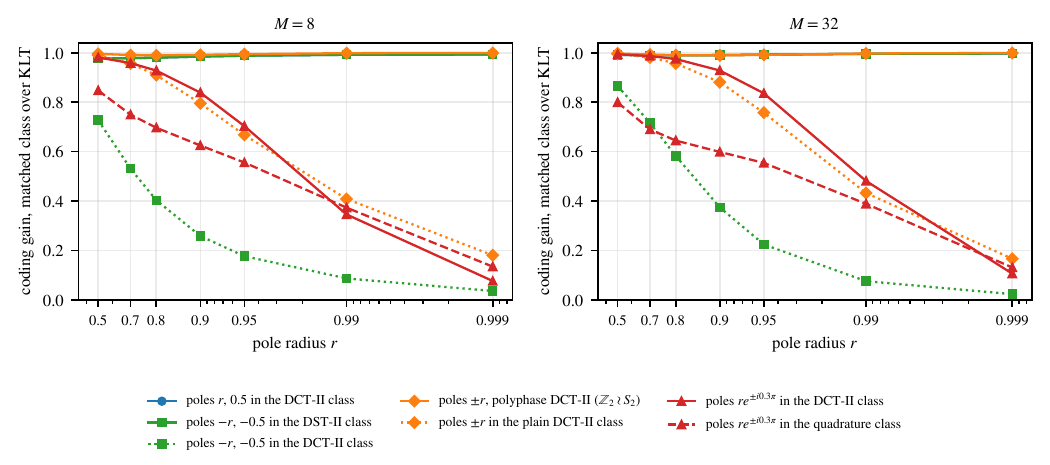}
\caption{Coding gain of each family's matched class relative to the KLT against the pole radius, at $M=8$ and $M=32$: real poles in the DCT-II class, negated poles in the DST-II class (and, dotted, in the DCT-II class), the two-phase family under its wreath group (polyphase DCT-II) and in the plain DCT-II class, and the resonant family in the DCT-II and the quadrature classes.}
\label{fig:classes}
\end{figure*}

\section{The Cosine Transform for Images}\label{sec:images}

The chain that attaches the DCT-II to image and video coding has one modeling step and several proved steps. The modeling step is that the rows and columns of a natural image block are stationary first-order Markov windows with a pole near $0.95$, between about $0.9$ and $0.98$ in the measurements collected in~\cite{clarke1985transform,jain1989fundamentals,rao1990discrete}. The model is separable and its support is empirical, and the review of~\cite{netravali1980picture} already notes that the Markov conclusions hold for the model and not for real images at small block sizes. No theorem replaces this step. The proved steps are classical for $p=1$: the KLT is a basis of single sinusoids whose frequencies move onto the DCT-II grid as the pole rises (Theorem~\ref{thm:levels}(c), from~\cite{ray1970further,clarke1981relation,flickner1982derivation}), the covariance is a rank-two perturbation of a member of the DCT-II class with the perturbation $\rho(1-\rho)$ on the two end samples (Remark~\ref{rem:firstorder}, from~\cite{jain1979sinusoidal,reznik2026exact}), and the normalized covariance tends to the rank-one member of that class (Proposition~\ref{prop:limits}(a)). The member $\bR_c$ of Corollary~\ref{cor:identity} is the covariance of the boundary-modified Markov process for which the DCT-II is the exact KLT in~\cite{jain1979sinusoidal}, so the corner matrix $\bD$ is the exact precision-domain departure of the stationary window from that process, and for $p=1$ the exact transform is the DCT-II followed by the corrections of~\cite{reznik2026exact}.

What the group adds is the reason and the order. Section~\ref{sec:klt} identifies the DCT-II class as the commutant of the dihedral group of the even extension, which is why one fixed basis diagonalizes the whole class and not only the limit; Theorem~\ref{thm:departure} gives the departure of a window of any order as a corner matrix of rank at most $2p$ in closed form, Corollary~\ref{cor:identity} the interlacing at every length, and Section~\ref{sec:fast} the exact transform of order $p$ as the DCT-II cores plus a correction of rank at most $p$ per parity branch, equal to $p$ for a minimal AR($p$) model, one with $\phi_p\ne0$. For a separable block the covariance is $\bR_{\mathrm v}\otimes\bR_{\mathrm h}$, the two-dimensional DCT-II is the matched transform of $D_{2M_1}\times D_{2M_2}$ acting on the doubly even extension, the departure of the precision from the class is supported on the rows and columns of the $p$ border pixels on each side of the block, and the departure of the covariance has rank at most $2p(M_1+M_2)-4p^2$, which is $2(M_1+M_2)-4$ for the Markov-1 model. For an $8\times8$ block with pole $0.95$ the two-dimensional DCT-II reaches \sepGainNineFive{} of the KLT coding gain (\sepDctDbNineFive{} dB against \sepKltDbNineFive{} dB) with residual \sepResNineFive{} and departure of rank \sepRank. At pole $0.9$ the ratio is \sepGainNine. Theorem~\ref{thm:gain}(d) bounds the loss without computing the KLT: the coding-gain ratio of the block is the square of the one-dimensional ratio, $\cgSepRatioNineFive$ at pole $0.95$, and the square of the correlation bound~\eqref{eq:corrbound}, computed from the DCT-II coefficient correlations of the row model alone, is $\cgSepCorrNineFive$. The two-dimensional DCT-II is therefore within \cgSepCorrLossPctNineFive{} percent, \cgSepCorrDbNineFive{} dB, of the KLT coding gain of the $8\times8$ Markov-1 block at pole $0.95$, with \cgSepLossPctNineFive{} percent measured, and within \cgSepCorrLossPctNine{} percent at pole $0.9$. The rank bound~\eqref{eq:gainbounds} gives $\cgSepJensenNineFive$ at this length, loose because the trace of $\bUps$, $-2\rho/(1+\rho)$, is spread over eight coefficients, and it tightens as $M$ grows (Section~\ref{sec:num}). This is the statement the chain lacked: with the Markov model granted, the loss of the DCT-II against the optimal transform of the block is bounded by a quantity computed from the model with no eigendecomposition, and the group reading says why the bound is small: one fixed basis diagonalizes the whole class, and every row and column of the block is two ranks away from it. For an $8\times8$ block whose rows and columns are AR(2) windows with poles $0.9$ and $0.5$ the ratio is \sepTwoGainRealPos{} (\sepTwoDctDbRealPos{} dB against \sepTwoKltDbRealPos{} dB) with departure of rank \sepTwoRankRealPos{} against the bound \sepTwoRankBound, and with poles $0.9$ and $-0.5$ it is \sepTwoGainMixed. For the resonant AR(2) block, poles $0.9e^{\pm i0.3\pi}$, the ratio falls to \sepTwoGainRes{} (\sepTwoDctDbRes{} dB against \sepTwoKltDbRes{} dB) with residual \sepTwoResRes, which is the oriented-texture case where the dihedral group is the wrong match and the resonant classes of Section~\ref{sec:groups} apply. The same mechanism, a boundary condition selecting the type, is the one by which an intra-predicted residual is matched by a sine type rather than a cosine type~\cite{han2012jointly,sullivan2012overview}, and by which a negative dominant pole selects the DST-II here (Proposition~\ref{prop:nyquist}). What the finite-window results add to the classical chain~\cite{clarke1981relation,flickner1982derivation,jain1979sinusoidal,yemini1979asymptotic,unser1984approximation,reznik2026exact} is therefore the closed-form departure at every order, the interlacing at every window length and order, the bound on the coding-gain loss, the order-$p$ transform in fast form, and the group that explains why the class, and not only the limit, is diagonalized.

\section{A Fast Transform of Any Order from the Two Groups}\label{sec:fast}

The matched transform of a group that is not multiplicity-free fixes a block structure and leaves the basis inside each block to the data~\cite{thornton2026unification}. For the window this happens twice, and both stages that it fixes are butterflies, the two-point sum and difference $(x,y)\mapsto(x+y,x-y)/\sqrt2$ of the fast Fourier transform: the exact group $\Z_2$ of Theorem~\ref{thm:exact}(a) splits the window into its even and odd parts by one two-point butterfly per mirrored pair of samples, and inside each part the dihedral closure, matched only approximately, supplies the half-length cosine cores, the DCT-II and DCT-IV of length $M/2$ that sit between the fold and the correction, which are built from the same butterfly together with plane rotations. What the two groups leave to the data is a correction of rank at most $p$ per part, equal to $p$ for a minimal AR($p$) model, and Section~\ref{sec:cascade} writes it as a cascade of first-order stages, one per unit of rank, so that the order enters the signal flow only as the number of stages. This section derives the three parts in turn, for AR($p$) windows, for rational spectra and for co-channel sums, draws the result as a flow graph parametrized by the order, and counts its operations. Throughout, $M=2L$ is even (the odd case is Proposition~\ref{prop:fast}(f)); $\bF$ is the orthonormal fold with rows $(\be_n+\be_{M-1-n})/\sqrt2$ for $0\le n<L$ followed by $(\be_n-\be_{M-1-n})/\sqrt2$; $\bK_{\mathrm{II}}$ and $\bK_{\mathrm{IV}}$ are the DCT-II and DCT-IV matrices of length $L$ with the basis vectors as rows, $\bK_{\mathrm{II}}[j,n]=\gamma_j\sqrt{2/L}\cos(\pi j(n+\tfrac12)/L)$ and $\bK_{\mathrm{IV}}[j,n]=\sqrt{2/L}\cos(\pi(j+\tfrac12)(n+\tfrac12)/L)$; and $\bD_1$ is the top-left $p\times p$ block of $\bD$, given by~\eqref{eq:departure}.

\subsection{The Butterfly from the Exact Group and the Cores from Its Closure}\label{sec:butterfly}

The reflection $\bJ$ generates $\Z_2$ acting on the window by permuting its samples. Its orbits are the $L$ mirrored pairs $\{n,M-1-n\}$, so the permutation representation is the direct sum of $L$ copies of the regular representation of $\Z_2$, and the matched transform of the regular representation is the character table of the group, the two-point transform $\frac1{\sqrt2}\bigl[\begin{smallmatrix}1&1\\1&-1\end{smallmatrix}\bigr]$: a butterfly. The matched transform of the exact group is therefore $\bF$, one butterfly between the samples $n$ and $M-1-n$ for each pair, and nothing more, because each of the two characters occurs $L$ times and the group fixes only the split into its two isotypic components, the even and the odd part. This stage is the same for every window of this paper, AR($p$), ARMA($p,q$) or a co-channel sum: by Theorem~\ref{thm:exact}(a) every stationary window is persymmetric, and the group does not depend on the order. Inside each component the dihedral closure of Section~\ref{sec:groups}, the group of the even extension, is matched only approximately. Its matched transform restricted to the two components gives the two cores, and the cores are built from the same butterfly together with plane rotations: the DCT-II class member of length $L$ is persymmetric, so the reflection of the half window folds it once more into a DCT-II and a DCT-IV core of length $L/2$, and so on down the chain of nested reflections, which is the group reading of the recursion of~\cite{chen1977fast}; the DCT-IV core does not fold, since the reflection maps its cosine kernel onto a sine kernel, and is realized through the cyclic group as a Fourier transform of length $L/2$ between two diagonal stages of twiddle factors~\cite{wang1984fast,rao1990discrete}. Proposition~\ref{prop:fast} states the result for the AR($p$) window, where the corner block gives the correction in closed form, and Proposition~\ref{prop:rational} for rational spectra, where the folded departure gives it.

\begin{proposition}\label{prop:fast}
Let $M=2L\ge2p$. (a) [exact stage] $\bF\bQ\bF^\T=\bQ_{\mathrm s}\oplus\bQ_{\mathrm a}$, the two blocks of size $L$ being the restrictions of $\bQ$ to the even and odd isotypic components of $\Z_2$. The KLT is $\bF^\T(\bU_{\mathrm s}\oplus\bU_{\mathrm a})$ for any eigenbases $\bU_{\mathrm s}$, $\bU_{\mathrm a}$ of the two blocks. (b) [cores] The class member splits the same way, $\bF(\bT_c+\bH_c)\bF^\T=\bGam_{\mathrm s}\oplus\bGam_{\mathrm a}$, and $\bK_{\mathrm{II}}\bGam_{\mathrm s}\bK_{\mathrm{II}}^\T=\bLam_{\mathrm{II}}=\operatorname{diag}(c(\pi j/L))$, $\bK_{\mathrm{IV}}\bGam_{\mathrm a}\bK_{\mathrm{IV}}^\T=\bLam_{\mathrm{IV}}=\operatorname{diag}(c(\pi(2j+1)/2L))$, $j=0,\dots,L-1$. The rows of $\bK_{\mathrm{II}}$ and $\bK_{\mathrm{IV}}$ are $\sqrt2$ times the even- and odd-indexed columns of the DCT-II of length $M$ restricted to $n<L$: the two cores are the matched transform of the dihedral group restricted to the two isotypic components of its subgroup $\Z_2$. (c) [corner blocks] $\sigma^2\bQ_{\mathrm s}=\bGam_{\mathrm s}-\bDel_1$ and $\sigma^2\bQ_{\mathrm a}=\bGam_{\mathrm a}-\bDel_1$, where $\bDel_1$ is the $L\times L$ matrix with $\bD_1$ in its first $p$ rows and columns and zeros elsewhere. (d) [correction] With $\bW_{\mathrm s}=\bK_{\mathrm{II}}[:,0{:}p]$ and $\bW_{\mathrm a}=\bK_{\mathrm{IV}}[:,0{:}p]$, the $L\times p$ matrices of the first $p$ coordinates of the core vectors,
\begin{equation}\label{eq:branch}
\bK_{\mathrm{II}}\,\sigma^2\bQ_{\mathrm s}\bK_{\mathrm{II}}^\T=\bLam_{\mathrm{II}}-\bW_{\mathrm s}\bD_1\bW_{\mathrm s}^\T,
\end{equation}
and likewise for the odd branch with $\bK_{\mathrm{IV}}$, $\bLam_{\mathrm{IV}}$, $\bW_{\mathrm a}$. Each branch is a symmetric perturbation of rank at most $p$ of a diagonal matrix. Its eigenvalues $\lambda$ that are not diagonal entries are the roots of the $p\times p$ secular determinant, the equation that locates the eigenvalues of a diagonal matrix plus a low-rank perturbation,
\begin{equation}\label{eq:secular}
\det\bigl(\bI_p-\bD_1\bW^\T(\bLam-\lambda\bI)^{-1}\bW\bigr)=0,
\end{equation}
and the eigenvector at such a root is $\bv=(\bLam-\lambda\bI)^{-1}\bW\bnu$ with $\bnu$ in the null space of the matrix in~\eqref{eq:secular}, that is, $v_j=\bm w_j^\T\bnu/(\Lambda_{jj}-\lambda)$ with $\bm w_j^\T$ the $j$th row of $\bW$: a sum of $p$ terms of Cauchy type, each a constant over the difference $\Lambda_{jj}-\lambda$ between a diagonal entry and the eigenvalue, the form of the entries of a Cauchy matrix. The KLT eigenvalue is $\mu=\sigma^2/\lambda$. Let $(n_+,n_-)$ be the inertia of $\bD_1$, its numbers of positive and negative eigenvalues, $n_++n_-\le p$, and let $\Lambda_1\le\dots\le\Lambda_L$ and $\lambda_1\le\dots\le\lambda_L$ be the core grid and the branch eigenvalues in increasing order. Then
\begin{equation}\label{eq:inertia}
\Lambda_{i-n_+}\le\lambda_i\le\Lambda_{i+n_-},\qquad\sum_{i=1}^{L}(\lambda_i-\Lambda_i)=-\operatorname{tr}(\bD_1\bW^\T\bW),
\end{equation}
indices outside $1,\dots,L$ dropped: within each branch the eigenvalues move at most $n_-$ places up and $n_+$ places down the core grid, at most $p$ places in all and jointly at most $2p$, as in Corollary~\ref{cor:identity}, and the total displacement is a closed form in the corner block. For $p=1$ and $0<\rho<1$, $\bD_1=-\rho(1-\rho)<0$, so $\Lambda_i\le\lambda_i\le\Lambda_{i+1}$: every branch eigenvalue of the precision moves up by less than one grid step, and for $-1<\rho<0$ down. (e) [polyphase stage] If $a(z)=b(z^m)$ with $m$ dividing $M$, the polyphase permutation of Theorem~\ref{thm:exact}(c) precedes the fold and reduces the transform to $m$ copies of the transform of length $M/m$ and order $p/m$. (f) [odd length] For $M=2L+1\ge2p+1$ the center sample is fixed by the reflection and joins the even component, of dimension $L+1$. With $\bF$ the fold that keeps the center sample unscaled, (a) to (d) hold with the even core of length $L+1$ equal to $\sqrt2$ times the even-indexed DCT-II columns of length $M$ restricted to $n\le L$, the center row divided by $\sqrt2$, which is the DCT-VI kernel $\cos(\pi(n+\tfrac12)j/(L+\tfrac12))$ with that weighting, and the odd core of length $L$ equal to $\sqrt2$ times the odd-indexed columns restricted to $n<L$, which is the DST-VII kernel $\sin(\pi(n+1)(j+\tfrac12)/(L+\tfrac12))$ up to the reversal of the index and signs, the parity split of the odd-length DCT-II of~\cite{reznik2013relationship}. The grids are $c(2\pi j/M)$ and $c(\pi(2j+1)/M)$, and the corrections are the same corner block $\bD_1$ in the first $p$ coordinates of each branch, of rank at most $p$, equal to $p$ for a minimal AR($p$) model.
\end{proposition}

\begin{proof}
\ifextended
(a) By Theorem~\ref{thm:exact}(a), $\bJ\bQ\bJ=\bQ$, so $\bQ$ commutes with the reflection and preserves its $+1$ and $-1$ eigenspaces, which the two row groups of $\bF$ span orthonormally. This is the isotypic decomposition of the permutation representation of $\Z_2$, with multiplicity $L$ in each component, so the group fixes the split and nothing inside it.

(b) $\bT_c+\bH_c$ is persymmetric as well. The DCT-II column $u_k$ of length $M$ satisfies
\[
u_k(M-1-n)=(-1)^ku_k(n),\qquad\text{since}\qquad \cos\Bigl(\frac{\pi k(M-n-\tfrac12)}{M}\Bigr)=\cos\Bigl(\pi k-\frac{\pi k(n+\tfrac12)}{M}\Bigr),
\]
so the even-indexed columns lie in the even component and the odd-indexed ones in the odd component. Restricted to $n<L$ and scaled by $\sqrt2$ they are orthonormal, and
\[
\sqrt2\,u_{2j}(n)=\gamma_j\sqrt{\tfrac2L}\cos\Bigl(\frac{\pi j(n+\tfrac12)}{L}\Bigr),\qquad
\sqrt2\,u_{2j+1}(n)=\sqrt{\tfrac2L}\cos\Bigl(\frac{\pi(j+\tfrac12)(n+\tfrac12)}{L}\Bigr)
\]
are the DCT-II and DCT-IV kernels of length $L$. Their eigenvalues under $\bT_c+\bH_c$ are $c(2\pi j/M)$ and $c(\pi(2j+1)/M)$.

(c) $\bD$ is supported on the top-left block $\bD_1$ and on the bottom-right block $\bJ_p\bD_1\bJ_p$, its mirror, which do not overlap when $M\ge2p$. For rows and columns $n,n'<p$ of either parity group the fold gives
\[
\tfrac12\bigl(D_{nn'}+D_{M-1-n,M-1-n'}\bigr)\pm\tfrac12\bigl(D_{n,M-1-n'}+D_{M-1-n,n'}\bigr).
\]
The first bracket is $2(\bD_1)_{nn'}$ by the mirror symmetry and the second vanishes, so each diagonal block of $\bF\bD\bF^\T$ is $\bDel_1$. The off-diagonal blocks, in which the first bracket enters with a difference, are zero.

(d) Equation~\eqref{eq:branch} is (b) and (c) in the core basis, with $\bK\bDel_1\bK^\T=\bW\bD_1\bW^\T$.

\emph{The secular equation.} For an eigenpair $(\lambda,\bv)$ with $\lambda$ not a diagonal entry,
\[
(\bLam-\lambda\bI)\bv=\bW\bD_1\bW^\T\bv.
\]
Setting $\bnu=\bD_1\bW^\T\bv\in\R^p$ gives $\bv=(\bLam-\lambda\bI)^{-1}\bW\bnu$, and substituting back,
\[
\bnu=\bD_1\bW^\T(\bLam-\lambda\bI)^{-1}\bW\bnu,
\]
which is~\eqref{eq:secular}. Conversely a null vector $\bnu$ of~\eqref{eq:secular} gives an eigenvector by the same two lines.

\emph{The inertia bounds.} The perturbation $-\bW\bD_1\bW^\T$ in~\eqref{eq:branch} is congruent to $-\bD_1$ on the range of $\bW$, whose columns are orthonormal, so by Sylvester's law of inertia it has $n_-$ positive and $n_+$ negative eigenvalues. Write it as the sum of its positive and its negative semidefinite parts, of ranks $n_-$ and $n_+$. For a positive semidefinite perturbation of rank $r$,
\[
\Lambda_i\le\lambda_i\le\Lambda_{i+r},
\]
its lower half from Weyl's inequalities and its upper half from the interlacing theorem for perturbations of rank at most $r$~\cite[Thms.~4.3.6 and~4.3.7]{horn1985matrix}. Applied to each part this gives the bounds of~\eqref{eq:inertia}, and the trace identity is
\[
\sum_{i=1}^{L}(\lambda_i-\Lambda_i)=\operatorname{tr}(\bLam-\bW\bD_1\bW^\T)-\operatorname{tr}\bLam=-\operatorname{tr}(\bD_1\bW^\T\bW).
\]

(e) is Theorem~\ref{thm:exact}(c).

(f) \emph{The fold at odd length.} For $M=2L+1$ the fold keeps the center sample unscaled,
\[
\bF=\begin{bmatrix}\tfrac1{\sqrt2}\bI_L&\bm0&\tfrac1{\sqrt2}\bJ_L\\ \bm0^\T&1&\bm0^\T\\ \tfrac1{\sqrt2}\bI_L&\bm0&-\tfrac1{\sqrt2}\bJ_L\end{bmatrix},
\]
with rows $(\be_n+\be_{M-1-n})/\sqrt2$ for $n<L$, then $\be_L$, then $(\be_n-\be_{M-1-n})/\sqrt2$. It is orthonormal and block-diagonalizes every persymmetric matrix as in (a), with an even component of dimension $L+1$ and an odd one of dimension $L$.

\emph{The cores.} The DCT-II columns of odd length satisfy $u_k(M-1-n)=(-1)^ku_k(n)$ as before, so the even-indexed columns lie in the even component and the odd-indexed ones in the odd component, where $u_{2j+1}(L)=0$. The fold maps them to
\[
\bF u_{2j}=\bigl(\sqrt2u_{2j}(0),\dots,\sqrt2u_{2j}(L-1),\,u_{2j}(L)\bigr)\oplus\bm0,\qquad
\bF u_{2j+1}=\bm0\oplus\bigl(\sqrt2u_{2j+1}(0),\dots,\sqrt2u_{2j+1}(L-1)\bigr),
\]
which are orthonormal because the columns are and the fold is.

\emph{The kernels.} With $2j/M=j/(L+\tfrac12)$ and $(2j+1)/M=(j+\tfrac12)/(L+\tfrac12)$ the even and odd kernels are
\[
\cos\Bigl(\frac{\pi(n+\tfrac12)j}{L+\tfrac12}\Bigr)\qquad\text{and}\qquad\cos\Bigl(\frac{\pi(n+\tfrac12)(j+\tfrac12)}{L+\tfrac12}\Bigr).
\]
Under $n\mapsto L-1-n$ the argument of the odd kernel becomes $\pi(j+\tfrac12)-\pi(n+1)(j+\tfrac12)/(L+\tfrac12)$, and $\cos(\pi(j+\tfrac12)-x)=(-1)^j\sin x$, so the odd kernel becomes
\[
(-1)^j\sin\Bigl(\frac{\pi(n+1)(j+\tfrac12)}{L+\tfrac12}\Bigr),
\]
the DST-VII kernel of length $L$.

\emph{Grids and corrections.} The eigenvalues are those of the DCT-II columns under $\bT_c+\bH_c$, $c(2\pi j/M)$ and $c(\pi(2j+1)/M)$. The corner blocks fold as in (c), since the center sample is at distance $L\ge p$ from either corner.
\fi
\end{proof}

For $p=1$, $\bD_1=-\rho(1-\rho)$, the perturbation in~\eqref{eq:branch} has rank one, \eqref{eq:secular} is a scalar secular equation and the eigenvectors are Cauchy vectors: this is the factorization of~\cite{reznik2026exact}, with its DCT-II and DCT-IV cores, and the recursive form of~\cite{reznik2026direct} is the same split applied to the half-length AR(1) matrix in place of the core. What the proposition adds is the order: the corner block of Theorem~\ref{thm:departure} is what each branch sees at its outer end, whatever $p$, and the only part of the transform that depends on the poles is the eigenbasis of a diagonal matrix plus one of rank at most $p$. The inertia form of the interlacing is his strict directional interlacing when $p=1$, and for $p\ge2$ it says that the corner block pushes the branch eigenvalues in both directions, $n_-$ places up and $n_+$ down: Section~\ref{sec:num} finds inertia $\inertiaArTwoReal$ for the AR(2) window with poles $0.9$ and $0.5$, $\inertiaArThreePair$ for the AR(3) window and $\inertiaArFourReal$ for the AR(4) window, with every eigenvalue inside its bounds, the interlacing strict and the trace identity exact. The group reading says why the cores are fixed: they are the matched transform of the dihedral closure, restricted along the exact subgroup $\Z_2$, and the correction is what the dihedral symmetry, approximate rather than exact, leaves to the data. At odd length the same two groups give the cores of types VI and VII, the parity split of the odd-length DCT-II that~\cite{reznik2013relationship} identified and that~\cite{reznik2026direct} reaches as the $\rho\to1$ limit of its odd-length recursion, so the direct form has fixed cores at every length. The bordered form with an arrowhead at the center (a diagonal matrix bordered by one full row and column), which~\cite{reznik2026direct} uses for $p=1$, belongs to the recursive form of Proposition~\ref{prop:recursive}(b), where the half-length window of an odd window carries the center sample as a border. One more regularity is measured but not proved: when all poles are real and of one sign, the KLT ordered by decreasing eigenvalue alternates between symmetric and antisymmetric vectors, starting with a symmetric one for positive poles and an antisymmetric one for a negative pole, as the first-order basis does~\cite{ray1970further}; for the mixed-sign and the resonant windows the alternation fails, since two branch eigenvalues of the same parity can then be adjacent.

\subsection{Rational Spectra and Sums}\label{sec:fastrational}

\begin{proposition}\label{prop:rational}
Let $\bR$ be a window with a rational spectrum of orders $(p,q)$ as in Corollary~\ref{cor:rational}, $M=2L$, and $\bZ=\bR-\bR_c(S)$. (a) The fold is exact, $\bF\bR\bF^\T=\bR_{\mathrm s}\oplus\bR_{\mathrm a}$, and $\bZ$ is persymmetric, so $\bF\bZ\bF^\T=\bZ_{\mathrm s}\oplus\bZ_{\mathrm a}$ with each block of rank at most $\max(p,q)$. (b) In the core bases
\begin{equation}\label{eq:branchcov}
\bK_{\mathrm{II}}\bR_{\mathrm s}\bK_{\mathrm{II}}^\T=\operatorname{diag}\bigl(S(2\pi j/M)\bigr)+\bK_{\mathrm{II}}\bZ_{\mathrm s}\bK_{\mathrm{II}}^\T,
\end{equation}
and likewise for the odd branch with $\bK_{\mathrm{IV}}$ and the samples $S(\pi(2j+1)/M)$: each branch is a diagonal matrix plus a symmetric matrix of rank at most $\max(p,q)$, whose eigenbasis is the correction, and Algorithm~\ref{alg:fast} applies with the design taken from~\eqref{eq:branchcov} in place of~\eqref{eq:branch}. The eigenvalues move at most $\max(p,q)$ places within a branch. For a sum of uncorrelated AR($k_1$) and AR($k_2$) windows the ranks are at most $k_1+k_2$ per branch.
\end{proposition}

\begin{proof}
\ifextended
(a) $\bR$ is symmetric Toeplitz and $\bR_c(S)$ is a class member, so both are persymmetric and so is $\bZ$. The fold block-diagonalizes a persymmetric matrix as in Proposition~\ref{prop:fast}(a). For the rank, the exponential part of $\bZ$ lies in $\bigoplus_j\operatorname{span}\{\bm u_j,\bm v_j\}^{\otimes2}$ by the proof of Corollary~\ref{cor:rational}, and
\[
\bm v_j=\zeta_j^{-(M-1)}\bJ\bm u_j,
\]
so $\operatorname{span}\{\bm u_j,\bm v_j\}=\operatorname{span}\{\bm u_j,\bJ\bm u_j\}$ folds into one even vector $\bm u_j+\bJ\bm u_j$ and one odd vector $\bm u_j-\bJ\bm u_j$. That is one dimension per pole in each branch. The two $\delta$ corners fold as the corner blocks in Proposition~\ref{prop:fast}(c), one block of rank at most $q-p$ in each branch.

(b) is (a) in the core bases, which diagonalize $\bR_c(S)$ with the samples of $S$ on the two grids, and the interlacing is as before.
\fi
\end{proof}

The design is then read from the covariance rather than the precision: the folded departure in the core bases is formed from the autocovariances and the samples of $S$, or in closed form from the poles by the formulas in the proof of Corollary~\ref{cor:rational}, and the apply stage is unchanged. In particular the exact transform of a co-channel sum of two autoregressive windows is computed by the same fold, the same two cores, and corrections of rank $k_1+k_2$ per branch. Table~\ref{tab:rational} measures the statement.

\begin{table*}[t]
\caption{Corollary~\ref{cor:rational} and Proposition~\ref{prop:rational} by window: ARMA orders $(p,q)$ (for a sum, the degrees of the ARMA representation it admits), length, rank of the departure from the DCT-II class against its bound, rank of the folded departure in the even and odd branches, off-diagonal fraction left by the covariance-side fast form, and the eigenvalue shift}
\label{tab:rational}
\centering\footnotesize
\setlength{\tabcolsep}{6pt}
\begin{tabular}{@{}lcccclcc@{}}
\toprule
Window & $(p,q)$ & $M$ & rank & bound & branches & off-diag. & shift\\
\midrule
\input{rational_table.tex} 
\bottomrule
\end{tabular}
\end{table*}

\subsection{The Correction as a Cascade of First-Order Stages}\label{sec:cascade}

Both branch matrices, \eqref{eq:branch} and~\eqref{eq:branchcov}, have the form $\bLam+\bm\Theta$ with $\bLam$ diagonal and $\bm\Theta$ symmetric of rank $\kappa$: $\bm\Theta=-\bW\bD_1\bW^\T$ with $\kappa\le p$ for the AR($p$) window, and $\bm\Theta=\bK\bZ_{\mathrm s}\bK^\T$, or its odd counterpart, with $\kappa\le\max(p,q)$ for a rational spectrum and $\kappa\le k_1+k_2$ for a sum. Write $\bm\Theta=\sum_{h=1}^{\kappa}\xi_h\bm\psi_h\bm\psi_h^\T$ with $(\xi_h,\bm\psi_h)$ its nonzero eigenpairs.

\begin{proposition}\label{prop:cascade}
Let $\bLam^{(0)}=\bLam$. For $h=1,\dots,\kappa$ let $\bm\psi_h'=(\bV^{(1)}\cdots\bV^{(h-1)})^\T\bm\psi_h$ and let $\bV^{(h)}$ be an orthonormal eigenbasis of the diagonal-plus-rank-one matrix $\bLam^{(h-1)}+\xi_h\bm\psi_h'\bm\psi_h'^{\T}$, with eigenvalues $\bLam^{(h)}$. Then $\bV=\bV^{(1)}\cdots\bV^{(\kappa)}$ is an eigenbasis of $\bLam+\bm\Theta$ with eigenvalues $\bLam^{(\kappa)}$, and every stage is a first-order stage: when the entries of $\bLam^{(h-1)}$ are distinct and those of $\bm\psi'_h$ nonzero, the eigenvalues of stage $h$ are the roots $\lambda$ of the scalar secular equation $1+\xi_h\sum_j\psi_{h,j}'^{\,2}/(\Lambda^{(h-1)}_{jj}-\lambda)=0$, one in each interval between consecutive entries of $\bLam^{(h-1)}$ and one beyond them on the side of the sign of $\xi_h$, and the eigenvector at $\lambda$ is the Cauchy vector $v_j=\psi'_{h,j}/(\Lambda^{(h-1)}_{jj}-\lambda)$ normalized; a zero entry of $\bm\psi'_h$ or a repeated entry of $\bLam^{(h-1)}$ leaves its coordinate fixed by the stage. The eigenvalues move at most one place at each stage, $\kappa$ places in all. The correction of order $p$ is thus the correction of order one taken $\kappa$ times in series, $p$ times for a minimal AR($p$) model, and for $p=1$ the single stage is the Cauchy correction of~\cite{reznik2026exact}.
\end{proposition}

\begin{proof}
\ifextended
Induction on $h$. Suppose $\bm Y_{h-1}=\bV^{(1)}\cdots\bV^{(h-1)}$ satisfies
\[
\bm Y_{h-1}^\T\Bigl(\bLam+\sum_{i<h}\xi_i\bm\psi_i\bm\psi_i^\T\Bigr)\bm Y_{h-1}=\bLam^{(h-1)}.
\]
Adding the term $h$ and conjugating by $\bm Y_{h-1}$ gives
\[
\bm Y_{h-1}^\T\Bigl(\bLam+\sum_{i\le h}\xi_i\bm\psi_i\bm\psi_i^\T\Bigr)\bm Y_{h-1}=\bLam^{(h-1)}+\xi_h\bm\psi_h'\bm\psi_h'^{\T},
\]
which $\bV^{(h)}$ diagonalizes by definition. So $\bm Y_h=\bm Y_{h-1}\bV^{(h)}$ diagonalizes the partial sum through the term $h$, and at $h=\kappa$ the partial sum is $\bLam+\bm\Theta$, which is the claim. The secular equation, the Cauchy form of the eigenvectors, the location of the roots and the deflation of zero entries and repeated diagonal entries are the rank-one modification theorem of~\cite{bunch1978rank}, and the interlacing is its one-place statement at each stage.
\fi
\end{proof}

In the group reading the stages are what the dihedral symmetry leaves undetermined, counted one unit of rank at a time: the cores diagonalize every class member, the departure from the class has rank $2\kappa$ by Corollaries~\ref{cor:identity} and~\ref{cor:rational}, the fold distributes it as $\kappa$ per branch, and each unit is one Cauchy stage. For the AR($p$) window $\kappa=\operatorname{rank}\bD_1$, since $\bW$ has orthonormal columns, so by~\eqref{eq:detcorner} $\kappa=p$ exactly when $\phi_p\ne0$, that is, for a minimal AR($p$) model, and $\kappa=p_0$ in general. The cascade and the single stage of Proposition~\ref{prop:fast}(d) are two factorizations of the same orthogonal matrix. What the cascade adds is the parametrization by the order, which the flow graph makes visible.

\subsection{The Flow Graph Parametrized by the Order}\label{sec:flowgraph}

\begin{figure*}[!t]
\centering
\resizebox{\textwidth}{!}{%
\begin{tikzpicture}[x=1cm,y=0.9cm,
  fgdot/.style={circle,fill=black,inner sep=0pt,minimum size=4.6pt},
  fgsdot/.style={circle,fill=black,inner sep=0pt,minimum size=2.8pt},
  fgw/.style={line width=0.75pt},
  fgneg/.style={line width=0.75pt,dashed},
  fgthin/.style={line width=0.3pt,black!70},
  fgrot/.style={draw,fill=white,rounded corners=1.5pt,minimum width=8mm,minimum height=13mm,inner sep=0pt,font=\footnotesize},
  fgbx/.style={draw,line width=0.6pt},
  fglbl/.style={font=\footnotesize,align=center}]
\fill[black!10] (-3.6,1.05) rectangle (-0.55,-8.1);
\node[fglbl,font=\small] at (-2.08,-1.6) {demodulation\\$\bm E_\omega^{*}$\\[2pt](rotated poles,\\Proposition~\ref{prop:modulated})};
\node[fglbl,font=\small] at (-2.08,-5.4) {polyphase\\permutation\\[2pt]($a(z)=b(z^m)$,\\Theorem~\ref{thm:exact}(c))};
\foreach \n in {0,...,7}{\node[fgdot] (i\n) at (0,-\n) {}; \node[left=1pt of i\n,font=\small] {$x_{\n}$}; \node[fgdot] (f\n) at (1.6,-\n) {};}
\foreach \n in {0,...,3}{\draw[fgw] (i\n) -- (f\n);}
\foreach \n in {4,...,7}{\draw[fgneg] (i\n) -- (f\n);}
\foreach \n in {0,...,7}{\pgfmathtruncatemacro{\m}{7-\n}\draw[fgw] (i\n) -- (f\m);}
\draw[fgbx] (2.05,1.05) rectangle (8.75,-3.5);
\node[font=\small] at (5.4,0.62) {DCT-II core, $L=4$};
\foreach \n in {0,...,3}{\node[fgdot] (a\n) at (2.55,-\n) {}; \draw[fgw] (f\n) -- (a\n); \node[fgdot] (b\n) at (4.0,-\n) {};}
\foreach \n in {0,1}{\draw[fgw] (a\n) -- (b\n);}\foreach \n in {2,3}{\draw[fgneg] (a\n) -- (b\n);}
\foreach \n in {0,...,3}{\pgfmathtruncatemacro{\m}{3-\n}\draw[fgw] (a\n) -- (b\m);}
\node[fgdot] (c0) at (5.45,0) {}; \node[fgdot] (c1) at (5.45,-1) {};
\draw[fgw] (b0) -- (c0); \draw[fgneg] (b1) -- (c1); \draw[fgw] (b0) -- (c1); \draw[fgw] (b1) -- (c0);
\node[fgrot] (r23) at (6.7,-2.5) {$-\tfrac{3\pi}{8}$};
\draw[fgw] (b2) -- (b2-|r23.west); \draw[fgw] (b3) -- (b3-|r23.west);
\node[fglbl,fill=white,inner sep=1pt] at (6.7,-1.5) {DCT-IV, $L=2$};
\node[fgdot] (s2) at (7.45,-2) {}; \node[fgdot] (s3) at (7.45,-3) {};
\draw[fgw] (r23.east|-b2) -- (s2); \draw[fgw] (r23.east|-b3) -- (s3);
\foreach \n in {0,...,3}{\node[fgdot] (d\n) at (8.3,-\n) {};}
\draw[fgw] (c0) -- (d0); \draw[fgw] (c1) -- (d1); \draw[fgw] (s2) -- (d2); \draw[fgw] (s3) -- (d3);
\draw[fgbx] (2.05,-3.62) rectangle (8.75,-8.1);
\node[font=\small] at (5.4,-7.62) {DCT-IV core, $L=4$};
\foreach \n in {4,...,7}{\node[fgdot] (p\n) at (2.55,-\n) {}; \draw[fgw] (f\n) -- (p\n); \node[fgdot] (q\n) at (3.35,-\n) {};}
\draw[fgw] (p7) -- (q4); \draw[fgw] (p4) -- (q5); \draw[fgw] (p5) -- (q6); \draw[fgw] (p6) -- (q7);
\node[fgrot] (t0) at (4.05,-4.5) {$-\tfrac{\pi}{16}$}; \node[fgrot] (t1) at (4.05,-6.5) {$-\tfrac{5\pi}{16}$};
\foreach \n/\t in {4/t0,5/t0,6/t1,7/t1}{\draw[fgw] (q\n) -- (q\n-|\t.west);}
\foreach \n in {4,...,7}{\node[fgdot] (g\n) at (4.8,-\n) {}; \node[fgdot] (h\n) at (5.95,-\n) {};}
\draw[fgw] (t0.east|-q4) -- (g4); \draw[fgw] (t0.east|-q5) -- (g5); \draw[fgw] (t1.east|-q6) -- (g6); \draw[fgw] (t1.east|-q7) -- (g7);
\draw[fgw] (g4) -- (h4); \draw[fgw] (g5) -- (h5); \draw[fgneg] (g6) -- (h6); \draw[fgneg] (g7) -- (h7);
\draw[fgw] (g4) -- (h6); \draw[fgw] (g6) -- (h4); \draw[fgw] (g5) -- (h7); \draw[fgw] (g7) -- (h5);
\node[fgrot] (t2) at (6.7,-6.5) {$-\tfrac{\pi}{4}$};
\draw[fgw] (h6) -- (h6-|t2.west); \draw[fgw] (h7) -- (h7-|t2.west);
\foreach \n in {4,...,7}{\node[fgdot] (k\n) at (7.45,-\n) {};}
\draw[fgw] (h4) -- (k4); \draw[fgw] (h5) -- (k5); \draw[fgw] (t2.east|-h6) -- (k6); \draw[fgw] (t2.east|-h7) -- (k7);
\foreach \n in {4,...,7}{\node[fgdot] (e\n) at (8.3,-\n) {};}
\draw[fgw] (k4) -- (e4); \draw[fgw] (k6) -- (e6); \draw[fgneg] (k7) -- (e5); \draw[fgneg] (k5) -- (e7);
\def\fgcauchy#1#2#3{
  \pgfmathsetmacro{\xl}{#1}\pgfmathsetmacro{\xr}{#1+1.9}\pgfmathsetmacro{\yt}{-#2+0.45}\pgfmathsetmacro{\yb}{-#2-3.45}
  \draw[fgbx,fill=white] (\xl,\yt) rectangle (\xr,\yb);
  \node[font=\scriptsize,below] at ({\xl+0.95},\yt) {#3};
  \foreach \j in {0,...,3}{\pgfmathsetmacro{\yy}{-#2-\j}
    \node[fgsdot] (cA\j) at ({\xl+0.35},\yy) {};
    \node[fgsdot] (cB\j) at ({\xr-0.35},\yy) {};}
  \foreach \j in {0,...,3}{\foreach \i in {0,...,3}{\draw[fgthin] (cA\j) -- (cB\i);}}}
\foreach \n in {0,...,7}{\draw[fgw] (8.3,-\n) -- (9.2,-\n);}
\fgcauchy{9.2}{0}{$\bm V^{(1)\mathsf T}$}\fgcauchy{11.45}{0}{$\bm V^{(2)\mathsf T}$}\fgcauchy{14.5}{0}{$\bm V^{(\kappa)\mathsf T}$}
\fgcauchy{9.2}{4}{$\bm V^{(1)\mathsf T}$}\fgcauchy{11.45}{4}{$\bm V^{(2)\mathsf T}$}\fgcauchy{14.5}{4}{$\bm V^{(\kappa)\mathsf T}$}
\foreach \n in {0,...,7}{\draw[fgw] (11.1,-\n) -- (11.45,-\n); \draw[fgw] (13.35,-\n) -- (13.75,-\n); \draw[fgw] (14.1,-\n) -- (14.5,-\n);}
\node at (13.92,-1.5) {$\cdots$}; \node at (13.92,-5.5) {$\cdots$};
\foreach \n in {0,...,7}{\draw[fgw] (16.4,-\n) -- (16.75,-\n); \draw[fgw] (17.95,-\n) -- (18.4,-\n); \node[fgdot] (o\n) at (18.4,-\n) {}; \node[right=1pt of o\n,font=\small] {$(\bm U^{\mathsf T}\bm x)_{\n}$};}
\draw[fgbx,fill=white] (16.75,0.6) rectangle (17.95,-7.6);
\node[fglbl,font=\small] at (17.35,-3.5) {order\\by\\eigen-\\value};
\node[fglbl,anchor=north] at (0.8,-8.25) {fold: $L$ butterflies\\exact group $\mathbb Z_2$,\\Theorem~\ref{thm:exact}(a)};
\node[fglbl,anchor=north,text width=6.6cm] at (5.4,-8.25) {cores, the dihedral closure, Proposition~\ref{prop:fast}(b). The DCT-II core folds again by the reflection of the half window. The DCT-IV core is a Fourier butterfly of length two between plane rotations.};
\node[fglbl,anchor=north] at (12.8,-8.25) {$\kappa$ Cauchy stages per branch, Proposition~\ref{prop:cascade}:\\$\kappa=p$ for AR($p$), $\kappa\le\max(p,q)$ for ARMA($p,q$),\\$\kappa\le k_1+k_2$ for a sum};
\node[fglbl,anchor=north] at (17.35,-8.25) {permutation};
\end{tikzpicture}
}
\caption{Flow graph of the exact transform of a window of length $M=8$ with a rational spectrum, parametrized by the order. Every butterfly is scaled by $1/\sqrt2$, and a dashed branch carries the weight $-1$. A box labeled $\theta$ is a plane rotation of the two wires it spans, $(u,v)\mapsto(u\cos\theta-v\sin\theta,\,u\sin\theta+v\cos\theta)$, the multiplication of $u+iv$ by $e^{i\theta}$. The DCT-II core of length four is drawn unfolded: the second column of butterflies is its fold, the butterfly on rows $0$ and $1$ is its DCT-II core of length two, and the rotation on rows $2$ and $3$ its DCT-IV core of length two, which receives the odd part in reverse order. Its outputs are the coefficients $0$, $2$, $1$ and $3$ on rows $0$ to $3$. The DCT-IV core of length four pairs its inputs, again in reverse order, into the real and imaginary parts of two complex numbers, rotates each, applies a Fourier butterfly of length two and a last rotation, and unpairs the result into its coefficients $0$ to $3$ on rows $4$ to $7$. Each opened Cauchy stage is the scaled Cauchy matrix of~\eqref{eq:stage}, drawn as its branches from every input to every output (Fig.~\ref{fig:cauchy}). The solid stages are fixed by the two groups and do not depend on the order. The order sets only the number $\kappa$ of Cauchy stages in each branch, each the eigenbasis of a diagonal-plus-rank-one matrix in the running basis (Proposition~\ref{prop:cascade}), and the model sets their entries. The shaded panel at the left holds the pre-stages, the exact symmetries beyond the reflection: the demodulation for rotated poles and the polyphase permutation for interleaved windows, after which the graph of length $M/m$ runs on each phase.}
\label{fig:butterfly}
\end{figure*}
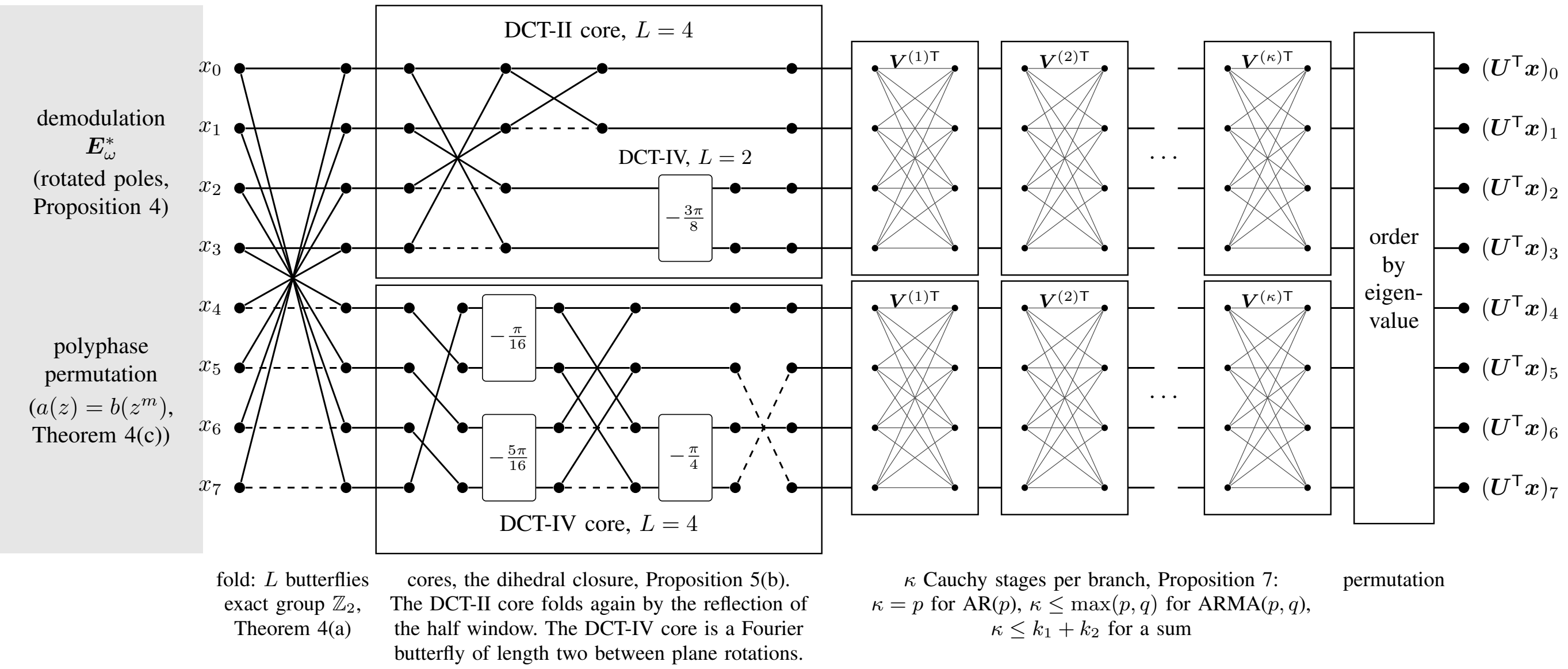

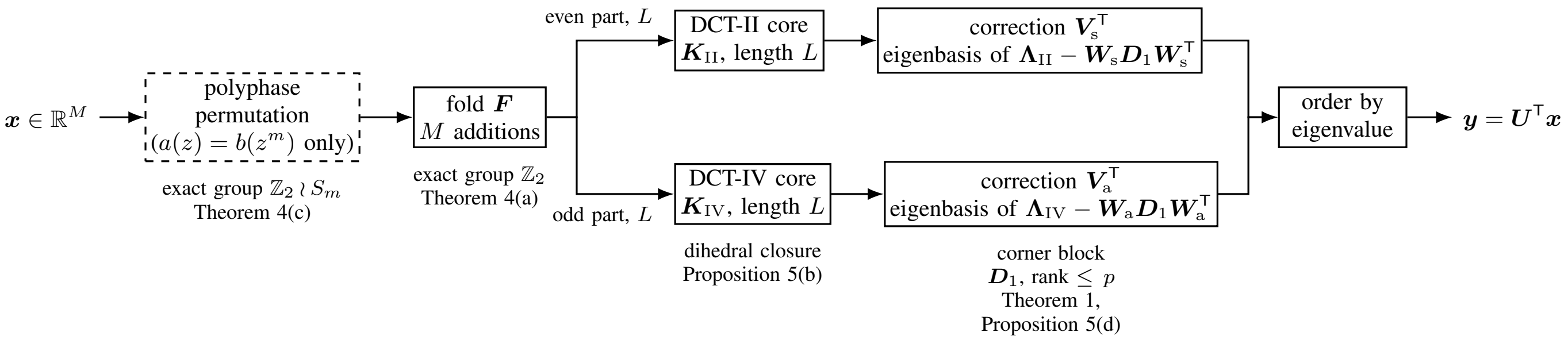
\begin{figure*}[!t]
\centering
\resizebox{\textwidth}{!}{%
\begin{tikzpicture}[>=Latex, node distance=5mm and 7mm, font=\footnotesize, line width=0.70pt,
  box/.style={draw, minimum height=8mm, minimum width=17mm, align=center, inner sep=2pt},
  stage/.style={draw=none, align=center, text width=24mm, font=\scriptsize}]
\node[box, dashed] (poly) {polyphase\\permutation\\($a(z)=b(z^m)$ only)};
\node[box, right=of poly] (fold) {fold $\bF$\\$M$ additions};
\node[box, above right=2mm and 17mm of fold] (k2) {DCT-II core\\$\bK_{\mathrm{II}}$, length $L$};
\node[box, below right=2mm and 17mm of fold] (k4) {DCT-IV core\\$\bK_{\mathrm{IV}}$, length $L$};
\node[box, right=of k2] (v2) {correction $\bV_{\mathrm s}^\T$\\eigenbasis of $\bLam_{\mathrm{II}}-\bW_{\mathrm s}\bD_1\bW_{\mathrm s}^\T$};
\node[box, right=of k4] (v4) {correction $\bV_{\mathrm a}^\T$\\eigenbasis of $\bLam_{\mathrm{IV}}-\bW_{\mathrm a}\bD_1\bW_{\mathrm a}^\T$};
\node[box, right=9mm of $(v2.east)!0.5!(v4.east)$] (merge) {order by\\eigenvalue};
\node[left=6mm of poly] (in) {$\bm x\in\R^M$};
\node[right=6mm of merge] (out) {$\bm y=\bU^\T\bm x$};
\draw[->] (in) -- (poly);
\draw[->] (poly) -- (fold);
\draw[->] (fold.east) -- ++(4mm,0) |- (k2.west);
\draw[->] (fold.east) -- ++(4mm,0) |- (k4.west);
\node[font=\scriptsize, anchor=south east, xshift=-1.5mm, yshift=0.3mm] at (k2.west) {even part, $L$};
\node[font=\scriptsize, anchor=north east, xshift=-1.5mm, yshift=-0.3mm] at (k4.west) {odd part, $L$};
\draw[->] (k2) -- (v2);
\draw[->] (k4) -- (v4);
\path let \p1=(v2.east), \p2=(v4.east) in coordinate (join) at ({max(\x1,\x2)+4mm},0);
\draw[->] (v2.east) -- (v2.east -| join) |- (merge.west);
\draw[->] (v4.east) -- (v4.east -| join) |- (merge.west);
\draw[->] (merge) -- (out);
\node[stage, below=1mm of poly] {exact group $\Z_2\wr S_m$\\Theorem~\ref{thm:exact}(c)};
\node[stage, below=1mm of fold] {exact group $\Z_2$\\Theorem~\ref{thm:exact}(a)};
\node[stage, below=1mm of k4] {dihedral closure\\Proposition~\ref{prop:fast}(b)};
\node[stage, below=1mm of v4] {corner block $\bD_1$, rank $\le p$\\Theorem~\ref{thm:departure}, Proposition~\ref{prop:fast}(d)};
\end{tikzpicture}}
\caption{Block form of the exact transform of an AR($p$) window of even length $M=2L$. The fold is the matched transform of the exact reflection. The two cores are the matched transform of the dihedral closure restricted to the two parity components. The two corrections carry all the dependence on the poles and are the eigenbases of diagonal matrices plus matrices of rank at most $p$ built from the corner block, each a cascade of Cauchy stages, $p$ for a minimal AR($p$) model. The dashed stage applies when the window interleaves shorter windows.}
\label{fig:fast}
\end{figure*}

Figure~\ref{fig:butterfly} draws the transform for $M=8$. Every solid part is determined by the two groups and is the same for every order and every window of this paper: the fold is four butterflies between mirrored samples; the DCT-II core of length four is, by the same reflection applied to the half window, a fold of two butterflies followed by a DCT-II core of length two, which is one more butterfly, and a DCT-IV core of length two; the DCT-IV core of length four is realized through the cyclic group. The order enters only as the number $\kappa$ of Cauchy stages in each branch: $\kappa=p$ for a minimal AR($p$) window, $\kappa\le\max(p,q)$ for an ARMA($p,q$) window, $\kappa\le k_1+k_2$ for a co-channel sum, and the first-order flow graph of~\cite{reznik2026exact} is the figure with one stage in each branch. The pre-stages in the shaded panel at the left are exact and apply when the window has more symmetry than the reflection: the polyphase permutation of Theorem~\ref{thm:exact}(c), after which the flow graph of length $M/m$ and order $p/m$ runs on each phase, and the demodulation $\bm E_\omega^*$ of Proposition~\ref{prop:modulated}, which carries a complex window with rotated poles onto the real one. Figure~\ref{fig:fast} is the same flow at the block level, and Algorithm~\ref{alg:fast} is the same flow as a procedure, with the design stage that computes the Cauchy stages once per model.

\begin{figure*}[!t]
\centering
\begin{tikzpicture}[
  font=\small,
  cdot/.style={circle,fill=black,inner sep=0pt,minimum size=4.2pt},
  csum/.style={circle,fill=black,inner sep=0pt,minimum size=4.2pt},
  cwire/.style={line width=0.6pt},
  cthin/.style={line width=0.35pt,black!70},
  clab/.style={font=\scriptsize,fill=white,inner sep=1pt},
  >={Stealth[length=4pt]}]
\def\xin{0}\def\xa{2.0}\def\xb{6.6}\def\xout{8.8}\def\dy{1.15}
\fill[black!8,rounded corners=2pt] (\xa+0.35,0.5) rectangle (\xb-0.35,-3*\dy-0.45);
\node[font=\small] at ({(\xa+\xb)/2},1.6) {Cauchy kernel: branch $j\to i$ carries $\dfrac{1}{\lambda^{(h-1)}_j-\lambda^{(h)}_i}$};
\foreach \j in {0,1,2,3}{
  \pgfmathsetmacro{\y}{-\j*\dy}
  \node[cdot] (in\j) at (\xin,\y) {};
  \node[left=2pt of in\j] {$u^{(h-1)}_{\j}$};
  \node[cdot] (a\j) at (\xa,\y) {};
  \draw[cwire,->] (in\j) -- (a\j) node[clab,midway,above=1pt] {$\psi'_{h,\j}$};
  \node[csum] (b\j) at (\xb,\y) {};
  \node[cdot] (out\j) at (\xout,\y) {};
  \draw[cwire,->] (b\j) -- (out\j) node[clab,midway,above=1pt] {$1/\lVert\tilde{\bm v}_{h,\j}\rVert$};
  \node[right=2pt of out\j] {$u^{(h)}_{\j}$};
}
\foreach \j in {0,1,2,3}{\foreach \i in {0,1,2,3}{\draw[cthin] (a\j) -- (b\i);}}
\draw[line width=0.9pt] (a3) -- (b0);
\node[clab] at ($(a3)!0.24!(b0)+(-0.1,-0.2)$) {$\dfrac{1}{\lambda^{(h-1)}_3-\lambda^{(h)}_0}$};
\node[font=\footnotesize] at ({(\xin+\xa)/2},-3*\dy-0.8) {scale input $j$};
\node[font=\footnotesize] at ({(\xa+\xb)/2},-3*\dy-0.8) {sum over $j$ at each output $i$};
\node[font=\footnotesize] at ({(\xb+\xout)/2},-3*\dy-0.8) {normalize};
\node[font=\footnotesize] at (\xin,0.95) {from stage $h-1$};
\node[font=\footnotesize] at (\xout+0.2,0.95) {to stage $h+1$};
\node[anchor=north west,align=left,draw=black!60,rounded corners=2pt,inner sep=6pt,text width=5.6cm] at (10.25,1.35) {%
  \textbf{Design, once per model}\\[4pt]
  The roots $\lambda^{(h)}_i$ of the secular equation\\[2pt]
  \hspace*{0.6em}$1+\xi_h\displaystyle\sum_{j}\frac{\psi_{h,j}'^{\,2}}{\lambda^{(h-1)}_j-\lambda}=0$\\[4pt]
  lie one in each gap of the running grid and one beyond it on the side of the sign of $\xi_h$:\\[6pt]
  \begin{tikzpicture}[baseline]
    \draw[line width=0.5pt,->] (0,0) -- (5.25,0);
    \foreach \x/\k in {0.4/0,1.6/1,2.8/2,4.0/3}{\draw (\x,0.12) -- (\x,-0.12); \node[font=\scriptsize,below] at (\x,-0.1) {$\lambda^{(h-1)}_{\k}$};}
    \foreach \x/\k in {1.0/0,2.2/1,3.4/2,4.7/3}{\fill (\x,0) circle (1.6pt); \node[font=\scriptsize,above] at (\x,0.08) {$\lambda^{(h)}_{\k}$};}
  \end{tikzpicture}\\[6pt]
  The roots become the running grid of stage $h+1$.\\[3pt]
  A zero $\psi'_{h,j}$ or a repeated $\lambda^{(h-1)}_j$ leaves coordinate $j$ fixed.};
\end{tikzpicture}
\caption{One Cauchy stage $\bV^{(h)\T}$ of Fig.~\ref{fig:butterfly}, opened for a branch of length $L=4$ and drawn for $\xi_h>0$. Left, the stage as a flow graph, the factorization~\eqref{eq:stage} read from left to right: each coordinate $u^{(h-1)}_j$ is scaled by $\psi'_{h,j}$, every branch from input $j$ to output $i$ of the shaded Cauchy kernel carries $1/(\lambda^{(h-1)}_j-\lambda^{(h)}_i)$, with one branch drawn heavy, and each sum is normalized by $1/\norm{\tilde{\bm v}_{h,i}}$. Right, the design of the stage, the secular equation whose roots $\lambda^{(h)}_i$ interlace the running grid $\lambda^{(h-1)}_j$ and are handed on to stage $h+1$. The notation is defined in Section~\ref{sec:flowgraph}.}
\label{fig:cauchy}
\end{figure*}
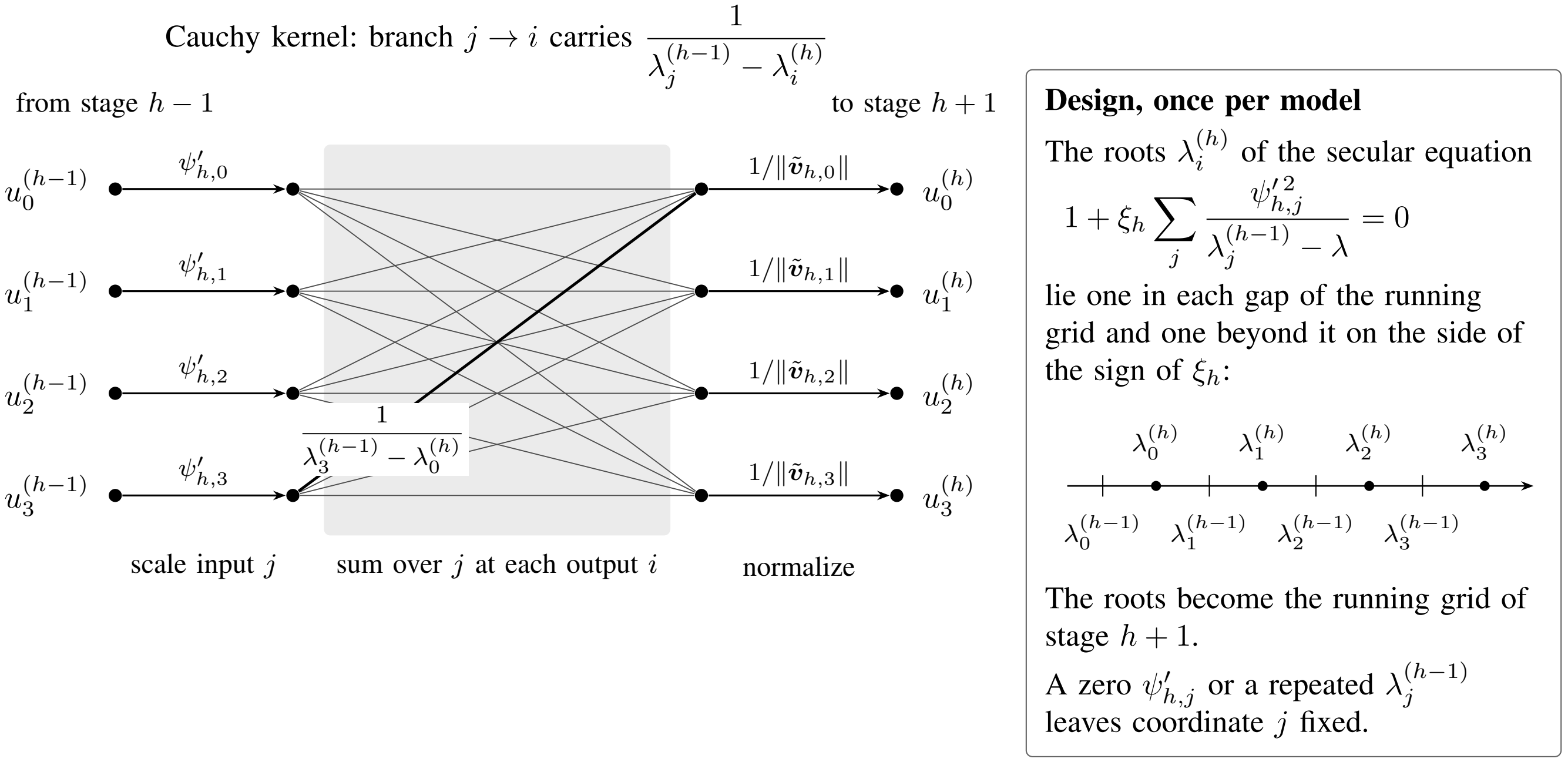

Fig.~\ref{fig:cauchy} opens one Cauchy stage of one branch, the part of the transform that depends on the model. Write $\bm u^{(0)}$ for the core output of the branch, $\bm u_{\mathrm s}$ or $\bm u_{\mathrm a}$ in Algorithm~\ref{alg:fast}, and $\bm u^{(h)}=\bV^{(h)\T}\bm u^{(h-1)}$ for its coordinates after stage $h$, so that $\bm u^{(\kappa)}$ is the branch output $\bm z_{\mathrm s}$ or $\bm z_{\mathrm a}$. Write $\lambda^{(h)}_i=\Lambda^{(h)}_{ii}$, $i=0,\dots,L-1$, for the diagonal entries after stage $h$ in increasing order, so that $\lambda^{(0)}_j$ is the core grid and $\lambda^{(h-1)}_j$ is the running grid that stage $h$ receives. By Proposition~\ref{prop:cascade} the $i$th column of $\bV^{(h)}$ is the Cauchy vector $\tilde{\bm v}_{h,i}$ divided by its norm, where
\[
(\tilde{\bm v}_{h,i})_j=\frac{\psi'_{h,j}}{\lambda^{(h-1)}_j-\lambda^{(h)}_i},\qquad j=0,\dots,L-1.
\]
The stage is therefore a Cauchy matrix between two diagonal scalings,
\begin{equation}\label{eq:stage}
\bV^{(h)\T}=\operatorname{diag}\bigl(\norm{\tilde{\bm v}_{h,i}}^{-1}\bigr)\Bigl[\frac{1}{\lambda^{(h-1)}_j-\lambda^{(h)}_i}\Bigr]_{ij}\operatorname{diag}(\bm\psi'_h),
\end{equation}
and these are the three columns of the figure. Coordinate $j$ is scaled by $\psi'_{h,j}$, every scaled coordinate reaches every output $i$ through a branch of weight $1/(\lambda^{(h-1)}_j-\lambda^{(h)}_i)$, and output $i$ is normalized by $1/\norm{\tilde{\bm v}_{h,i}}$.

The design of a stage needs only scalars. The roots $\lambda^{(h)}_i$ solve the secular equation
\[
1+\xi_h\sum_{j=0}^{L-1}\frac{\psi'^{\,2}_{h,j}}{\lambda^{(h-1)}_j-\lambda}=0,
\]
one in each gap of the running grid and one beyond it on the side of the sign of $\xi_h$, and the norms are
\[
\norm{\tilde{\bm v}_{h,i}}^2=\sum_{j=0}^{L-1}\frac{\psi'^{\,2}_{h,j}}{\bigl(\lambda^{(h-1)}_j-\lambda^{(h)}_i\bigr)^2}.
\]
The roots become the running grid of stage $h+1$, which is why each stage moves every eigenvalue at most one place. A zero entry of $\bm\psi'_h$ or a repeated entry of the running grid leaves its coordinate fixed by the stage, the deflation of Proposition~\ref{prop:cascade}. Applied densely a stage takes $L^2$ multiplications, and Section~\ref{sec:count} applies it in $O(L\log L)$ operations to a fixed accuracy by the fast multipole method.

\subsection{Recursive Form and Separation}\label{sec:fastmore}

\begin{figure*}[t]
\centering
\resizebox{\textwidth}{!}{\begin{tikzpicture}[font=\footnotesize,
  lab/.style={font=\scriptsize},
  plus/.style={draw,fill=black!25,minimum width=5pt,minimum height=5pt,inner sep=0pt},
  minus/.style={draw,fill=white,minimum width=5pt,minimum height=5pt,inner sep=0pt},
  zero/.style={draw=black!30,fill=white,minimum width=5pt,minimum height=5pt,inner sep=0pt},
  brace/.style={decorate,decoration={brace,amplitude=3pt}}]
\def\dx{0.4}
\begin{scope}
\node[anchor=west,font=\footnotesize] at (-0.3,2.75) {(a) the nested half-windows, $M=8$};
\foreach \n in {0,...,7}{\node[lab] at (\n*\dx,0) {$x_\n$};}
\draw[brace] (-0.15,0.25) -- (7*\dx+0.15,0.25) node[midway,above=3pt,lab] {the window: reflection $\Z_2$};
\foreach \a/\b in {0/3,4/7}{\draw[brace] (\a*\dx-0.12,0.95) -- (\b*\dx+0.12,0.95);}
\node[lab] at (3.5*\dx,1.42) {two halves, each with its reflection: $\Z_2\wr\Z_2$};
\foreach \a/\b in {0/1,2/3,4/5,6/7}{\draw[brace] (\a*\dx-0.1,1.72) -- (\b*\dx+0.1,1.72);}
\node[lab] at (3.5*\dx,2.2) {four quarters: $\Z_2\wr\Z_2\wr\Z_2$};
\end{scope}
\begin{scope}[yshift=-3.6cm,xshift=-2.3cm]
\node[anchor=west,font=\footnotesize] at (-0.3,2.75) {(b) the isotypic components, with the Haar basis (left) and the Walsh-Hadamard basis (right)};
\def\row#1#2#3#4{%
  \node[lab,anchor=east] at (2.35,#1) {#2};
  \foreach \v [count=\k from 0] in {#3}{\foreach \s [count=\n from 0] in \v{%
     \ifnum\s=1 \node[plus] at (2.6+\k*1.72+\n*0.19,#1) {};\fi \ifnum\s=-1 \node[minus] at (2.6+\k*1.72+\n*0.19,#1) {};\fi \ifnum\s=0 \node[zero] at (2.6+\k*1.72+\n*0.19,#1) {};\fi}}
  \foreach \v [count=\k from 0] in {#4}{\foreach \s [count=\n from 0] in \v{%
     \ifnum\s=1 \node[plus] at (9.75+\k*1.72+\n*0.19,#1) {};\fi \ifnum\s=-1 \node[minus] at (9.75+\k*1.72+\n*0.19,#1) {};\fi}}}
\row{2.1}{$V_0$, dim 1: the mean}{{1,1,1,1,1,1,1,1}}{{1,1,1,1,1,1,1,1}}
\row{1.5}{$W_1$, dim 1: half against half}{{1,1,1,1,-1,-1,-1,-1}}{{1,1,1,1,-1,-1,-1,-1}}
\row{0.9}{$W_2$, dim 2: quarter against quarter}{{1,1,-1,-1,0,0,0,0},{0,0,0,0,1,1,-1,-1}}{{1,1,-1,-1,-1,-1,1,1},{1,1,-1,-1,1,1,-1,-1}}
\row{0.3}{$W_3$, dim 4: sample against sample}{{1,-1,0,0,0,0,0,0},{0,0,1,-1,0,0,0,0},{0,0,0,0,1,-1,0,0},{0,0,0,0,0,0,1,-1}}{{1,-1,-1,1,1,-1,-1,1},{1,-1,1,-1,-1,1,-1,1},{1,-1,-1,1,-1,1,1,-1},{1,-1,1,-1,1,-1,1,-1}}
\node[lab,anchor=west] at (2.6,-0.3) {\tikz{\node[plus]{};}\ $+1$\quad \tikz{\node[minus]{};}\ $-1$\quad \tikz{\node[zero]{};}\ $0$\quad (each vector normalized)};
\end{scope}
\end{tikzpicture}}
\caption{The dyadic tower of the Haar basis for $M=8$. (a) The nested half-windows, whose reflections generate the automorphism group of the binary tree over the samples, the iterated wreath product. (b) Its isotypic components, the Haar multiresolution subspaces, with two bases adapted to them: the Haar basis keeps each vector on one block of the tree, and the Walsh-Hadamard basis is what the tower of folds of Proposition~\ref{prop:recursive}(d) produces when both branches are folded at every node. The AR($p$) window adds a correction of rank at most $p$ at every node.}
\label{fig:tower}
\end{figure*}
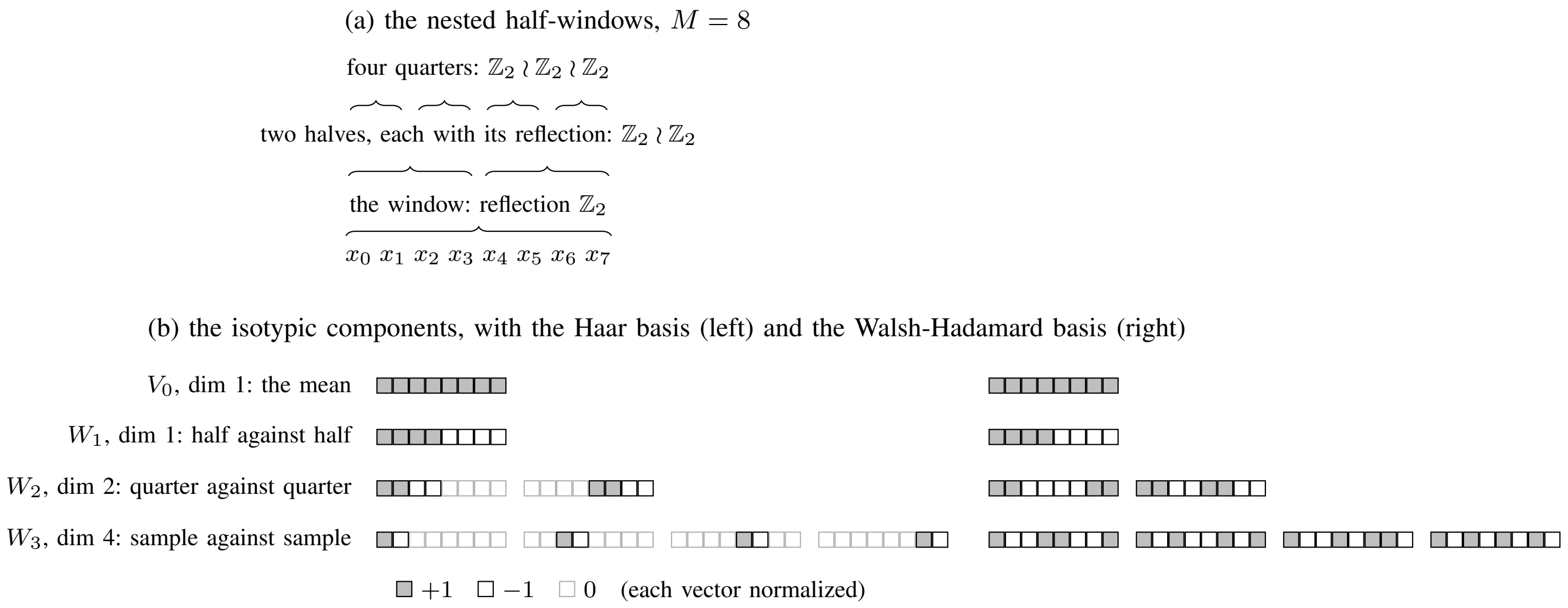

\begin{proposition}\label{prop:recursive}
Let $M\ge4p$, $L=\lfloor M/2\rfloor$, and let $\bQ^{(\ell)}$ be the precision of the window of length $\ell$ of the same AR($p$) process. (a) [even length] If $M=2L$, then $\sigma^2\bQ_{\mathrm s}=\sigma^2\bQ^{(L)}-\bm\Omega_{\mathrm s}$ and $\sigma^2\bQ_{\mathrm a}=\sigma^2\bQ^{(L)}-\bm\Omega_{\mathrm a}$ with $\bm\Omega_{\mathrm s}$, $\bm\Omega_{\mathrm a}$ supported on the last $p\times p$ block, the center of the window, of rank at most $p$. For $p=1$, $\bm\Omega_{\mathrm s}=\rho(1-\rho)\be_{L-1}\be_{L-1}^\T$ and $\bm\Omega_{\mathrm a}=-\rho(1+\rho)\be_{L-1}\be_{L-1}^\T$. (b) [odd length] If $M=2L+1$, the fold of Proposition~\ref{prop:fast}(f) puts the center sample in the even branch, of length $L+1$, and $\sigma^2\bQ_{\mathrm s}=\sigma^2\bQ^{(L+1)}-\bm\Omega_{\mathrm s}$ with $\bm\Omega_{\mathrm s}$ supported on the last $(p+1)\times(p+1)$ block, of rank at most $p+1$, while the odd branch, of length $L$, is as in (a). For $p=1$, $\bm\Omega_{\mathrm s}$ is $\bigl[\begin{smallmatrix}0&(\sqrt2-1)\rho\\(\sqrt2-1)\rho&-\rho^2\end{smallmatrix}\bigr]$ on the last two coordinates and $\bm\Omega_{\mathrm a}=-\rho^2\be_{L-1}\be_{L-1}^\T$. (c) [recursion] Hence the transform of length $M$ is the fold, followed by the transforms of lengths $\lceil M/2\rceil$ and $\lfloor M/2\rfloor$ on the two branches and a correction that diagonalizes a diagonal matrix plus one of rank at most $p$, or $p+1$ on the branch that holds the center sample of an odd window. The same step applies to every branch of length at least $4p$, and the recursion ends at branches of length between $2p$ and $4p-1$, which Proposition~\ref{prop:fast} transforms directly. Fig.~\ref{fig:recursion} draws the tree. (d) [dyadic tower] Every fold is of even length exactly when $M=m\cdot2^k$ with $2p\le m<4p$, and the recursion then has $k$ levels and $2^k$ leaves of length $m$. The full tower down to length $2p$ is the case $M=2p\cdot2^k$, the dyadic tower of the Haar basis: the folds are then the matched transform of the iterated wreath product $\Z_2\wr\Z_2\wr\cdots\wr\Z_2$ of $k$ factors acting on the nested half-windows, whose isotypic components are the Haar multiresolution subspaces~\cite{haar1910theorie,mallat1989theory,thornton2026unification}; the Haar basis and the Walsh-Hadamard basis (the orthogonal $\pm1$ basis of the Hadamard matrix of order $2^k$) are two bases adapted to those components, and the tower of folds, which folds both branches at every node, produces the Walsh-Hadamard one (Fig.~\ref{fig:tower}); the window adds a correction of rank at most $p$ at every node.
\end{proposition}

\begin{proof}
\ifextended
(a) \emph{Rows away from the center.} Rows $n<L-p$ of $\bQ_{\mathrm s}$ and of $\bQ^{(L)}$ coincide. Both carry the outer corner block of Theorem~\ref{thm:departure} and the Toeplitz band, and the fold changes nothing there because these rows have no entry in the second half of the window. The outer corner block occupies rows $n<p$, so it stays clear of the center when $L-p\ge p$, which is the hypothesis $M\ge4p$.

\emph{Rows at the center.} Rows $n\ge L-p$ of $\bQ$ couple to the coordinates $L,\dots,L+p-1$, which the fold maps onto $L-1,\dots,L-p$ with the sign of the branch, and $\bQ^{(L)}$ carries its inner corner block there. The difference is therefore confined to the last $p\times p$ block. For $p=1$ the center entry of $\sigma^2\bQ$ is $1+\rho^2$, the fold adds $\mp\rho$ from the coupling across the center, and the center entry of $\sigma^2\bQ^{(L)}$ is $1$, so
\[
\sigma^2(\bQ_{\mathrm s})_{L-1,L-1}=1+\rho^2-\rho,\qquad \sigma^2(\bQ_{\mathrm a})_{L-1,L-1}=1+\rho^2+\rho,\qquad \sigma^2Q^{(L)}_{L-1,L-1}=1,
\]
which gives the two values stated.

(b) \emph{Odd length.} With the fold of Proposition~\ref{prop:fast}(f) the odd branch does not contain the center sample and is treated exactly as in (a). In the even branch, rows $n<L-p$ of $\bQ_{\mathrm s}$ and of $\bQ^{(L+1)}$ coincide for the same reason. Rows $L-p\le n\le L$ couple to the center sample and, through the fold, to the coordinates $L+1,\dots,L+p$, which the fold maps onto $L-1,\dots,L-p$, and $\bQ^{(L+1)}$ carries its inner corner block in its last $p$ rows. The difference is therefore confined to the last $(p+1)\times(p+1)$ block, which bounds its rank by $p+1$. For $p=1$ the fold adds nothing to the entry $(L-1,L-1)$, whose mirror partner $L+1$ is two samples away, and it multiplies the coupling to the center by $\sqrt2$, so on the last two coordinates
\[
\sigma^2\bQ_{\mathrm s}=\begin{bmatrix}1+\rho^2&-\sqrt2\rho\\-\sqrt2\rho&1+\rho^2\end{bmatrix},\qquad
\sigma^2\bQ^{(L+1)}=\begin{bmatrix}1+\rho^2&-\rho\\-\rho&1\end{bmatrix},
\]
which gives $\bm\Omega_{\mathrm s}$. In the odd branch the entry $(L-1,L-1)$ is $1+\rho^2$ against the corner value $1$ of $\sigma^2\bQ^{(L)}$, which gives $\bm\Omega_{\mathrm a}$.

(c) \emph{One level.} Let $\bU^{(\ell)}$ be an orthonormal eigenbasis of $\bQ^{(\ell)}$, with $\bU^{(\ell)\T}\sigma^2\bQ^{(\ell)}\bU^{(\ell)}=\bLam^{(\ell)}$ diagonal. For each branch, with $\ell$ its length and $\bm\Omega$ its center block,
\[
\bU^{(\ell)\T}\sigma^2\bQ_{\mathrm{branch}}\bU^{(\ell)}=\bLam^{(\ell)}-\bU^{(\ell)\T}\bm\Omega\,\bU^{(\ell)},
\]
a diagonal matrix minus a symmetric matrix of rank at most $p$, or $p+1$ on the branch that holds the center sample. So the transform of length $M$ is the fold, then the transform of length $\ell$ on each branch, then the eigenbasis of this diagonal-minus-low-rank matrix, the correction.

\emph{The recursion.} $\bU^{(\ell)}$ is again the KLT of an AR($p$) window, of length $\lceil M/2\rceil$ or $\lfloor M/2\rfloor$, so the same step applies to it when its length is at least $4p$. By induction on the length the recursion ends at windows of length between $2p$ and $4p-1$. Each of these has length at least $2p$, and at least $2p+1$ when it is odd, so Proposition~\ref{prop:fast} gives its transform, through part (f) at odd length.

(d) \emph{When every fold is even.} A fold halves an even length, and folds are applied while the length is at least $4p$. So every fold is of even length exactly when $M,M/2,\dots,M/2^{k-1}$ are even and at least $4p$ and $m=M/2^k<4p$, that is, when $M=m\cdot2^k$ with $2p\le m<4p$. The recursion then has $k$ levels and $2^k$ leaves of length $m$, and the leaves have length $2p$ when $M=2p\cdot2^k$. The group generated by the reflections of the nested half-windows is then the automorphism group of the dyadic tree over the samples, the iterated wreath product $\Z_2\wr\Z_2\wr\cdots\wr\Z_2$. Its isotypic components are the Haar multiresolution subspaces: the constants, the difference between the two halves, the differences between the quarters within each half, and so on, of dimensions $1,1,2,4,\dots$ Within a component of dimension above one the group fixes the subspace, not the basis: the Haar basis keeps each vector on one block of the tree, while the tower of folds, which carries both the sum and the difference branch down to the leaves, produces the Walsh-Hadamard basis, each vector of which also lies in one component.
\fi
\end{proof}

Proposition~\ref{prop:recursive} is the group reading of the recursive factorization of~\cite{reznik2026direct}, which is its case $p=1$, and the reading of dilation and contraction for this window: for $M=2p\cdot2^k$ the window is matched to the dyadic tower of the Haar basis at every level up to a block of rank at most $p$ at the center of each node, and by Proposition~\ref{prop:cascade} each node correction is again a cascade of at most $p$ Cauchy stages. A fold of odd length is a step of the linear algebra only, with one more stage on the branch that holds the center sample. The next fold of that branch pairs the center sample with the end pair $(x_0+x_{M-1})/\sqrt2$, which no permutation of the samples does, so the group reading of part (d) needs every fold to be of even length. For operation counts Algorithm~\ref{alg:fast}, with fixed cores and one cascade per branch, is the realization with the smaller count. The recursive form puts $2^\ell$ corrections of size $M/2^\ell$ at level $\ell$, $2M^2$ multiplications over all levels when applied densely.

\begin{remark}\label{rem:wiener}
For a co-channel sum $\bm x=\bm u+\bm v$ of uncorrelated AR($k_1$) and AR($k_2$) windows with known models, the linear minimum mean-square estimate of $\bm u$ is $\bR_{\mathrm u}(\bR_{\mathrm u}+\bR_{\mathrm v})^{-1}\bm x$. In the DCT-II basis this separator is the diagonal spectral filter $S_{\mathrm u}/(S_{\mathrm u}+S_{\mathrm v})$ on the grid $\pi k/M$ plus a correction of rank at most $2\max(k_1,k_2)$. The separator equals $(\bQ_{\mathrm u}+\bQ_{\mathrm v})^{-1}\bQ_{\mathrm v}$, and with $\bm A$ and $\bm B$ the DCT-II class members of $\bQ_{\mathrm u}+\bQ_{\mathrm v}$ and of $\bQ_{\mathrm v}$, whose quotient $\bm A^{-1}\bm B$ is the diagonal filter, and $\bm E$, $\bm F$ their corner departures of Theorem~\ref{thm:departure}, the difference $(\bm A+\bm E)^{-1}(\bm B+\bm F)-\bm A^{-1}\bm B=(\bm A+\bm E)^{-1}(\bm F-\bm E\bm A^{-1}\bm B)$ has nonzero rows only among the first and last $\max(k_1,k_2)$. Separation therefore runs at the price of a fast cosine transform and a low-rank stage, and Section~\ref{sec:num} finds ranks of \wienerRankSumOneOne, \wienerRankSumTwoOne{} and \wienerRankSumTwoThree{} in three cases, each equal to the bound. The models are an input here. Their estimation from the sum, which is the identifiability of a sum of autoregressive processes from one second-order statistic, is not treated in this paper.
\end{remark}

\begin{figure}[t]
\centering
\resizebox{\ifextended 0.62\textwidth\fi}{!}{\begin{tikzpicture}[
  font=\footnotesize,
  win/.style={draw,rounded corners=1.5pt,minimum width=15pt,minimum height=11pt,inner sep=2pt,fill=white},
  oddfold/.style={win,line width=1.1pt},
  leaf/.style={win,fill=black!10},
  edge/.style={line width=0.5pt},
  rk/.style={font=\scriptsize,inner sep=1pt,fill=white},
  ctr/.style={circle,fill=black,inner sep=0pt,minimum size=3pt},
  lev/.style={font=\scriptsize,black!60,anchor=east}]
\begin{scope}
\node[font=\footnotesize,anchor=west] at (-3.5,0.8) {(a)\ \ $M=13$, $p=1$: the folds at $13$ and at $7$ are odd};
\node[oddfold] (a13) at (0,0) {$13$};
\node[oddfold] (a7)  at (-1.6,-1.0) {$7$};  \node[ctr] at ([xshift=-3pt,yshift=3pt]a7.south east) {};
\node[win]     (a6)  at ( 1.6,-1.0) {$6$};
\node[win]     (a4)  at (-2.4,-2.0) {$4$};  \node[ctr] at ([xshift=-3pt,yshift=3pt]a4.south east) {};
\node[leaf]    (a3a) at (-0.8,-2.0) {$3$};
\node[leaf]    (a3b) at ( 0.8,-2.0) {$3$};
\node[leaf]    (a3c) at ( 2.4,-2.0) {$3$};
\node[leaf]    (a2a) at (-2.8,-3.0) {$2$};
\node[leaf]    (a2b) at (-2.0,-3.0) {$2$};
\draw[edge] (a13) -- (a7) node[rk,midway,above left=-3pt] {rank $\le p{+}1$};
\draw[edge] (a13) -- (a6);
\draw[edge] (a7) -- (a4) node[rk,midway,above left=-3pt] {rank $\le p{+}1$};
\draw[edge] (a7) -- (a3a); \draw[edge] (a6) -- (a3b); \draw[edge] (a6) -- (a3c);
\draw[edge] (a4) -- (a2a); \draw[edge] (a4) -- (a2b);
\node[lev] at (-3.3,0) {$M\ge4p$: fold};
\node[lev] at (-3.3,-1.0) {$\lceil M/2\rceil,\ \lfloor M/2\rfloor$};
\node[lev,align=right] at (-3.3,-3.0) {leaves $2p\le\ell<4p$:\\ Prop.~\ref{prop:fast} directly};
\end{scope}
\begin{scope}[yshift=-4.5cm]
\node[font=\footnotesize,anchor=west] at (-3.5,0.8) {(b)\ \ $M=16=2p\cdot2^3$: every fold is even, the dyadic tower};
\node[win] (b16) at (0,0) {$16$};
\node[win] (b8a) at (-1.6,-1.0) {$8$}; \node[win] (b8b) at ( 1.6,-1.0) {$8$};
\node[win] (b4a) at (-2.4,-2.0) {$4$}; \node[win] (b4b) at (-0.8,-2.0) {$4$};
\node[win] (b4c) at ( 0.8,-2.0) {$4$}; \node[win] (b4d) at ( 2.4,-2.0) {$4$};
\foreach \i/\x in {1/-2.8,2/-2.0,3/-1.2,4/-0.4,5/0.4,6/1.2,7/2.0,8/2.8}{\node[leaf] (b2\i) at (\x,-3.0) {$2$};}
\draw[edge] (b16) -- (b8a); \draw[edge] (b16) -- (b8b);
\draw[edge] (b8a) -- (b4a); \draw[edge] (b8a) -- (b4b); \draw[edge] (b8b) -- (b4c); \draw[edge] (b8b) -- (b4d);
\draw[edge] (b4a) -- (b21); \draw[edge] (b4a) -- (b22); \draw[edge] (b4b) -- (b23); \draw[edge] (b4b) -- (b24);
\draw[edge] (b4c) -- (b25); \draw[edge] (b4c) -- (b26); \draw[edge] (b4d) -- (b27); \draw[edge] (b4d) -- (b28);
\node[lev] at (-3.3,0) {$\Z_2$};
\node[lev] at (-3.3,-1.0) {$\Z_2\wr\Z_2$};
\node[lev] at (-3.3,-2.0) {$\Z_2\wr\Z_2\wr\Z_2$};
\node[lev] at (-3.3,-3.0) {leaves $2p$};
\end{scope}
\end{tikzpicture}}
\caption{The recursion of Proposition~\ref{prop:recursive}. Each node is a window of length $\ell$, and its two children are the branches of its fold, each a diagonal matrix minus a block of rank at most $p$ in the eigenbasis of the shorter window. (a) $M=13$, $p=1$: the folds at $13$ and at $7$ are odd (heavy outline). The center sample goes to the branch marked with a dot, whose block has rank at most $p+1$. The leaves (shaded) have length between $2p$ and $4p-1$ and are transformed directly by Proposition~\ref{prop:fast}, through part (f) when the length is odd. (b) $M=16=2p\cdot2^3$: every fold is even, the recursion is the dyadic tower of the Haar basis of Fig.~\ref{fig:tower}, and its levels carry the iterated wreath product.}
\label{fig:recursion}
\end{figure}
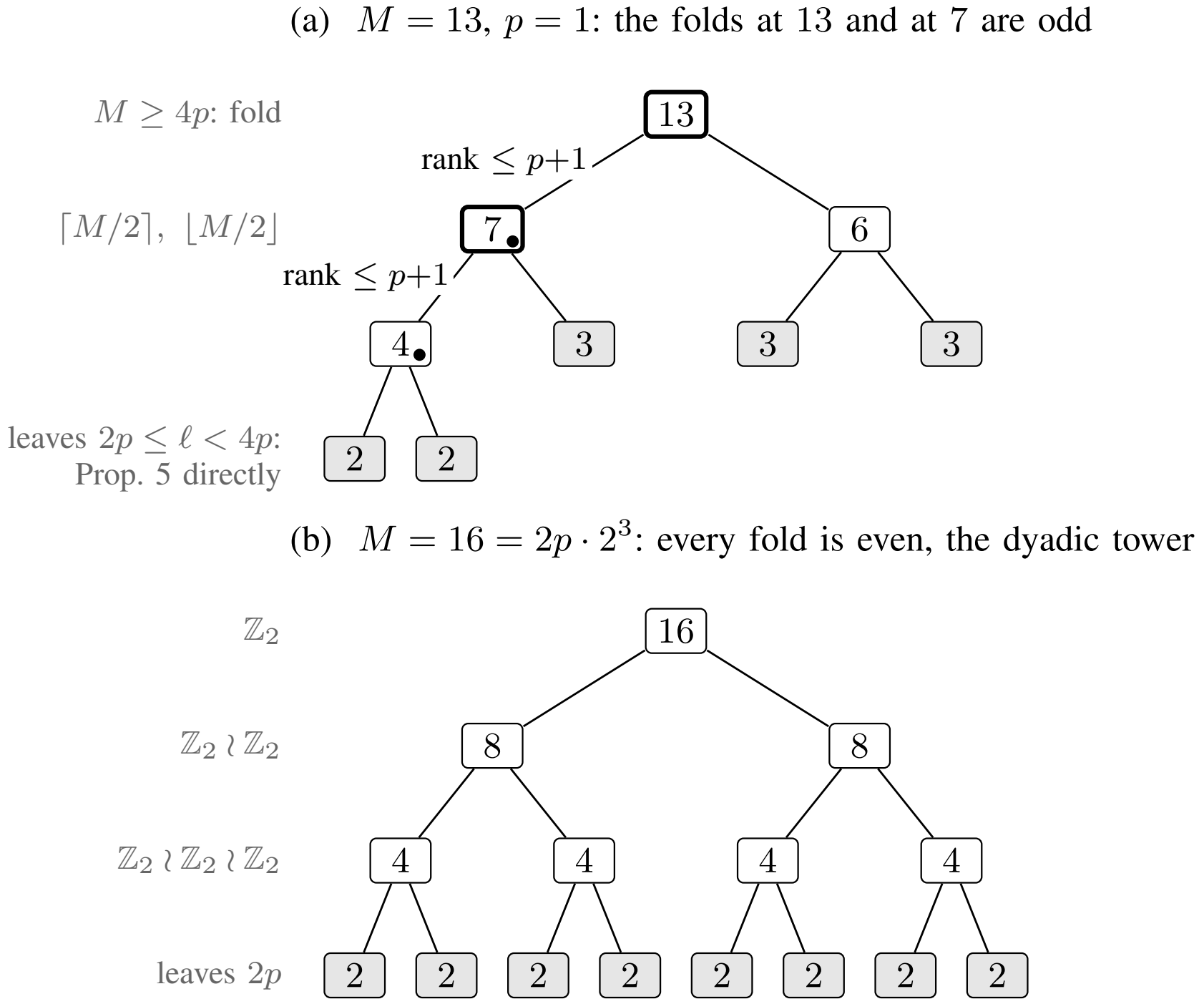

\begin{algorithm}[t]
\caption{Exact KLT of a window of length $M=2L$ with a rational spectrum}
\label{alg:fast}
\begin{algorithmic}[1]
\REQUIRE the model (AR coefficients or poles; the MA part or the summands for a rational spectrum), $\sigma^2$, $M=2L$
\STATE \textbf{Design} (once per model), for each branch:
\STATE\quad AR($p$): $\bD_1$ from~\eqref{eq:departure}; $\bLam=\operatorname{diag}(c(\pi j/L))$ or $\operatorname{diag}(c(\pi(2j+1)/2L))$; $\bm\Theta=-\bW\bD_1\bW^\T$ with $\bW=\bK[:,0{:}p]$
\STATE\quad rational spectrum or sum: $\bLam=\operatorname{diag}(S)$ on the branch grid and $\bm\Theta=\bK\bZ\bK^\T$ from~\eqref{eq:branchcov}
\STATE\quad the eigenpairs $(\xi_h,\bm\psi_h)$ of $\bm\Theta$, $\kappa=\operatorname{rank}\bm\Theta$. The stages $\bV^{(1)},\dots,\bV^{(\kappa)}$ of Proposition~\ref{prop:cascade}, each by one scalar secular equation and its Cauchy vectors, or their product $\bV$ at once from~\eqref{eq:secular}. The eigenvalues $\bLam^{(\kappa)}$, and the KLT eigenvalues $\sigma^2/\lambda$ (precision side) or $\lambda$ (covariance side)
\STATE \textbf{Apply} (per window $\bm x$):
\STATE\quad if the poles are rotated by $\omega$: $\bm x\gets\bm E_\omega^*\bm x$. If $a(z)=b(z^m)$: permute into phases and run the steps below on each phase with $M/m$, $p/m$
\STATE\quad $\bm y\gets\bF\bm x$ \hfill $L$ butterflies
\STATE\quad $\bm u_{\mathrm s}\gets\bK_{\mathrm{II}}\bm y[0{:}L]$, $\bm u_{\mathrm a}\gets\bK_{\mathrm{IV}}\bm y[L{:}M]$ \hfill fast DCT-II and DCT-IV
\STATE\quad $\bm z_{\mathrm s}\gets\bV^{(\kappa)\T}\cdots\bV^{(1)\T}\bm u_{\mathrm s}$, and $\bm z_{\mathrm a}$ likewise \hfill $\kappa$ Cauchy stages per branch
\ENSURE $\bm z_{\mathrm s}$, $\bm z_{\mathrm a}$ ordered by eigenvalue: the KLT coefficients of $\bm x$
\end{algorithmic}
\end{algorithm}

\begin{table*}[t]
\caption{Proposition~\ref{prop:fast} by pole pattern: order, length, rank of the correction term in the even and odd branches, the secular determinant at the branch eigenvalues (largest value), the off-diagonal fraction left by the assembled transform, the eigenvalue shift within a branch and jointly, and the largest deviation of a correction from the identity}
\label{tab:fast}
\centering\footnotesize
\setlength{\tabcolsep}{6pt}
\begin{tabular}{@{}lccclccc@{}}
\toprule
Pattern & $p$ & $M$ & rank & secular & off-diag. & shift & $\max|\bV-\bI|$\\
\midrule
\input{fast_table.tex} 
\bottomrule
\end{tabular}
\end{table*}

\subsection{Operation Count}\label{sec:count}

Let $\mathrm{ops}(M)$ be the number of arithmetic operations of the apply stage. The fold is $L$ butterflies, $M$ additions, with the scaling absorbed into the next stage. The DCT-II core of length $L$ is again a persymmetric class member, so the fold applies to it a second time and splits it into a DCT-II and a DCT-IV core of length $L/2$~\cite{chen1977fast}. The DCT-IV core is a Fourier transform of length $L/2$ with twiddle factors before and after~\cite{wang1984fast,rao1990discrete}, so both cores take $O(L\log L)$ operations, and the recursion $\mathrm{ops}_{\mathrm{II}}(L)=L+\mathrm{ops}_{\mathrm{II}}(L/2)+\mathrm{ops}_{\mathrm{IV}}(L/2)$ with $\mathrm{ops}_{\mathrm{IV}}(L)=O(L\log L)$ gives $O(M\log M)$ for the fixed part. In the group reading the fixed part is fast for the reason the classical transforms are fast~\cite{thornton2026unification}: the matched transform of the dihedral group factors along a tower of subgroups, here the reflections of the nested half-windows for the DCT-II chain and the cyclic tower of the Fourier realization for the DCT-IV stage, and every level is a sum and a difference per sample. The correction is the part that is not group-determined, and the order enters the count only through it. Applied as one dense matrix per branch it takes $2L^2=M^2/2$ multiplications whatever $\kappa$, since the product $\bV^{(1)}\cdots\bV^{(\kappa)}$ is formed at design time. This bounds the exact transform by $O(M^2)$ with a constant of one half, against the $M^2$ of a direct eigenbasis. Applied stage by stage, each $\bV^{(h)\T}$ is a Cauchy matrix $[\,\psi'_{h,j}/(\Lambda^{(h-1)}_{jj}-\lambda_i)\,]$ between two diagonal scalings, which the fast multipole method~\cite{greengard1987fast} evaluates in $O(L\log L)$ operations to a fixed accuracy, as~\cite{reznik2026exact} does for the single stage of $p=1$, so the whole transform runs in $O(\kappa M\log M)$ operations to that accuracy: $O(pM\log M)$ for the AR($p$) window, $O(\max(p,q)M\log M)$ for an ARMA($p,q$) window and $O((k_1+k_2)M\log M)$ for a sum. For a fixed order this is the $O(M\log M)$ of the classical fast transforms, with a constant that grows linearly in the order. In the worst case over models of a given order nothing else enters, since neither the fold, the cores nor the number of stages depends on the poles. The factor $\kappa$ is not inherent, as a measurement of the merged correction shows. A Cauchy matrix with interlaced nodes and poles is evaluated fast because its off-diagonal blocks, in the order of the nodes, have small numerical rank at every level of the hierarchical partition into halves, quarters and so on, which is what the multipole method uses. The merged correction $\bV=\bV^{(1)}\cdots\bV^{(\kappa)}$ of one branch is a product of $\kappa$ such matrices, and Section~\ref{sec:num} finds that it has the same property with ranks within one of a single stage's. For the windows with real poles of orders one, two and four, at $M=256$ and $512$, the off-diagonal blocks of the merged correction have rank at most \epsLevelSixMaxReal{} at absolute accuracy $10^{-6}$ and \epsLevelTenMaxReal{} at $10^{-10}$ at every level, within one of the first-order values at the same length. For the resonant window, whose core grid is not monotone in the index, the same holds in the order of the grid values with ranks at most \epsLevelSixMaxRes{} and \epsLevelTenMaxRes. Applied in that hierarchical form the merged correction of the windows measured takes $O(rM\log M)$ operations with $r$ set by the accuracy and not by the order, the count of one Cauchy stage: the exact transform of order $p$ is then, as measured, as compressible as the first-order one, and the order enters the apply count through the design alone. This is a measurement on the windows of Section~\ref{sec:num}, not a theorem, and whether the hierarchical ranks of the merged correction stay within one of a single stage's for every order and every pole pattern is open. The polyphase pre-stage divides the count by running $m$ graphs of length $M/m$, and the demodulation adds $M$ complex multiplications. The design stage runs once per model: $\kappa$ scalar secular equations per branch, each with $L$ roots bracketed by the running diagonal and each evaluation taking $O(L)$ operations, so $O(\kappa L^2)$ per branch by bisection to machine precision, the same order as the $p\times p$ determinant route of~\eqref{eq:secular}, or $O(L^3)$ by a direct eigendecomposition of the branch matrix.

\subsection{Complex Data and the Cyclic Class}\label{sec:complexdata}

Proposition~\ref{prop:modulated} treats a complex window whose poles are the real ones rotated by a
common carrier, which the modulation diagonal conjugates back to the real case. A window with
arbitrary complex coefficients is not of that form, and for it the construction of
Section~\ref{sec:butterfly} breaks at its first step.

The reason is that the map which commutes with a Hermitian Toeplitz matrix is
$\bx\mapsto\bJ\overline{\bx}$, and that map is antilinear. An antilinear map has no eigenspace
decomposition of the kind the parity fold rests on, so $\C^M$ does not split into two invariant
subspaces under it and the butterfly of the exact reflection is unavailable. What remains matched
is the cyclic group $\Z_M$, whose matched transform is the unitary discrete Fourier transform
$\bF_M$ with entries $M^{-1/2}e^{2\pi i nk/M}$ and whose invariant class is the circulants. The
class member is the circulant of the symbol, and the question is what the window precision departs
from it by.

\begin{proposition}\label{prop:complex}
Let $a(z)=1-\sum_{j=1}^{p}\phi_jz^j$ have complex coefficients, no root on the unit circle and
$a_p\ne0$, let the window be of length $M>2p$, and write $c_d=\sum_k\overline{a_k}a_{k+d}$ for the
symbol coefficients, so that $c_{-d}=\overline{c_d}$. Then $\sigma^2\bQ$ is the banded Hermitian
Toeplitz matrix of $c$ with the two $p\times p$ corner blocks
\begin{equation}\label{eq:complexcorner}
  (\bG)_{uv}=\sum_{s\le\min(u,v)}a_{u-s}\,\overline{a_{v-s}},\qquad 0\le u,v<p,
\end{equation}
in the leading position and the doubly reversed conjugate of $\bG$ in the trailing one. Writing
$\bC$ for the circulant with first column $c_{d\bmod M}$, the departure $\sigma^2\bQ-\bC$ is
supported on the first $p$ and the last $p$ coordinates, that is, it vanishes outside the four
$p\times p$ corner blocks, and its rank is exactly $2p$.
\end{proposition}

\ifextended
\begin{IEEEproof}
Two steps, the identity $\bQ\bR=\bI$ and then the structure of the departure.

\emph{The precision.} Let $\bQ$ be the matrix the statement defines and let $r_m$ be the
autocovariance of the stationary solution, so that $r_{-m}=\overline{r_m}$ and
$r_m=\sum_{j=1}^{p}\phi_jr_{m-j}$ for $m\ge1$. Take an interior row $u$ with $p\le u<M-p$. There
the row of $\bQ$ is the full symbol band and
\begin{equation}
  \sigma^2(\bQ\bR)_{uv}=\sum_{d=-p}^{p}c_d\,r_{u-d-v}
    =\sum_{k,\ell}\overline{a_k}\,a_\ell\,r_{u-\ell+k-v},
\end{equation}
which is the Yule-Walker identity for the filtered process: $\sum_\ell a_\ell x_{t-\ell}$ is the
innovation sequence, whose covariance with $x_{t-m}$ vanishes for $m>0$ by causality and equals
$\sigma^2$ at $m=0$, so the sum is $\sigma^2\delta_{uv}$. Take now a border row $u<p$. The full
symbol band would reach indices below zero, and the terms it would place there are exactly the
terms of the sum above with $k>u$. Replacing the leading block by $\bG$ deletes those terms, and
the one-sided relations $r_m=\sum_j\phi_jr_{m-j}$, valid for $m\ge1$, make the deleted terms sum to
zero against the columns of $\bR$ at $v\ge u$ while restoring the value at $v=u$. The trailing
border is the same computation on the reversed conjugated window, which is the stationary window of
the model with coefficients $\overline{a}$, so its block is the doubly reversed conjugate of $\bG$.
Hence $\bQ\bR=\bI$ and $\bQ$ is the precision.

\emph{The departure.} For $M>2p$ the matrices $\sigma^2\bQ$ and $\bC$ carry the same symbol band in
every entry whose row and column indices are not both within $p$ of an end, so
$\sigma^2\bQ-\bC$ vanishes outside the four $p\times p$ corner blocks, which is the stated support.
For the rank, write $\bC=\bm A_c^{\herm}\bm A_c$ with $\bm A_c$ the circulant that applies
the filter $a$ cyclically, and $\sigma^2\bQ=\bm L^{\herm}\bm L$ with $\bm L$ equal to $\bm A_c$
except in its first $p$ rows, which are the rows of a Cholesky factor of the inverse covariance of
the first $p$ samples, the exact start. The two products differ only through those rows, so
$\sigma^2\bQ-\bC=\bm S^{\herm}\operatorname{diag}(\bI_p,-\bI_p)\bm S$ with $\bm S$ the
$2p\times M$ matrix that stacks the first $p$ rows of $\bm L$ over the first $p$ rows of $\bm A_c$.
On the columns $0,\dots,p-1$ and $M-p,\dots,M-1$ the matrix $\bm S$ is block triangular: the rows
from $\bm L$ have a nonsingular triangular block on the first columns and are zero on the last
ones, since $M>2p$, and the wrap of the rows from $\bm A_c$ is a triangular block on the last
columns with $a_p$ on its diagonal, nonsingular under the hypothesis $a_p\ne0$. So $\bm S$ has
full row rank $2p$, and with $\operatorname{diag}(\bI_p,-\bI_p)$ nonsingular the rank of the
departure is exactly $2p$.
\end{IEEEproof}
\fi

The same border support settles how much a cyclic-class basis can be behind the matched one, and at
what window length. The answer is carried by the $2p$ border eigenvalues alone, which makes it a
quantity an application can be screened by before any data is taken.

\begin{proposition}\label{prop:borderlik}
Let the window be of length $M>2p$ with $a_p\ne0$ as above, let $\bR$ be its covariance, let $\bF$
be the unitary discrete Fourier transform, let $\bC$ be the cyclic class member whose symbol is the
spectral density sampled on the $M$-point grid, and let
$\bLam=\operatorname{diag}\operatorname{diag}(\bF\bR\bF^{\herm})$ be the moment-matched diagonal
model in that basis. Write $\bDel=\bR^{-1}-\bC^{-1}$, let $B$ be the border index set of the first
and last $p$ coordinates, and let $\mu_1,\dots,\mu_{2p}$ be the eigenvalues of
$\bR_B\bDel_B$. Then
\begin{enumerate}
\item $\bDel$ vanishes off $B\times B$ and has rank exactly $2p$;
\item with $h(\mu)=-\mu-\log(1-\mu)$, the divergence of the class member from the truth is
  \begin{equation}\label{eq:borderlik}
    D(\bR\,\|\,\bC)=\sum_{i=1}^{2p}h(\mu_i);
  \end{equation}
\item the per-sample divergence of the best Fourier-diagonal model obeys
  $D(\bR\,\|\,\bLam)/M\le M^{-1}\sum_i h(\mu_i)$, and the sum tends as $M$ grows to a limit
  $K_p(a,\sigma^2)$ fixed by the model and not by the window length, so that per-sample quantity is
  $O(1/M)$;
\item for $p=1$ with $a(z)=1-\phi z$ the eigenvalues tend, as $M$ grows, to $|\phi|/(1+|\phi|)$ and $-|\phi|/(1-|\phi|)$, and
  \begin{equation}\label{eq:kone}
    K_1(\phi)=\frac{2|\phi|^2}{1-|\phi|^2}+\log\!\left(1-|\phi|^2\right).
  \end{equation}
\end{enumerate}
\end{proposition}

\ifextended
\begin{IEEEproof}
Part 1 is Proposition~\ref{prop:complex} and the form of $\bC$. By that proposition $\sigma^2\bR^{-1}$
is the banded Toeplitz matrix of the symbol with the two corner blocks
of~\eqref{eq:complexcorner}, and $\sigma^2\bC^{-1}$ is the same band closed cyclically, which adds
the wrap in the two off-diagonal corners and completes the band in the two diagonal corners. The two
agree on every interior row, so the difference vanishes off $B\times B$, and on $B$ it is the corner
correction against the wrap, of rank exactly $2p$ when $a_p\ne0$ by the same argument that fixes the
rank there.

For part 2 put $\bA=\bR^{-1}$, so $\bC^{-1}=\bA-\bDel$ and
$\operatorname{tr}(\bC^{-1}\bR)=M-\operatorname{tr}(\bR\bDel)$, while
$\log\det\bC-\log\det\bR=-\log\det(\bI-\bR\bDel)$. The circular complex divergence of $\bC$ from
$\bR$ is $\operatorname{tr}(\bC^{-1}\bR)-M+\log\det\bC-\log\det\bR$, which is therefore
\begin{equation}
  D(\bR\,\|\,\bC)=-\operatorname{tr}(\bR\bDel)-\log\det(\bI-\bR\bDel).
\end{equation}
Because $\bDel$ is supported on $B\times B$, the nonzero spectrum of $\bR\bDel$ is that of
$\bR_B\bDel_B$, which has at most $2p$ members. Writing the trace and the determinant over that
spectrum gives $\sum_i(-\mu_i-\log(1-\mu_i))$, which is~\eqref{eq:borderlik}. No factor of one half
appears, the data being circular complex rather than real.

For part 3, among models diagonal in $\bF$ the divergence
$D(\bR\,\|\,\bF^{\herm}\operatorname{diag}(d)\bF)=\sum_k g_k/d_k-M+\sum_k\log d_k-\log\det\bR$,
with $g_k=(\bF\bR\bF^{\herm})_{kk}$, is stationary only at $d_k=g_k$ and is convex in $\log d$,
so $\bLam$ is its minimizer and in particular $D(\bR\,\|\,\bLam)\le D(\bR\,\|\,\bC)$. The limit
exists because $\bDel_B$ is fixed by $a$ and $\sigma^2$ through~\eqref{eq:complexcorner} and the
wrap, while $\bR_B$ has the two diagonal $p\times p$ blocks equal to the Toeplitz matrix of the
autocovariance, independent of $M$, and the two cross-border blocks decay with the impulse response
of $1/a$, so $\bR_B\bDel_B$ converges and with it the sum.

For part 4, with $a(z)=1-\phi z$ and $r=|\phi|$ the matrix $\sigma^2\bR^{-1}$ is tridiagonal with $1+r^2$ on the
interior diagonal, $1$ at the two ends of the diagonal and $-\overline{\phi}$ above the diagonal,
while $\sigma^2\bC^{-1}$ carries $1+r^2$ at every diagonal entry and the wrap $-\phi$ in the two
far corners. Hence on $B=\{1,M\}$,
\begin{equation}
  \sigma^2\bDel_B=\begin{bmatrix}-r^2&\phi\\ \overline{\phi}&-r^2\end{bmatrix},\qquad
  \bR_B=\frac{\sigma^2}{1-r^2}\begin{bmatrix}1&\phi^{M-1}\\ \overline{\phi}^{\,M-1}&1\end{bmatrix}.
\end{equation}
The cross terms vanish as $M$ grows, leaving $\bR_B\bDel_B\to(1-r^2)^{-1}\sigma^2\bDel_B$, whose eigenvalues
are $(-r^2\pm r)/(1-r^2)$, that is $r/(1+r)$ and $-r/(1-r)$. Then
$h(r/(1+r))=-r/(1+r)+\log(1+r)$ and $h(-r/(1-r))=r/(1-r)+\log(1-r)$,
and adding them gives $2r^2/(1-r^2)+\log(1-r^2)$, which is~\eqref{eq:kone}. At a finite length the
eigenvalues differ from these limits through the cross terms, by a relative amount of order $r^{M-1}$.
\end{IEEEproof}
\fi

Two consequences belong apart, because they pull in opposite directions and an application
feels both. The margin a Fourier-class basis concedes is a border effect spread over $M$ samples, so
it decays like $1/M$ with a numerator fixed by $2p$ numbers: a long transform leaves little of it to
collect however colored the process is, and the quantity to compute before choosing a window is
$K_p(a,\sigma^2)/M$. The margin a cosine class member concedes on complex data is not of that kind.
By Proposition~\ref{prop:complex} the conjugate reversal is antilinear there, so no cosine class
member is matched to the window at all, and the resulting gap is a property of the class rather than
of the border and does not decay with $M$. Section~\ref{sec:radar} measures both on recorded data
and finds them separating by the predicted amount.

A caution on reading~\eqref{eq:borderlik} as a bound rather than as the identity it is. As a root of
$a$ approaches the unit circle some $\mu_i$ approaches $1$ and $h$ diverges, while
$D(\bR\,\|\,\bLam)/M$ does not, so the inequality in part 3 becomes loose exactly where the spectrum
is sharpest. It is a ceiling to screen with, tight for moderate roots and conservative for sharp
ones.

The support is known in advance, so the complex construction needs no sampling: the $2p$ border
rows and columns are formed directly from~\eqref{eq:complexcorner} and the wrap of the circulant,
the Hermitian eigendecomposition of that $2p\times2p$ block supplies at most $2p$ generators, the
generators are the corresponding columns of $\bF_M^{\herm}$, and the cascade that follows is the
Hermitian counterpart of Proposition~\ref{prop:cascade}, with real eigenvalues and complex Cauchy
vectors.

\begin{remark}\label{rem:complexsplit}
Two consequences are facts about the realization rather than about the structure. The conjugation
in~\eqref{eq:complexcorner} sits on the second index. Placing it on the first leaves a Hermitian
matrix that satisfies none of the identities, and the check $\bQ\bR=\bI$ is what exposes that. And
a complex window whose coefficients are real splits into two independent real transforms, one on
the real part of the data and one on the imaginary part, which uses the fold and two cores of
length $M/2$ in place of one core of length $M$. The released code takes that route whenever the
coefficients are real.
\end{remark}

\subsection{Windows on the Covariance Side}\label{sec:covside}

Corollary~\ref{cor:rational} bounds the departure of an ARMA($p,q$) window or a co-channel sum from
its DCT-II class member, and Corollary~\ref{cor:ma} gives the moving-average case in the two Hankel
corners. Those bounds hold on the covariance, and unlike the autoregressive case there is no closed
form for the generator, since the covariance of a mixed window is not banded and its precision is
not. The construction therefore works on the covariance side, and it forms nothing of the window's
size at any point.

\begin{proposition}\label{prop:covside}
Let $\bA$ be the branch block, in core coordinates, of the departure of a window covariance from
its class member, symmetric of rank at most $r$. One application of $\bA$ to a vector takes
$O(M\log M)$ operations and forms neither the covariance nor $\bA$: carry the vector back to window
coordinates through the core, unfold it, multiply by the Toeplitz covariance through a circulant
embedding of length $2M$, fold the result, take it through the core again, and subtract the
diagonal that the class member becomes in core coordinates. Let $\bOm$ hold $r+\ell$ independent
Gaussian columns, $\ell\ge1$, let $\bm Y$ be an orthonormal basis of the range of $\bA\bOm$, and let
$\bm V_{\!A}\bm\Lambda_A\bm V_{\!A}^\T$ be the eigendecomposition of $\bm Y^\T\bA\bm Y$. Then with probability one
$\bA=(\bm Y\bm V_{\!A})\bm\Lambda_A(\bm Y\bm V_{\!A})^\T$ exactly, up to rounding, and these eigenpairs are the generators of
the cascade.
\end{proposition}

\ifextended
\begin{IEEEproof}
The application: the fold and the core are the operators of Proposition~\ref{prop:fast} and each
takes $O(M\log M)$ operations. A Toeplitz matrix of size $M$ multiplies a vector through a
circulant embedding of size $2M$ in the same order. The class member is diagonal in core
coordinates by definition of the class, so subtracting it is $O(M)$. The exactness: $\bA$ is
symmetric with $\operatorname{rank}\bA=r'\le r$, so its range is an $r'$-dimensional subspace, and
$\bA\bOm$ spans it unless the $r+\ell$ Gaussian columns meet the kernel of $\bA$ in a nontrivial
way, an event of probability zero since $r+\ell>r'$. With $\operatorname{range}\bm Y
=\operatorname{range}\bA$ we have $\bm Y\bm Y^\T\bA=\bA$ and $\bA\bm Y\bm Y^\T=\bA$ by symmetry, so
$\bm Y(\bm Y^\T\bA\bm Y)\bm Y^\T=\bA$, and diagonalizing the small matrix $\bm Y^\T\bA\bm Y$
gives the eigenpairs of $\bA$ on its range, the rest being zero.
\end{IEEEproof}
\fi

The randomized range finder used here is the one of~\cite{halko2011finding}, with the difference
that the rank is known in advance from Corollary~\ref{cor:rational}, so the step is exact rather
than approximate and the oversampling $\ell$ guards only against a degenerate draw.

A co-channel sum of uncorrelated components takes the same path without forming the mixed
representation that Corollary~\ref{cor:rational} invokes. The spectral density and the
autocovariance of a sum are the sums of the components', so the class member of the sum is the sum
of the class members and the departure of the sum is the sum of the departures. The bound is the one
the corollary gives for the mixed representation the sum admits, and the construction uses only the
component autocovariances.

Integrated windows are the one family in which the real and the complex cases sit on different
classes. Proposition~\ref{prop:integrated} puts a real integrated window on the DST-VII class, with
its departure supported on the first $p+d-1$ and last $p+d$ coordinates, so the border block is
formed entrywise and no sampling is needed. A complex integrated window is not near that class in
general, because the class is real symmetric and a Hermitian window that is not real symmetric is
not. The cyclic class carries it instead.

\begin{proposition}\label{prop:complexintegrated}
Let $a(z)$ have complex coefficients with $a_p\ne0$ and no root on the unit circle, let the
integrated window of difference order $d\ge1$ be pinned at its start as in
Proposition~\ref{prop:integrated}, and let $M>2(p+d)$. Then the precision of the integrated window
is the banded Hermitian Toeplitz matrix of the symbol $c(\omega)(2-2\cos\omega)^d$ with corner
blocks in its first and last $p+d$ coordinates, and its departure from the circulant of the same
symbol is supported on those coordinates with rank at most $2(p+d)$.
\end{proposition}

\ifextended
\begin{IEEEproof}
Differencing $d$ times convolves the symbol of the Toeplitz interior with
$|1-e^{i\omega}|^{2}=2-2\cos\omega$ each time and widens the band by one each time, as in the proof
of Proposition~\ref{prop:integrated}(b), so the interior is the band of
$c(\omega)(2-2\cos\omega)^d$ and the rows touched by the corner blocks and by the pinning are the
first and last $p+d$. The circulant of the same symbol agrees with that band wherever the wrap does
not reach, which for $M>2(p+d)$ is every entry outside the four $(p+d)\times(p+d)$ corner blocks,
and the rank bound is the size of the border those blocks occupy.
\end{IEEEproof}
\fi

\begin{remark}\label{rem:factored}
The symbol of an integrated window is evaluated in the factored form
$c(\omega)\,(4\sin^2(\omega/2))^d$ rather than as a sum over the differenced coefficient sequence.
The two agree in exact arithmetic. The sum cancels to zero at $\omega=0$ and leaves a rounding
error the size of its own terms there, and for $d\ge2$ the smallest eigenvalues of the integrated
precision are of order $M^{-2d}$, so the largest eigenvalues of the covariance are read from
exactly the part of the grid where that cancellation would dominate. A realization that evaluates
the differenced sum instead returns a largest eigenvalue wrong by orders of magnitude at $d=2$.
\end{remark}

\subsection{The Two Cores at Odd Length}\label{sec:oddcores}

Part~(f) of Proposition~\ref{prop:fast} gives the odd-length fold and names its two cores. The
realization needs them as explicit orthonormal matrices with a one-transform evaluation, and it is
useful to record that the pair is forced rather than chosen.

For $M=2L+1$ the reflection fixes the center sample $m=L$, so the fold has three kinds of row: the
symmetric pairs for $n<L$, the center sample alone, and the antisymmetric pairs. The center sample
lies in the symmetric subspace, which therefore has dimension $L+1$ against $L$ for the
antisymmetric one.

\begin{proposition}\label{prop:oddcores}
Let $M=2L+1$. Among the families $\cos(2\pi(m+\alpha)(k+\beta)/M)$ the only ones carried into
themselves up to a sign by the reflection $m\mapsto M-1-m$ are those with $\alpha=\tfrac12$, and
among those the choice $\beta=0$ gives the symmetric family and $\beta=\tfrac12$ the antisymmetric
one. With the fold weighting the $L$ mirrored pairs by $\sqrt2$ and the center sample by one, the
two branch cores are
\begin{align}
  \bK_{\mathrm s}[k,n]&=\sqrt{4/M}\,\cos\bigl(2\pi(n+\tfrac12)k/M\bigr),\ n<L,\label{eq:kcoress}\\
  \bK_{\mathrm a}[k,n]&=\sqrt{4/M}\,\cos\bigl(2\pi(n+\tfrac12)(k+\tfrac12)/M\bigr),\label{eq:kcoresa}
\end{align}
with $\sqrt{2/M}\cos(\pi k)$ in the center column of $\bK_{\mathrm s}$ and its row $k=0$ divided by
$\sqrt2$. Both are orthonormal, of sizes $(L+1)\times(L+1)$ and $L\times L$, and the branch grids
are $c(2\pi k/M)$ and $c(\pi(2k+1)/M)$, which together are the grid of spacing $\pi/M$ split by the
parity of the index.
\end{proposition}

\ifextended
\begin{IEEEproof}
Apply the reflection to the candidate. Its argument becomes $2\pi(M-1-m+\alpha)(k+\beta)/M$, which
differs from $-2\pi(m+\alpha)(k+\beta)/M$ by $2\pi(M-1+2\alpha)(k+\beta)/M$. The cosine is even, so
invariance up to a sign for every $k$ asks that $(M-1+2\alpha)(k+\beta)/M$ be an integer or a
half-integer independent of $k$, and with $M$ odd that forces $M-1+2\alpha\equiv0$, hence
$\alpha=\tfrac12$. With that offset the reflected column equals the original times
$\cos(2\pi(k+\beta))$, which is $+1$ for $\beta=0$ and $-1$ for $\beta=\tfrac12$, giving the two
families and no other. The second vanishes at the center sample, as the antisymmetric subspace
requires. Orthonormality is the discrete orthogonality of the two families on $M$ points with the
stated weights, and the dimensions follow from the center sample lying in the symmetric subspace.
The grids are the samples of the symbol at the two sets of frequencies, which are the even-indexed
and the odd-indexed points of the grid of spacing $\pi/M$.
\end{IEEEproof}
\fi

\begin{remark}\label{rem:oddnames}
In the numbering of~\cite{martucci1994symmetric} the matrices~\eqref{eq:kcoress}
and~\eqref{eq:kcoresa} are the DCT-VII of $L+1$ points and the DCT-VIII of $L$ points, read with the frequency as the row
index. Their transposes, with the basis vectors as columns, are the DCT-VI and, up to the reversal of
the index and the signs, the DST-VII named in Proposition~\ref{prop:fast}(f). Each is realized by
one complex transform of length $M$ with twiddles before and after. For the transpose of the first
the kernel is antiperiodic with period $M$ as a function of the frequency index, so extending the
coefficient vector evenly in that index cancels the sum rather than doubling it, and the sum is
written instead as a shifted transform of the zero-padded sequence. Since $M$ is odd the transform
length is odd and the Bluestein route~\cite{bluestein1970linear} carries it. The departure of each
folded branch from its class member is supported on its first $p$ rows and columns with rank
exactly $p$ and the same corner block $\bD_1$ as at even length, so the cascade and its realization in
Sections~\ref{sec:fmmreal} and~\ref{sec:secularreal} are unchanged.
\end{remark}

\subsection{The Polyphase Route in the Realized Form}\label{sec:polyreal}

\begin{remark}\label{rem:polymix}
By Theorem~\ref{thm:exact}(c) the polyphase permutation turns the covariance into
$\bI_m\otimes\bR_b$, so every eigenvalue of $\bR$ carries $m$ times its multiplicity in $\bR_b$ and
the eigenvectors are determined only up to an orthonormal mixing of the $m$ phases. The identity is
as valid a choice of that mixing as any other, so the phases need no mixing at all and the
transform is a permutation followed by $m$ independent transforms of length $M/m$ and order $p/m$.
That also fixes how the route is to be tested: agreement of eigenvectors with the route switched
off is the wrong thing to ask for and would fail for a correct implementation, so the released
suite checks orthonormality, exact diagonalization of the covariance, and agreement of the
eigenvalue multiset. Detection has to be exact in both directions, since a model whose support is
not entirely on multiples of $m$, or whose $m$ does not divide $M$, must not take the route. The
admissible $m$ are the divisors of the greatest common divisor of the support of $\phi$, and the
one taken is the largest of those that also divides $M$.
\end{remark}

\subsection{Fast Evaluation of the Cauchy Stages}\label{sec:fmmreal}

A stage diagonalizes $\bLam+\xi\,\bm\psi\bm\psi^\T$ with $\bLam=\operatorname{diag}(\Lambda_j)$ ascending, the notation of Proposition~\ref{prop:cascade} with the stage index dropped. Its
eigenvalues $\lambda_k$ interlace the $\Lambda_k$ and solve the secular equation
\begin{equation}\label{eq:secularstage}
  f(\lambda)=1+\xi\sum_j\frac{w_j}{\Lambda_j-\lambda}=0,\qquad w_j=\psi_j^2,
\end{equation}
and its eigenvector matrix is a Cauchy matrix $V_{jk}\propto\hat\psi_j/(\Lambda_j-\lambda_k)$ with column
norms fixed by the Loewner weights. Forming $\bV$ densely takes $O(n^2)$ in storage and in
arithmetic, so the operation count of Section~\ref{sec:count} rests on evaluating the five sums the
stage needs without forming it: the application of $\bV$ and of its transpose, the column norms,
the secular function with its derivative, and the Loewner weights. All five live on one
hierarchical tree over the sorted nodes, in the manner of~\cite{greengard1987fast}. Three points of
that evaluation are what separate the measured growth from the predicted one, and each is stated
here as it is realized.

\emph{Local shifting.} The near field includes the diagonal term $\Lambda_i-\lambda_i$ and the two
adjacent terms, which are differences of nearly equal numbers. The implementation stores the shift
$\delta_i=\lambda_i-\Lambda_i$ and the grid gaps and forms the near-field denominators from them, so the
quantities that cancel are not formed as differences at all. With that in place the hierarchical
evaluation is more accurate than a dense evaluation of the same sums, which a validation has to
respect: a dense reference that forms $\Lambda_i-\lambda_i$ by subtraction reports the fast evaluation as
wrong by about $10^{-11}$ when the fast evaluation is the more accurate of the two.

\emph{Admissibility, and why a uniform tree is not enough.} A source box may be replaced by its
interpolant only when the target is far enough from the box in the value variable, and the test
used is that the distance from the target interval to the box is at least $\eta$ times the box's own
value width, with $\eta=0.75$ by measurement. Index-balanced boxes are value-adaptive but not
uniform, so two boxes far apart in index can be close in value, and a method that admits them looks
accurate on a random charge vector while losing three or four digits on a unit vector, which is the
case that decides whether the eigenvector columns come out orthonormal. The test alone is not
enough. With the level by level interaction list of a uniform tree, a pair that fails the test is
evaluated directly at its own level, and at a coarse level that is the product of two large box
sizes. On these node sets that fallback is a term proportional to $n^2$, measured in
Table~\ref{tab:implplan} at between \planUniMin{} and \planUniMax{} of $n^2$ per pass, the fraction
growing with the order of the model. Deferring a pair that fails at its own level to its four
children and testing it one level down, and evaluating directly only the pairs that survive to the
deepest level, leaves \planAdaptive{} direct evaluations at every length measured. Since the test
depends on positions alone, the structure is built once per stage and shared by every kernel and
every sweep, with the target intervals widened to the enclosing grid cells so that one plan stays
valid while the iterates move inside their brackets.

\emph{The Loewner weights.} The column norms require
$\sum_{j\ne i}\log\lvert1+\delta_j/(\Lambda_j-\Lambda_i)\rvert$ for every $i$, and computing it as a difference
of two large logarithm sums is unstable. Expanding $\log(1+x)$ in the far field turns each order
into a Cauchy-type sum with charges $\delta_j^k$ and kernel $(\Lambda_j-\Lambda_i)^{-k}$ on the same tree, so
the two large sums are not formed. The expansion converges while the largest shift in a source box
is well inside the distance to the target box, and the largest root's shift is bounded only by
$\xi\sum_jw_j$, so a pair is admitted here only when that holds as well. Where the
shifts are too large for the expansion the sum is evaluated on the target side, as
Section~\ref{sec:secularreal} describes. The interpolation order sets the accuracy floor of the
design stage, at about $2\times10^{-12}$ absolute at the lengths where the hierarchical stages are
selected.

\subsection{The Secular Equation in the Realized Form}\label{sec:secularreal}

Each sweep of the hierarchical branch is one pass over every iterate at once, so the number of
sweeps, and not the number of roots, sets the expense of the design stage. A profile at $L=8192$
puts the sweeps at 81 percent of it. The schedule inherited from the dense branch is a fixed number
of bisections followed by safeguarded Newton steps, and it requires \sweepNewtonArOneZeroNine{}
sweeps to reach the evaluation floor. The convergence is bisection-like, which says the Newton step
is being rejected and replaced by a bisection most of the time, and the mechanism is the one the
literature on this equation addresses~\cite{li1993secular,melman1998}: a root commonly sits very
close to one of the two poles $\Lambda_k$ and $\Lambda_{k+1}$ that bound its interval, a Newton step taken from
the interior overshoots past that pole, and the safeguard catches it.

The remedy carries both bounding poles in closed form. With $\tau=\lambda-\Lambda_k$ and
$\Delta_k=\Lambda_{k+1}-\Lambda_k$, write
\begin{equation}\label{eq:split}
  f(\tau)=h(\tau)-\frac{\xi w_k}{\tau}+\frac{\xi w_{k+1}}{\Delta_k-\tau},
\end{equation}
where $h$ collects every term whose pole lies outside the interval and so varies slowly across it.
Fitting each one-sided sum with a pole of its own at the end it faces, matched to the sum and its
derivative, gives
\begin{align}
  \alpha_-&=\xi(F_-'\tau^2+w_k),\qquad \beta_-=\xi(F_-+F_-'\tau),\\
  \alpha_+&=\xi\bigl(F_+'(\Delta_k-\tau)^2+w_{k+1}\bigr),\\
  \beta_+&=\xi\bigl(F_+-F_+'(\Delta_k-\tau)\bigr),
\end{align}
with $\beta_0=1+\beta_-+\beta_+$, where $F_-$ and $F_+$ are the sums $\sum_jw_j/(\Lambda_j-\lambda)$ over the poles below and above the interval, and the next
iterate is the root in the current bracket of
\begin{equation}\label{eq:middlequad}
  \beta_0\tau^2-(\beta_0\Delta_k+\alpha_-+\alpha_+)\,\tau+\alpha_-\Delta_k=0 .
\end{equation}
This is the middle way of~\cite{li1993secular}, and three points of its arithmetic decide whether
it is usable. Written on the full sums, $\beta_-$ and $\beta_+$ are differences of nearly equal large
quantities, since the pole term dominates $F_-$ and $F_+$ near a root, so the hierarchical
evaluation is given a variant that omits the terms $j=k$ and $j=k+1$ from every sum and leaves the
caller to add them from the shift and the gap. On those split sums the coefficients come out clean,
because each pole's own contribution to its fit is exactly the weight that was held back. The
smaller root of~\eqref{eq:middlequad} is taken from the product of the roots rather than from the
differencing branch of the quadratic formula, since for a root glued to the left pole $4\beta_0\alpha_-\Delta_k$ is
negligible beside the square of the middle coefficient and the differencing branch returns noise.
And a root is frozen once its scaled residual reaches the evaluation floor, with the best iterate
seen retained, since the far end of a bracket can be several iterates stale and a midpoint fallback
against that stale end moves a converged root back off its solution. With those three in place the
schedule reaches a scaled residual of $10^{-12}$ in \sweepMiddleArOneZeroNine{} and
\sweepMiddleArThree{} sweeps on the two windows of Table~\ref{tab:implsolver}.

\emph{Roots held from the nearer pole.} The shift $\delta_k=\lambda_k-\Lambda_k$ makes a denominator
$\Lambda_j-\lambda_k=(\Lambda_j-\Lambda_k)-\delta_k$ accurate only while the root is nearer $\Lambda_k$ than $\Lambda_j$ is to
$\Lambda_{k+1}$. When the weight $w_{k+1}$ is very small the root lies within rounding of $\Lambda_{k+1}$, and
$\Lambda_{k+1}-\lambda_k$ formed from the gap and the shift is a difference of two numbers that agree in
all their digits, while the entry it divides is the dominant one of that eigenvector. The same is
true of $\Lambda_{k+2},\Lambda_{k+3},\dots$ whenever those lie closer to $\Lambda_{k+1}$ than the rounding of the gap
below it, as they do when a cluster of grid values follows a wide gap, and there the difference can
round to zero. Each root is therefore held as a base and an offset, $(k,\delta_k)$ in the lower half
of its gap and $(k+1,-\eta_k)$ with $\eta_k=\Lambda_{k+1}-\lambda_k$ in the upper half, the half being
decided by the sign of $f$ at the midpoint, and it is solved for in that variable, so that the small
quantity is the unknown and not a difference. Every difference between a grid value and a root is
then formed as the difference of two grid values, which floating-point subtraction rounds
correctly, less the offset. The eigenvector entries, the column norms and the Loewner weights all
take their differences in this form, in the dense and in the hierarchical realization alike. Grids
with such roots arise whenever a class spectrum has a floor, as that of a signal in white noise
does, and at a switch between regimes.

\emph{What the hierarchical form needs on clustered grids.} Three further points decide the accuracy
of the hierarchical form when the grid is clustered, and each was found to be the limit in turn. The
schedule above brings nearly every root to the evaluation floor, but on an irregular grid a few
fall back on halving and are left far from it after a fixed number of sweeps, so a root whose scaled
residual is above $10^{-13}$ at the end is finished on its own by the dense iteration, at $O(n)$
operations each. The expansion of the Loewner sums in powers of $\delta_j/(\Lambda_j-\Lambda_i)$ reaches
rounding only while that ratio is a few hundredths, so a pair with larger shifts is evaluated on the
target side instead: the sum over the source box is a smooth function of the target position, it is
computed exactly at the Chebyshev nodes of the target box and interpolated to the targets, and the
target box is halved until it is far enough away. And the tree is held in coordinates local to each
box, because a cluster of grid values a relative $10^{-12}$ apart gives boxes narrower than the
spacing of floating-point numbers at their own magnitude, in which interpolation nodes have no
exact position. With these in place, for a pulse-shaped signal in white noise at variance between
\apAccNoiseLo{} and \apAccNoiseHi{} of the signal power, at length 2048, the round trip error of the
transform is at most \apAccDenseHi{} with dense stages and at most \apAccFastHi{} with hierarchical
ones.

\subsection{Two Models on One Window}\label{sec:pencil}

Let a signal and a noise be autoregressive of orders $p_{\mathrm s}$ and $p_{\mathrm n}$ with innovation variances $\sigma_{\mathrm s}^2$ and $\sigma_{\mathrm n}^2$, window covariances $\bR_{\mathrm s}$ and $\bR_{\mathrm n}$ and precisions $\bQ_{\mathrm s}$ and $\bQ_{\mathrm n}$. The modes of the window ordered by signal to noise ratio are the generalized eigenvectors
\begin{equation}\label{eq:pencil}
\bR_{\mathrm s}\bv=\lambda\bR_{\mathrm n}\bv,\quad \bV^\T\bR_{\mathrm n}\bV=\bI,\quad \bV^\T\bR_{\mathrm s}\bV=\operatorname{diag}(\lambda_j).
\end{equation}
In the coefficients $\bV^\T\bx$ the noise is white and the signal is uncorrelated with variances $\lambda_j$, so the Wiener estimate of the signal from the sum, the likelihood ratio of the sum against the noise alone, and the truncation to the modes of largest ratio are diagonal operations. When the noise is white, $\bV$ is the transform of Proposition~\ref{prop:fast}. When it is not, $\bV$ is not orthogonal and is the eigenbasis of neither covariance. The two models nevertheless belong to one class, and the transform has the same fast form with a cascade for each model.

With $\bm y=\bR_{\mathrm s}\bv$ the problem \eqref{eq:pencil} reads $\bQ_{\mathrm n}\bm y=\lambda\,\bQ_{\mathrm s}\bm y$. By Proposition~\ref{prop:fast} the fold and the core of a branch bring both precisions to the form \eqref{eq:branch} on the same core, $\sigma_{\mathrm n}^2\bQ_{\mathrm n}\mapsto\bLam_{\mathrm n}-\bW_{\mathrm n}\bD_{1,\mathrm n}\bW_{\mathrm n}^\T$ and $\sigma_{\mathrm s}^2\bQ_{\mathrm s}\mapsto\bLam_{\mathrm s}-\bW_{\mathrm s}\bD_{1,\mathrm s}\bW_{\mathrm s}^\T$, where $\bLam_{\mathrm n}$ and $\bLam_{\mathrm s}$ hold the two symbols on the core grid and $\bW_{\mathrm n}$, $\bW_{\mathrm s}$ are the first $p_{\mathrm n}$ and $p_{\mathrm s}$ columns of the core. The two corrections sit on the same leading coordinates, so for every $\lambda$ the branch of the pencil differs from the diagonal pencil $(\bLam_{\mathrm n},\bLam_{\mathrm s})$ by a matrix of rank at most $\max(p_{\mathrm s},p_{\mathrm n})$.

\begin{proposition}\label{prop:pencil}
On a branch with core $\bK$, let $\bV_1$ be the cascade of Proposition~\ref{prop:cascade} for
\[
\bLam_{\mathrm n}\bLam_{\mathrm s}^{-1}-\bZ_{\mathrm n}\bD_{1,\mathrm n}\bZ_{\mathrm n}^\T,\qquad \bZ_{\mathrm n}=\bLam_{\mathrm s}^{-1/2}\bW_{\mathrm n},
\]
with eigenvalues $\bLam_1$, and let $\bV_2$ be the cascade for
\[
\bLam_1^{-1}-\bZ_{\mathrm s}\bD_{1,\mathrm s}\bZ_{\mathrm s}^\T,\qquad \bZ_{\mathrm s}=\bLam_1^{-1/2}\bV_1^\T\bLam_{\mathrm s}^{-1/2}\bW_{\mathrm s},
\]
with eigenvalues $\bLam_2$. Then $\bLam_1$ and $\bLam_2$ are positive, the generalized eigenvalues of the branch are $\lambda_j=(\sigma_{\mathrm s}^2/\sigma_{\mathrm n}^2)/\Lambda_{2,jj}$, and with $\bT=\bLam_{\mathrm s}^{-1/2}\bV_1\bLam_1^{-1/2}\bV_2$ the branch of the transform and of its inverse transpose are
\begin{equation}\label{eq:penciltransform}
\bV^\T=\sigma_{\mathrm n}^{-1}\,\bV_2^\T\bLam_1^{1/2}\bV_1^\T\bLam_{\mathrm s}^{1/2}\bK,\qquad \bR_{\mathrm n}\bV=\sigma_{\mathrm n}\,\bK^\T\bT ,
\end{equation}
each preceded or followed by the fold. The first cascade has at most $p_{\mathrm n}$ stages and the second at most $p_{\mathrm s}$, so one application takes two fast cores and at most $p_{\mathrm s}+p_{\mathrm n}$ stages per branch, $O((p_{\mathrm s}+p_{\mathrm n})M\log M)$ operations with the stages of Section~\ref{sec:fmmreal}.
\end{proposition}

\begin{proof}
Write $\bm A=\bLam_{\mathrm n}-\bW_{\mathrm n}\bD_{1,\mathrm n}\bW_{\mathrm n}^\T$ and $\bm B=\bLam_{\mathrm s}-\bW_{\mathrm s}\bD_{1,\mathrm s}\bW_{\mathrm s}^\T$, both positive definite. The first matrix of the statement is $\bLam_{\mathrm s}^{-1/2}\bm A\bLam_{\mathrm s}^{-1/2}$, so $\bLam_1$ is positive and $\bm X_1=\bLam_{\mathrm s}^{-1/2}\bV_1\bLam_1^{-1/2}$ gives $\bm X_1^\T\bm A\bm X_1=\bI$. The second matrix is $\bm X_1^\T\bm B\bm X_1$, since $\bm X_1^\T\bLam_{\mathrm s}\bm X_1=\bLam_1^{-1}$, so $\bLam_2$ is positive, and $\bT=\bm X_1\bV_2$ gives $\bT^\T\bm A\bT=\bI$ and $\bT^\T\bm B\bT=\bLam_2$. Hence $\bm A\bm t_j=\Lambda_{2,jj}^{-1}\bm B\bm t_j$ for the columns of $\bT$, which with the two variances is the value of $\lambda_j$. With $\bm Y=\bK^\T\bT\bLam_2^{-1/2}$ on the branch, $\bm Y^\T\sigma_{\mathrm s}^2\bQ_{\mathrm s}\bm Y=\bI$, and $\bv_j=\bR_{\mathrm s}^{-1}\bm y_j$ scaled to satisfy \eqref{eq:pencil} gives $\bV^\T=\sigma_{\mathrm n}^{-1}\bT^{-1}\bK$, which is the first expression of \eqref{eq:penciltransform} because $\bT^{-1}=\bV_2^\T\bLam_1^{1/2}\bV_1^\T\bLam_{\mathrm s}^{1/2}$. The second is its inverse transpose, and $\bR_{\mathrm n}\bV=\bV^{-\T}$ by \eqref{eq:pencil}. The ranks of the two corrections are at most $p_{\mathrm n}$ and $p_{\mathrm s}$, which bounds the stages.
\end{proof}

The diagonal the first cascade starts from, $\bLam_{\mathrm n}\bLam_{\mathrm s}^{-1}$, is the ratio of the two spectra on the core grid, $S_{\mathrm s}(\omega)/S_{\mathrm n}(\omega)$ up to the variances. Ordering the coefficients of the core by that ratio is the answer of the class alone, and the two cascades are the correction for the ends of the window, in the same sense as for one model. Neither transform of \eqref{eq:penciltransform} solves a system. The count $p_{\mathrm s}+p_{\mathrm n}$ is not the rank bound $\max(p_{\mathrm s},p_{\mathrm n})$ of the paragraph before the proposition. A cascade of $\max(p_{\mathrm s},p_{\mathrm n})$ stages that update both sides at once exists when the two corner blocks, the smaller one padded with zeros, are diagonalized by one real congruence. The corner blocks are indefinite in general, and such a congruence need not exist. The statement covers two autoregressive models. The windows of Sections~\ref{sec:fastrational} and~\ref{sec:covside}, whose corrections sit on the covariance side, are not covered by it.

\subsection{Realization}\label{sec:realization}

The statements of Sections~\ref{sec:complexdata} to~\ref{sec:pencil} are realized in a
reference implementation released with this paper, in Python and in C, in one and two dimensions,
for the autoregressive, moving-average, mixed, co-channel and integrated classes with real and with
complex data, and for the vector, periodic and cyclostationary classes of Section~\ref{sec:block}, and for the two-model transform of Section~\ref{sec:pencil} with real autoregressive models. Section~\ref{sec:bench} reports its accuracy and its operation count. The quantity
that carries the order claim there is the number of kernel evaluations the hierarchical stages
perform, which does not depend on the machine a measurement is taken on, and the timings are
reported beside it as an illustration rather than as the evidence.

Two limitations are stated rather than left implicit. The accuracy floor of the design stage is the
one Section~\ref{sec:fmmreal} gives for the Loewner expansion. And an integrated window of
difference order two or more has, at the lengths measured, a precision conditioned at about $10^{8}$, so the largest
eigenvalues of its covariance are resolved to about $10^{-7}$ relative by any construction of this
kind, which is what Table~\ref{tab:implranks} and the integrated rows of
Table~\ref{tab:implaccuracy} show.

\section{Vector, Periodic and Cyclostationary Windows}\label{sec:block}

The classes that supply the transforms so far, the cosine, sine and circulant classes, come from
multiplicity-free actions, so the group fixes one transform. A window of several channels, and a window of a process whose statistics
vary periodically, have as matched group the cyclic group of shifts by one block. That action is not
multiplicity-free. Its commutant is the class of block circulants, and the group fixes the transform
only up to a small unitary matrix at each frequency, which the model supplies. The departure of the
window from the class is again of low rank and on the border, so the cascade of
Section~\ref{sec:cascade} applies without change. This section states the class, the rank of the
departure for the models the implementation carries, and the reduction of a complex stage to a real
one. Complex data is admitted throughout, and real data is the special case of real blocks.

\subsection{The Block Circulant Class}\label{sec:blockclass}

Let $M=NP$ and index the window by a block $n\in\{0,\dots,N-1\}$ and a position within the block.
The shift by one block generates $\Z_N$, and a Hermitian matrix commutes with it exactly when its
$P\times P$ blocks satisfy $\bC_{(i,j)}=\bC_{(i-j)\bmod N}$ with $\bC_{-d}=\bC_d^{\herm}$. Each
character of $\Z_N$ occurs $P$ times in the representation, so the isotypic components have
dimension $P$ and the Reynolds projection onto the class is the average over block shifts.

\begin{proposition}\label{prop:blockcore}
Let $\widehat{\bC}_k=\sum_{d=0}^{N-1}\bC_d\,e^{-2\pi ikd/N}$ be the $P\times P$ Hermitian symbol at
frequency $k$, with eigendecomposition $\widehat{\bC}_k=\bW_k\operatorname{diag}(\lambda_{k,1},\dots,\lambda_{k,P})\bW_k^{\herm}$.
The $M$ vectors whose block $n$ is $N^{-1/2}e^{2\pi ikn/N}\bW_k\be_r$, for $0\le k<N$ and
$1\le r\le P$, are an orthonormal eigenbasis of the block circulant, with eigenvalues
$\lambda_{k,r}$. The transform is a discrete Fourier transform of length $N$ at each of the $P$
positions followed by $\bW_k^{\herm}$ at each frequency, $O(M\log N+MP)$ operations once the $N$
eigenproblems of size $P$ are solved.
\end{proposition}

\ifextended
\begin{IEEEproof}
Block $i$ of the product of the block circulant with the vector is
$N^{-1/2}\sum_j\bC_{i-j}e^{2\pi ikj/N}\bW_k\be_r=N^{-1/2}e^{2\pi iki/N}\widehat{\bC}_k\bW_k\be_r$,
by the substitution $d=i-j$, and $\widehat{\bC}_k\bW_k\be_r=\lambda_{k,r}\bW_k\be_r$. Vectors at
different $k$ are orthogonal by the orthogonality of the characters and vectors at one $k$ by the
unitarity of $\bW_k$. The count is that of $P$ transforms of length $N$ and $N$ products of size $P$.
\end{IEEEproof}
\fi

The class is therefore matched by the group and its basis by the model: the group supplies the
Fourier transform across blocks, and within each isotypic component the basis is the KLT of a
$P\times P$ symbol. For $P=1$ the class is the circulants of Section~\ref{sec:complexdata}.

\subsection{Windows with a Finite-Dimensional Realization}\label{sec:realizationrank}

\begin{proposition}\label{prop:realization}
Let $\bm y_t\in\C^m$ be stationary with autocovariance
$\bGam_h=E\,\bm y_{t+h}\bm y_t^{\herm}$, and suppose that $\bGam_h=\bH\bA^{h-1}\bG$ for every
$h\ge1$, with $\bA$ of size $s\times s$ and spectral radius below one. Let $\bR$ be the covariance
of a window of $N$ samples, the block Toeplitz matrix with blocks $\bGam_{i-j}$, and let $\bC$ be
the block circulant with blocks $\sum_{\ell\in\Z}\bGam_{d+\ell N}$, the periodized autocovariance.
Then $\bR-\bC$ has rank at most $2s$.
\end{proposition}

\ifextended
\begin{IEEEproof}
Block $(i,j)$ of $\bC-\bR$ is $\sum_{\ell\ge1}\bGam_{i-j+\ell N}+\sum_{\ell\ge1}\bGam_{i-j-\ell N}$.
In the first sum every lag is at least one, so
$\bGam_{i-j+\ell N}=\bH\bA^{i}\bA^{(\ell-1)N}\bA^{N-1-j}\bG$ and the sum over $\ell$ is
$\bX_i\bm Y_j$ with $\bX_i=\bH\bA^{i}$ and $\bm Y_j=(\bI-\bA^N)^{-1}\bA^{N-1-j}\bG$. The matrix with
these blocks is the product of one with $s$ columns and one with $s$ rows, of rank at most $s$. The
second sum is its conjugate transpose, since $\bGam_{-h}=\bGam_h^{\herm}$.
\end{IEEEproof}
\fi

The proposition covers four of the model classes at once. A vector autoregression of order $p$ in
$m$ channels has the companion realization, $s=mp$, so its departure has rank at most $2mp$. A
vector ARMA($p,q$) window has $s=m\max(p,q)$: its autocovariance satisfies the autoregressive
recurrence $\bGam_h=\sum_{k=1}^{p}\bm\Phi_k\bGam_{h-k}$ for every $h>q$, hence from
$h=\max(p,q)+1$ on, and a sequence that satisfies a block recurrence of order $r$ from index $r+1$
on is realized by the block companion matrix acting on $(\bGam_1,\dots,\bGam_r)$. A linear
state-space model observed without noise has $s$ equal to its state dimension, and white
observation noise adds a matrix that lies in the class. A product of two such kernels has the
Kronecker product of the two realizations, $s=s_1s_2$. The scalar case $m=1$ is the statement of
Corollary~\ref{cor:rational} for the cyclic class. Table~\ref{tab:blockranks} reports the rank found
against these bounds.

For the vector autoregression the same bound follows on the precision side, where the construction
is closed form. The window precision is the sum over $t=p,\dots,N-1$ of the outer products of the
rows of the recursion, weighted by the inverse innovation covariance, plus the inverse of the
stationary covariance of the first $p$ samples. The class member is the same sum taken cyclically
over all $N$ rows. The two differ in the $p$ wrapped rows and in the start block, each supported on
the first and last $p$ blocks.

\subsection{Periodic Autoregression}\label{sec:periodicar}

\begin{proposition}\label{prop:periodicar}
Let $x_t=\sum_{k=1}^{p}\phi_k(t)\,x_{t-k}+\varepsilon_t$ with complex coefficients of period $P$ in
$t$, uncorrelated innovations of variance $\sigma^2(t)$ of the same period, and a stable recursion.
Let $M=NP>2p$, and let $\bm a_t$ be the row with entries $1,-\phi_1(t),\dots,-\phi_p(t)$ at the
positions $t,t-1,\dots,t-p$. The window precision is
\begin{equation}\label{eq:periodicprecision}
  \bQ=\sum_{t=p}^{M-1}\sigma^{-2}(t)\,\bm a_t^{\herm}\bm a_t+\bGam_0^{-1}\oplus\bm 0,
\end{equation}
with $\bGam_0$ the covariance of $(x_0,\dots,x_{p-1})$. The matrix
$\bC=\sum_{t=0}^{M-1}\sigma^{-2}(t)\,\bm a_t^{\herm}\bm a_t$ with the positions taken modulo $M$ is
a positive definite block circulant with $P\times P$ blocks, and $\bQ-\bC$ is supported on the first
$p$ and the last $p$ coordinates. Its rank is at most $2p$, whatever the period.
\end{proposition}

\ifextended
\begin{IEEEproof}
The window is a unit lower triangular function of $(x_0,\dots,x_{p-1})$ and the innovations
$\varepsilon_p,\dots,\varepsilon_{M-1}$, which are mutually uncorrelated and uncorrelated with the
first $p$ samples, and the rows of the inverse of that triangular matrix are the unit vectors of the
first $p$ coordinates and the $\bm a_t$. That gives~\eqref{eq:periodicprecision}. The cyclic sum is
invariant under the shift by $P$ because the coefficients and variances have period $P$ and $P$
divides $M$. The difference $\bQ-\bC$ consists of the start block, on the first $p$ coordinates,
and of the rows $t=0,\dots,p-1$ taken cyclically, whose positions $t-p,\dots,t$ modulo $M$ lie among
the first $p$ and the last $p$ coordinates. A matrix supported on $2p$ coordinates has rank at
most $2p$.
\end{IEEEproof}
\fi

Writing the process as a vector autoregression over blocks of one period gives, through
Proposition~\ref{prop:realization}, the bound $2P\lceil p/P\rceil$, which grows with the period. The
scalar band of the precision gives $2p$. The start block $\bGam_0$ depends on the phase at which
the window begins. It is the stationary covariance of the state at phase $(p-1)\bmod P$, which
solves a Lyapunov equation of the recursion over one period.

\subsection{Cyclostationary Windows on the Covariance Side}\label{sec:cyclocov}

A process is cyclostationary with period $P$ when $r(t,u)=E\,x_t\overline{x_u}$ satisfies
$r(t+P,u+P)=r(t,u)$~\cite{gardner2006cyclo}.

\begin{proposition}\label{prop:finitelag}
Let $r(t,u)=0$ for $|t-u|>q$ and let $M=NP>2q$. The matrix $\bC$ with entries $r(t,u)$ for
$|t-u|\le q$, $r(t+M,u)$ for $u-t\ge M-q$ and $r(t,u+M)$ for $t-u\ge M-q$ is a block circulant
with $P\times P$ blocks, and $\bC-\bR$ vanishes outside the block $\bB$ that joins the first $q$
coordinates to the last $q$ and its conjugate transpose. Its rank is $2\operatorname{rank}\bB\le2q$.
\end{proposition}

\ifextended
\begin{IEEEproof}
$\bC$ is the covariance with its lags taken cyclically, which the condition $M>2q$ makes well
defined, and the periodicity of $r$ with $P$ dividing $M$ makes it invariant under the shift by $P$.
The entries of $\bC$ that $\bR$ lacks are those at cyclic lag at most $q$ and true lag at least
$M-q$, which join the first $q$ coordinates to the last $q$. A Hermitian matrix with a single
off-diagonal block $\bB$ and its conjugate transpose has rank $2\operatorname{rank}\bB$.
\end{IEEEproof}
\fi

The case that matters most in practice has far smaller rank than $2q$.

\begin{proposition}\label{prop:modulation}
Let $x_t=\sum_{n}a_n\,g_{t-nP}+w_t$ with uncorrelated symbols of unit variance, a pulse $g$ supported
on $0,\dots,L-1$, and white noise of variance $\sigma_w^2$ uncorrelated with the symbols, and let
$M=NP\ge2L-1$. Then $\bR=\bG\bG^{\herm}+\sigma_w^2\bI$, with the columns of $\bG$ the pulse shifted by
every multiple of $P$ and cut to the window, the class member of Proposition~\ref{prop:finitelag} is
$\bC=\bG_c\bG_c^{\herm}+\sigma_w^2\bI$ with the shifts taken cyclically, and
\begin{equation}\label{eq:modulationdeparture}
  \bC-\bR=\sum_{j=1}^{K}\bigl(\bm t_j\bm h_j^{\herm}+\bm h_j\bm t_j^{\herm}\bigr),\qquad K=\Bigl\lfloor\frac{L-1}{P}\Bigr\rfloor,
\end{equation}
where $\bm t_j$ is the part of the pulse of the $j$th symbol before the end of the window that lies
inside the window and $\bm h_j$ is the part that the cyclic shift carries to its start. The
departure has rank at most $2K$, twice the number of symbol periods the pulse extends beyond its
first, and the number of samples per symbol does not enter.
\end{proposition}

\ifextended
\begin{IEEEproof}
A symbol at $nP$ with $0\le n\le N-K-1$ has its whole pulse inside the window and contributes the
same column to $\bG$ and to $\bG_c$. A symbol at $(N-j)P$ with $1\le j\le K$, which are the $j$
with $jP\le L-1$, contributes $\bm t_j+\bm h_j$ to $\bG_c$. In $\bG$ it contributes $\bm t_j$, and
$\bm h_j$ is the column of the symbol at $-jP$, whose pulse enters the window from before its start.
Hence $\bG_c\bG_c^{\herm}-\bG\bG^{\herm}=\sum_j(\bm t_j+\bm h_j)(\bm t_j+\bm h_j)^{\herm}-\bm t_j\bm t_j^{\herm}-\bm h_j\bm h_j^{\herm}$,
which is~\eqref{eq:modulationdeparture}.
\end{IEEEproof}
\fi

A root raised cosine pulse spanning eight symbol periods, $8P+1$ samples, therefore needs sixteen stages at any
sampling rate, where Proposition~\ref{prop:finitelag} alone would allow $2(L-1)$. The transform
takes the pulse as received, the symbol timing, the noise level and the lack of correlation between
symbols as known, and it does not use the symbol alphabet.

\begin{remark}\label{rem:matchedfilter}
The $2K$ of Proposition~\ref{prop:modulation} counts the pulses that the two ends of the window cut,
$K$ at each end. The eigenvectors of $\bR$ with eigenvalues above $\sigma_w^2$ span the columns of $\bG$.
With $\bG^{\herm}\bG=\bm Y\operatorname{diag}(g_i)\bm Y^{\herm}$ the eigendecomposition of the Gram matrix of the cut pulses, the vectors
$\bG\bm y_i/\sqrt{g_i}$ are those eigenvectors, with eigenvalues $g_i+\sigma_w^2$, and the coefficients of a
window $\bx$ on them are $\operatorname{diag}(g_i)^{-1/2}\bm Y^{\herm}\bG^{\herm}\bx$. The vector $\bG^{\herm}\bx$ is the output of
a filter matched to the pulse, sampled at the symbol instants, so the signal coefficients of the
transform are the matched-filter samples rotated by a unitary matrix and scaled. On a record without ends the Gram matrix
is Toeplitz, a Fourier transform across the symbols diagonalizes it, and for a Nyquist pulse it is
the identity. The transform of a window therefore differs from a matched filter followed by a Fourier
transform across the symbols only through the $2K$ cut pulses, and the two agree in the limit of a
long window, with a difference of relative order $K/N$.
\end{remark}

\begin{remark}\label{rem:improper}
With real symbols, as in binary phase-shift keying, $E\,x_tx_u$ does not vanish and the Hermitian
covariance does not describe the window. The real vector that interleaves the real and imaginary
parts of the samples is a linearly modulated real signal with the interleaved pulse and $2P$
samples per symbol, so Proposition~\ref{prop:modulation} applies to it with the same $K$. The
transform is a real orthogonal one of length $2M$. For real blocks the symbol of the class member at
the frequency $-k$ is the conjugate of the symbol at $k$, so its eigenvectors come in conjugate
pairs, and each pair is replaced by its real and imaginary parts at no added Fourier transform. The
departure is then real symmetric and the stages are real.
\end{remark}

\subsection{A Complex Stage Is a Real Stage and a Phase}\label{sec:phasestage}

Each stage of a complex cascade diagonalizes $\bLam+\xi\,\bm\psi\bm\psi^{\herm}$ with $\bLam$ real diagonal,
$\xi$ real and $\bm\psi$ complex. Let $\bm E_{\bm\psi}$ be the diagonal matrix of the phases of the entries of
$\bm\psi$, with one where an entry vanishes, and let $|\bm\psi|$ hold the moduli. Then
\begin{equation}\label{eq:phasestage}
  \bLam+\xi\,\bm\psi\bm\psi^{\herm}=\bm E_{\bm\psi}\bigl(\bLam+\xi\,|\bm\psi||\bm\psi|^{\T}\bigr)\bm E_{\bm\psi}^{\herm},
\end{equation}
since $\bm E_{\bm\psi}$ commutes with $\bLam$. The eigenvalues are those of the real problem and the
eigenvectors are the real ones multiplied by $\bm E_{\bm\psi}$. A complex cascade therefore needs no
complex secular solver, and the statements of Sections~\ref{sec:fmmreal}
and~\ref{sec:secularreal} carry over as they stand, each application being two real applications
on the real and imaginary parts.

\subsection{Other Classes in the Implementation}\label{sec:otherclasses}

The same two ingredients, a class transform and a border departure of low rank, give several
further windows, each in one or two sentences. \emph{Bridge.} The samples between two pinned groups
of $p$ samples have as precision the corresponding block of the joint precision, which by
Theorem~\ref{thm:departure} is the banded Toeplitz matrix of the symbol with no corner correction.
The member of the type-I sine class with eigenvalues $c(\pi k/(M+1))$, $k=1,\dots,M$, differs from it by a Hankel
term supported on the first and last $p-1$ coordinates, so the rank is at most $2(p-1)$ and the
first-order bridge lies in the class. \emph{Switching at known times.} A window whose coefficients
change at known samples departs from the direct sum of the windows of its regimes by a matrix
supported on the $2p$ coordinates around each switch, by the argument of
Proposition~\ref{prop:periodicar}. \emph{Mat\'ern.} At smoothness $\nu=k+\tfrac12$ the sampled
process has an autocovariance that is a polynomial of degree $k$ in the lag times a decaying
exponential, which satisfies a recurrence of order $k+1$, so
Corollary~\ref{cor:rational} applies with $p=k+1$. \emph{Grid fields.} A precision that is a
function of two commuting one-dimensional difference operators is diagonalized exactly by the
product of their transforms, with no cascade. \emph{Long memory.} Fractional noise has no finite
realization and its departure has no bounded rank. The implementation truncates it at a stated
tolerance, and the result is an approximation and not the KLT.

\begin{table}[t]
\caption{Departure rank found by the reference implementation against the bound of its statement.
Dense stages.}
\label{tab:blockranks}
\centering
\begin{tabular}{@{}llcc@{}}
\toprule
window & bound & cases & at the bound \\
\midrule
\input{apps_tab_ranks.tex} 
\bottomrule
\end{tabular}
\end{table}

In all \apRkCases{} cases of Table~\ref{tab:blockranks} the rank found is at or below the bound,
and in \apRkAtBound{} of them it equals the bound.

\ifextended
\section{Carath\'eodory's Representation}\label{sec:cara}

A positive semidefinite Toeplitz matrix of size $M$ and rank $d<M$ is uniquely a sum of $d$ rank-one sinusoidal atoms, $\sum_kw_k\be(\omega_k)\be(\omega_k)^{\herm}$ with distinct frequencies and positive weights~\cite{caratheodory1907variabilitaet,caratheodory1911variabilitaet,caratheodory1911fejer,toeplitz1911theorie,grenander1958toeplitz}, the representation behind harmonic retrieval~\cite{pisarenko1973retrieval,stoica2005spectral}. Since the eigenvectors of an AR($p$) window are also sums of sinusoids, the two decompositions invite comparison.

\begin{proposition}\label{prop:cara}
(a) For every $\tau$ with $|\tau|=1$ the polynomial $z^Ma(1/z)+\tau a(z)$ has $M$ simple zeros $e^{i\omega_k}$ on the unit circle, the solutions of $M\omega-2\arg a(e^{i\omega})\equiv\arg(-\tau)\pmod{2\pi}$, and there are positive weights $w_k$ with $\bR=\sum_{k=1}^Mw_k\be(\omega_k)\be(\omega_k)^{\herm}$. (b) The sinusoids of an eigenvector of $\bR$ at eigenvalue $\mu$ sit where $|a(e^{i\omega})|^2=\sigma^2/\mu$ (Theorem~\ref{thm:levels}). (c) In the resonant limit of Proposition~\ref{prop:limits}(c) the limiting covariance has the unique two-atom representation at $\pm\omega_0$, and its eigenvectors with nonzero eigenvalue are sinusoids at $\omega_0$, so the two frequency sets coincide in the limit.
\end{proposition}

\begin{proof}
\ifextended
(a) The monic orthogonal polynomials $\Phi_n$ of the measure $S(\omega)\,d\omega/2\pi$ satisfy
\[
\Phi_n(z)=z^{n-p}\Phi_p(z)\ \ \text{for }n\ge p,\qquad \Phi_p(z)=z^pa(1/z),
\]
because the partial autocorrelations of an AR($p$) process, the Verblunsky coefficients of the Szeg\H{o} recursion, vanish beyond lag $p$~\cite{brockwell1991time,simon2005orthogonal}. The para-orthogonal polynomial
\[
\Phi_M(z)+\tau\Phi_M^*(z),\qquad \Phi_M^*(z)=z^M\Phi_M(1/z)=a(z),
\]
has $M$ simple zeros on the circle, and they are the nodes of the $M$-point Szeg\H{o} quadrature, with positive weights, exact for trigonometric polynomials of degree at most $M-1$~\cite{jones1989moment}. Applied to $e^{ij\omega}$ for $|j|\le M-1$ this gives
\[
r_j=\sum_kw_ke^{ij\omega_k},\qquad |j|\le M-1.
\]
On the circle $z^Ma(1/z)=e^{iM\omega}\,\overline{a(e^{i\omega})}$ for real coefficients, which gives the phase form.

(b) is Theorem~\ref{thm:levels}(a), with $S=\sigma^2/|a|^2$.

(c) The limiting covariance
\[
\bC(\omega_0)=\tfrac12\bigl(\be(\omega_0)\be(\omega_0)^{\herm}+\be(-\omega_0)\be(-\omega_0)^{\herm}\bigr)
\]
has rank two, the case $d<M$ of the theorem, and its range is the span of the cosine and the sine at $\omega_0$.
\else
(a) The monic orthogonal polynomials $\Phi_n$ of the measure $S(\omega)\,d\omega/2\pi$ satisfy $\Phi_n(z)=z^{n-p}\Phi_p(z)$ for $n\ge p$ with $\Phi_p(z)=z^pa(1/z)$, because the partial autocorrelations of an AR($p$) process, the Verblunsky coefficients of the Szeg\H{o} recursion, vanish beyond lag $p$~\cite{brockwell1991time,simon2005orthogonal}. The para-orthogonal polynomial $\Phi_M(z)+\tau\Phi_M^*(z)$, with $\Phi_M^*(z)=z^M\Phi_M(1/z)=a(z)$, has $M$ simple zeros on the circle, and they are the nodes of the $M$-point Szeg\H{o} quadrature, with positive weights, exact for trigonometric polynomials of degree at most $M-1$~\cite{jones1989moment}. Applied to $e^{ij\omega}$ for $|j|\le M-1$ this gives $r_j=\sum_kw_ke^{ij\omega_k}$. On the circle $z^Ma(1/z)=e^{iM\omega}\overline{a(e^{i\omega})}$ for real coefficients, which gives the phase form. (c) $\bC(\omega_0)=\tfrac12(\be(\omega_0)\be(\omega_0)^{\herm}+\be(-\omega_0)\be(-\omega_0)^{\herm})$ has rank two, the case $d<M$ of the theorem, and its range is the span of the cosine and the sine at $\omega_0$.
\fi
\end{proof}

The two sums are therefore dual and distinct. Carath\'eodory's atoms are a quadrature of the spectrum fixed by the covariance, at frequencies set by the phase of $a(e^{i\omega})$ (a warping of the uniform grid $2\pi k/M$ by twice the phase response of the whitening filter), with one free unimodular parameter. The eigenvector sinusoids move with the eigenvalue, at frequencies set by the magnitude of $a(e^{i\omega})$, and the boundary rows select them. For each of the six pole patterns of the basis experiment of Section~\ref{sec:num} at $M=32$, some level-set frequency lies at least \caraGapMinThirtyTwo{} grid steps $\pi/M$ from every atom. The two descriptions meet at the endpoint of the family, where the poles reach the circle, the limiting covariance loses rank, and the uniqueness of the theorem takes over. That endpoint is the rank-two limit of Proposition~\ref{prop:limits}(c), in the cyclic class exactly when its atoms are Fourier vectors of the $M$-cycle. The AR($p$) window and a line spectrum with at most $p+1$ atoms are also the two ends of the Carath\'eodory-Toeplitz extension set of the same $p+1$ autocovariances, the maximum-entropy extension and an extreme one~\cite{landau1987maximum}, which places the stable window at the center of the set the theorem describes and the resonant limit on its boundary.
\fi

\section{Numerical Illustration}\label{sec:num}

\emph{Exact symmetry.} Table~\ref{tab:groups} lists, for ten pole patterns, the order of the group of permutations commuting with the window covariance, found by exhaustive search over all permutations of the index set ($M=8$, and $M=9$ for the three-phase pattern). The \nGenericPatterns{} patterns with distinct autocovariances have a group of order two containing the reflection, including the resonant and the mixed-sign windows. The \nPolyPatterns{} polyphase patterns have orders $\orderPhaseTwo$, $\orderPhaseThree$ and $\orderPhaseFour$ for $m=2,3,4$, equal to $2^mm!$, with the phase permutations and the per-phase reflections present, the polyphase block form $\bI_m\otimes\bR_b$ holding to $\maxBlockErr$ and the $m$-fold eigenvalue multiplicity to $\maxMultErr$.

\begin{table}[t]
\caption{Exact permutation symmetry of the window covariance by exhaustive search: distinct autocovariances, group order, group, and dimension of the symmetric commutant against that of a symmetric matrix}
\label{tab:groups}
\centering\footnotesize
\setlength{\tabcolsep}{3pt}
\begin{tabular}{@{}lcccll@{}}
\toprule
Pattern & $M$ & distinct & order & group & dimension\\
\midrule
\input{group_table.tex} 
\bottomrule
\end{tabular}
\end{table}

\emph{The basis.} For the first-order window the roots of~\eqref{eq:rayd} at $M=8$ deviate from the DCT-II grid by at most $\rdDevDctTwoNineEight$ grid steps at pole $0.9$ and $\rdDevDctTwoNNEight$ at $0.99$, and at pole $\rdSmallRho$ they lie within $\rdDevDstOneSmallEight$ grid steps of the DST-I grid (Fig.~\ref{fig:frequencies}(a)). At $M=8$ and $32$ every crossing carries a sinusoid in all six windows: the component of a unit eigenvector at a crossing has norm at least $\actMinRealPos$ for the windows with real positive poles of orders one to three, whose eigenvectors carry one sinusoid each, at least $\actMinOthers$ for the mixed-sign and the resonant AR(2) windows, and down to $\actMinPairRealThirtyTwo$ for the AR(3) window with a pair and a real pole at $M=32$. The resonant AR(2) window at $M=32$ has $\nTwoSinResThirtyTwo$ eigenvectors with two sinusoids and $\nOneSinResThirtyTwo$ with one, and the AR(3) window with a pair and a real pole has eigenvectors with up to $\maxSinPairRealThirtyTwo$. The identity~\eqref{eq:identity} holds to $10^{-9}$ with rank at most $2p$ for every pattern, and the eigenvalues interlace the spectrum samples with an observed shift of at most $\shiftMaxObserved$. The exact rank of Theorem~\ref{thm:departure} is checked at $M=\rankM$ on \rankNumAr{} AR windows, \rankNumZeroPole{} of them declared at an order above their own by zero poles: the corner block factors as $\bH_a(\bH_a+\bT_a)$ to $\rankFactMaxErr$, its determinant equals~\eqref{eq:detcorner} to $\rankDetMaxErr$, and $\bD$ and $\bR-\bR_c$ both have rank $2p_0$ in every case; for ARMA windows with $q\le p$ the departure has rank $2p$ with uncancelled poles and loses two for each pole cancelled by a zero (\rankArmaCancelled{} for the ARMA(2,1) window whose zero cancels the pole $0.5$, \rankArmaWhite{} for the ARMA(1,1) window whose zero cancels its pole, which is white noise), the ARMA(1,2) window reaches its bound \rankArmaQgtP, and the sum of AR(1) and AR(2) windows has rank \rankSumDisjoint{} with disjoint poles and \rankSumShared{} when they share one.

\emph{Classes.} Table~\ref{tab:classes} gives, at $M=8$, the residual and the coding-gain ratio of each window in the DCT-I, DCT-II, DST-I, DST-II and cyclic classes, with the polyphase or quadrature construction where it applies.\ifextended{} Table~\ref{tab:classes32} repeats the measurement at $M=32$.\fi{} At $M=8$, for the windows with real positive poles, of orders one to three, the DCT-II has the smallest residual and the largest coding-gain ratio ($\resDctTwoArOneEight$ and $0.99$ for AR(1)). For the negated poles the DST-II takes its place with identical numbers, as Proposition~\ref{prop:nyquist} requires (conjugation error $\maxNyquistErr$). A mixed sign pattern follows its dominant pole. The two-phase windows are matched by their wreath group: the polyphase DCT-II for poles $\pm0.9$ and the polyphase DST-II for $\pm0.9i$ reach residual $\resPolyPosEight$ and coding-gain ratio $\gainPolyPosEight$, while for $\pm0.9$ the best plain class reaches $\resPlainPolyPosEight$ and $\gainPlainPolyPosEight$. The resonant window is matched by no class: the best plain residual is $\resBestPlainResEight$ with the DCT-II giving the best coding-gain ratio, $\gainDctTwoResEight$, and the quadrature class, spanned by the two carriers at the resonance times half-length envelopes, holds more of the energy (residual $\resQuadResEight$) at a lower coding-gain ratio ($\gainQuadResEight$).

\ifextended
\begin{table*}[t]
\caption{The measurement of Table~\ref{tab:classes} at $M=32$}
\label{tab:classes32}
\centering\footnotesize
\setlength{\tabcolsep}{5pt}
\begin{tabular}{@{}lcccccl@{}}
\toprule
Pattern & DCT-I & DCT-II & DST-I & DST-II & cyclic (Hartley) & polyphase or quadrature\\
\midrule
\input{class_table_32.tex} 
\bottomrule
\end{tabular}
\end{table*}
\fi

\emph{The coding-gain loss.} Table~\ref{tab:bounds} measures Theorem~\ref{thm:gain} on \cgNumWindows{} windows of orders one to four from $M=\cgSmallestM$ to $\cgLargestM$. The identity~(a) holds to $10^{-9}$, the rank of $\bUps$ is $2p$ and its trace equals $\sigma^{-2}\operatorname{tr}(\bD\bR)$ in every case ($\cgTraceArOneNine$ for the pole $0.9$, which is $-2\rho/(1+\rho)$), and the chain~\eqref{eq:gainbounds} holds throughout. The correlation bound~\eqref{eq:corrbound} applies to the \cgCorrApplies{} windows with real poles, where $\norm{\bX}_2<1$ at every length, and exceeds the ratio by at most \cgCorrGapMax. For the three windows with a resonant pair $\norm{\bX}_2$ exceeds one and the rank bound is the one that applies. The first-order window with pole $0.95$ loses \cgRatioArOneNineFiveEight{} at $M=8$ and \cgRatioArOneNineFiveOneTwentyEight{} at $M=128$ against the rank bound \cgJensenArOneNineFiveEight{} and \cgJensenArOneNineFiveOneTwentyEight. The AR(2) window with poles $0.9$ and $0.5$ loses \cgRatioArTwoRealEight{} and \cgRatioArTwoRealOneTwentyEight. The resonant window loses \cgRatioArTwoResEight{} and \cgRatioArTwoResOneTwentyEight, and the AR(4) window \cgRatioArFourRealEight{} and \cgRatioArFourRealOneTwentyEight. At $M=128$ every ratio and every rank bound is within $5$ percent of one and the rank bound is within $1$ percent of the ratio: the loss is set by $2p$ numbers spread over $M$ coefficients, as the theorem says, and the rank bound is loose only at short lengths, where the trace is spread over few coefficients. For the separable $8\times8$ block the two-dimensional ratio equals the square of the one-dimensional one, $\cgSepRatioNineFive$ at pole $0.95$ and $\cgSepRatioNine$ at $0.9$, with the squared correlation bounds $\cgSepCorrNineFive$ and $\cgSepCorrNine$ (Section~\ref{sec:images}).

\begin{table*}[t]
\caption{Theorem~\ref{thm:gain} by window and length: the coding-gain ratio $\cg_{\mathrm{KLT}}/\cg_{\mathrm{DCT}}$, the rank bound~\eqref{eq:gainbounds} (middle form) and the correlation bound~\eqref{eq:corrbound}, written ratio / rank bound / correlation bound; the correlation bound is omitted where $\norm{\bX}_2\ge1$ ($M=64$ omitted)}
\label{tab:bounds}
\centering\footnotesize
\setlength{\tabcolsep}{4pt}
\begin{tabular}{@{}lcccc@{}}
\toprule
Window & $M=8$ & $M=16$ & $M=32$ & $M=128$\\
\midrule
\input{bounds_table.tex} 
\bottomrule
\end{tabular}
\end{table*}

\emph{Fast structure.} Table~\ref{tab:fast} checks Proposition~\ref{prop:fast} on \fastNumPatterns{} pole patterns of orders one to four at $M=16$ and $32$. The fold block-diagonalizes the precision to $\fastMaxCross$, each branch equals its core block minus the corner block to $\fastMaxBranch$, the two cores diagonalize the class blocks to $\fastMaxCore$ with eigenvalues on the two grids, the correction term has rank exactly $p$ in every branch, the secular determinant~\eqref{eq:secular} is at most $\fastMaxSecularEig$ at the branch eigenvalues and at least \fastMinSecularMid{} midway between consecutive ones, no branch eigenvalue falls on the core grid, the Cauchy-type vectors reproduce the eigenvectors to $\fastMaxEigvec$, and the assembled transform $\bF^\T(\bK_{\mathrm{II}}^\T\bV_{\mathrm s}\oplus\bK_{\mathrm{IV}}^\T\bV_{\mathrm a})$ diagonalizes the covariance with off-diagonal fraction at most $\fastMaxKlt$. The eigenvalues move at most \fastShiftBranchMax{} places within a branch and \fastShiftJointMax{} jointly, within the bounds $p$ and $2p$, and, on the seven patterns of Table~\ref{tab:fast} together with the first-order window with pole $-0.7$, within the inertia bounds~\eqref{eq:inertia}: the corner block has inertia $(n_+,n_-)=\inertiaArOnePos$ for the pole $0.9$ and $\inertiaArOneNeg$ for the pole $-0.7$, $\inertiaArTwoReal$ for the three AR(2) windows, $\inertiaArThreePair$ for the two AR(3) windows and $\inertiaArFourReal$ for the AR(4) window, the observed displacements lie between $\placesMinObserved$ and $\placesMaxObserved$ places, the interlacing is strict in every branch, and the trace identity holds to rounding. On the same eight windows the KLT ordered by eigenvalue alternates in parity for the five whose poles are real and of one sign, starting with a symmetric vector for positive poles and an antisymmetric one for the pole $-0.7$, and does not alternate for the mixed-sign, the resonant and the pair-plus-real windows. The last column measures how far each correction is from the identity, after ordering its columns by their largest entry: \fastDevArOneSixteen{} for the first-order window at $M=16$, \fastDevArTwoSixteen{} for the AR(2) window with real poles, and \fastDevResSixteen{} for the resonant window, whose basis the cores do not approximate. The polyphase stage holds exactly for $a(z)=b(z^m)$ with $m=2$ and $3$ (cross-phase blocks $\fastPolyMaxCross$, phase blocks equal to the shorter window to $\fastPolyMaxBlock$).

\emph{Rational spectra, the recursive form, modulated windows.} Table~\ref{tab:rational} checks Corollary~\ref{cor:rational} and Proposition~\ref{prop:rational} on \ratNumCases{} windows with rational spectra, five ARMA windows of orders $(1,1)$, $(1,2)$, $(2,2)$, $(3,1)$ and $(0,2)$ and three co-channel sums, at $M=16$ and $32$: the departure from the class has rank exactly $2\max(p,q)$, or $2(k_1+k_2)$ for the sums, in every case, the fold is exact to $\ratMaxCross$, each branch carries at most $\max(p,q)$, the covariance-side design diagonalizes the window with off-diagonal fraction at most $\ratMaxKlt$, and the eigenvalues move at most \ratShiftMax{} places. For the recursive form, each folded block of the AR($p$) precision equals the half-length precision up to a block confined to the center $p\times p$ coordinates (entries outside it at most $\recMaxOutside$) of rank exactly $p$ for the seven patterns of Table~\ref{tab:fast}, with the first-order center blocks $-\bm\Omega_{\mathrm s}$ and $-\bm\Omega_{\mathrm a}$ of Proposition~\ref{prop:recursive}, $-\rho(1-\rho)$ and $\rho(1+\rho)$, to $\recCenterErr$. For three complex first-order windows (poles $0.9e^{0.37\pi i}$, $0.95e^{0.5\pi i}$, $0.7e^{0.9\pi i}$) the covariance equals the conjugated real window to $\modMaxConj$, the modulated DCT-II diagonalizes the modulated class member to $\modMaxOffdiag$, and the departure has rank two. The Wiener separator of the three sums in the DCT-II basis is the diagonal spectral filter plus a correction of rank \wienerRankSumOneOne, \wienerRankSumTwoOne{} and \wienerRankSumTwoThree{} (bound \wienerBoundSumTwoThree{} in the last case), of relative size \wienerSizeSumTwoThree{} in the last case.

\emph{The cascade.} Proposition~\ref{prop:cascade} is checked on the \cascCases{} windows of Tables~\ref{tab:fast} and~\ref{tab:rational}, the seven AR($p$) patterns and the eight rational-spectrum windows at $M=16$ and $32$: the number of stages in each branch equals its bound in every case, $p$ for the AR($p$) patterns, $\max(p,q)$ for the ARMA windows and $k_1+k_2$ for the sums (at most \cascMaxStages{} stages, for the sum of the AR(2) and AR(3) windows), every stage is a Cauchy matrix in the running basis (largest deviation of a column from its Cauchy vector $\cascMaxCauchy$), and the cascade is orthonormal and diagonalizes the branch with off-diagonal fraction at most $\cascMaxOffdiag$ and eigenvalues agreeing with the direct ones to $\cascMaxEig$.

\emph{Odd length and compressibility.} Proposition~\ref{prop:fast}(f) is checked on \oddNumCases{} cases, one window of each order one to three (the AR(1), the AR(2) with real poles and the pair-plus-real AR(3) windows of Table~\ref{tab:fast}) at $M=9$, $17$ and $33$: the fold with the center sample block-diagonalizes the precision to $\oddMaxCross$, the restricted DCT-II columns are orthonormal and equal the DCT-VI and DST-VII kernels to $\oddMaxKernel$, they diagonalize the class blocks to $\oddMaxCore$, the corrections have rank exactly $p$ in both branches and occupy the first $p$ coordinates, and the assembled transform diagonalizes the covariance with off-diagonal fraction at most $\oddMaxKlt$. For the compressibility of the merged correction (Section~\ref{sec:count}), the first-order window with pole $0.9$ at $M=256$ and $512$ has a top-level off-diagonal block of rank \epsRankSixTwoFiftySix{} and \epsRankSixFiveTwelve{} at absolute accuracy $10^{-6}$ and \epsRankTenTwoFiftySix{} and \epsRankTenFiveTwelve{} at $10^{-10}$ (with the threshold relative to the largest singular value of the block, \epsRelSixTwoFiftySix, \epsRelSixFiveTwelve, \epsRelTenTwoFiftySix{} and \epsRelTenFiveTwelve, the values reported for the first-order correction in Section~6 of~\cite{reznik2026exact}). Over the orders one to four the top-level ranks lie between \epsRankSixMin{} and \epsRankSixMax{} at $10^{-6}$ and between \epsRankTenMin{} and \epsRankTenMax{} at $10^{-10}$, within one of the first-order values, while the individual stages have ranks \epsStageMin{} to \epsStageMax{} at $10^{-6}$. Over the \epsNumLevelsFiveTwelve{} levels of the hierarchical partition at $M=512$ the ranks of the merged correction stay at or below \epsLevelSixMaxReal{} and \epsLevelTenMaxReal{} for the three windows with real poles (orders one, two and four). The resonant window is the exception that explains the rule: its core grid $c(\pi j/L)$ is not monotone in $j$, so in the index order the level ranks reach \epsLevelSixMaxResIndex{} for the merged correction and \epsLevelSixMaxResStage{} for a single stage, while in the order of the grid values, which is the order of the Cauchy structure, the single stage has ranks at most \epsLevelSixMaxResStageSorted{} and the merged two-stage correction at most \epsLevelSixMaxRes{} and \epsLevelTenMaxRes{} at the two accuracies, above the single stage but far below the block size.

\emph{Moving-average and integrated windows.} Corollary~\ref{cor:ma} is checked on MA windows of orders one to three at $M=16$ and $32$: the covariance has bandwidth $q$, the class member built from the autocovariance equals the member of Corollary~\ref{cor:rational} built from the spectrum samples, the departure equals minus the two Hankel corners to rounding, and its rank is exactly $2q$. The residual of the MA(2) window in the DCT-II class is \maResMaTwoThirtyTwo{} at $M=32$. Proposition~\ref{prop:integrated} is checked on ten integrated windows, the random walk, the integrated AR windows of orders one to three with $d=1$, and windows with $d=2$ and $3$: the precision has bandwidth $p+d$ and the integrated symbol in its interior, the DST-VII member equals the form of (a) to rounding, the departure is supported on the first $p+d-1$ and the last $p+d$ rows and columns with rank $2(p+d-1)$ in all cases but ARIMA(1,3,0), where it is $\intRankOneThreeZero$, the random walk has the DST-VII as its KLT with the closed-form eigenvalues to machine precision, and no integrated window has the reflection (relative asymmetry at least \intAsymMin). The residual of the integrated AR(1) window with pole $0.9$ in the DST-VII class is \intResOneOneZeroSixteen{} at $M=16$ and \intResOneOneZeroThirtyTwo{} at $M=32$, that of the integrated AR(2) window with real poles \intResTwoOneZeroThirtyTwo{} and of the resonant one \intResResThirtyTwo{} at $M=32$, against \intResDTwoThirtyTwo{} for the twice-integrated AR(1) window. On the covariance side the integrated ARMA windows with $d=1$ of orders $(1,1)$, $(0,1)$, $(0,2)$ and $(2,2)$ differ from the DST-VII member with coefficient variances $S_y(\omega_k)/(2-2\cos\omega_k)$ by rank \intCovRanks{} respectively, at most $2\max(p,q)$, with residuals \intCovResOneOneOneThirtyTwo{} and \intCovResZeroOneOneThirtyTwo{} for the first two at $M=32$. The sixteen trigonometric types scored on the ten integrated windows at $M=16$ and $32$ give the DST-VII the largest coding-gain ratio for every once-integrated AR window and a residual above that of the DST-IV (DST-III for the resonant window), the random walk zero residual in the DST-VII class alone, and for $d\ge2$ no residual below $0.1$ and no coding-gain ratio above $0.61$ in any type, the \intBestTypeDTwo{} being nearest. Theorem~\ref{thm:gain} with the DST-VII on \intGainNumCases{} once-integrated windows at $M=16$ and $32$: the identity holds to $10^{-9}$, the rank of $\bUps$ equals the covariance-side departure rank, the rank bound holds throughout and the correlation bound for the \intGainCorrApplies{} windows with a small correlation norm (ratio \intGainRatioZeroOneOneThirtyTwo{} against bound \intGainCorrZeroOneOneThirtyTwo{} for ARIMA(0,1,1) at $M=32$). The integrated ARMA(2,1,2) window loses most, ratio \intGainRatioTwoOneTwoThirtyTwo{} against the rank bound \intGainJensenTwoOneTwoThirtyTwo.

\emph{Discovery.} The reading of the exact group from sample covariances (Section~\ref{sec:groups}) was run on \discNumPatterns{} patterns, the AR(1) window, the resonant AR(2) window and the two-, three- and four-phase windows of Table~\ref{tab:groups} (groups of orders \discOrderArOne, \discOrderRes, \discOrderPhaseTwo, \discOrderPhaseThree{} and \discOrderPhaseFour), with $N$ from \discSmallestN{} to \discLargestN{} Gaussian windows and \discTrials{} trials at each $N$. The nearest permutation outside the group has resolving sample size $N^\star=\discNstarArOne$, \discNstarRes, \discNstarPhaseTwo, \discNstarPhaseThree{} and \discNstarPhaseFour{} windows respectively, and the reading equals the group in every trial from $N=\discExactFromArOne$, \discExactFromRes, \discExactFromPhaseTwo, \discExactFromPhaseThree{} and \discExactFromPhaseFour{} onward, the three-phase window at $M=9$ among them. At $N=\discLargestN$ the median of the largest statistic on the group is \discMaxInLastArOne, \discMaxInLastRes, \discMaxInLastPhaseTwo, \discMaxInLastPhaseThree{} and \discMaxInLastPhaseFour{} (expectation one), the median of the smallest statistic outside it is \discMinOutLastArOne, \discMinOutLastRes, \discMinOutLastPhaseTwo, \discMinOutLastPhaseThree{} and \discMinOutLastPhaseFour, and the separation ratio grows in proportion to $N$. A reading is certified when the ratio is at least $3$, every statistic of the reading is below $3$ and the reading is closed under composition. Of the \discCertifiedTotal{} certified readings at $N\ge5000$, \discCertifiedWrongWords{} was wrong, a proper subgroup of the two-phase group read at $N=10000$ when the statistics of one coset fluctuated upward, and the rest were the exact group. The calibrated pencil over the $M(M-1)/2$ skew-symmetric candidates counts, in the median over ten trials, \skewCountFivePhaseTwo, \skewCountFivePhaseThree{} and \skewCountFivePhaseFour{} directions at $N=5000$ for the three interleaved windows against the kernel dimensions \skewDimPhaseTwo, \skewDimPhaseThree{} and \skewDimPhaseFour, and \skewCountLastArOne{} and \skewCountLastRes{} at $N=\skewLastN$ for the two generic windows, whose kernel is empty. The common kernel of three two-phase windows with inner poles $0.9$, $0.5$ and $-0.7$ has dimension \skewFamilyDim{} against \skewFamilySingle{} for each window alone, and their common permutation group has order \skewFamilyGroupOrder. For the complex window with pole $0.9e^{i\carWzero\pi}$ the modulated-class residual is minimized at $\carMinEight\pi$ ($M=8$) and $\carMinThirtyTwo\pi$ ($M=32$) with residuals \carResEight{} and \carResThirtyTwo, equal to those of the real AR(1) window in the DCT-II class, whose own minimizer over the family is $\omega=\carRealMinEight$ with residual \carRealAtWEight{} at the carrier ($M=8$). The minimizer read from $N$ sample windows (\carTrials{} trials) has root-mean-square error $\carRmseFirstEight\pi$ and $\carRmseFirstThirtyTwo\pi$ at $N=\carNfirst$ and $\carRmseLastEight\pi$ and $\carRmseLastThirtyTwo\pi$ at $N=\carNlast$ for $M=8$ and $32$. For the real resonant window the complex family is minimized at $\omega=\carResonantComplexMinEight$ (residual \carResonantComplexResEight{} against \carResonantComplexAtWEight{} at the resonance, $M=8$) while the quadrature class is minimized at $\carResonantQuadMinEight\pi$ and $\carResonantQuadMinThirtyTwo\pi$ for $M=8$ and $32$.

\emph{Limits.} Fig.~\ref{fig:classes} follows the matched classes as the pole radius rises from $0.5$ to $0.999$. The DCT-II keeps at least $\gainRealPosDctTwoMinEight$ of the KLT coding gain for the real-pole family at $M=8$ and the DST-II the same for the negated family, while the wrong sign class falls to $\gainRealNegDctTwoNNNEight$ at radius $0.999$. The polyphase DCT-II keeps at least $\gainPolyPosMinEight$ for the two-phase family, where the plain DCT-II falls to $\gainPolyPosPlainNNEight$ at radius $0.99$. For the resonant family the ratio falls in every class, to $\gainResDctTwoNNEight$ (DCT-II), $\gainResDstTwoNNEight$ (DST-II) and $\gainResQuadNNEight$ (quadrature) at radius $0.99$, while the quadrature residual stays at $\resResQuadNNNEight$ at radius $0.999$ against $\resResDctTwoNNNEight$ for the DCT-II. The limit matrix $\bC(0.3\pi)$ itself has residual at least $\limMinPlainThirtyTwoOff$ in every class of the catalog at $M=32$ and $\limQuadThirtyTwoOff$ in the quadrature class. At $\omega_0=3\pi/8$, which is $2\pi\cdot6/32$, it is a circulant and the cyclic class holds it exactly ($\limHartleyThirtyTwoGrid$), while the extension classes leave at least $\limMinExtThirtyTwoGrid$.

\ifextended
\emph{The two measures.} The residual and the coding gain can disagree. At $M=32$ the DCT-I residual is below the DCT-II residual for every real-pole window ($\resDctOneArOneThirtyTwo$ against $\resDctTwoArOneThirtyTwo$ for AR(1)) while its coding-gain ratio is lower ($\gainDctOneArTwoThirtyTwo$ against $\gainDctTwoArTwoThirtyTwo$ for the poles $0.9$ and $0.5$). The residual is a second-moment distance dominated by the largest eigenvalues. The coding gain is a geometric mean that rewards the nulls of the small coefficients. Matched-group selection for transform coding should be scored by the second. The class member $\bR_c$ of Corollary~\ref{cor:identity} is a poor proxy for the nearest member at a short window: for the AR(1) window with pole $0.9$ its relative distance from $\bR$ is $\covDepArOneEight$ at $M=8$ and $\covDepArOneThirtyTwo$ at $M=32$, against a precision departure of $\precDepArOneEight$ and $\precDepArOneThirtyTwo$, because inversion amplifies the corner blocks where the spectrum is large; the Reynolds projection is the nearest member, and the residual it leaves is the one tabulated.
\fi

\section{Reference Implementation: Accuracy and Operation Count}\label{sec:bench}

This section reports the measurements of the reference implementation described in
Sections~\ref{sec:complexdata} to~\ref{sec:realization}. Every value in it is produced by a script
in the released bundle and written to a result file, and the manuscript reads that file rather than
carrying typed values, as the eight experiments of Section~\ref{sec:num} do for the rest of the
paper. The machine is recorded as \texttt{\machine}, with Python \pyVersion{} and NumPy
\npVersion. A rerun on other hardware changes the seconds of Table~\ref{tab:impltiming} and
Fig.~\ref{fig:impltiming} and leaves the kernel evaluation counts, the ranks, the sweep counts and
the accuracy figures as they are, which is the reason both are reported and the reason the claim
about the order rests on the counts.

\subsection{The Operation Count}\label{sec:benchcount}

\begin{figure*}[!t]
  \centering
  \includegraphics[width=\textwidth]{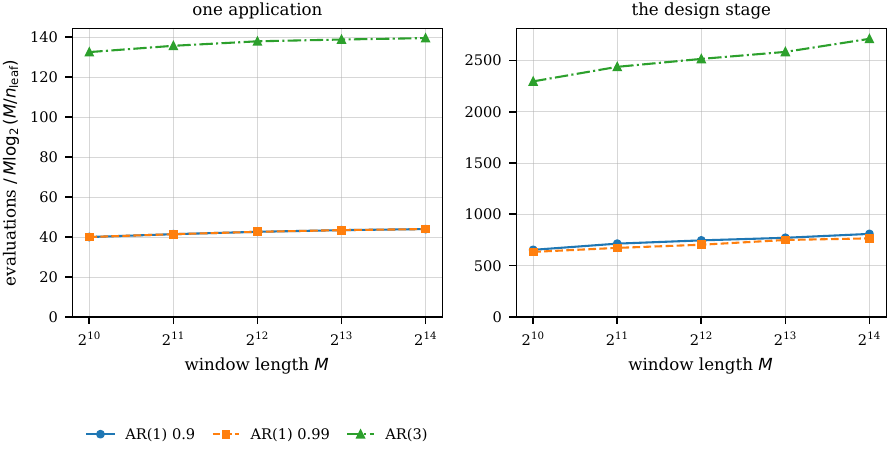}
  \caption{Kernel evaluations the hierarchical stages perform, divided by
  $M\log_2(M/n_{\mathrm{leaf}})$. The tree stops at boxes of $n_{\mathrm{leaf}}$ nodes, so that
  product is the number of levels carrying far-field work and a flat curve is the order claim of
  Section~\ref{sec:count}. These counts do not depend on the machine.}
  \label{fig:implcounts}
\end{figure*}

Fig.~\ref{fig:implcounts} and Table~\ref{tab:implcounts} report the kernel evaluations the
hierarchical stages perform, for one design and one application, over $M=\countMfirst$ to
$\countMlast$. The normalization is $M\log_2(M/n_{\mathrm{leaf}})$, since the tree stops at boxes
of $n_{\mathrm{leaf}}$ nodes and that product is the number of levels carrying far-field work;
flat in $M$ is then the order claim of Section~\ref{sec:count}.

On an AR(3) window the normalized application count rises by \applyFlatArThree{} percent over a
\countRange-fold range of length, with increments of \applyIncFirstArThree{} falling to
\applyIncLastArThree{} at successive doublings. On an AR(1) window the rise is
\applyFlatArOneZeroNine{} percent. The design counts behave the same way, rising by
\designFlatArThree{} percent on the AR(3) window. The residual rise is the near field at the
leaves, whose share of the total falls as $M$ grows, which is why the increments shrink rather than
settle at a constant offset.

\subsection{Timing}\label{sec:benchtiming}

\begin{figure*}[!t]
  \centering
  \includegraphics[width=\textwidth]{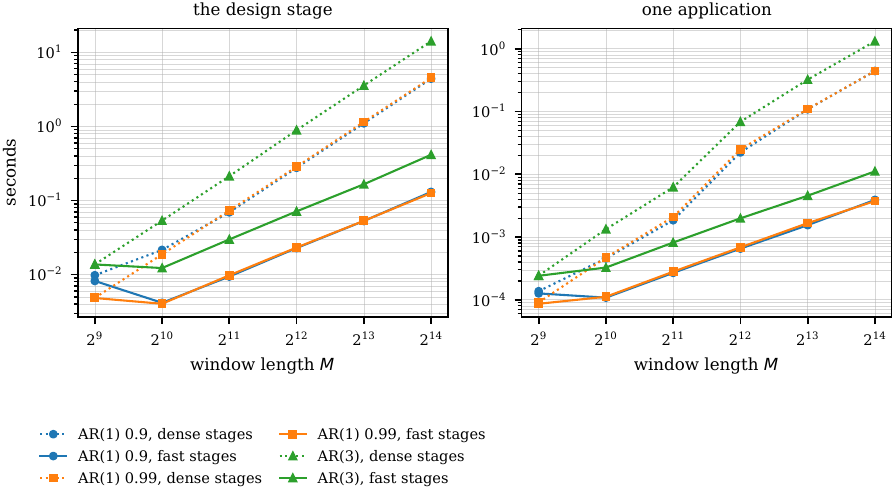}
  \caption{Seconds for the design stage and for one application, with the Cauchy stages formed
  densely and with the hierarchical evaluation. Timings move between runs
  in a way the counts of Fig.~\ref{fig:implcounts} do not.}
  \label{fig:impltiming}
\end{figure*}

Table~\ref{tab:impltiming} and Fig.~\ref{fig:impltiming} give seconds for the design stage and for
one application, with the Cauchy stages formed densely and with the hierarchical evaluation. At the
longest length measured one application with hierarchical stages is
\applyRatioLastArOneZeroNine{} times faster than with dense stages on the AR(1) window and
\applyRatioLastArThree{} times on the AR(3) window, and the design stage is
\designRatioLastArThree{} times. Fitted over the lengths at which the hierarchical stages are
selected, the application with dense stages grows as $M^{\fitApplyDenseArThree}$ on the AR(3)
window and with hierarchical stages as $M^{\fitApplyFastArThree}$, against a model
$t=M(a\log_2M+b)$ whose worst deviation is \devApplyFastArThree{} percent. These figures carry less
than the counts do, since timings on a desktop machine move
between runs, which is visible in the fitted deviations and not in the counts.

\subsection{Accuracy}\label{sec:benchacc}

Table~\ref{tab:implaccuracy} compares the construction against a dense eigendecomposition of the
window covariance. Over the window families and lengths measured the eigenvalues agree to
$\accEig$ relative, analysis and synthesis are inverse to $\accRound$, and the basis is orthonormal
to $\accOrth$. The hierarchical stages and the dense stages reach the same figures, which is the
point of the local shifting of Section~\ref{sec:fmmreal}.

\subsection{The Model Classes}\label{sec:benchranks}

Table~\ref{tab:implranks} reports, for \nRankCases{} windows across the classes of
Sections~\ref{sec:complexdata} and~\ref{sec:covside}, the rank the departure is found to have
against the bound its statement gives. Every case sits at or below its bound: the complex
autoregressive and complex mixed windows at $2p$ exactly, as
Proposition~\ref{prop:complex} states. The real mixed and co-channel windows at or below the bound
of Corollary~\ref{cor:rational}. The integrated windows at or below
Proposition~\ref{prop:integrated} in the real case and Proposition~\ref{prop:complexintegrated} in
the complex case. For the integrated windows the departure away from the border is $\offBorder$
relative, which is the numerical form of the statement that the support is exactly the border.

\subsection{The Interaction Rule and the Secular Step}\label{sec:benchplan}

\begin{figure}[!t]
  \centering
  \includegraphics[width=0.526\textwidth]{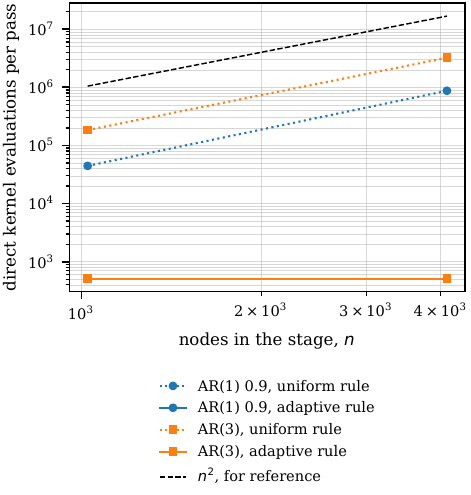}
  \caption{Direct kernel evaluations one pass leaves under the two interaction rules, against the
  number of nodes in the stage. The uniform rule leaves a fixed fraction of $n^2$, and the fraction
  grows with the order of the model. The deferred rule leaves a count that does not grow with $n$.}
  \label{fig:implplan}
\end{figure}

Table~\ref{tab:implplan} and Fig.~\ref{fig:implplan} give the direct kernel evaluations one pass
leaves under the two interaction rules of Section~\ref{sec:fmmreal}. Under the uniform rule the
direct work per pass is between \planUniMin{} and \planUniMax{} of $n^2$ and the fraction grows
with the order of the model. Under the deferred rule the same count is \planAdaptive{} evaluations
at every length measured, which at $n=\planNbig$ is $\planFracAtBig$ of $n^2$.

\begin{figure}[!t]
  \centering
  \includegraphics[width=0.539\textwidth]{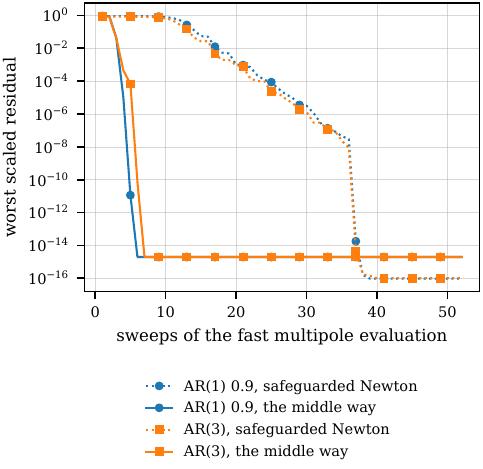}
  \caption{The secular equation. A sweep is one pass of the hierarchical evaluation over every
  iterate, so a sweep is the same expense under either schedule. Safeguarded Newton is rejected and
  replaced by a bisection most of the time, because a root commonly sits close to one of the two
  poles bounding its interval.}
  \label{fig:implsolver}
\end{figure}

Table~\ref{tab:implsolver} and Fig.~\ref{fig:implsolver} compare the two secular schedules of
Section~\ref{sec:secularreal} on the node sets the design stage meets. Both are the released
schedules and a sweep is one pass of the hierarchical evaluation in both, so the comparison is of
the step and not of the work per sweep. Safeguarded Newton reaches a scaled residual of $10^{-12}$
after \sweepNewtonArOneZeroNine{} sweeps on the AR(1) window and \sweepNewtonArThree{} on the AR(3)
window. The middle way reaches it after \sweepMiddleArOneZeroNine{} and \sweepMiddleArThree.

\subsection{The Polyphase Route and Odd Lengths}\label{sec:benchpoly}

Table~\ref{tab:implpoly} reports the polyphase route of Remark~\ref{rem:polymix} against the same
transform with the route switched off, at $M=\polyM$. The design stage is between \polyDesignMin{}
and \polyDesignMax{} times faster and the two routes agree on the eigenvalue multiset. The number
of stages is the same either way, since $m$ phases carrying $p/m$ stages per branch is the same
count as one window carrying $p$, and what changes is the dimension each stage acts on.

Table~\ref{tab:imploddcores} and Table~\ref{tab:imploddtransform} report odd lengths. The two cores
of Proposition~\ref{prop:oddcores} are orthonormal and their one-transform realizations agree with
the explicit matrices to $\oddCoreWorst$ relative. The fold leaves $\oddLeakWorst$ between the two
branches, the departure of each branch is the grid less the same corner block to $\oddGenWorst$,
its rank is exactly $p$ in every case measured, and the transform reaches the same accuracy as at
even length.

\section{Experiments on Recorded Data}\label{sec:recorded}

This section applies the construction to recorded data: a microphone array in a reverberant room,
the receive array of a 77 GHz automotive radar, and the fast-time axis of the same radar. The first
two ask whether the transform improves a direction-finding pipeline, and the third measures the
likelihood margins that Propositions~\ref{prop:complex} and~\ref{prop:borderlik} predict.

\subsection{Direction Finding on Real Array Data}\label{sec:df}

The measurements of the earlier sections are on the models the construction is built for. This subsection reports what
the transform does on a real recording, in the application the exact whitening was meant to serve.
The corpus is the LOCATA development set~\cite{evers2020locata}, the dicit array: thirteen
collinear microphones at plus and minus $\{0, 4, 8, 16, 32, 64, 96\}$ cm, recorded at 48 kHz in a
reverberant room and decimated here to 16 kHz, with per-sample voice activity for every source,
which gives both the number of active sources and the true cone angle of each from the array axis
at every evaluation time. Seven recordings are used. A frame counts as noise only when no source is
active in it, widened by 100 ms on each side so that onsets and tails do not enter the noise model,
and every figure below is reported on frames or blocks the model was not fitted on.

\subsubsection{The transform on the clock}\label{sec:dftime}

Room noise is strongly colored in time, so the first question is how well each transform
decorrelates it. A Gaussian model that is diagonal in a transform domain is a proper density, so
the comparison is by the log-likelihood a model gives held-out noise frames, paired frame by frame.
That statistic uses $N M$ scalars where a sample covariance uses $M^2$ entries, which matters here:
at $M = 256$ the off-diagonal energy of a sample covariance built from the noise frames of one
recording has a floor near $0.55$, well above the $4 \times 10^{-2}$ the cosine transform leaves on
the model, so the covariance route cannot resolve the comparison with the noise this corpus
provides.

Table~\ref{tab:dflikelihood} gives the result beside what the theory predicts. The prediction needs
no new argument: a diagonal model in an orthonormal transform loses, in expectation, exactly the
Kullback-Leibler divergence from the window law to that model, which for the fitted AR($p$) model
is available in closed form,
\begin{equation}\label{eq:dfkl}
  \frac{1}{M}\,\mathrm{KL}\bigl(\bR \,\|\, \bK^{\T}\bLam\bK\bigr)
  = \frac{\operatorname{tr}(\bLam^{-1}\bK\bR\bK^{\T}) - M
          + \log\det\bLam - \log\det\bR}{2M},
\end{equation}
with $\bK$ the transform and $\bLam$ its variances. The KLT of the fitted model has
$\mathrm{KL} = \dfKltKl$ per sample by construction, so the quantity in~\eqref{eq:dfkl} for the
cosine transform is the gain the KLT is predicted to show.

Over the \dfCells{} cells of Table~\ref{tab:dflikelihood} the KLT leads the cosine transform in
every one, by \dfGainMin{} to \dfGainMax{} nats per sample, between \dfTmin{} and \dfTmax{}
standard errors. The measured gain divided by the prediction of~\eqref{eq:dfkl} runs from
\dfPredRatioMin{} to \dfPredRatioMax, median \dfPredRatioMed, so on real room noise the gain is the
one the theory gives, to within the error of the fitted model. Against the real Fourier basis, the
transform the Whittle approximation uses, the measured gain is larger and the agreement with the
prediction looser, \dfPredRatioFourierMin{} to \dfPredRatioFourierMax{} times, which is consistent
with the Fourier model being the further of the two from the window law at these lengths. The gain
falls as the window grows, from \dfGainAtShort{} nats per sample at $M = 64$ to \dfGainAtLong{} at
$M = 1024$: the departure of Theorem~\ref{thm:departure} is a border effect of fixed rank, so its
share of a window falls as the window lengthens, and the new transform earns its keep on short
windows. For scale, a white model is \dfWhiteGap{} nats per sample behind the KLT on the same
frames, so the cosine transform recovers most of what there is to recover and the KLT recovers the
rest.

\subsubsection{The transform on the array axis}\label{sec:dfspace}

The direction-finding use is on the array axis rather than on the clock. At each frequency the
ambient field across the seven uniformly spaced microphones is a Hermitian Toeplitz covariance, so
a complex AR($p$) model of the spatial sequence is the object of Proposition~\ref{prop:complex} and
its KLT whitens that model exactly. The window is seven long, so nothing here is about operation
count. It is about exactness and about how few noise frames the model needs. Four whiteners are
built from the same noise-only frames at each frequency bin: none, the circulant class member of
the fitted model whitened by the Fourier basis, the exact KLT of the fitted complex AR model, and
the unstructured sample covariance.

Table~\ref{tab:dfspatial} reports the eigenvalue spread each leaves on held-out noise frames, which
is one when the whitening is exact. Over \dfSpatialRows{} recording and noise-count pairs the
AR-KLT whitener leaves a smaller spread than the circulant one in \dfSpatialWins, by a median
factor of \dfSpatialRatioMed{} and up to \dfSpatialRatioMax. The unstructured estimate is useless
below about eight noise frames, where it is singular on a seven-element array, and overtakes both
structured models above about thirty-two. That is the same crossing the window-length comparison of
Section~\ref{sec:dftime} shows, in the variable that matters for an array: a structured model with
a handful of parameters is what to use when noise-only data is scarce.

\subsubsection{A Caution: the Transform as a Spatial Prewhitener}\label{sec:dfdecisions}

Why a whitener is contemplated at all should be stated before the measurement. MUSIC separates a
signal subspace from a noise subspace by the orthogonality of the two, and the rule that finds the
split treats the smaller eigenvalues of the spatial covariance as a flat floor: the noise
eigenvalues are assumed equal, so that everything above them is signal. A noise field that is not
spatially white breaks that assumption at the root, since its own eigenvalues are spread and the
split is read in the wrong place. The standard remedy is to whiten the noise first, which is why a
practitioner reaches for a noise model, and why a transform that whitens a modeled field exactly
is a candidate for the job. Section~\ref{sec:dfspace} shows the transform does whiten the measured
field better than the alternatives. This section reports what that bought downstream, and the
answer turns on which downstream method is used.

The two measurements above are about the transform as a model of noise, which is what it is. This
section reports a third measurement, of a use we tried and do not recommend, and reports it as a
caution rather than as a result.

Enumeration first. Table~\ref{tab:dfdecisions} applies the minimum description length rule to the
whitened eigenvalues over \dfBlocks{} blocks of about 0.26 s with a known source count. No whitener
gives a usable count: the best accuracy over all methods and noise-frame counts is \dfCountBest,
and the rule overcounts throughout. Its assumption, that the whitened noise is exactly white, is
not met by a residual spread of six to eleven.

Bearing is where the caution lies. Table~\ref{tab:dfbearing} estimates it by MUSIC with whitened
steering vectors, block by block over the \dfBearBlocks{} blocks with exactly one active source.
Whitening by an AR-KLT fitted to the silent frames raises the median absolute error from
\dfBearMedNone{} degrees to \dfBearMedArklt, and lowers the fraction of estimates within ten
degrees from \dfBearHitNone{} to \dfBearHitArklt. Anyone prewhitening array data this way should
know that number.

Two things say that this is a misuse rather than a property of the transform. The first is that a
whitener fitted to the silent frames is then applied to active blocks, and the noise in an active
block is not the noise the silent frames carry. The
right-hand columns of Table~\ref{tab:dfbearing} remove the dominant eigendirection from each active
block and measure the spread of what remains: it is \dfResidRatioMin{} to \dfResidRatioMax{} times
the spread the same whitener leaves on held-out silent frames. What is present during speech, after
the direct path is taken out, still carries strong directional structure, which is what
reverberation in a room looks like. A room response is linear and time invariant, so the
difficulty is not nonlinearity but non-stationarity, with coherence beside it: the field during
speech is produced by the source, so it is signal dependent rather than a persistent floor, and it is
spatially coherent rather than a diffuse field with a rational spectrum.
Section~\ref{sec:dfadesprit} measures how far toward coherent and how far toward spread this
particular field sits. The second is that the unstructured sample covariance, which
whitens the silent-frame noise better than the AR-KLT does, \dfSilentSample{} against
\dfSilentArklt, loses nothing on the bearing at all, with a median of \dfBearMedSample{} degrees.
So the harm does not follow from whitening, or from whitening strongly.

Fitting the model where the decision stands recovers part of it and not all of it.
Table~\ref{tab:dffitted} repeats the bearing with the noise model fitted to each block's own
remainder, the block less its dominant direction, with the dominant direction taken from the
unwhitened block so that the comparison is one refinement step against the estimate it refines.
The AR-KLT improves from \dfBearMeanArklt{} degrees mean to \dfFitMeanArklt, which is the size of
the silent-frame misuse. It does not reach the unwhitened estimate, \dfFitMedArklt{} degrees median
against \dfFitMedNone, and among the four models fitted to the same remainder the two that assert a
rational spatial spectrum do worse on every statistic than the two that do not, worst plain median
\dfFitPlainWorstMed{} against best rational \dfFitRationalBestMed. A rational spectrum of order two
across seven elements is a strong assertion about a reverberant field, and where it is wrong its
inverse amplifies particular spatial frequencies, which is a direction-dependent distortion of the
array manifold. We give that as the reading we find consistent with the measurement, not as a
result.

What this section establishes, then, is a boundary rather than a benefit: on this corpus, with this
pipeline, the transform is a good model of the noise and a poor spatial prewhitener for MUSIC, and
neither fitting it to silence nor fitting it to the active block makes it a good one. Two
properties of the pipeline bound how far that carries. The bearing estimator is MUSIC with whitened
steering vectors, incoherently averaged over the band, rather than a steered-response method with a
phase transform, which is what published results on this corpus use. And seven elements with a
rational model of order two is the smallest case the theory admits, since the implementation asks
for $p < \lfloor M/2 \rfloor$. A longer array, or a field closer to the model, may place the
transform differently, and nothing measured here speaks to its use on the clock, where
Section~\ref{sec:dftime} finds the gain the theory predicts.

\subsubsection{ESPRIT, which rests on something else}\label{sec:dfesprit}

MUSIC is not the only choice, and the alternative changes the question. ESPRIT does not use the
orthogonality of the two subspaces at all. It uses the shift invariance of a uniform array: the
same signal subspace seen by two subarrays displaced by one element differs by a diagonal of phase
factors, and the bearings are read from those. Nothing in that argument asks the noise to be white,
so there is a reasonable expectation that prewhitening is unnecessary, and a reason to suspect it
is harmful: whitening by a general matrix $\bW$ sends the manifold $\mathbf{a}(\theta)$ to
$\bW\mathbf{a}(\theta)$, which is no longer a geometric sequence in the element index, so the
relation ESPRIT exploits does not hold on whitened data.

Table~\ref{tab:dfesprit} measures both, over the same \dfBearBlocks{} one-source blocks, with
bearings read from the phase of the LS-ESPRIT eigenvalue on the two six-element subarrays of the
seven-element uniform run and combined across bins by the median. Whitening the data and running
ESPRIT on it, which is the form that breaks the invariance, raises the median error from
\dfEspNoneMed{} degrees to \dfEspSilentNaiveMed, a factor of \dfEspNaivePenalty, and halves the
fraction within ten degrees, \dfEspNoneHit{} to \dfEspSilentNaiveHit. Using the whitener only to
estimate the signal subspace and then mapping that subspace back through the inverse of the
whitener, so that the invariance is intact when it is used, leaves ESPRIT where it started:
\dfEspSilentSubMed{} degrees median against \dfEspNoneMed, and \dfEspSilentSubHit{} within ten
against \dfEspNoneHit.

Two things follow. The first is a statement about how to use a whitener with any method built on
array invariance rather than on subspace orthogonality: apply it to the subspace estimate and not
to the data. The penalty for the other order is the factor of \dfEspNaivePenalty{} above, and it
has nothing to do with which transform supplies the whitener. The second is that even applied
correctly, whitening neither helps nor hurts ESPRIT here, which is consistent with its argument
not needing a white noise floor in the first place. On this corpus MUSIC without whitening remains
the better of the two estimators, \dfMusicNoneMed{} degrees median against \dfEspNoneMed.

\subsubsection{Coherent arrivals, and a dilated estimator built for them}\label{sec:dfadesprit}

Section~\ref{sec:dfdecisions} argued that a whitener fitted to silence addresses the wrong term
because the interference during speech is reverberation. That argument has a second consequence,
and it concerns the estimator rather than the whitener. A reverberant arrival is a delayed and
scaled copy of the direct path, so it is coherent with it. Coherence is what defeats a subspace
estimator built on averaged snapshots: when the arrivals are coherent the source covariance is rank
one however many snapshots enter the average, so the one signal eigenvector recovered spans a
mixture of the arrival bearings rather than the direct path, and no amount of further averaging
separates them.

An estimator that forms a group average over shifts of a single snapshot, rather than an average
over snapshots, does not meet that collapse. The dilated average of one snapshot is the Toeplitz
matrix of its biased autocorrelation, and the true bearings are a fixed point of a cascade of exact
one-component steps on it for any number of snapshots and any source coherence. That construction, the dilated cascade, is the natural thing to try on this data, since the
property it is built for is the property the room supplies. Its implementation here passes a
self-test before any of what follows runs: it reproduces a noiseless single bearing to machine
precision and reproduces the rank-one collapse of the averaged estimators on synthetic coherent
arrivals.

Table~\ref{tab:dfadesprit} applies it to the same \dfAdBlocks{} one-source blocks, one estimate per
snapshot and per bin, combined by the median. The result is a negative one and is reported as such.
At one component the median error is \dfAdKOneMed{} degrees against \dfMusicNoneMed{} for
snapshot-averaged MUSIC, and the fraction within ten degrees is \dfAdKOneHit{} against
\dfMusicNoneHit. Allowing two or three components, which is what a direct path with one or two
specular reflections would ask for, moves the median by \dfAdOrderSpread{} degrees in total and not
in one direction. Summing the dilated average over the sixteen snapshots of a block, which restores
the variance reduction of averaging while keeping the Toeplitz group structure, recovers most of the
difference and reaches \dfAdSumMed{} degrees. Applying the AR($p$) KLT whitener to each snapshot
before the estimator, which breaks the invariance the method rests on in the same way
Section~\ref{sec:dfesprit} describes for ESPRIT, raises the median to \dfAdWhiteMed{} degrees, and
is recorded here as a caution of the same kind rather than as a candidate.

The certification statistics of the cascade say something about the data. At one component the
whiteness test on the residual accepted at the first stage in \dfAdRouteFirst{} of
\dfAdSnapshots{} snapshots, and at two components the test failed through every stage in only
\dfAdRouteBest. A single snapshot of this recording therefore looks like one component and a noise
floor, not like a small number of discrete coherent arrivals. That is the first indication of why an
estimator built for discrete coherent arrivals finds nothing here to use.

\begin{table*}[!t]
\caption{Bearing error on the \dfAdBlocks{} one-source blocks, snapshot-averaged estimators against
the dilated cascade. Errors in degrees against the recorded source position. The last row breaks the
invariance the cascade rests on and is included as a caution.}
\label{tab:dfadesprit}
\centering
\footnotesize
\setlength{\tabcolsep}{4pt}
\begin{tabular}{lcccc}
\toprule
Estimator & Mean & Median & $\le 10^\circ$ & $\le 20^\circ$ \\
\midrule
\input{fo3_tab_adesprit} 
\bottomrule
\end{tabular}
\end{table*}

A negative result on recorded speech does not settle whether the immunity is real, because the
recording holds every effect at once and none of them is adjustable. Three controlled measurements
on the geometry of this same uniform subarray, seven elements at $0.32$~m read at
\dfCohFreq{}~Hz, separate them.

The first establishes that the immunity is real and large. A direct path at \dfCohDirect{} degrees
and one fully coherent image arrival at \dfCohImage{} degrees, over \dfCohTrials{} trials per cell,
give the sweep \ifextended{}of Table~\ref{tab:dfcoherence}\fi. With the image at $0.7$ of the direct amplitude, least-squares
ESPRIT holds a median error of \dfCohEspFloor{} degrees that is flat in the signal to noise ratio
across the range tried, which is the signature of a bias and not of noise, while the cascade
continues to fall and reaches \dfCohAdHigh{} degrees at the same operating point, a factor of
\dfCohRatio. At $0.3$ of the amplitude the ESPRIT floor is \dfCohEspFloorMid{} degrees against
\dfCohAdHighMid{} for the cascade. The dilated average summed over snapshots keeps a floor of its
own, \dfCohSumFloor{} degrees, which is the noise-free bias of that average under two coherent
components and is the reason the cascade and not the summed average is the construction to use when
coherence is the binding effect.

The second rules out the explanation first offered for the recorded deficit. A peak search compares
magnitudes over a grid of assumed steering vectors and a rotation read takes a phase difference
between subarrays, so a departure of the real array from the assumed one need not touch them
equally. Section~\ref{sec:dfcalib} shows that the residual on this array is a deterministic
manifold error rather than noise. A second sweep\ifextended{}, Table~\ref{tab:dfmanifold},\fi{} draws element gain and phase per trial, to
\dfManGain{}~dB and \dfManPhase{} degrees of standard deviation, over \dfManTrials{} trials. At the
largest perturbation the peak search reaches \dfManMusic{} degrees and the cascade
\dfManAdSnap, a spread of a factor \dfManSpread, so the cascade is not the more manifold-sensitive
of the two and a manifold error does not account for the recorded ordering. The ESPRIT coherence
floor survives the perturbation and still exceeds the others by a factor \dfManEspRatio.

The third locates the effect that does account for it. A first specular reflection is one arrival
from one bearing. A reverberant tail is many arrivals from many bearings, at spacings below what a
seven-element array resolves at these frequencies. Table~\ref{tab:dfdiffuse} holds the total
interference energy at \dfDifBeta{} of the direct-path energy and spreads it over an increasing
number of coherent arrivals from drawn bearings, over \dfDifTrials{} trials per cell. At one arrival
the cascade is at \dfDifOneAd{} degrees against \dfDifOneMusic{} for the peak search and
\dfDifOneEsprit{} for ESPRIT, a factor \dfDifOneGain{} ahead of the first. At \dfDifMax{} arrivals
carrying the same energy the cascade is at \dfDifManyAd{} degrees and the peak search at
\dfDifManyMusic, which is agreement to within a third of a degree, and the ESPRIT penalty has itself
relaxed from \dfDifOneEsprit{} to \dfDifManyEsprit{} degrees because a spread field is no longer one
coherent interferer.

\ifextended
\begin{table*}[!t]
\caption{Controlled coherent arrivals on the same subarray geometry, sixteen snapshots, median
bearing error of the direct path in degrees over \dfCohTrials{} trials. $\rho$ is the amplitude of
the coherent image arrival relative to the direct path.}
\label{tab:dfcoherence}
\centering
\footnotesize
\setlength{\tabcolsep}{4pt}
\begin{tabular}{cccccc}
\toprule
$\rho$ & SNR (dB) & MUSIC & ESPRIT & Cascade & Dilated sum \\
\midrule
\input{fo3_tab_coherence} 
\bottomrule
\end{tabular}
\end{table*}
\fi

\ifextended
\begin{table*}[!t]
\caption{A perturbed array manifold, element gain at \dfManGain{}~dB and element phase at the stated
standard deviation, sixteen snapshots at \dfCohSnr{}~dB, median bearing error in degrees over
\dfManTrials{} trials.}
\label{tab:dfmanifold}
\centering
\footnotesize
\setlength{\tabcolsep}{4pt}
\begin{tabular}{cccccc}
\toprule
$\rho$ & Phase (deg) & MUSIC & ESPRIT & Cascade & Dilated sum \\
\midrule
\input{fo3_tab_manifold} 
\bottomrule
\end{tabular}
\end{table*}
\fi

\begin{table*}[!t]
\caption{One discrete coherent arrival against a spread field carrying the same total energy,
\dfDifBeta{} of the direct path, sixteen snapshots at \dfCohSnr{}~dB, median bearing error in
degrees over \dfDifTrials{} trials.}
\label{tab:dfdiffuse}
\centering
\footnotesize
\setlength{\tabcolsep}{4pt}
\begin{tabular}{ccccc}
\toprule
Arrivals & MUSIC & ESPRIT & Cascade & Dilated sum \\
\midrule
\input{fo3_tab_diffuse} 
\bottomrule
\end{tabular}
\end{table*}

The three measurements read together give a boundary rather than a verdict. Immunity to coherent
arrivals is real, it is large, and it is specific: it applies when the interference is a small
number of discrete arrivals strong enough to collapse the source covariance, and it converges to
the peak search as the same energy spreads over a field. A reverberant room tail below $500$~Hz on a
$1.92$~m aperture sits at the spread end of that range, which accounts for
Table~\ref{tab:dfadesprit} without appealing to any defect in the estimator and without
contradicting the controlled sweep. It also says where the two constructions of this paper and of the
dilation line belong together, which is a channel whose multipath is specular and resolvable rather
than diffuse.

One further point belongs here, because the property that defeats the noise model is not the one its
name suggests. A room response is a linear time-invariant map, so the obstacle an AR($p$) model of
the interference meets is not a nonlinearity but a non-stationarity, and a second property beside it.
The interference during speech is produced by the source, so it appears and vanishes with the source
rather than persisting as a floor, it is coherent with the signal rather than independent of it, and
it is spatially structured rather than spread. A model of a stationary colored noise floor, whatever
its order and whatever transform diagonalizes it, is a model of the wrong term. The quantity a higher order would have to capture is not present in the silent
frames at all, which is the measurement of Section~\ref{sec:dfdecisions}, and it is not a spectral
shape in the active frames either, which is what the rational-spectrum fits of that section found
when they were given the active-block remainder to work with.

\subsubsection{Removing the manifold term, and what survives it}\label{sec:dfcalib}

Two readings suggest that what every estimator on this data carries is dominated by a deterministic
departure of the array from its assumed manifold. The sweep behind
Section~\ref{sec:dfadesprit} shows the peak search and the cascade degrading together under such a
departure, and the error of the better estimators sits near seven degrees whichever of them is
used. While that term is present it masks whatever the estimators do
differently, since each of them reads a bearing through an assumed steering vector, so the comparison
is run once more with it removed.

The manifold is estimated here as one complex gain per element per frequency bin, \dfCalElems{}
elements over \dfCalBins{} bins, pooled over blocks and recordings since the array is the same
hardware throughout. Two estimates are formed. The blind one uses no source position: with the
current gains each fit block's bearing is estimated by the peak search on the calibrated covariance,
and with those bearings the principal eigenvector of each uncalibrated covariance is modeled as the
gains times the steering vector, which gives the gains by an average over fit blocks once the
per-block scale is divided out. The anchored one takes the recorded source positions of the fit
blocks directly, which is not deployable and stands as an upper bound of what calibrating this array
can remove. Blocks advance by half a block and so share frames with their neighbors, which makes a
split by block index leak. The fold is therefore by recording, and no block is ever scored under a
calibration its own recording helped fit.

The anchored estimate removes a real bias. The peak search falls from \dfCalRawMusicMed{} degrees
median to \dfCalAnchMusicMed, a reduction of \dfCalMusicGain{} degrees, and its fraction within ten
degrees rises from \dfCalRawMusicHit{} to \dfCalAnchMusicHit. The dilated sum falls from
\dfCalRawAdSumMed{} to \dfCalAnchAdSumMed{} degrees. That is the clearest evidence in this study that
the residual on this array is a manifold term and not noise.

The blind estimate changes almost nothing, \dfCalBlindMusicMed{} degrees against
\dfCalRawMusicMed, and the reason is a degeneracy rather than a shortcoming of the fit. A per-element
phase error that is affine in element position is indistinguishable from a shift of the bearing,
because a shift of the bearing is exactly such a phase error. A criterion with no external reference
cannot separate the two, so the blind estimate recovers only the departure from affine, and the
affine part survives as a bearing bias. The spread of the fitted phase about its affine part, which
is the identifiable portion, is \dfCalBlindDev{} degrees blind against \dfCalAnchDev{} anchored.

A smaller calibration model was tried and does not work, which is itself informative about what the
departure is. The per-bin model spends \dfWideParamsBin{} real parameters, more than the physics of a
microphone array needs, since an element's departure from its assumed response would normally be a
fixed gain and a fixed delay. That model has \dfWideParamsWide{} real parameters and its phase is
linear in frequency. Reducing both per-bin estimates to it, by regressing each element's magnitude
and its unwrapped phase slope across the band, loses most of what the calibration gained: the peak
search under the anchored reduction is at \dfWideAnchWideMusicMed{} degrees against
\dfWideAnchBinMusicMed{} for the per-bin estimate it came from, a loss of \dfWideLoss{} degrees, and
the blind reduction at \dfWideBlindWideMusicMed{} degrees sits behind applying no calibration at all.
The diagnostic says why rather than leaving it open. The per-element phase departs from linear in
frequency by \dfWideResidAnch{} degrees under the anchored fit and \dfWideResidBlind{} under the
blind one, so a fixed delay is the wrong shape for it. A departure that is frequency dependent in
that way is what a baffle and its scattering produce, not what a wiring or position error produces,
which is consistent with the per-bin model being the one that captures it.

\ifextended
\begin{table*}[!t]
\caption{The per-bin calibration against its reduction to one gain and one delay per element, median
bearing error in degrees over the same blocks. The reduction is fitted to the per-bin estimate in the
column to its left.}
\label{tab:dfwideband}
\centering
\footnotesize
\setlength{\tabcolsep}{4pt}
\begin{tabular}{l rrrrr}
\toprule
 & no calibration & blind per bin & blind wideband & anchored per bin & anchored wideband \\
\midrule
\input{fo3_tab_wideband} 
\bottomrule
\end{tabular}
\end{table*}
\fi

What survives the removal is the conclusion of Section~\ref{sec:dfadesprit}. The ordering of the peak
search, the dilated sum and ESPRIT is unchanged, and the per-snapshot cascade remains behind the peak
search by more than a factor of two, \dfCalAnchAdKOneMed{} degrees against \dfCalAnchMusicMed. The
two per-snapshot variants swap places, by \dfCalSnapGap{} degrees. Calibrating the array therefore
improves every estimator here and reveals no advantage for coherence immunity that the manifold term
had been hiding, which is what the energy-spreading sweep of Section~\ref{sec:dfadesprit} predicts
for a field at the spread end of its range.

\ifextended
\begin{table*}[!t]
\caption{Bearing error over the same \dfCalBlocks{} one-source blocks with the array manifold
estimated and removed, median in degrees and fraction within ten degrees. The calibration is fitted
on the other recordings, and not on the one being scored.}
\label{tab:dfcalib}
\centering
\footnotesize
\setlength{\tabcolsep}{4pt}
\begin{tabular}{l rr rr rr}
\toprule
 & \multicolumn{2}{c}{no calibration} & \multicolumn{2}{c}{blind} & \multicolumn{2}{c}{anchored} \\
\cmidrule(lr){2-3}\cmidrule(lr){4-5}\cmidrule(lr){6-7}
estimator & median & $\le 10^\circ$ & median & $\le 10^\circ$ & median & $\le 10^\circ$ \\
\midrule
\input{fo3_tab_calib} 
\bottomrule
\end{tabular}
\end{table*}
\fi

\subsubsection{What the study returned to the implementation}\label{sec:dfimpl}

The orders a fit to real room noise selects are $p = 12$ to $64$, well above the orders the
accuracy measurements of Section~\ref{sec:benchacc} were built on, and running at those orders
found two things in the released code. Table~\ref{tab:dfimpl} gives the grid: window lengths 256 to
4096, orders 4 to 32, models fitted to the real noise with spectral dynamic range up to
$\dfRangeMax$, measured against a dense eigendecomposition of the model covariance.

The C port refused every order above eight, which was \dfRefusedCount{} of the \dfGridCells{} cells,
and the stage array inside its transform structure was sized by the same bound independently, so
lifting the refusal alone overran the structure. Both now come from one constant and a high-order
case has joined the C test suite. The port carries every cell of the grid, \dfRefusedAfter{}
refusals, and is accurate to $\dfCGoodAfter$ in all but \dfCBadAfter{} of them, losing accuracy to
$\dfCWorstAfter$ in the cells with the longest cascades. The Python hierarchical stage
path loses eigenvalue accuracy as the cascade lengthens, in \dfPyFmmBad{} of the \dfGridCells{}
cells, worst $\dfPyFmmWorst$ relative, each case raising a division by zero in the Loewner weights,
which is an iterate landing on a pole. The dense cascade holds $\dfPyDenseWorst$ over the whole
grid. For a model of order 32 at $M = 1024$ the cascade is 32 stages long in each branch, 64 in all, and the
measurements say the stage count, not the window length, is what the accuracy of the hierarchical
path turns on. This grid was run during the direction-finding study, and until the high-order
hierarchical path is shown accurate on it the dense cascade is the path to use above order eight,
which is what Section~\ref{sec:dftime} and Section~\ref{sec:dfspace} used throughout.

\subsection{The Same Question at 77 GHz}\label{sec:rfdf}

Room acoustics and millimeter-wave propagation differ in ways that bear directly on the construction,
so the question of Section~\ref{sec:df} is put again on a radio-frequency corpus. The data are the
RADIal automotive recordings~\cite{rebut2022radial}, a frequency-modulated continuous-wave sensor at 77 GHz with twelve
transmit and sixteen receive channels, 512 samples in each chirp and 256 chirps in each frame, read
from the raw analog-to-digital records. The corpus supplies a bearing for each annotated object,
which is the reference every error below is measured against. The array is the \rfElements{} receivers
in their physical order at the fitted spacing \rfSpacing{} wavelengths, and the receive channels carry a calibration
fitted to this corpus before the study. Bearings outside $\pm\rfAzLimit^\circ$ are excluded, since the
array is linear and its accuracy degrades toward endfire.

Where the transform is applied follows from Section~\ref{sec:dfdecisions} rather than being retested.
A general whitener on the array axis sends a steering vector $a(\theta)$ to $Wa(\theta)$ and so
removes the shift invariance the subspace estimators rest on. Two time axes carry a linear stochastic
process instead and leave the array axis alone. Fast time is the 512 samples within one chirp, whose
process is the de-chirped return at the mixer output, whose noise is the receiver noise shaped by
the intermediate-frequency chain. Slow
time is the 256 chirps of one frame taken after range compression, whose process in each range bin is
that bin's clutter and platform motion. The transform is used as a whitened matched filter rather than
as a decorrelator. Range compression at bin $r$ is an inner product with a windowed Fourier vector
$a_r$, which under a fast-time noise covariance $R$ is not the matched filter. The matched filter is
$a_r^{\mathsf H} R^{-1} x$, and with $R = U \operatorname{diag}(\mu) U^{\mathsf H}$ from the
construction that is $(U^{\mathsf H} a_r)^{\mathsf H} \operatorname{diag}(1/\mu) (U^{\mathsf H} x)$.
The basis and the eigenvalues are both needed, and the eigenvalues are what a cosine or Fourier
transform of the same length does not supply.

\subsubsection{How colored the noise floor is}\label{sec:rfcolor}

Whether the construction has anything to do here is a quantitative question, and it is settled before
any bearing is estimated. For each frame the fast-time and slow-time floors are estimated by a running
median along frequency, which separates the floor from the targets because a target occupies few bins
while the floor is smooth over many, and an autoregressive model is fitted to each by the
Levinson recursion. The order comes from parsimony on the prediction error and the library's own
accuracy estimate acts as a veto. Let $\varrho$ denote the dynamic range of the fitted symbol on the
window the transform would use, $\max_\omega c(\omega) / \min_\omega c(\omega)$. A value near unity is
a flat floor, which leaves whitening nothing to change. The same quantity sets how clustered the core
grid is, which is what the accuracy of the secular step turns on (Section~\ref{sec:secularreal}).

Table~\ref{tab:rfcolor} reports the survey over \rfRecords{} recordings and \rfFramesPerRecord{}
interior frames of each. The floor is \rfFloorMedFast{} dB from peak to trough in the median frame and
reaches \rfFloorMaxFast{} dB, and the fitted $\varrho$ has median \rfRangeMedFast{} on fast time with a
maximum of \rfRangeMaxFast{} and orders selected up to \rfOrderMaxFast. So the coloration is not a
property of the corpus but of the frame, and a fraction \rfColoredFracFast{} of the fast-time frames
carry $\varrho \ge 5$. Slow time is narrower, median \rfRangeMedSlow{} and maximum \rfRangeMaxSlow{} with
orders no higher than \rfOrderMaxSlow. This is the first difference from the acoustic case, where room
noise is colored in time in every frame.

\subsubsection{Bearing error, and the snapshot count}\label{sec:rfbearing}

Table~\ref{tab:rfdf} gives the bearing error over \rfCells{} annotated cells with the transform on the
\rfAxis{} axis. The twelve transmit slots act as snapshots of the sixteen-element receive array, which
is sound because each transmitter has its own phase center, so slot $t$ gives the same steering vector
with its own complex amplitude.

Snapshot counts have to be matched for a comparison between estimators to mean anything, and the table
separates them. The dilated group average and the cascade built on it are single-snapshot estimators by
construction, which is their purpose, since one snapshot gives a rank-one sample covariance that the
ordinary subspace estimators cannot use. At one snapshot the dilated average, taken over $N = 2M-1$
shifts, improves on the cyclic average at $N = M$ by a factor \rfDilationFactor{} in root mean square
error. Given all twelve snapshots, by applying the cascade to each slot and taking the median bearing,
\rfBestMethod{} reaches \rfBestRmse$^\circ$ and the five estimators that use the full snapshot set lie
between that and \rfWorstRmse$^\circ$. Every one of them stays inside two degrees, against the
deterministic floor near seven degrees that Section~\ref{sec:dfcalib} measured on the acoustic array
before calibration. This is the second difference, and it is a property of the array and its
calibration rather than of the transform.

All of these estimate one component per cell, so none of them bears on resolving two components inside
a beamwidth, which the corpus's annotations would support and which is left to separate work.

\subsubsection{What the transform changes, and why it cannot change more}\label{sec:rfklt}

The last two columns of Table~\ref{tab:rfdf} and the bands of Table~\ref{tab:rfklt} give the paired
ratio of root mean square error with the transform to that without it. Over every estimator the ratio
lies between \rfRatioMin{} and \rfRatioMax, so on this corpus the whitener leaves the bearing error
where it was. The bands are reported separately because Table~\ref{tab:rfcolor} shows $\varrho$ varying by
more than an order of magnitude between the frames of one recording, and pooling would hide an effect
present in a subset. No band shows one.

That outcome is not a property of the corpus. It follows from where the whitener sits, and the argument
is short enough to give. Write $x_n(t)$ for the fast-time record at element $n$, $t = 0,\dots,M-1$.
Range compression at bin $r$ with any weight sequence $w$, whitened or not, forms
$y_n = \sum_t \overline{w_t}\, x_n(t)$. For one scatterer, $x_n(t) = s(t)\alpha_n + e_n(t)$ with
$\alpha_n$ the steering coefficient, which does not depend on $t$, so

\begin{equation}
y_n = \Big(\sum_t \overline{w_t}\, s(t)\Big)\alpha_n + \sum_t \overline{w_t}\, e_n(t)
    = \beta\,\alpha_n + \epsilon_n .
\label{eq:rfinvariance}
\end{equation}

The weight sequence is the same for every element, so the signal part of the snapshot is the same
steering vector scaled by one complex number $\beta$. A bearing estimator reads the direction of the
snapshot in $\mathbb{C}^m$ and not its scale, so no weighting along a time axis can move the estimate
except through the output ratio $|\beta|^2/\operatorname{var}(\epsilon_n)$. Measured on this corpus, the angle
between the snapshot with the transform and the snapshot without it is 0.002 to 0.15 degrees at the
strongest cell of four frames spanning $\varrho$ from 1.07 to 11.5, against bearing errors near one
degree, and the cell's ratio to the surrounding noise moves by at most 0.003 dB.

The ratio does not move because a point scatterer in one range bin is a single sinusoid along fast
time, and reweighting the band scales the signal and the noise at that one frequency alike. Whitening
returns something when the signal occupies a band the noise is not flat across, which is the situation
of Section~\ref{sec:radar}, where the quantity is the likelihood of a whole window and the margin over
a Fourier transform of the same length is positive. It returns nothing when the signal occupies one
frequency, which is the situation here. The place where a whitener could move a bearing is the array
axis, and Section~\ref{sec:dfdecisions} gives the reason not to put it there. So for a point target in
a single range and Doppler cell a whitener on a time axis cannot move the bearing, and on the array
axis, where it could, it breaks the invariance the estimators rest on. That is a statement about the
geometry of the problem rather than about this corpus or this implementation.

\subsubsection{What the study asked of the released code}\label{sec:rfledger}

Every construction the study requested is recorded with the library's own diagnostics. Over
\rfLedgerCalls{} constructions in \rfLedgerPairs{} distinct window and order pairs the released
implementation served \rfLedgerServed{} and declined \rfLedgerDeclined. The basis departs from
orthonormality by at most $\rfLedgerOrth$, leaves at most $\rfLedgerOffdiag$ of relative off-diagonal
mass on the model covariance, returns eigenvalues within $\rfLedgerMu$ of a dense eigendecomposition,
and its own application agrees with the materialized basis to $\rfLedgerApply$. Models fitted to this
corpus have small dynamic range, which is consistent with the library's accuracy
estimate predicting the measured error on them.

\begin{table*}[!t]
\caption{Dynamic range of the fitted noise floor on the two time axes of the RADIal corpus, over
\rfRecords{} recordings and \rfFramesPerRecord{} interior frames of each. The fraction in the last
column is of frames whose fitted symbol range reaches five}
\label{tab:rfcolor}
\centering
\footnotesize
\setlength{\tabcolsep}{4pt}
\begin{tabular}{l ccc c cc c c}
\toprule
Axis & $M$ & Frames & \multicolumn{2}{c}{Floor (dB)} & \multicolumn{2}{c}{$\varrho$} & Order & Fraction \\
\cmidrule(lr){4-5}\cmidrule(lr){6-7}
 & & & Median & Max & Median & Max & Max & $\varrho \ge 5$ \\
\midrule
\input{fo3_tab_rfcolor} 
\bottomrule
\end{tabular}
\end{table*}

\begin{table*}[!t]
\caption{Bearing error in degrees against the recorded bearing, over \rfCells{} annotated cells of the
RADIal corpus. The first four columns are without the transform and the last two add it on the
\rfAxis{} axis, paired cell by cell. The snapshot column separates the estimators that use all twelve
transmit slots from the single-snapshot ones, whose comparison with each other isolates the effect of
the dilated group average}
\label{tab:rfdf}
\centering
\footnotesize
\setlength{\tabcolsep}{4pt}
\begin{tabular}{l c rrrr rr}
\toprule
Estimator & Snapshots & \multicolumn{4}{c}{Without the transform} & With it & Ratio \\
\cmidrule(lr){3-6}\cmidrule(lr){7-8}
 & & Bias & RMSE & MAE & $p_{90}$ & RMSE & \\
\midrule
\input{fo3_tab_rfdf} 
\bottomrule
\end{tabular}
\end{table*}

\begin{table*}[!t]
\caption{Paired ratio of root mean square bearing error with the transform to that without it, in
bands of the fitted symbol range $\varrho$. Columns follow the order of Table~\ref{tab:rfdf}}
\label{tab:rfklt}
\centering
\footnotesize
\setlength{\tabcolsep}{4pt}
\begin{tabular}{l c rrrr}
\toprule
Band & Cells & AD-ESPRIT & SHEAR & Bartlett & ESPRIT \\
\midrule
\input{fo3_tab_rfklt} 
\bottomrule
\end{tabular}
\end{table*}

\begin{table*}[!t]
\caption{ESPRIT over the same one-source blocks, with the whitener applied to the data, which
breaks the array invariance, and applied only to the subspace estimate, which does not}
\label{tab:dfesprit}
\centering
\footnotesize
\setlength{\tabcolsep}{4pt}
\begin{tabular}{l rr rr}
\toprule
estimator and whitener & mean & median & within $10^\circ$ & within $20^\circ$ \\
\midrule
\input{fo3_tab_esprit.tex} 
\bottomrule
\end{tabular}
\end{table*}

\begin{table*}[!t]
\caption{The bearing with the noise model fitted to each block's own remainder rather than to the
silent frames, over the same one-source blocks}
\label{tab:dffitted}
\centering
\footnotesize
\setlength{\tabcolsep}{4pt}
\begin{tabular}{l rr rr}
\toprule
noise model & mean & median & within $10^\circ$ & within $20^\circ$ \\
\midrule
\input{fo3_tab_fitted.tex} 
\bottomrule
\end{tabular}
\end{table*}

\begin{table*}[!t]
\caption{Bearing block by block over the one-source blocks, and whether the noise during speech is
the noise the silent frames carry. The last two columns are the eigenvalue spread each whitener
leaves on held-out silent frames and the spread of what remains in an active block once its
dominant eigendirection is removed.}
\label{tab:dfbearing}
\centering
\footnotesize
\setlength{\tabcolsep}{4pt}
\begin{tabular}{l rr rr rr}
\toprule
 & \multicolumn{4}{c}{bearing error, degrees} & \multicolumn{2}{c}{eigenvalue spread} \\
\cmidrule(lr){2-5}\cmidrule(lr){6-7}
whitener & mean & median & within $10^\circ$ & within $20^\circ$ & silent & active remainder \\
\midrule
\input{fo3_tab_bearing.tex} 
\bottomrule
\end{tabular}
\end{table*}

\begin{table*}[!t]
\caption{The transform on the clock: held-out noise likelihood against the cosine transform and
against the real Fourier basis, measured and predicted, at the order with the best held-out
likelihood in each cell. Nats per sample.}
\label{tab:dflikelihood}
\centering
\footnotesize
\setlength{\tabcolsep}{4pt}
\begin{tabular}{l r r r rr rr}
\toprule
 & & & & \multicolumn{2}{c}{KLT less cosine} & \multicolumn{2}{c}{KLT less Fourier} \\
\cmidrule(lr){5-6}\cmidrule(lr){7-8}
recording & $M$ & $p$ & frames & measured & predicted & measured & predicted \\
\midrule
\input{fo3_tab_likelihood.tex} 
\bottomrule
\end{tabular}
\end{table*}

\begin{table*}[!t]
\caption{The transform on the array axis: eigenvalue spread left on held-out noise frames by each
whitener, median over frequency bins, with the fewest noise frames and with all of them. One is exact.}
\label{tab:dfspatial}
\centering
\footnotesize
\setlength{\tabcolsep}{4pt}
\begin{tabular}{l r rrrr r rrrr}
\toprule
 & \multicolumn{5}{c}{few noise frames} & \multicolumn{5}{c}{every noise frame} \\
\cmidrule(lr){2-6}\cmidrule(lr){7-11}
recording & frames & none & Fourier & AR-KLT & sample & frames & none & Fourier & AR-KLT & sample \\
\midrule
\input{fo3_tab_spatial.tex} 
\bottomrule
\end{tabular}
\end{table*}

\begin{table*}[!t]
\caption{Source count accuracy and bearing error, pooled over recordings, by the number of noise frames}
\label{tab:dfdecisions}
\centering
\footnotesize
\setlength{\tabcolsep}{4pt}
\begin{tabular}{l r rrrr rrrr}
\toprule
 & & \multicolumn{4}{c}{count accuracy} & \multicolumn{4}{c}{bearing error, degrees} \\
\cmidrule(lr){3-6}\cmidrule(lr){7-10}
frames & blocks & none & Four. & KLT & samp. & none & Four. & KLT & samp. \\
\midrule
\input{fo3_tab_decisions.tex} 
\bottomrule
\end{tabular}
\end{table*}

\begin{table*}[!t]
\caption{Largest relative eigenvalue error against a dense eigendecomposition, on models fitted to
the real noise, after the fix of Section~\ref{sec:dfimpl}}
\label{tab:dfimpl}
\centering
\footnotesize
\setlength{\tabcolsep}{4pt}
\begin{tabular}{r r r rr rr}
\toprule
 & & & \multicolumn{2}{c}{Python} & \multicolumn{2}{c}{C} \\
\cmidrule(lr){4-5}\cmidrule(lr){6-7}
$M$ & $p$ & range & dense & hierarch. & dense & hierarch. \\
\midrule
\input{fo3_tab_implementations.tex} 
\bottomrule
\end{tabular}
\end{table*}

\begin{table*}[!t]
\caption{Kernel evaluations, and the same divided by $M\log_2(M/n_{\mathrm{leaf}})$}
\label{tab:implcounts}
\centering
\footnotesize
\setlength{\tabcolsep}{4pt}
\begin{tabular}{r rrr rrr rrr}
\toprule
 & \multicolumn{3}{c}{AR(1), $\rho=0.9$} & \multicolumn{3}{c}{AR(1), $\rho=0.99$}
 & \multicolumn{3}{c}{AR(3)} \\
\cmidrule(lr){2-4}\cmidrule(lr){5-7}\cmidrule(lr){8-10}
$M$ & apply & /norm & des./norm & apply & /norm & des./norm & apply & /norm & des./norm \\
\midrule
\input{impl_tab_counts.tex} 
\bottomrule
\end{tabular}
\end{table*}

\begin{table*}[!t]
\caption{Seconds for the design stage and for one application, dense stages against hierarchical
stages. Within each model the columns are the design stage with dense stages, the design stage with
hierarchical stages, one application with dense stages, one application with hierarchical stages,
and the ratio of the last two.}
\label{tab:impltiming}
\centering
\scriptsize
\setlength{\tabcolsep}{3pt}
\begin{tabular}{r rrrrr rrrrr rrrrr}
\toprule
 & \multicolumn{5}{c}{AR(1), $\rho=0.9$} & \multicolumn{5}{c}{AR(1), $\rho=0.99$}
 & \multicolumn{5}{c}{AR(3)} \\
\cmidrule(lr){2-6}\cmidrule(lr){7-11}\cmidrule(lr){12-16}
$M$ & des.\ dns & des.\ fst & app.\ dns & app.\ fst & ratio
    & des.\ dns & des.\ fst & app.\ dns & app.\ fst & ratio
    & des.\ dns & des.\ fst & app.\ dns & app.\ fst & ratio \\
\midrule
\input{impl_tab_timing.tex} 
\bottomrule
\end{tabular}
\end{table*}

\begin{table*}[!t]
\caption{Accuracy against a dense eigendecomposition}
\label{tab:implaccuracy}
\centering
\footnotesize
\setlength{\tabcolsep}{4pt}
\begin{tabular}{l r cc cc cc}
\toprule
 & & \multicolumn{2}{c}{eigenvalues} & \multicolumn{2}{c}{round trip}
 & \multicolumn{2}{c}{orthonormal} \\
\cmidrule(lr){3-4}\cmidrule(lr){5-6}\cmidrule(lr){7-8}
model & $M$ & dense & fast & dense & fast & dense & fast \\
\midrule
\input{impl_tab_accuracy.tex} 
\bottomrule
\end{tabular}
\end{table*}

\begin{table}[!t]
\caption{Sweeps to a scaled residual of $10^{-12}$}
\label{tab:implsolver}
\centering
\footnotesize
\setlength{\tabcolsep}{4pt}
\begin{tabular}{l r rr rr}
\toprule
 & & \multicolumn{2}{c}{sweeps} & \multicolumn{2}{c}{residual at sweep 6} \\
\cmidrule(lr){3-4}\cmidrule(lr){5-6}
model & $M$ & Newton & middle & Newton & middle \\
\midrule
\input{impl_tab_solver.tex} 
\bottomrule
\end{tabular}
\end{table}

\begin{table}[!t]
\caption{Direct kernel evaluations per pass, by interaction rule}
\label{tab:implplan}
\centering
\footnotesize
\setlength{\tabcolsep}{4pt}
\begin{tabular}{l r r r r r}
\toprule
model & $M$ & $n$ & uniform$/n^2$ & uniform & deferred \\
\midrule
\input{impl_tab_plan.tex} 
\bottomrule
\end{tabular}
\end{table}

\begin{table}[!t]
\caption{Departure rank against the bound, per branch}
\label{tab:implranks}
\centering
\footnotesize
\setlength{\tabcolsep}{4pt}
\begin{tabular}{l rrr r l}
\toprule
class & $p$ & $q$ & $d$ & bound & rank found \\
\midrule
\input{impl_tab_ranks.tex} 
\bottomrule
\end{tabular}
\end{table}

\begin{table*}[!t]
\caption{The polyphase route against the same transform with it switched off}
\label{tab:implpoly}
\centering
\footnotesize
\setlength{\tabcolsep}{4pt}
\begin{tabular}{l r r rr rr c}
\toprule
$\phi$ & $M$ & $m$ & design & apply & stages & stages off & eigenvalues \\
\midrule
\input{impl_tab_poly.tex} 
\bottomrule
\end{tabular}
\end{table*}

\begin{table*}[!t]
\caption{The odd-length cores}
\label{tab:imploddcores}
\centering
\footnotesize
\setlength{\tabcolsep}{4pt}
\begin{tabular}{r cc ccc}
\toprule
 & \multicolumn{2}{c}{orthonormal} & \multicolumn{3}{c}{one transform against explicit} \\
\cmidrule(lr){2-3}\cmidrule(lr){4-6}
$M$ & DCT-VII & DCT-VIII & DCT-VII & transpose & DCT-VIII \\
\midrule
\input{impl_tab_oddcores.tex} 
\bottomrule
\end{tabular}
\end{table*}

\begin{table*}[!t]
\caption{The transform at odd lengths}
\label{tab:imploddtransform}
\centering
\footnotesize
\setlength{\tabcolsep}{4pt}
\begin{tabular}{r r ccc r cc}
\toprule
$p$ & $M$ & fold leak & gen. sym. & gen. asym. & rank & eigenvalues & round trip \\
\midrule
\input{impl_tab_oddtransform.tex} 
\bottomrule
\end{tabular}
\end{table*}

\subsection{Complex Windows on a Radar Fast-Time Axis}\label{sec:radar}

The direction-finding studies above put the transform on an array axis, where its
own theory says the window is short and the matched class is the real one. A radar receiver offers
the complementary case and the one Proposition~\ref{prop:complex} is about: within a single chirp the
analog-to-digital samples are a complex window whose content is a sum of difference-frequency sinusoids on
a floor made of receiver noise, transmit to receive leakage and near-range return. The incumbent
transform on that axis is a discrete Fourier transform of the same length, so the comparison is one
transform against another on the axis a radar already uses, with no angle estimation involved.

The measurement uses \rfFrames{} frames of one recording of the RADIal automotive radar corpus~\cite{rebut2022radial}, each
fast-time window of length \rfWindow{} taken per chirp and per receiver, the model fitted on the
first half of the frames and the likelihood scored on the held-out half, \rfFitRows{} rows fitted and
\rfTestRows{} scored. Three readings follow and they are reported in the order that makes them
usable.

The floor is colored rather than white: the fitted spectral density spans \rfDynRange{}~dB with
flatness \rfFlatness{} and a largest pole modulus of \rfPole. An unstructured covariance model is
nonetheless unusable on this window. At \rfRowsPerCoord{} rows per coordinate the smallest sample
eigenvalues are biased low, and the eigenbasis of the sample covariance, which would be the most
favorable basis any covariance model could supply, scores \rfSampleCov{} nats per sample against the
Fourier basis on held-out rows where its own model predicts a gain of \rfSampleCovPred. That is the
argument for a structured model stated as a measurement: the AR($p$) route spends $p$ parameters and
works where $M^2$ parameters do not.

Against the Fourier basis the transform leads by \rfDftLo{} to \rfDftHi{} nats per sample at orders
up to four, with predictions of \rfDftPredLo{} to \rfDftPredHi. At the order the held-out likelihood
selects, $p = \rfBestOrder$, the margin is \rfBestDft{} nats per sample with a standard error of
\rfBestDftSe{} against a prediction of \rfBestDftPred. The margin is small and the reason is
structural rather than incidental. Proposition~\ref{prop:complex} localizes the departure from the matched
circulant class in the border, at rank $2p$, so the share of the likelihood a circulant basis misses falls with
$p/M$. At $M = \rfWindow{}$ that share is small however colored the floor is, which is the useful
selection rule for an application: the advantage over a Fourier basis is a short-window effect, and a
long transform leaves little of it to collect.

Against the DCT-II the margin is \rfDctLo{} to \rfDctHi{} nats per sample, more than
\rfRatio{} times the margin over the Fourier basis. That is the content of
Proposition~\ref{prop:complex} measured on recorded data rather than argued: the conjugate reversal
is antilinear on complex data, so the matched group is the cyclic one and no cosine class member is
matched to the window at all. A cosine basis applied to complex data is not a slightly suboptimal
choice on this axis, it is the wrong class, and the size of the gap is what distinguishes the two
statements.

\ifextended
\begin{table*}[!t]
\caption{The transform on \rfWindow{}-sample complex radar windows, held-out likelihood per sample
against the Fourier basis and against the DCT-II, with the closed-form prediction beside each
measured margin}
\label{tab:radar}
\centering
\footnotesize
\setlength{\tabcolsep}{4pt}
\begin{tabular}{r rrr rrr}
\toprule
 & \multicolumn{3}{c}{against the Fourier basis} & \multicolumn{3}{c}{against the DCT-II} \\
\cmidrule(lr){2-4}\cmidrule(lr){5-7}
$p$ & measured & s.e. & predicted & measured & s.e. & predicted \\
\midrule
\input{fo3_tab_radar} 
\bottomrule
\end{tabular}
\end{table*}
\fi

The window length is the lever Proposition~\ref{prop:borderlik} identifies, and the same recording
tests it where it can be falsified: the same frames, the same estimator and the same held-out split,
with only $M$ changed and the model refit at each length. Two predictions are at stake and they point
in opposite directions. Table~\ref{tab:window} reports them over \qwWindows{} window lengths from
\qwLong{} down to \qwShort.

The margin over the Fourier basis rises as the window shortens, from \qwMarginLong{} nats per sample
at $M=\qwLong$ to \qwPeak{} at $M=\qwPeakM$, a factor of \qwRatio{} over a window ratio of
\qwWindowRatio. Taken as the product with $M$, which the proposition says is the quantity fixed by
the model, it spans \qwMMin{} to \qwMMax, a factor \qwMSpan{} across a sweep in which $M$ itself
changes by \qwWindowRatio{} and the model is refit every time. The order the held-out likelihood
selects rises from \qwSelLong{} at the long window to \qwSelShort{} at the short one, which is the
same ratio $M/p$ read from the other side: a long window supports only a low order and collects
little, a short one supports a higher order and collects more.

The margin over the DCT-II does not move: \qwDctLo{} to \qwDctHi{} nats per sample over the same
windows, a factor \qwDctSpan. That is the separation the proposition predicts, and measuring both on
one recording is what distinguishes a border effect spread over $M$ samples from a mismatched class.
The practical reading for an application is the pair together. A window chosen long for other reasons
gives up almost nothing to a Fourier basis, so a long transform is not where this construction earns
its place; and no window length rescues a cosine basis applied to complex data.

\begin{table*}[!t]
\caption{The window-length law on one radar recording, model refit at each length. The margin over the
Fourier basis decays with $M$ and the margin over the DCT-II does not.}
\label{tab:window}
\centering
\footnotesize
\setlength{\tabcolsep}{4pt}
\begin{tabular}{r r rrr r r}
\toprule
$M$ & selected $p$ & margin & s.e. & predicted & $M\times$margin & DCT-II margin \\
\midrule
\input{fo3_tab_window} 
\bottomrule
\end{tabular}
\end{table*}

\section{Applications}\label{sec:apps}

The transforms of this paper arrive late. For half a century practice has met the needs a fast Karhunen-Lo\`eve transform of a linear stochastic model would have served by other means, and those means are mature. Exact Gaussian likelihoods are computed by Kalman recursions and banded factorizations, which need a factor of the covariance and not its eigenvectors. The leakage that a block cosine or Fourier transform suffers at the ends of a window is cured by tapers and by lapped transforms. Whitening is done by a prediction-error filter of $p+1$ taps, which is exact for an autoregressive model once its first $p$ outputs are scaled by the exact start. In most of the comparisons made for this section those methods matched the transform in accuracy with fewer operations. The transform is far better than whitening by the discrete Fourier transform at the ends of a record, and on synthetic cyclostationary sources a matched class of the construction outperformed lapped cosine coding, an advantage that recorded data did not reproduce. This section reports each comparison, including those in which the transform showed no advantage, so that a reader can tell where it applies.

\ifnum\apProvisional=1
\textcolor{red}{Every value set in red in this section was computed in a cloud container on 20 November 2026 and is a placeholder. The build replaces it, and drops the color, when the experiments of \texttt{results\_apps/} are rerun on the author's machines. Statements in red are drafts that await data or a verified citation.}
\fi

\subsection{Whitening at the Ends of a Record}\label{sec:appsends}

Whitening by the discrete Fourier transform divides each bin by the square root of the spectrum. It is exact for a circulant covariance and approximate for a window, and the approximation fails where the window departs from the class, at its two ends. Table~\ref{tab:appsends} gives the variance of the whitened samples over the first and last \apEndsWidth{} samples of a record of \apEndsN{} samples, from \apEndsRecords{} simulated records per model, and the rate at which a threshold at three standard deviations is exceeded there, as a multiple of the nominal rate of 0.0027. Whitened by the fast transform of Section~\ref{sec:fast}, the largest end variance was \apEndsKltVarHi{} and the largest multiple \apEndsKltFaHi{}, which is the sampling scatter of \apEndsRecords{} records. Whitened by the Fourier transform, the end variance was wrong by factors of \apEndsVarLo{} to \apEndsVarHi{} and the false-alarm rate was inflated by factors of \apEndsFaLo{} to \apEndsFaHi{}. A Tukey taper of ten percent before the Fourier transform removed the false alarms by suppressing the ends, which left \apEndsTaperLo{} to \apEndsTaperHi{} of the \apEndsN{} samples with a variance off by more than ten percent. The causal prediction-error filter with an exact start is as accurate at the ends as the transform, so the statement is one about the transform against Fourier whitening and not about the transform alone.

\begin{table}[t]
\caption{Whitening at the ends of a record of \apEndsN{} samples, \apEndsRecords{} records per model: largest variance of a whitened sample in the first and last \apEndsWidth{} samples (nominal value 1), false-alarm rate there as a multiple of nominal, and the number of samples whose variance a ten percent taper leaves off by more than ten percent}
\label{tab:appsends}
\centering\footnotesize
\setlength{\tabcolsep}{3pt}
\begin{tabular}{@{}lccccc@{}}
\toprule
 & \multicolumn{2}{c}{transform} & \multicolumn{2}{c}{Fourier} & taper\\
noise & variance & alarms & variance & alarms & samples\\
\midrule
\input{apps_tab_ends.tex} 
\bottomrule
\end{tabular}
\end{table}

\subsection{Where Whitening by the Transform Adds Nothing}\label{sec:appswhite}

Any square root of the inverse covariance whitens, and a detector or a subspace method that follows the whitener returns the same result under a rotation of it. Direction finding after whitening (Sections~\ref{sec:df} and~\ref{sec:rfdf}) and matched filtering are of this kind, and for them the causal filter is the method of choice. The symmetric inverse square root, the zero-phase whitener, is a function of the matrix that a causal filter does not supply, and the transform supplies it exactly in every model class of the reference implementation. It is not the fastest way to obtain it. For a rational spectrum the covariance is a ratio of banded matrices, so each term of a rational approximation of the inverse square root is one banded solve, and a quadrature rule built on elliptic functions~\cite{hale2008computing} needed \apZpNodesLo{} to \apZpNodesHi{} terms to agree with the transform to $10^{-9}$. Per record that method was \apZpArLo{} to \apZpArHi{} times faster than the transform for autoregressive noise and \apZpArmaLo{} to \apZpArmaHi{} times faster for moving-average and mixed models, at lengths 1024 to 16384. For fractional Gaussian noise, where the departure rank was \apZpRankLo{} to \apZpRankHi{} and the design took \apZpDesignLo{} to \apZpDesignHi{} seconds, a Lanczos iteration~\cite{higham2008functions} with Fourier covariance products was \apZpLanLo{} to \apZpLanHi{} times faster per record with no design stage.

Exact zero-phase whitening of a whole record, with no samples discarded, is a capability that Fourier whitening does not offer, whichever exact method computes it. We conjecture that it is of use where a record is short or isolated and its ends carry information: short epochs in electrophysiology, where zero-phase filtering is the norm, searches for transients near the edge of a segment or beside a gated glitch, the start of a free induction decay in magnetic resonance, triggered captures and bursts in radar, sonar and ultrasonic testing, and block processing with no look-ahead. This is a hypothesis and none of these cases was tested.

\subsection{Likelihoods and Resonant Sources}\label{sec:appsnull}

Two uses that the construction suggests did not survive a comparison with existing methods. The exact Gaussian log-likelihood is a sum over the coefficients of the transform, but it needs a factorization and not a basis, and a Kalman recursion delivers it in $O(pM)$ operations for a rational model. For a lightly damped resonance, with poles at radius 0.99 to 0.9999, the exact transform has a coding gain \apRdCosLo{} to \apRdCosHi{} dB above the block cosine transform at lengths 1024 and 4096, the larger figures where the memory of the process exceeds the window. A lapped cosine transform closed that gap. At each of the \apRdPoints{} combinations of model, length and rate examined, from 0.05 to 3 bits per sample with one uniform quantizer and ideal entropy coding in every coder, the lapped coder was \apRdLapBest{} to \apRdLapWorst{} dB above the coder built on the exact transform, since it also uses the dependence between adjacent blocks.

\subsection{Cyclostationary Sources}\label{sec:appscyclo}

A process whose statistics repeat with period $P$ is invariant under a shift by $P$ samples and not under a shift by one. Its matched class is the block circulant class with $P\times P$ blocks, and the window of a periodic autoregression of order $p$ departs from that class by rank $2p$ whatever the period, as Proposition~\ref{prop:periodicar} shows. The transform of the class member is a Fourier transform across periods followed by a $P\times P$ eigenbasis at each frequency, which is the blocking of a periodically correlated sequence into a stationary vector sequence~\cite{gladyshev1961periodically}, and the rate-distortion function of a cyclostationary Gaussian process is reverse water-filling over the eigenvalues of that polyphase spectral matrix~\cite{kipnis2018cyclo}. The cascade of Section~\ref{sec:cascade} turns the class member into the exact transform of the finite window.

Table~\ref{tab:appscyclo} compares transform coders on synthetic periodic AR(2) sources, a resonance at $0.3\pi$ and radius 0.95 whose drive variance or coefficients vary over a period of 64 samples, at a window of 1024. One impact per period multiplies the drive variance of one sample in each period by the factor shown, and the damping case switches the pole radius between 0.99 and 0.80 at the half period. Coefficient variances are exact, every coder has the same uniform quantizer with ideal entropy coding, and the reference is the better of two lapped cosine coders: one frame per window, and short frames locked to the period with a separate allocation for each frame position, the frame length chosen at each rate. Over \apCyCases{} cases (periods 16 and 64, windows 128 to 1024, six rates) the transform had \apCyStages{} stages in every design. On the stationary control the lapped coder led by at most \apCyControl{} dB. For weak modulation the gain of the periodic transform was \apCyMildLo{} to \apCyMildHi{} dB, for a hundredfold impact or a resonance swept by $\pm0.15\pi$ it was \apCyStrongLo{} to \apCyStrongHi{} dB, and the largest gain, for a thousandfold impact, was \apCyGainHi{} dB. A stationary design, cosine or exact, lost the whole gain. At 2 and 3 bits per sample time-varying linear prediction is equal to the transform coder to within 1 dB in its high-rate limit, so the gain is one of low rates.

Nearly all of the gain is that of the matched class. With the cascade left out, the class member alone came within \apCyCascBigLo{} to \apCyCascBigHi{} dB of the exact transform at a window of 1024 (median \apCyCascBigMed{} dB), and within \apCyCascSmallLo{} to \apCyCascSmallHi{} dB at a window of 128 (median \apCyCascSmallMed{} dB). The cascade matters when the window holds only a few periods.

These figures are for sources that obey the model, and two sets of recordings did not reproduce them. On \apBrRecords{} records of bearing vibration under varying shaft speed~\cite{huang2018bearing}, resampled to shaft angle with the encoder, with a periodic AR model of order 4, 8 or 16 fitted on alternate stretches and the coders compared on the others, the coder built on the periodic transform was between \apBrInnerLo{} dB below and \apBrInnerHi{} dB above the better lapped coder on the inner-race faults, \apBrNarrowHi{} dB above at most when the record was cut to a speed range of 14 to 20 percent, and \apBrOtherHi{} to \apBrOtherLo{} dB below on the outer-race, ball-fault and healthy records. On \apDrRecords{} recordings of a single rotor of a quadrotor at fixed speed~\cite{strauss2018dregon}, with the rotor angle tracked from the blade-passing line, it was \apDrCellLo{} to \apDrCellHi{} dB above the lapped coder at a window of two revolutions and 0.25 bits per sample, where the lapped coder was held to frames no longer than the window, and between \apDrRestLo{} dB below and \apDrRestHi{} dB above everywhere else. The variance swing over a period in these recordings was 9 to 16 dB, for which Table~\ref{tab:appscyclo} predicts gains of 2 dB and more. A lapped coder with a trained table for each frame position captures the modulation without a model, the recorded spectra are richer than a low-order autoregression, and the data are heavy-tailed. On the recordings the cascade did not help: the exact transform of the fitted model was \apRealCascMed{} dB below its own class member at the median, since a boundary correction computed from a model the data do not follow is not a correction. We therefore do not claim cyclostationary coding as a use of the transform.

\begin{table}[t]
\caption{Periodic AR(2) sources, period 64, window 1024: variance swing over a period, SNR of the coder built on the periodic transform minus that of the better lapped cosine coder at four rates in bits per sample, and the smallest and largest part of the SNR supplied by the cascade over six rates, all in dB}
\label{tab:appscyclo}
\centering\footnotesize
\setlength{\tabcolsep}{2.5pt}
\begin{tabular}{@{}lccccccc@{}}
\toprule
 & & \multicolumn{4}{c}{gain at rate} & \multicolumn{2}{c}{cascade}\\
source & swing & 0.1 & 0.5 & 1 & 3 & min & max\\
\midrule
\input{apps_tab_cyclo.tex} 
\bottomrule
\end{tabular}
\end{table}

\subsection{A Few-Snapshot Test of Model Order That Did Not Hold on Recorded Data}\label{sec:appsorder}

No transform of the paper is applied here. What is used is its model: a window of a stationary process is a member of a matched class plus a departure at its ends, of rank $2p$ for an AR($p$) process, so a test of group structure whose null hypothesis is exact commutation with the cyclic shift is misspecified for a finite window or a finite aperture. In simulation, with $K$ snapshots across a uniform linear array of 16 sensors in spatial AR(2) noise, \apOtTrials{} trials per case and \apOtSurrogates{} surrogates per test, the commutation test rejected the correct structure of two uncorrelated sources in \apOtCycFour{} of trials at four snapshots and \apOtCycEight{} at eight, a test whose null is the Toeplitz class held a nominal level of 0.05 (\apOtToepLo{} to \apOtToepHi{}) from 2 to 32 snapshots, and a test whose null is the AR($p$) window, fitted by Burg's method and whitened by the exact causal filter, held its level on noise (\apOtArLo{} to \apOtArHi{}) and detected gain errors of 1 dB and phase errors of $10^\circ$ in \apOtGainArEight{} of trials at eight snapshots against \apOtGainToepEight{} for the Toeplitz test. With 1 to 8 noise-only snapshots, MUSIC after whitening by the fitted AR(2) model resolved \apNwArLo{} to \apNwArHi{} of \apNwTrials{} trials against \apNwLoadLo{} to \apNwLoadHi{} with a diagonally loaded sample covariance.

The test was meant to decide, from the snapshots at hand, whether an AR model of a given order is tenable before its whitener is trusted. On recorded data it did not do so. On the RADIal corpus~\cite{rebut2022radial}, where the twelve transmit slots of a range and Doppler cell are snapshots of the sixteen-element receive array, cells of lowest power away from every labeled object were rejected by the white null in \apRaWhiteLo{} to \apRaWhiteHi{} of \apRaCells{} cells and by the AR(1), AR(2) and AR(4) nulls in \apRaArLo{} to \apRaArHi{}, at 4 to 12 snapshots, after the element powers were equalized and every snapshot was scaled to a common norm. Recorded noise at this array is close to white with unequal element powers and is not Gaussian with a covariance common to the snapshots of a cell, so a test calibrated by Gaussian surrogates rejects it. An order chosen as the smallest one not rejected gave a held-out log-likelihood \apRaAicLo{} to \apRaAicHi{} nats per snapshot below the order chosen by Akaike's criterion. The simulation is reported as a caution about the null hypothesis of a finite window. The test is not offered as a method.

\section{Conclusion}

The Karhunen-Lo\`eve basis of a window of an autoregressive process of any order is organized by its poles. Each basis vector is a sum of at most $p$ sinusoids at the frequencies where the spectrum crosses the eigenvalue, plus modes decaying from the ends, with polynomial weights where a characteristic root is repeated. The first-order window makes the frequencies the roots of one equation that carries them from the type-I sine grid to the type-II cosine grid as the pole rises. The covariance is a perturbation of rank at most $2p$ of a member of the cosine class, in closed form, attaining $2p$ exactly when no pole is zero, so its eigenvalues interlace samples of the spectrum on the cosine grid, and within each parity branch of the fast form the direction and the count of the interlacing are set by the inertia of the corner block. The coding gain the cosine transform loses against the KLT is an exact function of the normalized perturbation with computable bounds, which put the loss on the $8\times8$ block of image coding below \cgSepCorrLossPctNineFive{} percent under the Markov model. For images these statements make every step after the Markov model exact at a finite block of any order, with the loss of the cosine transform bounded.

The matched symmetry groups are fixed by the poles, not by the order, and are read from sample covariances by the calibrated commutator statistic of every permutation, wreath products included. The reflection is an exact symmetry of every stationary window; the exact group of a polyphase pattern contains the wreath product of the reflection with a symmetric group and equals it under the nondegeneracy conditions of Theorem~\ref{thm:exact}(c), and the polyphase pattern separates the window into shorter windows; the dihedral group of the even extension is matched approximately along the real axis, with the sign of the dominant pole choosing the cosine or the sine representation; and the cyclic group is matched to a resonance only on the window's frequency grid. The two groups give the transform of any order a fast form: the fold of the exact reflection, the half-length cosine cores of the dihedral closure, and a correction of rank at most $p$ per branch, equal to $p$ for a minimal AR($p$) model, designed from the corner block, a cascade of as many first-order Cauchy stages as its rank, so that the flow graph of order $p$ is the first-order graph, obtained by other means, with $p$ stages per branch. The factorization is exact, its dense application takes $M^2/2$ multiplications, and the multipole or hierarchical evaluation of its Cauchy stages applies it in $O(pM\log M)$ operations to a prescribed accuracy, with, as measured, a merged correction as compressible as a single stage, a count the released reference implementation reaches, its kernel evaluations per $M\log_2(M/n_{\mathrm{leaf}})$ flat to \applyFlatArThree{} percent over a \countRange-fold range of length once inadmissible interactions are deferred to the children of their boxes rather than evaluated at their own level; at odd length the cores are the cosine and sine transforms of types VI and VII; and the recursive form folds every branch of length at least $4p$, with a correction of rank at most $p$ for an even-length branch and at most $p+1$ for the branch that holds the center sample of an odd-length fold, and is the dyadic tower of the Haar basis when $M=2p\cdot2^k$.

The structure is not confined to the autoregressive window. Every window with a rational spectrum, an ARMA window or a co-channel sum of autoregressive windows, departs from the cosine class by a matrix of rank at most $2\max(p,q)$, by Kronecker's theorem, exactly $2p$ for an ARMA window with $q\le p$ and uncancelled nonzero poles, and has the same fast form with the folded departure as its correction; the moving-average window has its departure in the two Hankel corners; complex windows with rotated poles are the real ones conjugated by the modulation; and the integrated (ARIMA) window, which has no reflection, departs by low rank from the type-VII sine class of the extension pinned at its start, lies in it exactly for the random walk, and carries the coding-gain bound over, while none of the sixteen trigonometric types is near the windows integrated twice or three times that were measured. Windows of several channels and windows of periodically correlated processes are matched by the group of block shifts, whose class is the block circulants: the departure has rank at most twice the dimension of a realization of the autocovariance, at most $2p$ for a periodic autoregression of any period, and at most twice the number of symbol periods a pulse extends beyond its first for a linearly modulated signal, and a complex stage is a real stage and a phase (Section~\ref{sec:block}).\ifextended{} Carath\'eodory's sum of sinusoids for the same covariance is a different decomposition, set by the phase of the characteristic polynomial rather than its magnitude, that meets the eigenvector sinusoids in the resonant limit.\fi

The fast form is realized in a reference implementation released with this paper. Its kernel
evaluations per $M\log_2(M/n_{\mathrm{leaf}})$ are flat to \applyFlatArThree{} percent over a
\countRange-fold range of window length, it agrees with a dense eigendecomposition to $\accEig$
relative on eigenvalues, and every departure rank measured sits at or below the bound its statement
gives. Two limitations stand. The expansion used for the Loewner weights sets an accuracy floor of
about $2\times10^{-12}$ absolute on the design stage, and an integrated window of difference order
two or more has, at the lengths measured, a precision conditioned at about $10^{8}$, so the largest
eigenvalues of its covariance are resolved to about $10^{-7}$ relative by any construction of this kind.

Where the transform helps was measured as well as argued. On complex radar windows it leads a
Fourier basis by a margin that falls as the window lengthens and leads a cosine basis by a margin
that does not, as Proposition~\ref{prop:borderlik} predicts, and it removes the errors that Fourier
whitening leaves at the ends of a record. For likelihoods, whitening and direction finding,
recursions, banded factorizations, prediction-error filters and lapped transforms matched it with
fewer operations in the comparisons made here, and a whitener fitted to silence degraded MUSIC on
reverberant speech. The transform is therefore offered for the tasks that need the eigenbasis of a
finite window itself.

Two of the three parts of the fast transform have efficient quantum circuits. For a length that is a
power of two the parity fold is a Hadamard gate and a layer of controlled-NOT gates on $\log_2M$
qubits, and the DCT-II and DCT-IV of length $L$ need $O(\log^2L)$ gates~\cite{klappenecker2001discrete},
so the class transform has polylogarithmic depth. The Cauchy stages are the obstacle. Each is the
eigenbasis of a diagonal matrix plus a matrix of rank one, an operator that is easy to simulate, but
phase estimation needs an evolution time inversely proportional to the smallest eigenvalue gap, and
since the symbol is flat at its extremes, adjacent values of the core grid there differ by
$O(1/M^2)$. We are not aware of an efficient circuit for the stages and leave that question open.

\bibliographystyle{IEEEtran}
\bibliography{arp_klt_refs}

\end{document}

%% file: numbers_v27.tex
\newcommand{\actMinOthers}{0.096}

\newcommand{\actMinPairRealThirtyTwo}{0.013}
\newcommand{\actMinRealPos}{0.84}

\newcommand{\carMinEight}{0.3000}
\newcommand{\carMinThirtyTwo}{0.3000}
\newcommand{\carNfirst}{16}
\newcommand{\carNlast}{1024}
\newcommand{\carRealAtWEight}{0.562}

\newcommand{\carRealMinEight}{0.0000}

\newcommand{\carResEight}{0.0052}
\newcommand{\carResThirtyTwo}{0.0302}
\newcommand{\carResonantComplexAtWEight}{0.286}

\newcommand{\carResonantComplexMinEight}{0.0000}

\newcommand{\carResonantComplexResEight}{0.166}

\newcommand{\carResonantQuadMinEight}{0.3000}
\newcommand{\carResonantQuadMinThirtyTwo}{0.2972}

\newcommand{\carRmseFirstEight}{0.0132}
\newcommand{\carRmseFirstThirtyTwo}{0.0066}
\newcommand{\carRmseLastEight}{0.0015}
\newcommand{\carRmseLastThirtyTwo}{0.0011}
\newcommand{\carTrials}{50}
\newcommand{\carWzero}{0.3000}
\newcommand{\caraGapMinThirtyTwo}{0.96}

\newcommand{\cascCases}{30}
\newcommand{\cascMaxCauchy}{2 \times 10^{-15}}
\newcommand{\cascMaxEig}{2 \times 10^{-15}}
\newcommand{\cascMaxOffdiag}{3 \times 10^{-15}}
\newcommand{\cascMaxStages}{5}
\newcommand{\cgCorrApplies}{five}
\newcommand{\cgCorrGapMax}{0.012}

\newcommand{\cgJensenArOneNineFiveEight}{1.1158}
\newcommand{\cgJensenArOneNineFiveOneTwentyEight}{1.0028}

\newcommand{\cgLargestM}{128}
\newcommand{\cgNumWindows}{eight}
\newcommand{\cgRatioArFourRealEight}{1.3128}
\newcommand{\cgRatioArFourRealOneTwentyEight}{1.0427}

\newcommand{\cgRatioArOneNineFiveEight}{1.0047}
\newcommand{\cgRatioArOneNineFiveOneTwentyEight}{1.0022}

\newcommand{\cgRatioArTwoRealEight}{1.0151}
\newcommand{\cgRatioArTwoRealOneTwentyEight}{1.0031}

\newcommand{\cgRatioArTwoResEight}{1.1918}
\newcommand{\cgRatioArTwoResOneTwentyEight}{1.0215}

\newcommand{\cgSepCorrDbNineFive}{0.05}
\newcommand{\cgSepCorrLossPctNine}{2.1}
\newcommand{\cgSepCorrLossPctNineFive}{1.2}
\newcommand{\cgSepCorrNine}{1.0205}
\newcommand{\cgSepCorrNineFive}{1.0113}

\newcommand{\cgSepJensenNineFive}{1.2451}

\newcommand{\cgSepLossPctNineFive}{0.9}
\newcommand{\cgSepRatioNine}{1.0162}
\newcommand{\cgSepRatioNineFive}{1.0094}
\newcommand{\cgSmallestM}{8}
\newcommand{\cgTraceArOneNine}{-0.947}
\newcommand{\covDepArOneEight}{2.037}
\newcommand{\covDepArOneThirtyTwo}{0.433}
\newcommand{\departRealPosNNNEight}{0.061}

\newcommand{\dimGenericEight}{20}
\newcommand{\dimPhaseFour}{3}
\newcommand{\dimPhaseThree}{7}
\newcommand{\dimPhaseTwo}{9}
\newcommand{\discAllExactFromN}{20000}
\newcommand{\discCertifiedTotal}{354}

\newcommand{\discCertifiedWrongWords}{one}
\newcommand{\discExactFromArOne}{2000}
\newcommand{\discExactFromPhaseFour}{500}
\newcommand{\discExactFromPhaseThree}{5000}
\newcommand{\discExactFromPhaseTwo}{20000}
\newcommand{\discExactFromRes}{2000}
\newcommand{\discLargestN}{50000}
\newcommand{\discMaxInLastArOne}{1.01}
\newcommand{\discMaxInLastPhaseFour}{2.00}
\newcommand{\discMaxInLastPhaseThree}{1.56}
\newcommand{\discMaxInLastPhaseTwo}{1.21}
\newcommand{\discMaxInLastRes}{0.80}
\newcommand{\discMinOutLastArOne}{88}
\newcommand{\discMinOutLastPhaseFour}{1133}
\newcommand{\discMinOutLastPhaseThree}{39}
\newcommand{\discMinOutLastPhaseTwo}{60}
\newcommand{\discMinOutLastRes}{219}
\newcommand{\discNstarArOne}{573}
\newcommand{\discNstarPhaseFour}{43}
\newcommand{\discNstarPhaseThree}{1256}
\newcommand{\discNstarPhaseTwo}{832}
\newcommand{\discNstarRes}{230}
\newcommand{\discNumPatterns}{five}
\newcommand{\discOrderArOne}{2}
\newcommand{\discOrderPhaseFour}{384}
\newcommand{\discOrderPhaseThree}{48}
\newcommand{\discOrderPhaseTwo}{8}
\newcommand{\discOrderRes}{2}

\newcommand{\discSmallestN}{500}
\newcommand{\discTrials}{20}
\newcommand{\epsLevelSixMaxReal}{8}
\newcommand{\epsLevelSixMaxRes}{13}
\newcommand{\epsLevelSixMaxResIndex}{47}
\newcommand{\epsLevelSixMaxResStage}{46}
\newcommand{\epsLevelSixMaxResStageSorted}{8}
\newcommand{\epsLevelTenMaxReal}{13}
\newcommand{\epsLevelTenMaxRes}{20}
\newcommand{\epsNumLevelsFiveTwelve}{4}

\newcommand{\epsRankSixFiveTwelve}{7}
\newcommand{\epsRankSixMax}{8}
\newcommand{\epsRankSixMin}{6}
\newcommand{\epsRankSixTwoFiftySix}{6}
\newcommand{\epsRankTenFiveTwelve}{12}
\newcommand{\epsRankTenMax}{13}
\newcommand{\epsRankTenMin}{10}
\newcommand{\epsRankTenTwoFiftySix}{10}
\newcommand{\epsRelSixFiveTwelve}{9}
\newcommand{\epsRelSixTwoFiftySix}{8}
\newcommand{\epsRelTenFiveTwelve}{13}
\newcommand{\epsRelTenTwoFiftySix}{12}
\newcommand{\epsStageMax}{8}
\newcommand{\epsStageMin}{1}
\newcommand{\fastDevArOneSixteen}{0.099}
\newcommand{\fastDevArTwoSixteen}{0.096}

\newcommand{\fastDevResSixteen}{0.515}
\newcommand{\fastMaxBranch}{9 \times 10^{-16}}
\newcommand{\fastMaxCore}{2 \times 10^{-15}}
\newcommand{\fastMaxCross}{2 \times 10^{-15}}
\newcommand{\fastMaxEigvec}{1 \times 10^{-13}}
\newcommand{\fastMaxKlt}{5 \times 10^{-12}}
\newcommand{\fastMaxSecularEig}{5 \times 10^{-11}}
\newcommand{\fastMinSecularMid}{0.02}
\newcommand{\fastNumPatterns}{7}
\newcommand{\fastPolyMaxBlock}{3 \times 10^{-13}}
\newcommand{\fastPolyMaxCross}{0}
\newcommand{\fastShiftBranchMax}{2}
\newcommand{\fastShiftJointMax}{2}
\newcommand{\gainDctOneArTwoThirtyTwo}{0.79}

\newcommand{\gainDctTwoArTwoThirtyTwo}{0.99}
\newcommand{\gainDctTwoResEight}{0.84}

\newcommand{\gainPlainPolyPosEight}{0.95}

\newcommand{\gainPolyPosEight}{0.99}
\newcommand{\gainPolyPosMinEight}{0.989}

\newcommand{\gainPolyPosPlainNNEight}{0.41}

\newcommand{\gainQuadResEight}{0.63}

\newcommand{\gainRealNegDctTwoNNNEight}{0.04}

\newcommand{\gainRealPosDctTwoMinEight}{0.978}

\newcommand{\gainResDctTwoNNEight}{0.35}

\newcommand{\gainResDstTwoNNEight}{0.15}

\newcommand{\gainResQuadNNEight}{0.37}

\newcommand{\inactFreqA}{0.063}
\newcommand{\inactFreqB}{0.239}
\newcommand{\inactFreqC}{0.333}
\newcommand{\inactGap}{1.8}
\newcommand{\inactNorm}{3 \times 10^{-14}}
\newcommand{\inactNormB}{0.55}
\newcommand{\inactNormC}{0.48}
\newcommand{\inactR}{0.7785}
\newcommand{\inactResid}{2 \times 10^{-14}}
\newcommand{\inertiaArFourReal}{(2, 2)}
\newcommand{\inertiaArOneNeg}{(1, 0)}
\newcommand{\inertiaArOnePos}{(0, 1)}
\newcommand{\inertiaArThreePair}{(1, 2)}
\newcommand{\inertiaArTwoReal}{(1, 1)}
\newcommand{\intAsymMin}{0.88}
\newcommand{\intBestTypeDTwo}{DST-IV}
\newcommand{\intCovRanks}{2, 1, 3, 4}
\newcommand{\intCovResOneOneOneThirtyTwo}{0.0093}
\newcommand{\intCovResZeroOneOneThirtyTwo}{0.0001}
\newcommand{\intDstSevenGainMinDOneThirtyTwo}{0.777}
\newcommand{\intDstSevenGainOneOneZeroNineThirtyTwo}{0.927}
\newcommand{\intGainCorrApplies}{three}
\newcommand{\intGainCorrZeroOneOneThirtyTwo}{1.0034}
\newcommand{\intGainJensenOneOneZeroNineThirtyTwo}{1.0833}
\newcommand{\intGainJensenTwoOneTwoThirtyTwo}{1.2805}
\newcommand{\intGainLossPctOneOneZeroNineThirtyTwo}{7.3}
\newcommand{\intGainNumCases}{six}
\newcommand{\intGainRatioOneOneZeroNineThirtyTwo}{1.0785}
\newcommand{\intGainRatioTwoOneTwoThirtyTwo}{1.2715}
\newcommand{\intGainRatioZeroOneOneThirtyTwo}{1.0029}
\newcommand{\intMaxGainDTwoArThirtyTwo}{0.027}
\newcommand{\intMaxGainDTwoThirtyTwo}{0.57}
\newcommand{\intMinResDTwoThirtyTwo}{0.12}
\newcommand{\intRankOneThreeZero}{5}
\newcommand{\intResDTwoThirtyTwo}{0.19}
\newcommand{\intResDstFourOneOneZeroNineThirtyTwo}{0.0065}
\newcommand{\intResOneOneZeroSixteen}{0.0151}
\newcommand{\intResOneOneZeroThirtyTwo}{0.0092}
\newcommand{\intResResThirtyTwo}{0.0150}
\newcommand{\intResTwoOneZeroThirtyTwo}{0.0097}

\newcommand{\limHartleyThirtyTwoGrid}{0}
\newcommand{\limMinExtThirtyTwoGrid}{0.293}
\newcommand{\limMinPlainEightOff}{0.200}
\newcommand{\limMinPlainThirtyTwoOff}{0.228}
\newcommand{\limQuadEightOff}{0.055}
\newcommand{\limQuadThirtyTwoOff}{0.033}
\newcommand{\maResMaTwoThirtyTwo}{0.0108}
\newcommand{\maxBlockErr}{0}
\newcommand{\maxMultErr}{5 \times 10^{-15}}
\newcommand{\maxNyquistErr}{0}
\newcommand{\maxSinPairRealThirtyTwo}{3}
\newcommand{\modMaxConj}{3 \times 10^{-14}}
\newcommand{\modMaxOffdiag}{1 \times 10^{-15}}
\newcommand{\nGenericPatterns}{5}
\newcommand{\nOneSinResThirtyTwo}{18}
\newcommand{\nPolyPatterns}{5}
\newcommand{\nTwoSinResThirtyTwo}{14}
\newcommand{\oddMaxCore}{3 \times 10^{-13}}
\newcommand{\oddMaxCross}{1 \times 10^{-13}}
\newcommand{\oddMaxKernel}{4 \times 10^{-16}}
\newcommand{\oddMaxKlt}{9 \times 10^{-14}}
\newcommand{\oddNumCases}{nine}
\newcommand{\orderPhaseFour}{384}
\newcommand{\orderPhaseThree}{48}
\newcommand{\orderPhaseTwo}{8}
\newcommand{\placesMaxObserved}{2}
\newcommand{\placesMinObserved}{0}
\newcommand{\precDepArOneEight}{0.022}
\newcommand{\precDepArOneThirtyTwo}{0.010}

\newcommand{\rankArmaCancelled}{2}
\newcommand{\rankArmaQgtP}{4}
\newcommand{\rankArmaWhite}{0}
\newcommand{\rankDetMaxErr}{2 \times 10^{-16}}
\newcommand{\rankFactMaxErr}{2 \times 10^{-15}}
\newcommand{\rankM}{24}
\newcommand{\rankNumAr}{eleven}
\newcommand{\rankNumZeroPole}{three}
\newcommand{\rankSumDisjoint}{6}
\newcommand{\rankSumShared}{4}
\newcommand{\ratMaxCross}{1 \times 10^{-13}}
\newcommand{\ratMaxKlt}{3 \times 10^{-15}}
\newcommand{\ratNumCases}{8}
\newcommand{\ratShiftMax}{2}
\newcommand{\rdDevDctTwoNNEight}{0.13}

\newcommand{\rdDevDctTwoNineEight}{0.39}

\newcommand{\rdDevDstOneSmallEight}{0.006}
\newcommand{\rdSmallRho}{0.01}
\newcommand{\recCenterErr}{4 \times 10^{-16}}
\newcommand{\recMaxOutside}{2 \times 10^{-15}}
\newcommand{\resBestPlainResEight}{0.112}

\newcommand{\resDctOneArOneThirtyTwo}{0.023}
\newcommand{\resDctTwoArOneEight}{0.005}
\newcommand{\resDctTwoArOneThirtyTwo}{0.030}

\newcommand{\resPlainPolyPosEight}{0.010}

\newcommand{\resPolyPosEight}{0.004}

\newcommand{\resQuadResEight}{0.069}

\newcommand{\resResDctTwoNNNEight}{0.249}

\newcommand{\resResQuadNNNEight}{0.055}

\newcommand{\sepDctDbNineFive}{17.65}
\newcommand{\sepGainNine}{0.984}
\newcommand{\sepGainNineFive}{0.991}

\newcommand{\sepKltDbNineFive}{17.69}
\newcommand{\sepRank}{28}

\newcommand{\sepResNineFive}{0.003}
\newcommand{\sepTwoDctDbRealPos}{21.80}
\newcommand{\sepTwoDctDbRes}{8.61}
\newcommand{\sepTwoGainMixed}{0.978}
\newcommand{\sepTwoGainRealPos}{0.970}
\newcommand{\sepTwoGainRes}{0.704}
\newcommand{\sepTwoKltDbRealPos}{21.93}
\newcommand{\sepTwoKltDbRes}{10.13}
\newcommand{\sepTwoRankBound}{48}
\newcommand{\sepTwoRankRealPos}{48}

\newcommand{\sepTwoResRes}{0.304}
\newcommand{\shiftMaxObserved}{2}

\newcommand{\skewCountFivePhaseFour}{12}
\newcommand{\skewCountFivePhaseThree}{9}
\newcommand{\skewCountFivePhaseTwo}{4}

\newcommand{\skewCountLastArOne}{1}

\newcommand{\skewCountLastRes}{0}

\newcommand{\skewDimPhaseFour}{12}
\newcommand{\skewDimPhaseThree}{9}
\newcommand{\skewDimPhaseTwo}{4}

\newcommand{\skewFamilyDim}{2}
\newcommand{\skewFamilyGroupOrder}{8}
\newcommand{\skewFamilySingle}{4}
\newcommand{\skewLastN}{20000}

\newcommand{\wienerBoundSumTwoThree}{6}
\newcommand{\wienerRankSumOneOne}{2}
\newcommand{\wienerRankSumTwoOne}{4}
\newcommand{\wienerRankSumTwoThree}{6}
\newcommand{\wienerSizeSumTwoThree}{0.341}

%% file: numbers_impl.tex
\newcommand{\accEig}{6.55 \times 10^{-13}}
\newcommand{\accOrth}{3.13 \times 10^{-13}}
\newcommand{\accRound}{4.22 \times 10^{-15}}
\newcommand{\applyFlatArOneZeroNine}{9.9}

\newcommand{\applyFlatArThree}{5.3}

\newcommand{\applyIncFirstArThree}{3.15}

\newcommand{\applyIncLastArThree}{0.71}

\newcommand{\applyRatioLastArOneZeroNine}{113}

\newcommand{\applyRatioLastArThree}{118}
\newcommand{\countMfirst}{1024}
\newcommand{\countMlast}{16384}
\newcommand{\countRange}{16}

\newcommand{\designFlatArThree}{18}

\newcommand{\designRatioLastArThree}{34}

\newcommand{\devApplyFastArThree}{4}

\newcommand{\fitApplyDenseArThree}{2.55}

\newcommand{\fitApplyFastArThree}{1.27}

\newcommand{\machine}{macOS-\allowbreak{}15.3.2-\allowbreak{}arm64-\allowbreak{}arm-\allowbreak{}64bit-\allowbreak{}Mach-\allowbreak{}O}

\newcommand{\nRankCases}{14}
\newcommand{\npVersion}{2.5.1}
\newcommand{\oddCoreWorst}{1.54 \times 10^{-14}}
\newcommand{\oddGenWorst}{5.10 \times 10^{-14}}
\newcommand{\oddLeakWorst}{6.54 \times 10^{-17}}
\newcommand{\offBorder}{8.88 \times 10^{-16}}
\newcommand{\planAdaptive}{512}
\newcommand{\planFracAtBig}{3.1 \times 10^{-5}}
\newcommand{\planNbig}{4096}
\newcommand{\planUniMax}{0.192}
\newcommand{\planUniMin}{0.042}
\newcommand{\polyDesignMax}{52.9}
\newcommand{\polyDesignMin}{4.3}
\newcommand{\polyM}{4096}
\newcommand{\pyVersion}{3.13.12}

\newcommand{\sweepMiddleArOneZeroNine}{6}
\newcommand{\sweepMiddleArThree}{7}
\newcommand{\sweepNewtonArOneZeroNine}{37}
\newcommand{\sweepNewtonArThree}{37}

%% file: numbers_fo3.tex
\newcommand{\dfAdBlocks}{156}
\newcommand{\dfAdKOneHit}{0.36}

\newcommand{\dfAdKOneMed}{16.5}

\newcommand{\dfAdOrderSpread}{1.1}
\newcommand{\dfAdRouteBest}{0.040}
\newcommand{\dfAdRouteFirst}{0.895}
\newcommand{\dfAdSnapshots}{32448}

\newcommand{\dfAdSumMed}{10.0}

\newcommand{\dfAdWhiteMed}{27.8}
\newcommand{\dfBearBlocks}{156}
\newcommand{\dfBearHitArklt}{0.46}

\newcommand{\dfBearHitNone}{0.61}

\newcommand{\dfBearMeanArklt}{23.9}

\newcommand{\dfBearMedArklt}{12.0}

\newcommand{\dfBearMedNone}{7.5}
\newcommand{\dfBearMedSample}{7.3}
\newcommand{\dfBlocks}{415}
\newcommand{\dfCBadAfter}{5}
\newcommand{\dfCGoodAfter}{1.31 \times 10^{-10}}
\newcommand{\dfCWorstAfter}{1.26 \times 10^{-4}}

\newcommand{\dfCalAnchAdKOneMed}{14.4}

\newcommand{\dfCalAnchAdSumMed}{7.5}
\newcommand{\dfCalAnchDev}{31.5}

\newcommand{\dfCalAnchMusicHit}{0.74}
\newcommand{\dfCalAnchMusicMed}{4.9}

\newcommand{\dfCalBins}{13}

\newcommand{\dfCalBlindDev}{19.5}

\newcommand{\dfCalBlindMusicMed}{7.8}
\newcommand{\dfCalBlocks}{156}
\newcommand{\dfCalElems}{7}

\newcommand{\dfCalMusicGain}{2.6}

\newcommand{\dfCalRawAdSumMed}{10.0}

\newcommand{\dfCalRawMusicHit}{0.61}
\newcommand{\dfCalRawMusicMed}{7.5}

\newcommand{\dfCalSnapGap}{0.4}
\newcommand{\dfCells}{15}
\newcommand{\dfCohAdHigh}{0.08}
\newcommand{\dfCohAdHighMid}{0.07}
\newcommand{\dfCohDirect}{70}
\newcommand{\dfCohEspFloor}{11.4}
\newcommand{\dfCohEspFloorMid}{2.1}
\newcommand{\dfCohFreq}{400}
\newcommand{\dfCohImage}{110}
\newcommand{\dfCohRatio}{150}
\newcommand{\dfCohSnr}{12}
\newcommand{\dfCohSumFloor}{0.53}
\newcommand{\dfCohTrials}{200}
\newcommand{\dfCountBest}{0.036}
\newcommand{\dfDifBeta}{0.5}
\newcommand{\dfDifManyAd}{1.76}
\newcommand{\dfDifManyEsprit}{3.13}
\newcommand{\dfDifManyMusic}{1.75}
\newcommand{\dfDifMax}{32}
\newcommand{\dfDifOneAd}{0.31}
\newcommand{\dfDifOneEsprit}{8.63}
\newcommand{\dfDifOneGain}{4.0}
\newcommand{\dfDifOneMusic}{1.25}
\newcommand{\dfDifTrials}{300}

\newcommand{\dfEspNaivePenalty}{2.1}
\newcommand{\dfEspNoneHit}{0.44}

\newcommand{\dfEspNoneMed}{12.4}

\newcommand{\dfEspSilentNaiveHit}{0.19}

\newcommand{\dfEspSilentNaiveMed}{25.8}
\newcommand{\dfEspSilentSubHit}{0.45}

\newcommand{\dfEspSilentSubMed}{12.5}

\newcommand{\dfFitMeanArklt}{17.0}

\newcommand{\dfFitMedArklt}{12.2}

\newcommand{\dfFitMedNone}{7.5}

\newcommand{\dfFitPlainWorstMed}{8.3}
\newcommand{\dfFitRationalBestMed}{10.0}
\newcommand{\dfGainAtLong}{0.0026}
\newcommand{\dfGainAtShort}{0.0158}
\newcommand{\dfGainMax}{0.0203}
\newcommand{\dfGainMin}{0.0009}
\newcommand{\dfGridCells}{20}
\newcommand{\dfKltKl}{1.20 \times 10^{-13}}

\newcommand{\dfManAdSnap}{2.70}

\newcommand{\dfManEspRatio}{3.9}

\newcommand{\dfManGain}{0.5}
\newcommand{\dfManMusic}{2.50}
\newcommand{\dfManPhase}{30}
\newcommand{\dfManSpread}{1.08}
\newcommand{\dfManTrials}{300}
\newcommand{\dfMusicNoneHit}{0.61}

\newcommand{\dfMusicNoneMed}{7.5}
\newcommand{\dfPredRatioFourierMax}{1.57}
\newcommand{\dfPredRatioFourierMin}{0.55}
\newcommand{\dfPredRatioMax}{1.11}
\newcommand{\dfPredRatioMed}{0.96}
\newcommand{\dfPredRatioMin}{0.74}
\newcommand{\dfPyDenseWorst}{1.22 \times 10^{-10}}
\newcommand{\dfPyFmmBad}{7}
\newcommand{\dfPyFmmWorst}{0.179}
\newcommand{\dfRangeMax}{2.75 \times 10^{6}}
\newcommand{\dfRefusedAfter}{0}
\newcommand{\dfRefusedCount}{10}

\newcommand{\dfResidRatioMax}{9.9}
\newcommand{\dfResidRatioMin}{3.7}

\newcommand{\dfSilentArklt}{9.4}

\newcommand{\dfSilentSample}{6.6}
\newcommand{\dfSpatialRatioMax}{2.61}
\newcommand{\dfSpatialRatioMed}{1.84}

\newcommand{\dfSpatialRows}{49}
\newcommand{\dfSpatialWins}{48}
\newcommand{\dfTmax}{36}
\newcommand{\dfTmin}{5}
\newcommand{\dfWhiteGap}{1.47}

\newcommand{\dfWideAnchBinMusicMed}{4.9}

\newcommand{\dfWideAnchWideMusicMed}{7.3}

\newcommand{\dfWideBlindWideMusicMed}{10.6}
\newcommand{\dfWideLoss}{2.4}
\newcommand{\dfWideParamsBin}{182}
\newcommand{\dfWideParamsWide}{14}

\newcommand{\dfWideResidAnch}{27}
\newcommand{\dfWideResidBlind}{36}
\newcommand{\qwDctHi}{0.624}
\newcommand{\qwDctLo}{0.556}
\newcommand{\qwDctSpan}{1.12}
\newcommand{\qwLong}{512}
\newcommand{\qwMMax}{5.15}
\newcommand{\qwMMin}{1.70}
\newcommand{\qwMSpan}{3.04}
\newcommand{\qwMarginLong}{0.0099}

\newcommand{\qwPeak}{0.1270}
\newcommand{\qwPeakM}{32}
\newcommand{\qwRatio}{10.8}
\newcommand{\qwSelLong}{3}
\newcommand{\qwSelShort}{6}
\newcommand{\qwShort}{16}
\newcommand{\qwWindowRatio}{32}
\newcommand{\qwWindows}{6}

\newcommand{\rfBestDft}{0.0099}
\newcommand{\rfBestDftPred}{0.0111}
\newcommand{\rfBestDftSe}{0.0005}
\newcommand{\rfBestOrder}{3}
\newcommand{\rfDctHi}{0.64}
\newcommand{\rfDctLo}{0.18}
\newcommand{\rfDftHi}{0.0099}
\newcommand{\rfDftLo}{0.0021}
\newcommand{\rfDftPredHi}{0.0117}
\newcommand{\rfDftPredLo}{0.0035}
\newcommand{\rfDynRange}{24.1}
\newcommand{\rfFitRows}{2048}
\newcommand{\rfFlatness}{0.28}
\newcommand{\rfFrames}{8}
\newcommand{\rfPole}{0.893}
\newcommand{\rfRatio}{65}
\newcommand{\rfRowsPerCoord}{4.0}
\newcommand{\rfSampleCov}{-9.1}
\newcommand{\rfSampleCovPred}{0.55}
\newcommand{\rfTestRows}{2048}
\newcommand{\rfWindow}{512}

%% file: numbers_fo3_rf.tex
\newcommand{\rfAxis}{fast-time}
\newcommand{\rfAzLimit}{25}
\newcommand{\rfBestMethod}{ESPRIT}
\newcommand{\rfBestRmse}{1.433}
\newcommand{\rfCells}{80}
\newcommand{\rfColoredFracFast}{0.17}

\newcommand{\rfDilationFactor}{1.42}
\newcommand{\rfElements}{16}
\newcommand{\rfFloorMaxFast}{21.0}

\newcommand{\rfFloorMedFast}{8.9}

\newcommand{\rfFramesPerRecord}{8}
\newcommand{\rfLedgerApply}{2.5 \times 10^{-15}}
\newcommand{\rfLedgerCalls}{96}
\newcommand{\rfLedgerDeclined}{0}
\newcommand{\rfLedgerMu}{5.6 \times 10^{-15}}
\newcommand{\rfLedgerOffdiag}{1.0 \times 10^{-10}}
\newcommand{\rfLedgerOrth}{5.1 \times 10^{-14}}
\newcommand{\rfLedgerPairs}{5}
\newcommand{\rfLedgerServed}{96}

\newcommand{\rfOrderMaxFast}{8}
\newcommand{\rfOrderMaxSlow}{3}
\newcommand{\rfRangeMaxFast}{61.9}
\newcommand{\rfRangeMaxSlow}{6.6}
\newcommand{\rfRangeMedFast}{2.94}
\newcommand{\rfRangeMedSlow}{2.84}

\newcommand{\rfRatioMax}{1.000}

\newcommand{\rfRatioMin}{0.997}
\newcommand{\rfRecords}{8}
\newcommand{\rfSpacing}{0.494}
\newcommand{\rfWorstRmse}{1.473}

%% file: numbers_apps.tex
\newcommand{\apProvisional}{0}
\newcommand{\apAccDenseHi}{$2\times10^{-14}$}
\newcommand{\apAccFastHi}{$3\times10^{-12}$}

\newcommand{\apAccNoiseHi}{$10^{-2}$}
\newcommand{\apAccNoiseLo}{$10^{-8}$}
\newcommand{\apBrInnerHi}{0.6}
\newcommand{\apBrInnerLo}{1.6}
\newcommand{\apBrNarrowHi}{0.8}
\newcommand{\apBrOtherHi}{0.3}
\newcommand{\apBrOtherLo}{4.4}
\newcommand{\apBrRecords}{8}
\newcommand{\apCyCascBigHi}{0.94}
\newcommand{\apCyCascBigLo}{0.00}
\newcommand{\apCyCascBigMed}{0.05}
\newcommand{\apCyCascSmallHi}{3.87}
\newcommand{\apCyCascSmallLo}{0.02}
\newcommand{\apCyCascSmallMed}{0.37}
\newcommand{\apCyCases}{50}
\newcommand{\apCyControl}{0.17}
\newcommand{\apCyGainHi}{14.6}
\newcommand{\apCyMildHi}{1.33}
\newcommand{\apCyMildLo}{-0.01}
\newcommand{\apCyStages}{4}
\newcommand{\apCyStrongHi}{6.6}
\newcommand{\apCyStrongLo}{1.7}
\newcommand{\apDrCellHi}{2.2}
\newcommand{\apDrCellLo}{0.6}
\newcommand{\apDrRecords}{3}
\newcommand{\apDrRestHi}{0.3}
\newcommand{\apDrRestLo}{1.9}
\newcommand{\apEndsFaHi}{174}
\newcommand{\apEndsFaLo}{3.9}
\newcommand{\apEndsKltFaHi}{1.1}
\newcommand{\apEndsKltVarHi}{1.09}
\newcommand{\apEndsN}{2048}
\newcommand{\apEndsRecords}{2000}
\newcommand{\apEndsTaperHi}{218}
\newcommand{\apEndsTaperLo}{174}
\newcommand{\apEndsVarHi}{360}
\newcommand{\apEndsVarLo}{7.7}
\newcommand{\apEndsWidth}{64}
\newcommand{\apNwArHi}{0.98}
\newcommand{\apNwArLo}{0.68}
\newcommand{\apNwLoadHi}{0.79}
\newcommand{\apNwLoadLo}{0.46}
\newcommand{\apNwTrials}{500}
\newcommand{\apOtArHi}{0.097}
\newcommand{\apOtArLo}{0.027}
\newcommand{\apOtCycEight}{1.00}
\newcommand{\apOtCycFour}{0.53}
\newcommand{\apOtGainArEight}{0.95}

\newcommand{\apOtGainToepEight}{0.11}
\newcommand{\apOtSurrogates}{199}
\newcommand{\apOtToepHi}{0.073}
\newcommand{\apOtToepLo}{0.010}
\newcommand{\apOtTrials}{300}
\newcommand{\apRaAicHi}{1.45}
\newcommand{\apRaAicLo}{0.27}
\newcommand{\apRaArHi}{0.54}
\newcommand{\apRaArLo}{0.24}
\newcommand{\apRaCells}{200}
\newcommand{\apRaWhiteHi}{0.58}
\newcommand{\apRaWhiteLo}{0.30}
\newcommand{\apRdCosHi}{9.6}
\newcommand{\apRdCosLo}{0.04}
\newcommand{\apRdLapBest}{0.01}
\newcommand{\apRdLapWorst}{0.38}
\newcommand{\apRdPoints}{84}

\newcommand{\apRealCascMed}{0.26}
\newcommand{\apRkAtBound}{52}
\newcommand{\apRkCases}{52}
\newcommand{\apZpArHi}{5.8}
\newcommand{\apZpArLo}{2.6}
\newcommand{\apZpArmaHi}{3.9}
\newcommand{\apZpArmaLo}{1.1}
\newcommand{\apZpDesignHi}{17.9}
\newcommand{\apZpDesignLo}{1.0}
\newcommand{\apZpLanHi}{8.9}
\newcommand{\apZpLanLo}{1.5}
\newcommand{\apZpNodesHi}{30}
\newcommand{\apZpNodesLo}{17}
\newcommand{\apZpRankHi}{34}
\newcommand{\apZpRankLo}{25}

%% file: class_table.tex
AR(1), $0.9$ & 0.017/0.89 & 0.005/0.99 & 0.128/0.78 & 0.231/0.57 & 0.014/0.82 & -- \\
AR(2), $0.9$, $0.5$ & 0.019/0.61 & 0.004/0.99 & 0.140/0.50 & 0.249/0.26 & 0.012/0.53 & -- \\
AR(3), $0.9$, $0.7$, $0.5$ & 0.023/0.17 & 0.002/0.87 & 0.159/0.13 & 0.271/0.05 & 0.006/0.17 & -- \\
AR(1), $-0.9$ & 0.017/0.89 & 0.231/0.57 & 0.128/0.78 & 0.005/0.99 & 0.014/0.82 & -- \\
AR(2), $-0.9$, $-0.5$ & 0.019/0.61 & 0.249/0.26 & 0.140/0.50 & 0.004/0.99 & 0.012/0.53 & -- \\
AR(2), $0.9$, $-0.5$ & 0.017/0.96 & 0.005/0.99 & 0.125/0.85 & 0.223/0.75 & 0.012/0.93 & -- \\
AR(2), $0.9e^{\pm i0.3\pi}$ & 0.137/0.73 & 0.166/0.84 & 0.120/0.82 & 0.198/0.57 & 0.112/0.69 & quadrature 0.069/0.63 \\
AR(3), pair, $0.8$ & 0.056/0.52 & 0.079/0.81 & 0.036/0.64 & 0.098/0.35 & 0.020/0.48 & quadrature 0.321/0.24 \\
AR(4), two pairs & 0.080/0.86 & 0.096/0.91 & 0.091/0.92 & 0.154/0.75 & 0.121/0.79 & -- \\
AR(2), $\pm0.9$ & 0.021/0.95 & 0.113/0.80 & 0.130/0.77 & 0.113/0.80 & 0.010/0.90 & polyphase DCT-II 0.004/0.99 \\
AR(2), $\pm0.9i$ & 0.169/0.77 & 0.199/0.79 & 0.130/0.87 & 0.199/0.79 & 0.010/0.90 & polyphase DST-II 0.004/0.99 \\
AR(4), $b(z^2)$, $b$ resonant & 0.119/0.73 & 0.147/0.67 & 0.110/0.66 & 0.147/0.67 & 0.307/0.51 & polyphase DCT-II 0.115/0.81 \\

%% file: rational_table.tex
ARMA(1,1), pole $0.9$, zero $-0.5$ & (1,1) & 16 & 2 & $2\max(p,q)=2$ & 1/1 & $1 \times 10^{-15}$ & 1\\
ARMA(1,1), pole $0.9$, zero $-0.5$ & (1,1) & 32 & 2 & $2\max(p,q)=2$ & 1/1 & $2 \times 10^{-15}$ & 1\\
ARMA(1,2), pole $0.9$ & (1,2) & 16 & 4 & $2\max(p,q)=4$ & 2/2 & $1 \times 10^{-15}$ & 1\\
ARMA(1,2), pole $0.9$ & (1,2) & 32 & 4 & $2\max(p,q)=4$ & 2/2 & $2 \times 10^{-15}$ & 1\\
ARMA(2,2), poles $0.9$, $0.5$ & (2,2) & 16 & 4 & $2\max(p,q)=4$ & 2/2 & $1 \times 10^{-15}$ & 1\\
ARMA(2,2), poles $0.9$, $0.5$ & (2,2) & 32 & 4 & $2\max(p,q)=4$ & 2/2 & $1 \times 10^{-15}$ & 1\\
ARMA(3,1), pair at $0.3\pi$, $0.8$ & (3,1) & 16 & 6 & $2\max(p,q)=6$ & 3/3 & $9 \times 10^{-16}$ & 2\\
ARMA(3,1), pair at $0.3\pi$, $0.8$ & (3,1) & 32 & 6 & $2\max(p,q)=6$ & 3/3 & $2 \times 10^{-15}$ & 2\\
MA(2) & (0,2) & 16 & 4 & $2\max(p,q)=4$ & 2/2 & $9 \times 10^{-16}$ & 1\\
MA(2) & (0,2) & 32 & 4 & $2\max(p,q)=4$ & 2/2 & $2 \times 10^{-15}$ & 2\\
AR(1) $0.9$ + AR(1) $-0.7$ & (2,1) & 16 & 4 & $2(k+m)=4$ & 2/2 & $1 \times 10^{-15}$ & 2\\
AR(1) $0.9$ + AR(1) $-0.7$ & (2,1) & 32 & 4 & $2(k+m)=4$ & 2/2 & $2 \times 10^{-15}$ & 2\\
resonant AR(2) + AR(1) $0.95$ & (3,2) & 16 & 6 & $2(k+m)=6$ & 3/3 & $2 \times 10^{-15}$ & 2\\
resonant AR(2) + AR(1) $0.95$ & (3,2) & 32 & 6 & $2(k+m)=6$ & 3/3 & $2 \times 10^{-15}$ & 2\\
AR(2) $0.9$, $0.5$ + AR(3) pair, $0.8$ & (5,3) & 16 & 10 & $2(k+m)=10$ & 5/5 & $1 \times 10^{-15}$ & 1\\
AR(2) $0.9$, $0.5$ + AR(3) pair, $0.8$ & (5,3) & 32 & 10 & $2(k+m)=10$ & 5/5 & $2 \times 10^{-15}$ & 2\\

%% file: fast_table.tex
AR(1), pole $0.9$ & 1 & 16 & 1/1 & $8 \times 10^{-12}$ & $8 \times 10^{-16}$ & 1 / 1 & 0.099\\
AR(1), pole $0.9$ & 1 & 32 & 1/1 & $2 \times 10^{-11}$ & $2 \times 10^{-15}$ & 1 / 1 & 0.165\\
AR(2), poles $0.9$, $0.5$ & 2 & 16 & 2/2 & $3 \times 10^{-13}$ & $2 \times 10^{-15}$ & 1 / 1 & 0.096\\
AR(2), poles $0.9$, $0.5$ & 2 & 32 & 2/2 & $7 \times 10^{-12}$ & $2 \times 10^{-14}$ & 1 / 1 & 0.164\\
AR(2), poles $0.9$, $-0.5$ & 2 & 16 & 2/2 & $1 \times 10^{-13}$ & $7 \times 10^{-16}$ & 1 / 2 & 0.615\\
AR(2), poles $0.9$, $-0.5$ & 2 & 32 & 2/2 & $8 \times 10^{-13}$ & $2 \times 10^{-15}$ & 1 / 2 & 0.639\\
AR(2), poles $0.9e^{\pm i0.3\pi}$ & 2 & 16 & 2/2 & $5 \times 10^{-13}$ & $6 \times 10^{-15}$ & 1 / 2 & 0.515\\
AR(2), poles $0.9e^{\pm i0.3\pi}$ & 2 & 32 & 2/2 & $2 \times 10^{-12}$ & $8 \times 10^{-15}$ & 1 / 2 & 0.968\\
AR(3), poles $0.9$, $0.7$, $0.5$ & 3 & 16 & 3/3 & $8 \times 10^{-12}$ & $8 \times 10^{-14}$ & 1 / 1 & 0.179\\
AR(3), poles $0.9$, $0.7$, $0.5$ & 3 & 32 & 3/3 & $5 \times 10^{-11}$ & $4 \times 10^{-13}$ & 1 / 1 & 0.199\\
AR(3), pair at $0.3\pi$ and $0.8$ & 3 & 16 & 3/3 & $4 \times 10^{-11}$ & $5 \times 10^{-14}$ & 1 / 2 & 0.637\\
AR(3), pair at $0.3\pi$ and $0.8$ & 3 & 32 & 3/3 & $9 \times 10^{-12}$ & $5 \times 10^{-14}$ & 2 / 2 & 0.672\\
AR(4), poles $0.9$, $0.7$, $0.5$, $0.3$ & 4 & 16 & 4/4 & $4 \times 10^{-12}$ & $3 \times 10^{-13}$ & 1 / 1 & 0.224\\
AR(4), poles $0.9$, $0.7$, $0.5$, $0.3$ & 4 & 32 & 4/4 & $5 \times 10^{-11}$ & $4 \times 10^{-12}$ & 1 / 1 & 0.252\\

%% file: apps_tab_ranks.tex
vector AR & $2mp$ & 5 & 5 \\
vector ARMA & $2m\max(p,q)$ & 4 & 4 \\
state space & $2s$ & 4 & 4 \\
periodic AR, real & $2p$ & 6 & 6 \\
periodic AR, complex & $2p$ & 6 & 6 \\
modulated, circular symbols & $2\lfloor (L-1)/P\rfloor$ & 10 & 10 \\
modulated, real symbols & $2\lfloor (L-1)/P\rfloor$ & 6 & 6 \\
bridge & $2(p-1)$ & 4 & 4 \\
switching & $2p$ per switch & 4 & 4 \\
Mat\'ern, $\nu=k+\tfrac12$ & $2(k+1)$ & 3 & 3 \\

%% file: group_table.tex
AR(1), pole $0.9$ & 8 & yes & 2 & $\mathbb{Z}_2$ & 20 / 36 \\
AR(2), poles $0.9$, $0.5$ & 8 & yes & 2 & $\mathbb{Z}_2$ & 20 / 36 \\
AR(2), poles $0.9$, $-0.5$ & 8 & yes & 2 & $\mathbb{Z}_2$ & 20 / 36 \\
AR(2), poles $0.9e^{\pm i0.3\pi}$ & 8 & yes & 2 & $\mathbb{Z}_2$ & 20 / 36 \\
AR(3), pair at $0.3\pi$ and $0.8$ & 8 & yes & 2 & $\mathbb{Z}_2$ & 20 / 36 \\
AR(2), poles $\pm0.9$ & 8 & no & 8 & $\mathbb{Z}_2\wr S_2$ & 9 / 36 \\
AR(2), poles $\pm0.9i$ & 8 & no & 8 & $\mathbb{Z}_2\wr S_2$ & 9 / 36 \\
AR(4), $a(z)=b(z^2)$, $b$ resonant & 8 & no & 8 & $\mathbb{Z}_2\wr S_2$ & 9 / 36 \\
AR(3), poles $0.9e^{2\pi ij/3}$ & 9 & no & 48 & $\mathbb{Z}_2\wr S_3$ & 7 / 45 \\
AR(4), poles $0.9i^{\,j}$ & 8 & no & 384 & $\mathbb{Z}_2\wr S_4$ & 3 / 36 \\

%% file: class_table_32.tex
AR(1), $0.9$ & 0.023/0.96 & 0.030/0.99 & 0.091/0.89 & 0.121/0.74 & 0.076/0.85 & -- \\
AR(2), $0.9$, $0.5$ & 0.024/0.79 & 0.031/0.99 & 0.098/0.67 & 0.129/0.37 & 0.080/0.56 & -- \\
AR(3), $0.9$, $0.7$, $0.5$ & 0.025/0.25 & 0.032/0.94 & 0.115/0.18 & 0.147/0.06 & 0.084/0.14 & -- \\
AR(1), $-0.9$ & 0.023/0.96 & 0.121/0.74 & 0.091/0.89 & 0.030/0.99 & 0.076/0.85 & -- \\
AR(2), $-0.9$, $-0.5$ & 0.024/0.79 & 0.129/0.37 & 0.098/0.67 & 0.031/0.99 & 0.080/0.56 & -- \\
AR(2), $0.9$, $-0.5$ & 0.023/0.99 & 0.030/0.99 & 0.090/0.92 & 0.118/0.87 & 0.074/0.93 & -- \\
AR(2), $0.9e^{\pm i0.3\pi}$ & 0.100/0.84 & 0.112/0.93 & 0.088/0.91 & 0.114/0.72 & 0.107/0.80 & quadrature 0.068/0.60 \\
AR(3), pair, $0.8$ & 0.052/0.63 & 0.063/0.91 & 0.047/0.75 & 0.069/0.41 & 0.073/0.54 & quadrature 0.343/0.15 \\
AR(4), two pairs & 0.086/0.89 & 0.097/0.95 & 0.078/0.96 & 0.101/0.82 & 0.087/0.86 & -- \\
AR(2), $\pm0.9$ & 0.023/0.98 & 0.073/0.88 & 0.091/0.86 & 0.073/0.88 & 0.073/0.88 & polyphase DCT-II 0.029/0.99 \\
AR(2), $\pm0.9i$ & 0.104/0.87 & 0.117/0.88 & 0.091/0.95 & 0.117/0.88 & 0.073/0.88 & polyphase DST-II 0.029/0.99 \\
AR(4), $b(z^2)$, $b$ resonant & 0.155/0.81 & 0.169/0.70 & 0.147/0.70 & 0.169/0.70 & 0.269/0.65 & polyphase DCT-II 0.164/0.88 \\

%% file: bounds_table.tex
AR(1), pole $0.9$ & 1.0080 / 1.0619 / 1.0102 & 1.0082 / 1.0216 / 1.0111 & 1.0064 / 1.0102 / 1.0088 & 1.0023 / 1.0026 / 1.0031 \\
AR(1), pole $0.95$ & 1.0047 / 1.1158 / 1.0057 & 1.0052 / 1.0347 / 1.0068 & 1.0047 / 1.0121 / 1.0065 & 1.0022 / 1.0028 / 1.0030 \\
AR(1), pole $0.99$ & 1.0011 / 1.3199 / 1.0012 & 1.0013 / 1.1079 / 1.0015 & 1.0014 / 1.0353 / 1.0017 & 1.0011 / 1.0036 / 1.0015 \\
AR(2), poles $0.9$, $0.5$ & 1.0151 / 1.0811 / 1.0268 & 1.0128 / 1.0292 / 1.0236 & 1.0091 / 1.0136 / 1.0141 & 1.0031 / 1.0034 / 1.0045 \\
AR(2), poles $0.9$, $-0.5$ & 1.0111 / 1.0762 / 1.0153 & 1.0104 / 1.0266 / 1.0158 & 1.0078 / 1.0122 / 1.0116 & 1.0027 / 1.0030 / 1.0040 \\
AR(2), poles $0.9e^{\pm i0.3\pi}$ & 1.1918 / 1.2513 / -- & 1.1322 / 1.1647 / -- & 1.0769 / 1.0854 / -- & 1.0215 / 1.0221 / -- \\
AR(3), poles $0.9e^{\pm i0.3\pi}$, $0.8$ & 1.2401 / 1.3837 / -- & 1.1684 / 1.2291 / -- & 1.1011 / 1.1170 / -- & 1.0288 / 1.0299 / -- \\
AR(4), poles $0.9$, $0.7$, $0.5$, $0.3$ & 1.3128 / 1.5973 / -- & 1.2347 / 1.3202 / -- & 1.1451 / 1.1694 / -- & 1.0427 / 1.0445 / -- \\

%% file: fo3_tab_adesprit.tex
MUSIC, no whitening & 13.0 & 7.5 & 0.61 & 0.74 \\
ESPRIT, no whitening & 15.3 & 12.4 & 0.44 & 0.69 \\
AD-ESPRIT, one component & 18.6 & 16.5 & 0.36 & 0.59 \\
AD-ESPRIT, two components & 18.3 & 15.4 & 0.37 & 0.62 \\
AD-ESPRIT, three components & 19.6 & 16.0 & 0.34 & 0.58 \\
AD-ESPRIT, dilated start only & 17.9 & 14.5 & 0.37 & 0.62 \\
AD-ESPRIT, dilated sum over snapshots & 13.9 & 10.0 & 0.50 & 0.73 \\
AD-ESPRIT, snapshot whitened & 29.4 & 27.8 & 0.14 & 0.33 \\

%% file: fo3_tab_coherence.tex
0.0 & 0 & 0.75 & 0.82 & 0.85 & 0.65 \\
0.0 & 6 & 0.25 & 0.37 & 0.36 & 0.29 \\
0.0 & 12 & 0.25 & 0.18 & 0.20 & 0.15 \\
0.0 & 20 & 0.00 & 0.08 & 0.07 & 0.06 \\
0.3 & 0 & 0.75 & 2.00 & 0.98 & 0.68 \\
0.3 & 6 & 0.25 & 2.04 & 0.34 & 0.29 \\
0.3 & 12 & 0.25 & 2.07 & 0.22 & 0.18 \\
0.3 & 20 & 0.00 & 2.09 & 0.07 & 0.06 \\
0.7 & 0 & 0.75 & 11.36 & 2.35 & 1.08 \\
0.7 & 6 & 0.25 & 11.53 & 0.48 & 0.69 \\
0.7 & 12 & 0.25 & 11.37 & 0.23 & 0.54 \\
0.7 & 20 & 0.00 & 11.41 & 0.08 & 0.53 \\

%% file: fo3_tab_manifold.tex
0.0 & 0 & 0.25 & 0.18 & 0.22 & 0.14 \\
0.0 & 2 & 0.25 & 0.25 & 0.24 & 0.19 \\
0.0 & 5 & 0.25 & 0.41 & 0.34 & 0.32 \\
0.0 & 10 & 0.50 & 0.70 & 0.62 & 0.63 \\
0.0 & 20 & 1.25 & 1.53 & 1.20 & 1.16 \\
0.0 & 30 & 1.75 & 2.15 & 1.84 & 1.78 \\
0.7 & 0 & 0.25 & 11.40 & 0.25 & 0.50 \\
0.7 & 2 & 0.25 & 11.48 & 0.27 & 0.52 \\
0.7 & 5 & 0.25 & 11.56 & 0.40 & 0.65 \\
0.7 & 10 & 0.75 & 11.37 & 0.82 & 0.90 \\
0.7 & 20 & 1.50 & 11.83 & 1.59 & 1.83 \\
0.7 & 30 & 2.50 & 10.59 & 2.70 & 2.70 \\

%% file: fo3_tab_diffuse.tex
1 & 1.25 & 8.63 & 0.31 & 1.45 \\
2 & 1.50 & 5.06 & 1.11 & 1.62 \\
4 & 1.75 & 3.86 & 1.59 & 1.64 \\
8 & 1.62 & 3.57 & 1.72 & 1.78 \\
16 & 1.75 & 3.11 & 1.79 & 1.79 \\
32 & 1.75 & 3.13 & 1.76 & 1.91 \\

%% file: fo3_tab_wideband.tex
MUSIC & 7.5 & 7.8 & 10.6 & 4.9 & 7.3 \\
ESPRIT & 12.4 & 11.9 & 14.6 & 10.4 & 11.1 \\
AD-ESPRIT, one component & 16.5 & 15.1 & 16.7 & 14.4 & 14.1 \\
AD-ESPRIT, two components & 15.4 & 14.5 & 16.8 & 14.7 & 14.0 \\
AD-ESPRIT, dilated sum & 10.0 & 9.0 & 12.1 & 7.5 & 8.0 \\

%% file: fo3_tab_calib.tex
MUSIC & 7.5 & 0.61 & 7.8 & 0.62 & 4.9 & 0.74 \\
ESPRIT & 12.4 & 0.44 & 11.9 & 0.44 & 10.4 & 0.46 \\
AD-ESPRIT, one component & 16.5 & 0.36 & 15.1 & 0.35 & 14.4 & 0.31 \\
AD-ESPRIT, two components & 15.4 & 0.37 & 14.5 & 0.39 & 14.7 & 0.29 \\
AD-ESPRIT, dilated sum & 10.0 & 0.50 & 9.0 & 0.53 & 7.5 & 0.64 \\

%% file: fo3_tab_rfcolor.tex
Fast time, within one chirp & 512 & 64 & 8.9 & 21.0 & 2.94 & 61.9 & 8 & 0.17 \\
Slow time, across chirps & 256 & 64 & 8.9 & 21.1 & 2.84 & 6.6 & 3 & 0.08 \\

%% file: fo3_tab_rfdf.tex
AD-ESPRIT cascade, per slot & 12 & 0.217 & 1.454 & 0.595 & 0.924 & 1.454 & 1.000 \\
SHEAR & 12 & 0.214 & 1.469 & 0.618 & 1.020 & 1.469 & 1.000 \\
Bartlett on the calibrated array & 12 & 0.221 & 1.473 & 0.628 & 1.012 & 1.473 & 1.000 \\
ESPRIT & 12 & 0.113 & 1.433 & 0.679 & 1.181 & 1.432 & 1.000 \\
MUSIC & 12 & 0.229 & 1.473 & 0.626 & 1.012 & 1.473 & 1.000 \\
AD-ESPRIT cascade & 1 & 0.001 & 2.323 & 1.011 & 1.591 & 2.316 & 0.997 \\
ESPRIT, dilated average, $N = 2M-1$ & 1 & 0.001 & 2.323 & 1.011 & 1.591 & 2.316 & 0.997 \\
ESPRIT, cyclic average, $N = M$ & 1 & -0.158 & 3.301 & 2.343 & 3.921 & 3.301 & 1.000 \\

%% file: fo3_tab_rfklt.tex
Nearly flat, $1 \le \varrho < 2$ & 22 & 1.001 & 1.000 & 1.000 & 1.000 \\
Mildly colored, $2 \le \varrho < 5$ & 57 & 1.000 & 1.000 & 1.000 & 1.000 \\
Strongly colored, $\varrho \ge 5$ & 1 & 0.995 & 0.998 & 1.000 & 1.010 \\

%% file: fo3_tab_esprit.tex
MUSIC, no whitening & 13.0 & 7.5 & 0.61 & 0.74 \\
ESPRIT, no whitening & 15.3 & 12.4 & 0.44 & 0.69 \\
ESPRIT, data whitened, silence-fitted & 31.5 & 25.8 & 0.19 & 0.38 \\
ESPRIT, data whitened, block-fitted & 30.7 & 26.8 & 0.17 & 0.38 \\
ESPRIT, subspace only, silence-fitted & 14.8 & 12.5 & 0.45 & 0.69 \\
ESPRIT, subspace only, block-fitted & 16.5 & 13.8 & 0.40 & 0.64 \\
ESPRIT, subspace only, unstructured & 15.3 & 12.4 & 0.44 & 0.69 \\

%% file: fo3_tab_fitted.tex
no whitening & 13.0 & 7.5 & 0.61 & 0.74 \\
AR-KLT, fitted to silence & 23.9 & 12.0 & 0.46 & 0.60 \\
unstructured, fitted to the block & 14.4 & 8.3 & 0.60 & 0.70 \\
Toeplitz, fitted to the block & 14.1 & 8.1 & 0.55 & 0.71 \\
Fourier, fitted to the block & 15.1 & 10.0 & 0.51 & 0.71 \\
AR-KLT, fitted to the block & 17.0 & 12.2 & 0.42 & 0.63 \\

%% file: fo3_tab_bearing.tex
none & 13.0 & 7.5 & 0.61 & 0.74 & 68.6 & 368.5 \\
Fourier & 17.1 & 7.7 & 0.57 & 0.68 & 24.0 & 87.6 \\
AR-KLT & 23.9 & 12.0 & 0.46 & 0.60 & 9.4 & 71.2 \\
sample & 14.7 & 7.3 & 0.65 & 0.77 & 6.6 & 65.0 \\

%% file: fo3_tab_likelihood.tex
task1 rec1 & 64 & 12 & 9536 & $0.01477 \pm 0.00112$ & 0.01483 & $0.0836 \pm 0.0029$ & 0.1509 \\
task1 rec1 & 128 & 24 & 4479 & $0.01171 \pm 0.00121$ & 0.01373 & $0.0711 \pm 0.0032$ & 0.0981 \\
task1 rec1 & 256 & 48 & 2217 & $0.01083 \pm 0.00156$ & 0.01086 & $0.0452 \pm 0.0025$ & 0.0677 \\
task1 rec1 & 512 & 64 & 1092 & $0.00803 \pm 0.00145$ & 0.00923 & $0.0310 \pm 0.0030$ & 0.0396 \\
task1 rec1 & 1024 & 64 & 527 & $0.00350 \pm 0.00068$ & 0.00476 & $0.0151 \pm 0.0014$ & 0.0207 \\
task2 rec1 & 64 & 12 & 8346 & $0.02029 \pm 0.00215$ & 0.02312 & $0.0887 \pm 0.0077$ & 0.0963 \\
task2 rec1 & 128 & 24 & 4089 & $0.01514 \pm 0.00211$ & 0.01398 & $0.0541 \pm 0.0051$ & 0.0512 \\
task2 rec1 & 256 & 48 & 2028 & $0.01054 \pm 0.00179$ & 0.00949 & $0.0337 \pm 0.0032$ & 0.0288 \\
task2 rec1 & 512 & 64 & 1008 & $0.00560 \pm 0.00100$ & 0.00726 & $0.0126 \pm 0.0016$ & 0.0176 \\
task2 rec1 & 1024 & 64 & 494 & $0.00334 \pm 0.00074$ & 0.00317 & $0.0113 \pm 0.0019$ & 0.0084 \\
task3 rec1 & 64 & 12 & 35126 & $0.01236 \pm 0.00034$ & 0.01223 & $2.7715 \pm 0.0373$ & 3.5658 \\
task3 rec1 & 128 & 24 & 17245 & $0.00886 \pm 0.00032$ & 0.00888 & $2.2386 \pm 0.0481$ & 1.5403 \\
task3 rec1 & 256 & 48 & 8593 & $0.00505 \pm 0.00029$ & 0.00527 & $2.1551 \pm 0.0731$ & 1.3763 \\
task3 rec1 & 512 & 64 & 4271 & $0.00250 \pm 0.00019$ & 0.00262 & $0.5530 \pm 0.0268$ & 0.7179 \\
task3 rec1 & 1024 & 64 & 2119 & $0.00094 \pm 0.00011$ & 0.00124 & $0.3345 \pm 0.0237$ & 0.3270 \\

%% file: fo3_tab_spatial.tex
task1 rec1 & 2 & 20.6 & 22.2 & 15.4 & $1.59 \times 10^{12}$ & 63 & 20.6 & 20.0 & 8.2 & 4.1 \\
task1 rec2 & 2 & 41.4 & 48.8 & 25.4 & $2.20 \times 10^{12}$ & 41 & 41.4 & 18.6 & 9.1 & 6.6 \\
task1 rec3 & 2 & 45.3 & 33.7 & 17.3 & $2.42 \times 10^{12}$ & 31 & 45.3 & 21.3 & 8.8 & 10.4 \\
task2 rec1 & 2 & 199.7 & 150.0 & 71.2 & $3.37 \times 10^{12}$ & 35 & 199.7 & 103.1 & 46.5 & 30.6 \\
task2 rec2 & 2 & 1164.9 & 415.4 & 291.3 & $1.43 \times 10^{13}$ & 12 & 1164.9 & 323.1 & 183.9 & 770.2 \\
task3 rec1 & 2 & 30.7 & 21.6 & 11.8 & $2.08 \times 10^{12}$ & 127 & 30.7 & 11.2 & 5.7 & 4.7 \\
task4 rec1 & 2 & 29.7 & 19.4 & 21.7 & $1.40 \times 10^{12}$ & 125 & 29.7 & 10.3 & 8.3 & 2.8 \\

%% file: fo3_tab_decisions.tex
2 & 415 & 0.000 & 0.012 & 0.027 & 0.000 & 13.0 & 16.0 & 24.1 & 16.7 \\
4 & 415 & 0.000 & 0.027 & 0.036 & 0.000 & 13.0 & 15.7 & 23.7 & 21.0 \\
8 & 415 & 0.000 & 0.029 & 0.031 & 0.000 & 13.0 & 17.0 & 24.5 & 21.8 \\
16 & 415 & 0.000 & 0.031 & 0.027 & 0.019 & 13.0 & 16.8 & 22.4 & 18.6 \\
32 & 415 & 0.000 & 0.029 & 0.029 & 0.022 & 13.0 & 16.3 & 24.7 & 17.6 \\
64 & 415 & 0.000 & 0.029 & 0.022 & 0.014 & 13.0 & 15.9 & 24.5 & 14.9 \\
all & 415 & 0.000 & 0.031 & 0.019 & 0.010 & 13.0 & 17.1 & 23.9 & 14.7 \\

%% file: fo3_tab_implementations.tex
256 & 4 & $1.16 \times 10^{5}$ & $1.20 \times 10^{-11}$ & $1.20 \times 10^{-11}$ & $5.29 \times 10^{-12}$ & $5.29 \times 10^{-12}$ \\
256 & 8 & $4.58 \times 10^{5}$ & $9.51 \times 10^{-12}$ & $9.51 \times 10^{-12}$ & $2.00 \times 10^{-11}$ & $2.00 \times 10^{-11}$ \\
256 & 16 & $1.14 \times 10^{6}$ & $4.06 \times 10^{-11}$ & $4.07 \times 10^{-11}$ & $1.24 \times 10^{-10}$ & $1.24 \times 10^{-10}$ \\
256 & 32 & $1.61 \times 10^{6}$ & $8.20 \times 10^{-11}$ & $8.21 \times 10^{-11}$ & $1.31 \times 10^{-10}$ & $1.31 \times 10^{-10}$ \\
512 & 4 & $1.28 \times 10^{5}$ & $5.81 \times 10^{-12}$ & $5.81 \times 10^{-12}$ & $5.72 \times 10^{-12}$ & $5.72 \times 10^{-12}$ \\
512 & 8 & $5.58 \times 10^{5}$ & $2.81 \times 10^{-11}$ & $2.80 \times 10^{-11}$ & $2.36 \times 10^{-11}$ & $2.36 \times 10^{-11}$ \\
512 & 16 & $1.77 \times 10^{6}$ & $8.51 \times 10^{-11}$ & $8.52 \times 10^{-11}$ & $7.21 \times 10^{-11}$ & $7.21 \times 10^{-11}$ \\
512 & 32 & $2.75 \times 10^{6}$ & $6.20 \times 10^{-11}$ & $6.20 \times 10^{-11}$ & $3.53 \times 10^{-11}$ & $3.53 \times 10^{-11}$ \\
1024 & 4 & $1.19 \times 10^{5}$ & $1.22 \times 10^{-11}$ & $1.22 \times 10^{-11}$ & $6.12 \times 10^{-12}$ & $6.45 \times 10^{-12}$ \\
1024 & 8 & $4.75 \times 10^{5}$ & $2.35 \times 10^{-11}$ & $2.35 \times 10^{-11}$ & $1.90 \times 10^{-11}$ & $1.90 \times 10^{-11}$ \\
1024 & 16 & $1.42 \times 10^{6}$ & $4.20 \times 10^{-11}$ & $4.20 \times 10^{-11}$ & $1.04 \times 10^{-10}$ & $1.04 \times 10^{-10}$ \\
1024 & 32 & $1.97 \times 10^{6}$ & $8.14 \times 10^{-11}$ & $0.179$ & $1.43 \times 10^{-5}$ & $1.43 \times 10^{-5}$ \\
2048 & 4 & $1.20 \times 10^{5}$ & $1.08 \times 10^{-11}$ & $1.08 \times 10^{-11}$ & $5.82 \times 10^{-12}$ & $5.88 \times 10^{-12}$ \\
2048 & 8 & $4.81 \times 10^{5}$ & $2.77 \times 10^{-11}$ & $4.95 \times 10^{-3}$ & $2.38 \times 10^{-11}$ & $3.70 \times 10^{-11}$ \\
2048 & 16 & $1.51 \times 10^{6}$ & $4.71 \times 10^{-11}$ & $0.0904$ & $1.26 \times 10^{-4}$ & $1.26 \times 10^{-4}$ \\
2048 & 32 & $2.09 \times 10^{6}$ & $8.19 \times 10^{-11}$ & $0.0493$ & $2.39 \times 10^{-6}$ & $3.30 \times 10^{-6}$ \\
4096 & 4 & $1.20 \times 10^{5}$ & $1.02 \times 10^{-11}$ & $1.02 \times 10^{-11}$ & $5.33 \times 10^{-12}$ & $5.35 \times 10^{-12}$ \\
4096 & 8 & $4.83 \times 10^{5}$ & $2.79 \times 10^{-11}$ & $3.05 \times 10^{-3}$ & $2.74 \times 10^{-11}$ & $2.74 \times 10^{-11}$ \\
4096 & 16 & $1.53 \times 10^{6}$ & $5.68 \times 10^{-11}$ & $0.0459$ & $6.30 \times 10^{-5}$ & $6.30 \times 10^{-5}$ \\
4096 & 32 & $2.14 \times 10^{6}$ & $1.22 \times 10^{-10}$ & $0.0258$ & $2.69 \times 10^{-6}$ & $2.69 \times 10^{-6}$ \\

%% file: impl_tab_counts.tex
1024 & 204\,800 & 40.0 & 654 & 204\,800 & 40.0 & 634 & 678\,912 & 132.6 & 2295 \\
2048 & 508\,928 & 41.4 & 714 & 508\,928 & 41.4 & 673 & 1\,668\,096 & 135.8 & 2437 \\
4096 & 1\,221\,632 & 42.6 & 746 & 1\,221\,632 & 42.6 & 703 & 3\,956\,736 & 138.0 & 2514 \\
8192 & 2\,847\,744 & 43.5 & 771 & 2\,847\,744 & 43.5 & 750 & 9\,102\,336 & 138.9 & 2583 \\
16384 & 6\,484\,992 & 44.0 & 809 & 6\,484\,992 & 44.0 & 766 & 20\,585\,472 & 139.6 & 2710 \\

%% file: impl_tab_timing.tex
512 & 0.0098 & 0.00821 & 0.000137 & 0.000126 & 1 & 0.00485 & 0.00483 & 9.06e-05 & 8.63e-05 & 1 & 0.0137 & 0.0137 & 0.000242 & 0.000241 & 1 \\
1024 & 0.0215 & 0.00416 & 0.000457 & 0.000108 & 4 & 0.0188 & 0.00404 & 0.000471 & 0.000112 & 4 & 0.0536 & 0.0123 & 0.00133 & 0.000326 & 4 \\
2048 & 0.0694 & 0.00941 & 0.00185 & 0.000269 & 7 & 0.0732 & 0.00978 & 0.00209 & 0.00028 & 7 & 0.213 & 0.0298 & 0.0062 & 0.000817 & 8 \\
4096 & 0.277 & 0.0228 & 0.0223 & 0.000654 & 34 & 0.286 & 0.0232 & 0.0249 & 0.000685 & 36 & 0.89 & 0.0713 & 0.0685 & 0.00199 & 35 \\
8192 & 1.1 & 0.0529 & 0.109 & 0.00155 & 70 & 1.15 & 0.0532 & 0.109 & 0.00167 & 66 & 3.57 & 0.166 & 0.322 & 0.00455 & 71 \\
16384 & 4.42 & 0.131 & 0.442 & 0.0039 & 113 & 4.53 & 0.126 & 0.442 & 0.00374 & 118 & 14.1 & 0.415 & 1.31 & 0.0111 & 118 \\

%% file: impl_tab_accuracy.tex
AR(1) 0.9 & 64 & $3.22 \times 10^{-15}$ & $3.22 \times 10^{-15}$ & $7.77 \times 10^{-16}$ & $9.99 \times 10^{-16}$ & $1.22 \times 10^{-15}$ & $1.22 \times 10^{-15}$ \\
AR(1) 0.9 & 256 & $6.90 \times 10^{-15}$ & $6.90 \times 10^{-15}$ & $1.89 \times 10^{-15}$ & $1.11 \times 10^{-15}$ & $2.55 \times 10^{-15}$ & $1.18 \times 10^{-13}$ \\
AR(1) 0.9 & 1024 & $9.96 \times 10^{-15}$ & $9.96 \times 10^{-15}$ & $2.33 \times 10^{-15}$ & $1.55 \times 10^{-15}$ &  &  \\
AR(1) 0.99 & 64 & $3.03 \times 10^{-14}$ & $3.03 \times 10^{-14}$ & $6.66 \times 10^{-16}$ & $4.44 \times 10^{-16}$ & $9.99 \times 10^{-16}$ & $9.99 \times 10^{-16}$ \\
AR(1) 0.99 & 256 & $1.70 \times 10^{-13}$ & $1.70 \times 10^{-13}$ & $2.00 \times 10^{-15}$ & $1.11 \times 10^{-15}$ & $2.11 \times 10^{-15}$ & $9.49 \times 10^{-15}$ \\
AR(1) 0.99 & 1024 & $6.55 \times 10^{-13}$ & $6.55 \times 10^{-13}$ & $2.55 \times 10^{-15}$ & $1.94 \times 10^{-15}$ &  &  \\
AR(3) & 64 & $1.86 \times 10^{-15}$ & $1.86 \times 10^{-15}$ & $1.67 \times 10^{-15}$ & $6.66 \times 10^{-16}$ & $1.55 \times 10^{-15}$ & $1.78 \times 10^{-15}$ \\
AR(3) & 256 & $1.99 \times 10^{-15}$ & $1.99 \times 10^{-15}$ & $1.89 \times 10^{-15}$ & $1.89 \times 10^{-15}$ & $3.22 \times 10^{-15}$ & $3.13 \times 10^{-13}$ \\
AR(3) & 1024 & $3.98 \times 10^{-15}$ & $3.98 \times 10^{-15}$ & $4.22 \times 10^{-15}$ & $4.11 \times 10^{-15}$ &  &  \\

%% file: impl_tab_solver.tex
AR(1) 0.9 & 2048 & 37 & 6 & $0.90$ & $1.97 \times 10^{-15}$ \\
AR(3) & 2048 & 37 & 7 & $0.83$ & $7.21 \times 10^{-11}$ \\

%% file: impl_tab_plan.tex
AR(1) 0.9 & 2048 & 1024 & 0.042 & 44\,544 & 512 \\
AR(1) 0.9 & 8192 & 4096 & 0.052 & 871\,936 & 512 \\
AR(3) & 2048 & 1024 & 0.175 & 183\,808 & 512 \\
AR(3) & 8192 & 4096 & 0.192 & 3\,227\,136 & 512 \\

%% file: impl_tab_ranks.tex
ARMA & 1 & 1 & 0 & 2 & 1, 1 \\
ARMA & 2 & 1 & 0 & 4 & 2, 2 \\
ARMA & 0 & 2 & 0 & 4 & 2, 2 \\
ARMA & 1 & 2 & 0 & 4 & 2, 2 \\
sum & 2 & 0 & 0 & 4 & 2, 2 \\
sum & 3 & 0 & 0 & 8 & 4, 4 \\
integrated & 1 & 0 & 1 & 3 & 2 \\
integrated & 2 & 0 & 1 & 5 & 4 \\
integrated & 1 & 0 & 2 & 5 & 4 \\
complex ARMA & 1 & 1 & 0 & 2 & 2 \\
complex ARMA & 2 & 0 & 0 & 4 & 4 \\
complex ARMA & 0 & 1 & 0 & 2 & 2 \\
complex integrated & 1 & 0 & 1 & 4 & 4 \\
complex integrated & 1 & 0 & 2 & 6 & 6 \\

%% file: impl_tab_poly.tex
(0, 0.9) & 4096 & 2 & 4.6 & 4.4 & 4 & 4 & $1.89 \times 10^{-14}$ \\
(0, 0.5, 0, 0.3) & 4096 & 2 & 4.3 & 4.4 & 8 & 8 & $4.69 \times 10^{-15}$ \\
(0, 0, 0, 0.7) & 4096 & 4 & 52.9 & 19.8 & 8 & 8 & $3.68 \times 10^{-15}$ \\

%% file: impl_tab_oddcores.tex
33 & $1.55 \times 10^{-15}$ & $1.33 \times 10^{-15}$ & $1.93 \times 10^{-15}$ & $8.94 \times 10^{-16}$ & $1.81 \times 10^{-15}$ \\
129 & $4.66 \times 10^{-15}$ & $3.58 \times 10^{-15}$ & $3.89 \times 10^{-15}$ & $3.41 \times 10^{-15}$ & $2.98 \times 10^{-15}$ \\
513 & $2.09 \times 10^{-14}$ & $1.65 \times 10^{-14}$ & $1.54 \times 10^{-14}$ & $8.07 \times 10^{-15}$ & $7.93 \times 10^{-15}$ \\

%% file: impl_tab_oddtransform.tex
1 & 33 & $1.25 \times 10^{-17}$ & $3.17 \times 10^{-15}$ & $2.25 \times 10^{-15}$ & 1 & $1.63 \times 10^{-15}$ & $8.89 \times 10^{-16}$ \\
1 & 129 & $1.25 \times 10^{-17}$ & $8.69 \times 10^{-15}$ & $6.15 \times 10^{-15}$ & 1 & $8.87 \times 10^{-15}$ & $2.03 \times 10^{-15}$ \\
1 & 513 & $0.00$ & $4.09 \times 10^{-14}$ & $3.20 \times 10^{-14}$ & 1 & $5.27 \times 10^{-15}$ & $1.33 \times 10^{-15}$ \\
3 & 33 & $6.54 \times 10^{-17}$ & $3.19 \times 10^{-15}$ & $2.47 \times 10^{-15}$ & 3 & $1.89 \times 10^{-15}$ & $6.66 \times 10^{-16}$ \\
3 & 129 & $6.54 \times 10^{-17}$ & $9.30 \times 10^{-15}$ & $6.90 \times 10^{-15}$ & 3 & $2.14 \times 10^{-15}$ & $1.78 \times 10^{-15}$ \\
3 & 513 & $0.00$ & $5.10 \times 10^{-14}$ & $3.62 \times 10^{-14}$ & 3 & $3.13 \times 10^{-15}$ & $2.55 \times 10^{-15}$ \\

%% file: fo3_tab_radar.tex
1 & +0.0021 & 0.0001 & 0.0035 & +0.501 & 0.006 & 0.419 \\
2 & +0.0041 & 0.0002 & 0.0051 & +0.540 & 0.007 & 0.434 \\
3 & +0.0099 & 0.0005 & 0.0111 & +0.620 & 0.006 & 0.467 \\
4 & +0.0075 & 0.0006 & 0.0117 & +0.614 & 0.006 & 0.420 \\
6 & -0.0002 & 0.0007 & 0.0128 & +0.644 & 0.007 & 0.437 \\
8 & -0.0104 & 0.0010 & 0.0141 & +0.635 & 0.007 & 0.456 \\
12 & +0.0008 & 0.0011 & 0.0247 & +0.401 & 0.011 & 0.553 \\
16 & -0.0042 & 0.0013 & 0.0274 & +0.376 & 0.013 & 0.569 \\
24 & -0.0292 & 0.0026 & 0.0321 & +0.387 & 0.010 & 0.562 \\
32 & -0.0616 & 0.0039 & 0.0332 & +0.280 & 0.012 & 0.563 \\
48 & -0.0887 & 0.0047 & 0.0370 & +0.209 & 0.013 & 0.576 \\
64 & -0.1079 & 0.0053 & 0.0395 & +0.184 & 0.014 & 0.578 \\

%% file: fo3_tab_window.tex
16 & 6 & +0.1060 & 0.0033 & 0.0932 & 1.70 & +0.556 \\
32 & 12 & +0.1270 & 0.0038 & 0.1029 & 4.06 & +0.620 \\
64 & 12 & +0.0671 & 0.0034 & 0.0906 & 4.29 & +0.591 \\
128 & 3 & +0.0283 & 0.0014 & 0.0331 & 3.62 & +0.611 \\
256 & 3 & +0.0201 & 0.0011 & 0.0210 & 5.15 & +0.624 \\
512 & 3 & +0.0099 & 0.0005 & 0.0111 & 5.04 & +0.620 \\

%% file: apps_tab_ends.tex
AR(4) $r=0.95$ & 1.08 & 1.0 & 7.7 & 3.9 & 179 \\
AR(4) $r=0.99$ & 1.08 & 1.0 & 35.9 & 13.1 & 174 \\
AR(4) $r=0.999$ & 1.09 & 1.1 & 360 & 38.6 & 218 \\
ARMA(2,2) & 1.09 & 1.0 & 344 & 174 & 174 \\

%% file: apps_tab_cyclo.tex
stationary control & 0.0 & -0.03 & -0.02 & -0.02 & -0.01 & 0.02 & 0.11 \\
drive variance, index 0.5 & 3.3 & 0.12 & 0.20 & 0.20 & 0.28 & 0.02 & 0.09 \\
drive variance, index 0.9 & 6.8 & 0.51 & 0.81 & 0.91 & 1.14 & 0.03 & 0.10 \\
drive variance, index 0.99 & 7.8 & 0.64 & 1.06 & 1.31 & 1.76 & 0.04 & 0.11 \\
one impact, $10\times$ & 3.7 & 0.17 & 0.35 & 0.36 & 0.40 & 0.02 & 0.06 \\
one impact, $100\times$ & 12.0 & 2.91 & 2.84 & 2.65 & 2.29 & 0.01 & 0.05 \\
one impact, $1000\times$ & 21.8 & 9.51 & 7.70 & 6.57 & 5.20 & 0.01 & 0.06 \\
resonance swept $\pm0.05\pi$ & 1.6 & 1.21 & 0.87 & 0.67 & 0.57 & 0.02 & 0.09 \\
resonance swept $\pm0.15\pi$ & 5.5 & 2.06 & 3.19 & 2.23 & 1.89 & 0.02 & 0.08 \\
damping switched & 9.0 & 1.71 & 0.60 & 0.53 & 0.49 & 0.00 & 0.04 \\